\documentclass[11pt]{cernrep}
\usepackage{graphicx,wrapfig,epsfig,epsf}
\usepackage{amsmath,amssymb,citesort,enumerate,here}
\usepackage{hhline,multirow}

\begin{document}

\title{ Hard probes in heavy ion collisions at the LHC: \\
PDFs, SHADOWING AND $pA$ COLLISIONS}

\def\jyfl{1}
\def\hip{2}
\def\ames{3}
\def\strasbourg{4}
\def\notredame{5}
\def\columbia{6}
\def\heidelberg{7}
\def\cernth{8}
\def\cordoba{9}
\def\nikhef{10}
\def\slac{11}
\def\jlab{12}
\def\neviscol{13}
\def\kent{14}
\def\telaviv{15}
\def\duke{16}
\def\bochum{17}
\def\seattle{18}
\def\liverpool{19}
\def\cernepaip{20}
\def\psu{21}
\def\bnl{22}
\def\lbnl{23}
\def\davis{24}

\author{
{\bf Convenors:} K.~J.~Eskola$^{\jyfl,\hip}$, 
J.~w.~Qiu$^{\ames}$ {\rm (theory)}, 
W.~Geist$^{\strasbourg,\notredame}$ {\rm (experiment)}\\
{\bf Editor:} K. J. Eskola\\
{\bf Contributing authors:}
A.~Accardi$^{\columbia,\heidelberg}$,
N.~Armesto$^{\cernth,\cordoba}$,
M.~ Botje$^{\nikhef}$[{\small ALICE}],
S.~J.~Brodsky$^{\slac,\jlab}$,
B.~Cole$^{\neviscol}$[{\small ATLAS}],
K.~J.~Eskola$^{\jyfl,\hip}$, 
G.~Fai$^{\kent}$, 
L.~Frankfurt$^{\telaviv}$, 
R.~J.~Fries$^{\duke}$,
W.~Geist$^{\strasbourg,\notredame}$[{\small CMS}],
V.~Guzey$^{\bochum}$, 
H.~Honkanen$^{\jyfl,\hip}$, 
V.~J.~Kolhinen$^{\jyfl,\hip}$, 
Yu.~V.~Kovchegov$^{\seattle}$,
M.~McDermott$^{\liverpool}$, 
A.~Morsch$^{\cernepaip}$, 
J.~w.~Qiu$^{\ames}$,
C.~A.~Salgado$^{\cernth}$,
M.~Strikman$^{\psu}$,
H.~Takai$^{\bnl}$[{\small ATLAS}], 
S. Tapprogge$^{\hip}$[{\small ATLAS}],
R.~Vogt$^{\lbnl,\davis}$,
X.~f.~Zhang$^{\kent}$
}

\institute{~\\
$^{~\jyfl}$
Department of Physics, University of Jyv\"askyl\"a, Jyv\"askyl\"a, Finland\\
$^{~\hip}$
Helsinki Insititute of Physics, University of Helsinki, Finland\\
$^{~\ames}$
Department of Physics and Astronomy, Iowa State University, Ames, IA, USA\\
$^{~\strasbourg}$
Institut de Recherches Subatomiques, IN2P3-CNRS/Universit\'e 
Louis Pasteur, Strasbourg, France\\
$^{~\notredame}$
Physics Department, University of Notre Dame, Notre Dame, IN, USA\\
$^{~\columbia}$
Department of Physics, Columbia University, New York, NY, USA\\
$^{~\heidelberg}$ 
Institut f\"ur Theoretische Physik der Universit\"at Heidelberg, 
Heidelberg, Germany\\
$^{~\cernth}$
Theory Division, CERN, Geneva, Switzerland\\
$^{~\cordoba}$
Departamento de F\'{\i}sica, Universidad de C\'ordoba, C\'ordoba, Spain\\
$^{\nikhef}$
NIKHEF, Amsterdam, The Netherlands\\
$^{\slac}$
Stanford Linear Accelerator Center, Stanford University, Stanford, CA, USA\\
$^{\jlab}$
Thomas Jefferson National Accelerator Facility (JLab), Newport News, VA, USA\\
$^{\neviscol}$
Nevis Laboratories, Columbia University, New York, NY, USA\\
$^{\kent}$
Center for Nuclear Research, Department of Physics, Kent State University,
Kent, OH, USA\\
$^{\telaviv}$
Nuclear Physics Department, Tel Aviv University, Tel Aviv, Israel\\
$^{\duke}$
Physics Department, Duke University, Durham, NC, USA\\
$^{\bochum}$
Fakult\"at f\"ur Physik und Astronomie, 
Institut f\"ur Theor. Physik II, Ruhr-Univ. Bochum, Germany\\
$^{\seattle}$
Department of Physics, University of Washington, Seattle, WA, USA\\
$^{\liverpool}$
Department of Mathematical Sciences, Liverpool University, Liverpool, UK\\
$^{\cernepaip}$
Experimental Physics Division, CERN, Geneva, Switzerland\\
$^{\psu}$
Department of Physics, Pennsylvania State University, University Park, PA, USA\\
$^{\bnl}$
Brookhaven National Laboratory, Upton, NY, USA\\
$^{\lbnl}$
Lawrence Berkeley National Laboratory, Berkeley, CA, USA\\
$^{\davis}$
Physics Department, University of California at Davis, Davis, CA, USA\\
}

\maketitle 

\begin{flushright}
\vspace{-18cm}
HIP-2003-40/TH
\vspace{17cm}
\end{flushright} 

\begin{abstract}
This manuscript is the outcome of the subgroup ``PDFs, shadowing and
$pA$ collisions'' from the CERN workshop ``Hard Probes in Heavy Ion
Collisions at the LHC''. In addition to the experimental parameters
for $pA$ collisions at the LHC, the issues discussed are factorization
in nuclear collisions, nuclear parton distributions (nPDFs), hard
probes as the benchmark tests of factorization in $pA$ collisions at
the LHC, and semi-hard probes as observables with potentially large
nuclear effects. Also, novel QCD phenomena in $pA$ collisions at the
LHC are considered.  The importance of the $pA$ program at the LHC is
emphasized.

\end{abstract}

\newpage
\tableofcontents
\newpage
\section{WHY $pA$ COLLISIONS: INTRODUCTION AND SUMMARY}
\label{pA_SEC:intro}
{\em Kari J. Eskola, Walter Geist, and Jianwei Qiu}

\def\gsim{~\raise0.3ex\hbox{$>$\kern-0.75em\raise-1.1ex\hbox{$\sim$}}~}
\def\lsim{~\raise0.3ex\hbox{$<$\kern-0.75em\raise-1.1ex\hbox{$\sim$}}~}

Hard Probes \cite{Satz:cg,HardProbes2} are high-energy probes of the
Quark Gluon Plasma (QGP) which are produced in the primary partonic
collisions.  The production of hard probes involves a large transfer
of energy-momentum at a scale $Q\gg\Lambda_{\rm QCD}$. Such hard
probes include the production of Drell-Yan dileptons, massive gauge
bosons, heavy quarks, prompt photons, and high-$p_T$ partons observed
as jets and high-$p_T$ hadrons. In $pp$ collisions, it is well
established that the inclusive cross sections of these hard processes
can be computed through collinear factorization, i.e. using
short-distance cross sections of parton-parton scatterings and
well-defined universal parton distribution functions (PDFs). While the
partonic sub-cross sections and the scale evolution of the PDFs are
calculable in perturbative QCD (pQCD), the PDFs contain
nonperturbative information which must be extracted from the measured
cross sections of various hard processes.

Provided that leading-power collinear factorization is applicable also
in $AA$ collisions, the cross sections of hard probes can be used as
benchmark cross sections against which the signals and properties of
the QGP can be extracted.  Therefore, {\em it is of extreme importance
that the applicability of factorization be tested in $pA$
interactions}, especially at the LHC where truly hard probes will
finally become available for nuclear collisions.

In this document, we divide the hard probes into two classes according
to their scales: hard probes and semi-hard probes.  The hard probes
refer to observables with hard partonic subprocesses of energy
exchange $Q$ greater than several tens of GeV, up to 100 GeV and
higher.  For these hard probes, the process-dependent nuclear effects
(i.e., power corrections) should remain negligible.  The semi-hard
probes correspond to probes at moderate scales, $Q$ of few GeV up to
10\dots20 GeV, where the process-dependent nuclear effects may already
be sizable in $pA$ collisions and even larger in $AA$ collisions.
This document mainly addresses the hard and semi-hard probes in $pA$
collisions from the factorization (large scale) point of view but we
also discuss (Sec. 7) other interesting new physics which can be
explored in $pA$ collisions at the LHC.

{\bf Section 2} summarizes the main machine parameters for the $pA$
runs at the LHC as well as the acceptance regions of the ALICE, CMS
and ATLAS detectors.  In addition to protons, the LHC machine can
accelerate light beams as dilute as the deuteron and as compact as the
helium nucleus.  The machine can also accelerate a variety of heavier
nuclei ranging from oxygen to lead.  All three complementary
detectors, ALICE, CMS and ATLAS, have wide rapidity coverage for
multiplicity measurements, the capability of precision centrality
determination in $pA$ collisions and excellent jet reconstruction
efficiency for a range of transverse momentum, $p_T$, and
pseudorapidity, $\eta$.

Experimental data become more valuable and powerful when there are
theoretical calculations to compare with.  Because of the nature of
the strong interactions, our ability to do calculations and to make
predictions for hard probes in hadronic collisions completely relies
on factorization in QCD perturbation theory.  For the hard probes, the
benchmark tests, factorization can be expected to hold in $pp$, $pA$
and $AA$ collisions. For the semi-hard probes which are sensitive to
the properties of nuclear matter via multiple parton scatterings,
there is, however, a critical difference between $pA$ and $AA$
collisions: while factorization may be valid for such semi-hard
observables in $pA$ collisions, it can fail in $AA$ collisions.
Factorization in nuclear collisions is summarized in {\bf Section 3}.

Nuclear parton distribution functions (nPDFs) are an essential
ingredient in understanding the magnitude of the nuclear effects on
the factorized hard probe cross sections.  While $AA$ collisions
typically are too complex for detailed verification of the nPDF
effects on the semi-hard probe cross sections, {\em $pA$ collisions
provide a much cleaner environment for studies of the nPDFs}.  The
$pA$ collisions can also probe the spatial dependence of the nPDFs --
an important issue in the studies of hard probes in different $AA$
centrality classes as well as in understanding the origin of the
nuclear effects (shadowing) on the nPDFs.  We devote
{\bf Section 4} to discussion of several aspects of the nPDFs, ranging
from global DGLAP fits to studies of the origin of shadowing and
diffractive phenomena, and PDFs in the gluon saturation region at
small values of $x$ and $Q$.

{\bf Section 5}, ``The benchmark tests in $pA$'', deals with hard
probes for which the nonfactorizable nuclear effects, i.e.,
process-dependent power corrections, should remain negligible.  For
these large-$Q$ processes, the factorizable nuclear effects (nuclear
effects on the parton distributions) also remain small.  We emphasize
that it is very important to measure these processes in $pA$
collisions at the LHC: universal nuclear effects that are internal to
the nucleus would, if observed, confirm the practical applicability of
factorization in hard nuclear processes and would establish a reliable
reference baseline for $AA$.  It would be especially important to
verify at what scale the observed nuclear dependence becomes universal
and is completely determined by that of the nPDFs.  On the other hand,
observation of any significant nuclear effects in $pA$ would be a
truly interesting and important signal of sizable power corrections in
the cross sections and thereby also of new multiple collision dynamics
of QCD.

For nuclear collisions, the very large-scale processes with $Q\gsim
100$~GeV are in practice only available at the LHC, the ultimate hard
probe machine. As shown in Table~\ref{rates}, due to the high
cms-energy and luminosity -- higher than in $AA$ -- the counting rates
of the hard probes at the scale $Q\sim 100$~GeV in $pA$ collisions
will be large enough to give sufficient statistics for reliable
measurements of the nuclear effects. For example, in a month (assumed
to be $10^6$ s) we expect about $3.3\cdot 10^8$ jets with $E_T\ge
100$~GeV and $|\eta|\le2.5$ to be collected in $p$Pb collisions at
$\sqrt s=8.8$~GeV with the maximum machine luminosity
$1.4\cdot10^{30}$~cm$^{-2}$s$^{-1}$ and $2.6\cdot10^7$ jets with the
maximum ALICE luminosity $1.1\cdot 10^{29}$~cm$^{-2}$s$^{-1}$ (for the
luminosities, see Sec. 2.1).
About half of these jets fall into the interval $100\,{\rm GeV}\le
E_T\le 120$~GeV. The corresponding number of $Z^0$ bosons at $|y_Z|\le
2.4$ is $6.8\cdot10^6$ in a month with the maximum machine luminosity
and $5.4\cdot10^5$ with the maximum ALICE luminosity.

\begin{table}[thb]
\vspace{-0.3cm}
\begin{center}
\caption{Expected counting rates (in $1/$s, with two significant digits) of 
jets with transverse energy $E_T\ge 100$~GeV and rapidity $|\eta|\le
2.5$ (see Sec.~5.3),
and heavy gauge bosons at rapidities $|y|\le2.4$
in $p$Pb collisions at $\sqrt s=8.8$~TeV (see Sec.~5.1).
}
\vskip0.5cm

\begin{tabular}{c|c|c|c|c}
\hline\hline
luminosity (cm$^{-2}$s$^{-1}$) 	& jets$/$s& $Z^0/$s & $W^+/$s & $W^-/$s \\
\hline
$1.1\cdot 10^{29}$		& 26	& 0.54	  & 0.89    & 0.83    	\\
\hline
$1.4\cdot10^{30}$	    	& 330	& 6.8	  & 11	    & 11	\\
\hline\hline
\end{tabular}
\label{rates}

\end{center}
\end{table}

{\bf Section 6}, ``Processes with potentially large nuclear effects'',
discusses the physics of the semi-hard probes.  The theoretical
reference cross sections for these processes inevitably contain some
model dependence, causing unknown theoretical uncertainties in the
corresponding cross sections in $AA$ collisions.  For these semi-hard
probes, it is even more important to experimentally confirm the
reference cross sections in $pA$ collisions because of the substantial
non-universal nuclear dependence.  That is, the reliable reference
cross sections for semi-hard probes in $AA$ can be obtained {\it only
if the corresponding $pA$ measurements are performed}.  A concrete
example of this is the clear suppression of hadrons with $p_T\lsim10$
GeV recently observed at RHIC in central AuAu collisions at $\sqrt
s=200$~GeV relative to the $pp$ data
\cite{Adler:2003qi,Adams:2003kv,Back:2003qr}. Only with help of 
the $p(d)A$ data is it possible to verify that the origin of the
suppression indeed is the strongly interacting medium and not the
nuclear effects, the process-dependent power corrections and/or nPDFs,
in the production of $10\dots 20$~GeV partons. A significant
suppression in $p(d)A$ collisions would indicate that the nuclear
effects, the process-dependent power corrections in particular, would
play a major role in suppressing the hadron spectrum, especially so in
$AA$ collisions.  In this case, the observed suppression in AuAu
collisions would be only partly due to the dense medium.  On the other
hand, a significant enhancement (Cronin effect) in $p(d)A$ collisions
would also be a signal of sizable nuclear effects which would cause an
even stronger enhancement of $10\dots20$~GeV parton production in AuAu
collisions.  In this case, the suppression observed in the AuAu data
would, in fact, be stronger than expected based on the $pp$ reference
data.  The new data on single hadron $p_T$ distributions in $d$Au
collisions at $\sqrt{s}=200$~GeV at RHIC
\cite{Adler:2003ii,Adams:2003im,Back:2003ns} confirm the latter case
by showing the absence of suppression or even the appearance of a
Cronin-type enhancement up to $p_T\sim 8$~GeV, in contrast to the
strong suppression observed in AuAu collisions.  The experience at
RHIC clearly demonstrates that in order to measure the detailed
properties of the strongly interacting matter (Equation of State,
degrees of freedom, degree of thermalization, etc.) produced at the
LHC, one needs precise knowledge of the absolute reference cross
sections {\em where the nuclear effects are constrained by those
measured in $pA$ collisions}.

Apart from the connection to the QGP signals, we emphasize the
importance of measuring the magnitude of the nuclear effects on hard
and semi-hard probes in $pA$ relative to $pp$.  It is only through
these measurements, where we can separate the universal nuclear
dependence of the nPDFs from the process-dependent power corrections
due to rescattering in nuclear matter, that we can differentiate the
contributions to the nuclear dependence. Then it is possible to
determine when and where new QCD phenomena, such as gluon saturation
at the scale of a few GeV, set in.

Finally, in 
{\bf Section 7}, ``Novel QCD phenomena in $pA$ collisions at the
LHC'', we emphasize the fact that $pA$ collisions per se offer lots of
new interesting QCD phenomena to explore, such as the breaking down of
the leading twist QCD for a wide range of momentum transfer due to
increase of the gluon fields at small $x$, related phenomena of $p_T$
broadening of the forward spectra of leading jets and hadrons, strong
quenching of the low $p_T$ hadron production in the proton
fragmentation region, and coherent diffraction into three jets.

\vspace{0.3cm}

\noindent
{\bf In summary}, the $pA$ program at the LHC serves a dual role.  It
is needed to calibrate the $AA$ measurements for a sounder
interpretation. It also has intrinsic merits in the framework of a
more profound understanding of QCD (e.g. shadowing vs. diffraction,
nonlinear QCD and saturation, higher twists...). Thus, one should
foresee three key steps in this program: (1) ($pp$ and) $p$Pb runs
with reliable determination of centrality at the same collision energy
as PbPb interactions; (2) a systematic study of $pA$ collisions,
requiring a variety of collision energies and nuclei as well as a
centrality scan; and (3) an interchange of $p$- and $A$-beams for
asymmetric detectors. The expected physics output from measurements of
hard and semi-hard probes in $pA$ collisions at the LHC may be
summarized as follows. We are able to

\begin{itemize}
\item 
test the predictive power of QCD perturbation theory in nuclear
collisions by verifying the applicability of factorization theorems
and the universality of the nPDFs through the hardest probes,
only available at the LHC 
$\longrightarrow $ Secs. 3 and  5.

\item 
measure the nuclear effects (internal to the nucleus) in the nPDFs
over an unprecedented range of $x$ and $Q$; investigate the interplay
between the ``EMC'' effect, nuclear shadowing and saturation as well
as the transverse-coordinate dependence of the nPDFs; and possibly,
discover a new state of matter, the color glass condensate, by probing
very soft gluons in heavy nuclei through the rapidity dependence
$\longrightarrow $ Sec.~4.

\item  
determine the nuclear dependence of the cross sections of the
semi-hard probes in $pA$ collisions and study QCD multiple parton
scattering in nuclear matter and its corresponding dependences beyond
the universal nuclear effects included in the nPDFs $\longrightarrow $
Sec.~6.

\item 
extract excellent information on QCD dynamics in hadronization because
normal nuclear matter acts like a filter for color neutralization and
parton hadronization and explore potential new QCD phenomena in $pA$
collisions, such as diffraction into three jets, double PDFs,
etc. $\longrightarrow$ Sec.~7.

\item 
use the hard and semi-hard probe cross sections as references for the
QGP signals in $AA$ collisions: the hard probes set the benchmark of
the applicability of factorization while the semi-hard probes help to
understand the size of the nuclear modifications not caused by the
dense medium produced in $AA$ collisions.

\end{itemize}

\clearpage


\section{THE EXPERIMENTAL PARAMETERS FOR $pA$ AT THE LHC}
\label{pA_SEC:machine}

\subsection{Constraints from the LHC machine}
\label{subsec:morsch}
{\em Andreas Morsch}

\subsubsection{$pA$ collisions as part of the LHC ion programme}

The LHC Ion programme has been reviewed in a recent LHC Project Report
\cite{brandt}. 
Whereas collisions between light ions are already part of the LHC
Phase II programme, $pA$ and $pA$-like collisions (i.e. $pA$, $dA$,
$\alpha A$) are part of the LHC upgrade programme.  Collisions between
protons and ions are in principle possible since two independent RF
systems are already foreseen in the LHC nominal programme.  However,
$pA$ collisions also require the availability of two independent
timing systems. The necessary cabling is already included in the
baseline layout.  Acquisition of the dedicated electronics remains to
be discussed.

The question whether $dA$ collisions will be available in the LHC is still 
under investigation
because the same source has to be used for 
the production of both protons and deuterons. 
Since proton availability is required for other CERN programs in
parallel with the LHC $pA$ runs, the final answer will depend on the
time required for the source to switch back and forth between proton
and deuteron operation. Production of $\alpha$ particles in parallel
with proton operation is
simpler and can be considered as an interesting alternative to $pA$
and $dA$ collisions.

As an alternative, the PS division is presently studying a new layout
for Linac 3, with possibly two sources and a switch-yard allowing
simultaneous production of two different types of ions, including
deuterons and $\alpha$.

\subsubsection{Beam properties}

At the LHC, protons and ions have to travel in the same magnetic lattice and, 
hence, the two beams have the same rigidity. This has several consequences
for the physical properties of the collision system. The condition of having
the same rigidity results in beam momenta $p$ 
which depend on the atomic mass 
$A$ and charge
$Z$ of the beam particles. 
For the nominal LHC bending field one obtains
\begin{equation}
p = 7 \, {\rm TeV} \, \frac{Z}{A} \quad.
\label{morsch:eq_mom}
\end{equation}
The centre of mass energy is given by
\begin{equation}
\sqrt{s} = 14 \, {\rm TeV} \, \sqrt{\frac{Z_{1} Z_{2}}{A_{1} A_{2}}} \quad.
\label{morsch:eq_sqrts}
\end{equation}

Since, in general, the two beams have different momenta, the rest
system of the colliding particles does not coincide with the
laboratory system. The rapidity of the centre of mass system relative
to the laboratory is
\begin{equation}
y_{\rm cent} = \frac{1}{2} \, \ln{\frac{Z_{1} A_{2}}{Z_{2} A_{1}}}
\quad .\label{morsch:eq_y}
\end{equation}

In Table \ref{tab_1} the properties of several collision systems of interest
have been summarized together with their geometric cross sections.

\begin{table}
\begin{center}
\caption{Geometric cross section, maximum centre of mass energy and 
rapidity shift for several collision systems.}
\vspace{0.4cm}
\begin{tabular}{|c|c|c|c|}
\hline
System & $\sigma_{\rm tot}$ & $\sqrt{s}_{\rm max}$ & Rapidity \\
       & (barn)             &    (TeV)             &  Shift   \\
\hline
\hline
$p$O     & 0.39               &    9.9               &  0.35    \\
\hline
$p$Ar    & 0.72               &    9.4               &  0.40    \\
\hline
$p$Pb    & 1.92               &    8.8               &  0.47    \\
\hline
\hline
$d$O     & 0.66               &    7.0               &  0.00    \\
\hline
$d$Ar    & 1.10               &    6.6               &  0.05    \\
\hline
$d$Pb    & 2.58               &    6.2               &  0.12    \\
\hline
\hline
$\alpha$O & 0.76            &    7.0               &  0.00    \\
\hline
$\alpha$Ar& 1.22            &    6.6               &  0.05    \\
\hline
$\alpha$Pb& 2.75            &    6.2               &  0.12    \\
\hline
\end{tabular}
\label{tab_1}
\end{center}
\end{table}

\subsubsection{Luminosity considerations for $pA$ collisions}

Two processes 
limit the proton and ion beam lifetimes
and, hence, the maximum luminosity \cite{morsch1}:

\begin{itemize}
\item[1] Beam-beam loss is due to hadronic interactions between the beam
particles and proportional to the geometric cross section (see Table
\ref{tab_1}). For asymmetric beam intensities (different intensity 
proton and ion beams), the lifetime is proportional to the intensity of the
higher
intensity beam.

\item[2] Intra-beam scattering increases the beam emittance during the
run. Since 
the timescale of emittance growth is proportional to $AZ$, it mainly
affects the ion beam.
\end{itemize}

It can be shown that 
maximizing the beam lifetime
leads to solutions with asymmetric beam intensities (high-intensity
proton beam, low-intensity ion beam).  However, for the 
$pA$ luminosity requirements 
published by the ALICE collaboration
\cite{morsch2}, it is rather the source intensity than the luminosity
lifetime which 
limits the luminosity. Since high intensity
proton beams are readily available ($N_{\rm p}^{\rm max} =
10^{11}/{\rm bunch}$) but ion sources can deliver only relatively
small intensities, mainly limited by space charge effect in the SPS,
these conditions also push the realistic scenarios towards asymmetric
beams.

Table~\ref{tab_2} shows the maximum intensities per bunch for Pb, Ar,
and O.  {\bf It has to be emphasized that these maximum intensities
represent rough estimates and much more detailed studies are needed to
confirm them. They are by no means to be quoted as official LHC
baseline values.}  At these maximum intensities the transverse
emittance growth time amounts to about $35 \, {\rm h}$.  In the same
table, the proton intensities needed to obtain the luminosities
requested by ALICE are shown. In all cases these intensities are below
$N_{\rm p}^{\rm max}$.  Hence, higher luminosities from which CMS and
ATLAS could profit can be envisaged, as shown in the last column.


\begin{table}[b]
\begin{center}
\caption{ Maximum ion intensity per bunch, ALICE luminosity request, the
proton intensity needed to fulfill the ALICE request and the maximum
luminosity for $N_{\rm p} = 10^{11}$ for three different $pA$ systems.}
\vspace{0.4cm}
\begin{tabular}{|c|c|c|c|c|c|}
\hline
System & $N_{\rm ion}^{\rm max}$ & Luminosity (ALICE) & $N_{\rm p}$ &
       max. Luminosity \\ & ($10^{10}$) & (${\rm cm}^{-2}{\rm
       s}^{-1}$) & ($10^{10}$) & $({\rm cm}^{-2}{\rm s}^{-1}$) \\
\hline
\hline
$p$Pb     &  0.007  & $1.1 \cdot 10^{29}$   & 0.8   &   $1.4 \cdot 10^{30}$ \\
\hline
$p$Ar     &  0.055  & $3.0 \cdot 10^{29}$   & 0.27  &   $1.1 \cdot 10^{31}$ \\
\hline
$p$O      &  0.1    & $5.5 \cdot 10^{29}$   & 0.27  &   $2.0 \cdot 10^{31}$ \\
\hline
\end{tabular}
 \label{tab_2}
\end{center}
\end{table}

\eject

\subsection{The ALICE detector}
\label{subsec:botje}
{\em Michiel Botje}
%
%




\subsubsection{Introduction}\label{se:introduction}

To study the properties of hadronic matter under conditions of extreme
energy density the ALICE collaboration has designed a general purpose
detector optimised to measure a large variety of observables in heavy
ion collisions at the LHC~\cite{mbref:alicetp}. The apparatus will
detect and identify hadrons, leptons and photons over a wide range of
momenta. The requirement to study the various probes of interest in a
very high multiplicity environment, which may be as large 8000 charged
particles per unit of rapidity in central PbPb collisions, imposes
severe demands on the tracking of charged particles. This requirement
has led to a design based on high granularity, but slow, drift
detectors placed in a large volume solenoidal magnetic field.
Additional detectors augment the identification capabilities of the
central tracking system which covers the pseudorapidity range
$|\eta| \le 0.9$. A forward muon spectrometer ($2.5 < \eta < 4.0$)
provides measurements of the quarkonia state. Forward detectors provide
global event properties by measurements of photon and charged particle
multiplicities and forward calorimetry.

In addition to running with Pb beams at the highest available energy
of $\sqrt{s} = 5.5$~TeV, ALICE will take proton-proton ($pp$) and $pA$
($p$Pb, maybe $d$Pb or $\alpha$Pb) reference data.  Note that there
are restrictions imposed on the $pA$ programme by both the LHC machine
and the ALICE detector. First, the maximum luminosity is limited
because of pile-up in the TPC to typically
$(10^{30},10^{29},10^{27})$~cm$^{-2}$s$^{-1}$ for $pp$, $p$Pb and PbPb
collisions, respectively.  Second, the energy per nucleon of ions in the
LHC bending field scales to the nominal proton beam energy (7~TeV)
with the ratio $Z/A$ so that the centre of mass energy per nucleon and
the rapidity shift of the centre of mass with respect to the LHC
system are 
$\sqrt{s} \approx (9,6,6)$~TeV and $\Delta y \approx
(0.5,0,0)$ for $p$Pb, $d$Pb and $\alpha$Pb collisions, respectively,
as given by Eqs. \ref{morsch:eq_sqrts} and \ref{morsch:eq_y}. A
comparison of nominal $p$Pb data with $pp$ (14~TeV) or PbPb (5.5~TeV)
therefore needs a good understanding of the energy
dependence. Furthermore, the rapidity shift of $\Delta y \approx 0.5$
in $p$Pb collisions needs a dedicated study of the ALICE
acceptance. Whereas detailed simulations of PbPb and $pp$ collisions
are already well under way, a comprehensive study of the ALICE $pA$
acceptance is not yet available.

In the following we will briefly describe the main components of the
ALICE detector. For a recent, more detailed review,
see Ref.~\cite{mbref:paolo} and references therein.

\subsubsection{The ALICE detector}\label{se:detector}

\begin{figure}[tbh]
\begin{center}
\hspace{3.0cm}\includegraphics[width=14.5cm]{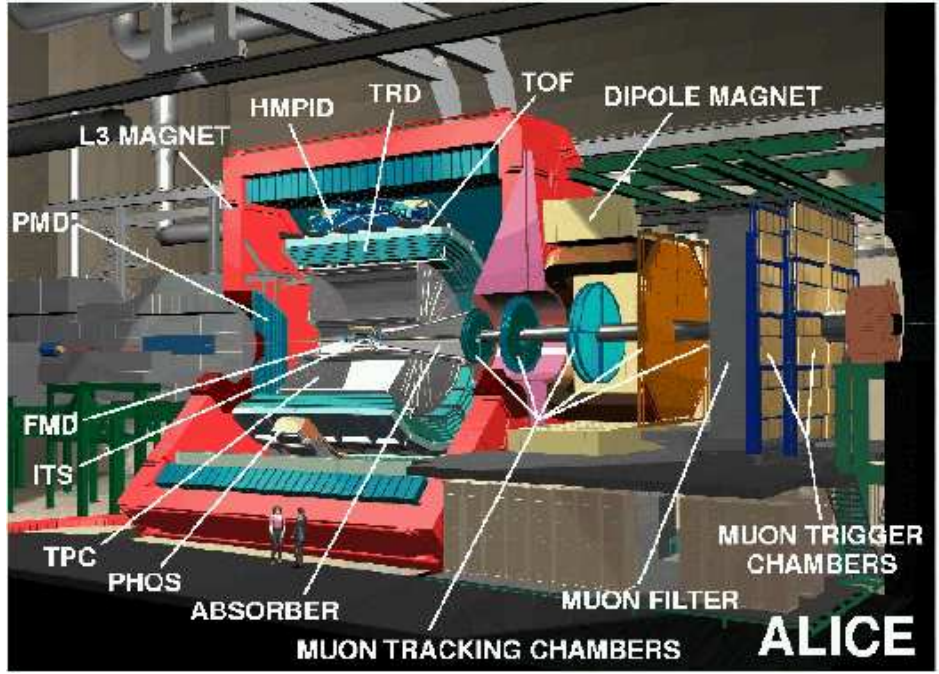} 
\caption{Layout of the ALICE detector.}
\label{fig:mbalice}
\end{center}
\end{figure}

The layout of the ALICE detector is shown in Fig.~\ref{fig:mbalice}.
The central barrel is placed inside the L3 magnet which provides a
solenoidal field of up to 0.5~T. The choice of field is a compromise
between low momentum acceptance ($\sim 100$~MeV/$c$), momentum
resolution and tracking and trigger efficiency. ALICE intends to run
mostly with a higher field of 0.5~T to study high $p_T$ probes but
also with a lower field configuration of $0.2$~T which is optimal for
soft hadronic physics.

The central tracking detector covers a range of $|\eta| < 0.9$ and the
full range in azimuth.  It consists of an inner tracking system (ITS)
with six layers of high resolution silicon detectors located at a
radial distance from the beam axis of $r = 4$, 7, 15, 24, 39 and
44~cm, respectively. The ITS provides tracking and identification of
low $p_T$ particles and improved momentum resolution for particles at
higher momenta which traverse the TPC. An impact parameter resolution
of better than 100~$\mu$m allows for vertex reconstruction of hyperons
and heavy quarks.

A time projection chamber (TPC) of 0.9 (2.5)~m inner (outer) radius
provides track finding, momentum measurement and particle
identification via $dE/dx$ with a resolution of better than 10\%.
Hadrons are identified in the TPC and ITS in the range
$\sim$100--550~MeV/$c$ and up to 900~MeV/$c$ for protons.  The overall
efficiency of the ALICE tracking is estimated to be better than 90\%
independent of $p_T$ down to 100~MeV/$c$.

One of the unique features of ALICE is its particle identification
capability. First of all, a transition radiation detector (TRD) with
an inner (outer) radius of 3 (3.5)~m provides electron identification.
The TRD allows for the measurement of light and heavy meson
resonances, the dilepton continuum and open charm and beauty via their
semileptonic decay channels. The fast tracking capability of the TRD
can be used to trigger on high $p_T$ leptons and hadrons, 
providing a leading-particle trigger for jets.

Time-of-flight counters (TOF), covering the TPC geometrical
acceptance, are installed at a radius of 3.5~m and have a time
resolution of better than 150~ps. This detector will be able to
identify $\pi$, $K$ and protons in the semi-hard regime up to about
3~GeV/$c$.

A single arm, high resolution electromagnetic calorimeter (PHOS) is
located 5~m below the interaction point and covers a range of $|\eta|
\le 0.12$ in pseudorapidity and $\Delta\phi = 100^{\circ}$ in
azimuth. The calorimeter is optimised to measure $\pi^0$, $\eta$ and
prompt photons in the energy range of one to several tens of~GeV.

A RICH detector (HMPID) is positioned on top of the ALICE detector at
about 4.5~m from the beam axis covering a range of $|\eta| < 0.5$ and
$\Delta \phi = 57^{\circ}$. The RICH will extend the particle
identification capabilities of the central barrel to $\approx
3$~GeV/$c$ for $\pi/K$ and $\approx 5$~GeV/$c$ for $K/p$ separation.

The forward muon spectrometer consists of an absorber, starting at 1~m
from the vertex, followed by a dipole magnet with 3~Tm field integral,
placed outside the central magnet, and 10 planes of high granularity
tracking stations. The last section of the muon arm consists of a
second absorber and two more tracking planes for muon identification
and triggering. This detector is designed to measure the decay of
heavy quark resonances ($J/\psi$, $\psi'$, $\Upsilon$, $\Upsilon'$ and
$\Upsilon''$) with sufficient mass resolution to separate the
different states. The coverage is $2.5 < \eta < 4$ with $p_{T}(\!\mbox{
single muon)} > 1$~GeV$/c$.

Several smaller detector systems located at small angles will measure
global event characteristics. In particular, the FMD detectors measure
the charged particle multiplicity in the ranges $-5.1 < \eta < -1.7$
and $1.7 < \eta < 3.4$ and full azimuth. The FMD and ITS combined will
then cover most of the phase space ($-5.1 < \eta < 3.4$) for
multiplicity measurements. Initial Monte Carlo studies indicate that a
centrality determination in $pA$ collisions with a resolution of about
2~fm might be within reach from a measurement of the total charged
multiplicity.

\begin{figure}[tbh]
\begin{center}
\vspace{1cm}
\includegraphics[width=12.5cm]{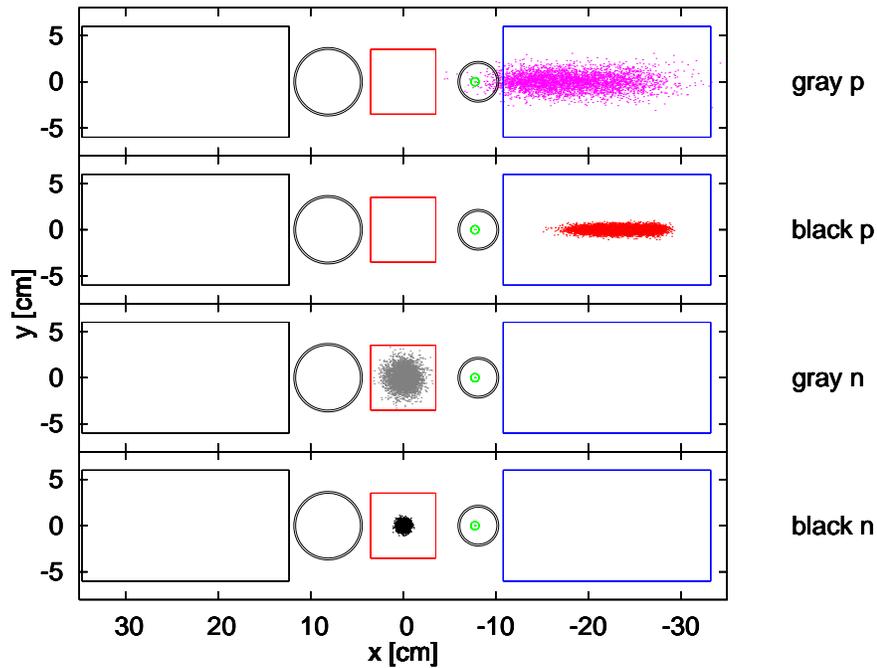} 
\caption{Hit distribution of slow nucleons from $p$Pb collisions at the
  front face of the ZDC. The squares at 0 ($-20$)~cm indicate the
  neutron (proton) calorimeter. The Pb beam is visible inside one of
  the beam pipes (shown as circles) at about $x = -9$~cm.}
\label{fig:zdchits}
\end{center}
\end{figure}

To measure and to trigger on the impact parameter in heavy ion
collisions, the neutron and proton spectators will be measured by zero
degree calorimetry. The neutron and proton spectators are separated in
space by the first LHC dipole and measured by a pair of neutron and
proton calorimeters installed at about 90~m both left and right of the
interaction region. A centrality determination in $pA$ collisions can
be provided by measuring black ($p < 250$~MeV, where $p$ is the momentum
in the rest system of nucleus $A$) and grey ($p$ = 250--1000~MeV) protons 
and neutrons which are
correlated with the impact parameter or, equivalently, with the number
of collisions~\cite{mbref:e910}. Initial simulations show that the ZDC
acceptance for slow nucleons in $pA$ collisions is quite large, see
Fig.~\ref{fig:zdchits}, so that such a centrality measurement in ALICE
might be feasible. More detailed studies are currently under way.
%

\clearpage

\subsection{The CMS detector}
\label{subsec:geist}
{\em Walter Geist}

The CMS detector at the LHC was conceived for optimal measurements of
hard processes in $pp$ collisions at a center-of-mass energy $\sqrt s
= 14$~TeV and at a very high luminosity. Among the rare and hard
processes are e.g. the production of Higgs- and SUSY-particles
corresponding to mass/energy scales beyond 100 GeV; their decays yield
charged leptons, photons, and weakly interacting particles as well as jets
of hadrons. The CMS design is therefore driven by the need for 
\begin{itemize}

\item[{\em (i)}] a superb muon-system, 

\item[{\em (ii)}] the best possible electromagnetic calorimeter (ECAL) 
compatible with {\em (i)}, and 

\item[{\em (iii)}] a high-performance tracking detector.  
The set-up is completed by

 \item[{\em (iv)}] a large acceptance hadronic calorimeter (HCAL).
\end{itemize}

\begin{figure}[tbh]
\begin{center}
\vspace{1cm}
\includegraphics[width=14cm]{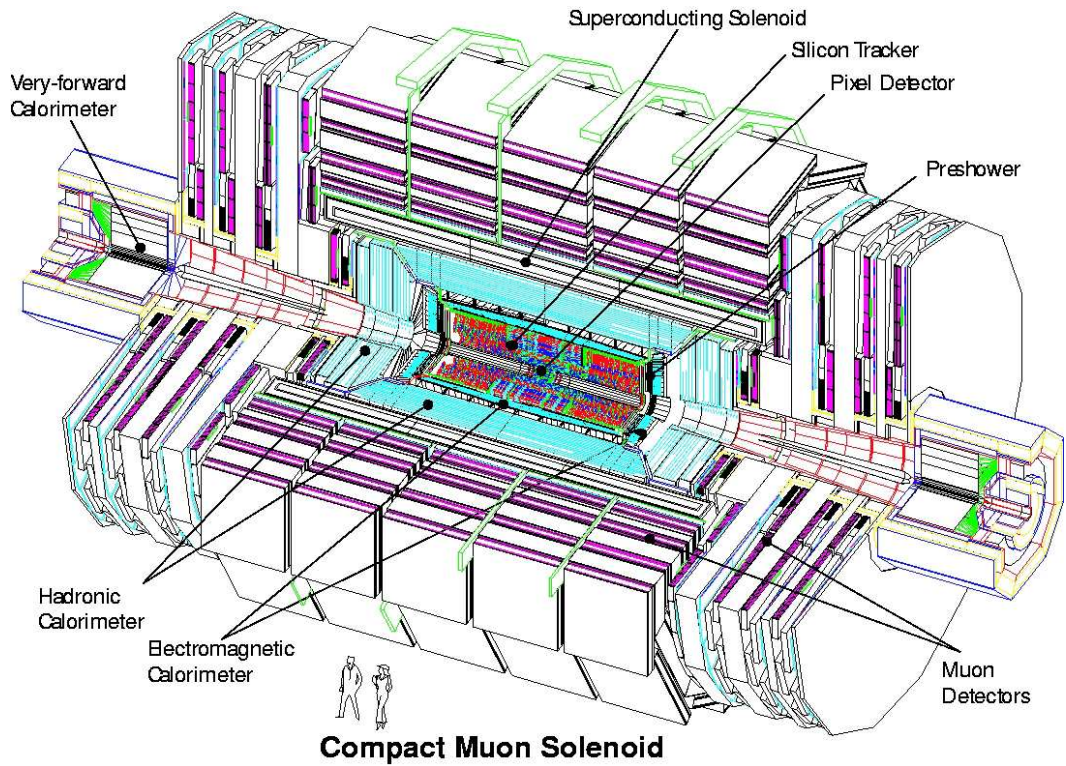} 
\caption{Layout of the CMS detector.}
\label{fig:cmsfig}
\end{center}
\end{figure}

\paragraph{Magnet.}
A direct consequence of these scientific goals is a strong (4 T)
compact solenoid centered at the beam crossing; it bends the
trajectories of charged particles in the plane perpendicular to the
beam direction and thus facilitates the use of the position of the
small diameter beam for establishing primary and secondary
vertices. Its length makes it possible to reconstruct tracks with good
momentum resolution up to pseudorapidities $|\eta| \le 2.4$. The field
integral leads to transverse momentum resolutions $\delta p_T/p_T$ smaller
than about 15\% at $p_T = 1$~TeV$/c$. At the same time, particles with
transverse momenta less than about 1 GeV$/c$ spiral close to the beam
line thereby reducing detector occupancies farther away.

\paragraph{Tracker.}
The all-Si tracker system of CMS consists of 13 cylindrical layers (3
layers of pixel detectors and 10 layers of strip-modules) as well as
of forward-backward discs (2 pixel discs and 12 Si-strip discs)
perpendicular to the beams. They complete the geometric acceptance for
tracks up to $|\eta| = 2.4$.  The Si-detectors are characterized by
fast response and very good spatial resolution. Four Si-strip layers
are made of back-to-back (`double sided') modules for
quasi-3-dimensional space points. Good spatial resolution and low
occupancies ($< 1$\%) in $pp$ collisions at high luminosity result in
track reconstruction efficiencies usually better than 98\%, and better
than 95\% inside jets. The relative resolution for transverse momenta
is expected to be $\delta p_T/p_T = (15p_T+ 5)$\%, with $p_T$ in units
of TeV, for $|\eta|\le 1.6$.  Furthermore, this system, in particular
the pixel part, provides a reliable separation of primary vertices and
secondary vertices due to $B$-decays.

\paragraph{Calorimeters.}
The electromagnetic and hadronic calorimeters surround the
tracker system hermetically. They are positioned inside the coil
of the magnet. Their large rapidity coverage ensures efficient
detection of `missing' transverse energy in the case of
weakly-interacting particle production.  
\begin{itemize}

\item ECAL: The barrel section of ECAL covers the rapidity range 
up to $|\eta|= 1.48$.  The forward part extends the acceptance to $|\eta|=
5$, where preshower detectors are also foreseen. The fine segmentation
of the PbWO$_4$ crystals in rapidity and azimuth matches the
transverse shower size from single $e^{\pm}$ and photons.

\item
HCAL: the hadronic calorimeter is also divided into a
barrel part using copper-plastic tiles for $|\eta| \le 3$ and a forward
calorimeter which increases the acceptance up to $|\eta|= 5$ and is
based upon a fibre concept.
\end{itemize}

\paragraph{Muon detection.}
The muon chambers outside the magnet coil are grouped into stations of
gaseous detector layers interleaved with the iron of the return yoke. They
are protected by an absorber thickness of 10 interaction lengths in
front of the first station and an additional 10 interaction lengths
before the last station. Also, the muon system is organized into
cylindrical and forward disc parts. It is a stand-alone system,
capable of determining the muon momenta to very good precision.

\paragraph{Trigger.}
The CMS triggers select rare processes from the overwhelming rate of
soft reactions at very high luminosities. The total rate of events to
be transferred to a storage medium is limited to about 100~Hz. In order
to cope with this low rate, the rate of individual triggers must be
limited, typically by imposing a transverse momentum/energy cutoff.

\subsubsection{CMS and $pA$ collisions}
While the first detailed investigations of CMS performance with PbPb
collisions have been undertaken, this is not the case for $pA$
collisions.
However, no major problems are anticipated \cite{Geist00}.
There are no a priori technical obstacles to running the CMS
experiment at the lower $pA$- and $AA$-luminosities (see
Sec.~\ref{subsec:morsch}) with e.g. lower $p_T$/energy thresholds
and/or a lower magnetic field, as long as occupancy problems do not
obstruct track reconstruction substantially. 
Global measurements corresponding to an increased acceptance at lower
transverse momenta are very useful, given that the $pA$-collision
energy will exceed that previously available by more than order of
magnitude.

It should be emphasized that the TOTEM experiment will take data in
conjunction with CMS. TOTEM needs a geometrical acceptance for secondary
particles up to $|\eta|=7$. Furthermore, a `zero-degree' calorimeter covering 
the range of beam rapidities is considered for CMS heavy ion runs.

\clearpage

\subsection{The ATLAS detector}
\label{subsec:takai}
{\em Brian Cole, Helio Takai, and Stefan Tapprogge}

\subsubsection{The ATLAS detector}

\begin{figure}[htbp]
\begin{center}
\mbox{\epsfig{file=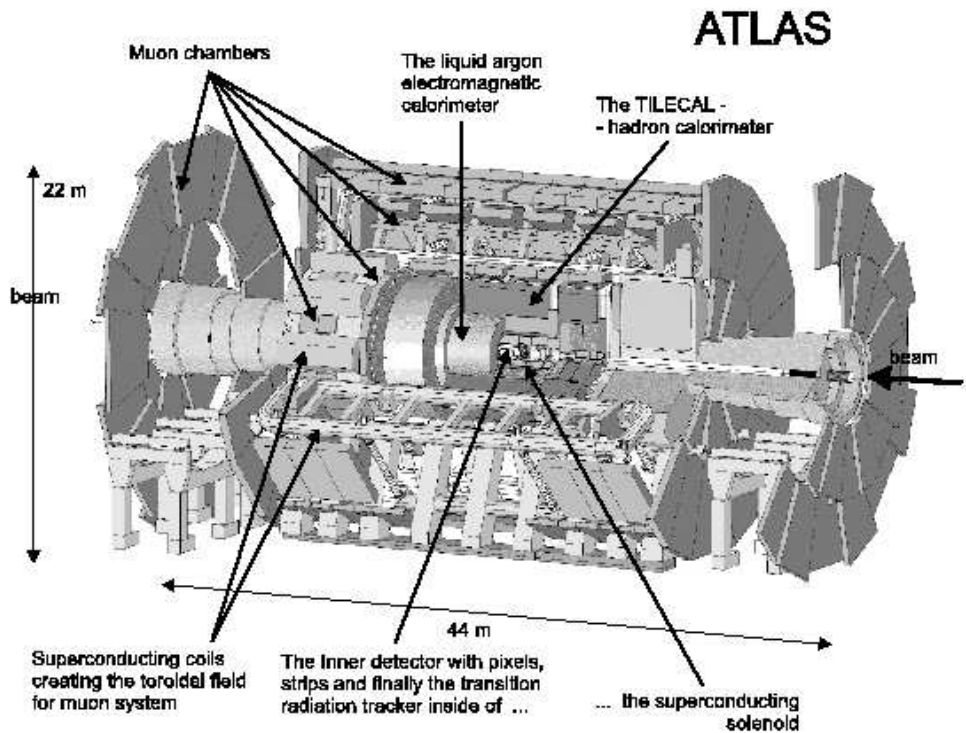,width=13cm}}
\caption{The overall layout of the ATLAS detector}
\label{atlasfig}
\end{center}
\end{figure}

The ATLAS detector is designed to study proton-proton collisions at
the LHC design energy of 14~TeV in the center of mass.  The extensive 
physics program pursued by the collaboration
includes the Higgs boson search,
searches for SUSY and other scenarios beyond the Standard Model, as
well as precision measurements of processes within (and possibly beyond)
the Standard Model. To achieve these goals at a full machine
luminosity of $10^{34}$~cm$^{-2}$s$^{-1}$,
 ATLAS will have a precise
tracking system (Inner Detector) for charged particle measurements, an
as hermetic as possible calorimeter system, which has an extremely
fine grain segmentation, and a stand-alone muon system. An overview of
the detector is shown in Fig.~\ref{atlasfig}.

The Inner Detector is composed of (1) a finely segmented silicon pixel
detector, (2) silicon strip detectors (Semiconductor Tracker (SCT))
and (3) the Transition Radiation Tracker (TRT).  The segmentation is
optimized for proton-proton collisions at design luminosity.
The inner detector system is designed to cover a pseudorapidity of
$|\eta| < 2.5$ and is located inside a 2~T solenoid magnet.

The calorimeter system in the ATLAS detector surrounding the solenoid
magnet is divided into electromagnetic and hadronic sections and
covers pseudorapidity $| \eta | < 4.9$. The EM calorimeter is
an accordion liquid argon device and is finely segmented
longitudinally and transversely for $| \eta | \le 3.1$.  The first
longitudinal segmentation has a granularity of $0.003 \times 0.1$ $(\Delta
\eta \times \Delta \phi)$ in the barrel and slightly coarser in the
endcaps.  The second longitudinal segmentation is composed of $\Delta
\eta \times \Delta \phi = 0.025 \times 0.025$ cells and the last
segment $\Delta \eta \times \Delta \phi = 0.05 \times 0.05$ cells.  In
addition, a finely segmented $(0.025 \times 0.1)$ pre-sampler system is
present in front of the electromagnetic (EM) calorimeter.  The overall
energy resolution of the EM calorimeter determined experimentally is
$10\%/\sqrt{E} \oplus 0.5\%$. The calorimeter also has good pointing
resolution, $60\, {\rm mrad}/\sqrt{E}$ for photons, and a timing resolution
of better than 200 ps for energy showers larger than 20 GeV.

The hadronic calorimeter is also segmented longitudinally and
transversely.  Except for the endcaps and the forward calorimeters,
the technology utilized for the calorimeter is a lead-scintillator
tile structure with a granularity of $\Delta \eta \times \Delta \phi =
0.1 \times 0.1$.  In the endcaps the hadronic calorimeter is
implemented in liquid argon technology for radiation hardness with the
same granularity as the barrel hadronic calorimeter.  The energy
resolution for the hadronic calorimeters is $50\%/\sqrt{E} \oplus 2\%$
for pions.  The very forward region, up to $\eta =4.9$, is covered by
the Forward Calorimeter implemented as an axial drift liquid argon
calorimeter.  The overall performance of the calorimeter system is
described in Ref.~\cite{atlas:ref1}.

The muon spectrometer in ATLAS is located behind the calorimeters,
thus shielded from hadronic showers. The spectrometer is implemented
using several technologies for tracking devices and a toroidal magnet
system which provides 
a 4~T field to have an independent
momentum measurement outside the calorimeter volume.  Most of the
volume is covered by Monitored Drift tubes (MDTs). In the forward region,
where the rate is high, Cathode Strip Chamber technology is chosen.
The stand-alone muon spectrometer momentum resolution is of the order
of $2\%$ for muons with $p_T$ in the range $10\dots 100$~GeV. The muon
spectrometer coverage is $| \eta | < 2.7$.

The trigger and data acquisition system of ATLAS is a multi-level
system which has to reduce the beam crossing rate of 40~MHz to an
output rate to mass storage of $\mathcal{O}(100)$~Hz. The first stage
(LVL1) is a hardware based trigger, making use of coarse 
granularity calorimeter data and dedicated muon trigger chambers only,
to reduce the output rate to about 75~kHz, within a maximum latency
of 2.5~$\mu$s.

The performance results mentioned here were obtained using a detailed
full simulation of the ATLAS detector response with GEANT and have 
been validated by an extensive program of test beam measurements of 
all components.

\subsubsection{Expected detector performance for $pA$}

The proton-nucleus events are asymmetric, with the central rapidity 
shifted by almost 0.5 units.  In proton-nucleus runs it is
expected that all of the subsystems will be available for data analysis.
For the expected luminosity of $10^{29}$~cm$^{-2}$s$^{-1}$ in $p$Pb 
collisions, the occupancy of the detectors will be less than or at most
comparable to the high luminosity proton-proton runs. The tracking
will also benefit from not having up to  25 collisions (as in
 $pp$ interactions at the design luminosity of $10^{34}$~cm$^{-2}$s$^{-1}$)
in the same bunch crossing, but only a single event vertex. 
The detector acceptance for a $pA$ run is
the same as for $pp$  and is shown in Table~\ref{atlas:table1}
(in the detector frame. See Sec.~\ref{subsec:morsch} for the
rapidity shifts in the cms frame of the $pA$ collisions).

\begin{table}
\begin{center}
\caption{ATLAS detector acceptance for $pA$ runs}
\vskip 0.2cm
\begin{tabular} {l|c}
\hline
Detector System  & $\eta_{\rm max}$ \\
\hline\hline
Pixel &  2.5 \\
SCT &  2.5 \\
TRT &  2.5 \\
EM Calorimeter &  3.1 \\
Hadronic Calorimeter  & 4.9 \\
Muon Spectrometer & 2.7\\
\hline\hline
\end{tabular}
\label{atlas:table1}
\end{center}
\end{table}

The muon spectrometer will also benefit from the low luminosity
proton-nucleus mode. In the high luminosity proton-proton runs, a
large number of hits in the spectrometer comes from slow neutrons from
previous interactions. These are not present in low luminosity
runs. The only expected backgrounds are muons from $\pi $ and $K$
decays in flight. As has been shown for $pp$, a substantial
number of these background muons are expected to be rejected when 
matching the track from the stand-alone muon system to a track in
the Inner Detector is required.

\subsubsection{The physics potential of ATLAS in $pA$}

Measurements of proton-nucleus collisions at the LHC will not only
provide essential control of the hard 
and semi-hard
processes 
in heavy ion collisions, but these
measurements can also address physics that is, in its own right, of
fundamental interest. The ATLAS detector provides an unprecedented
opportunity to study proton-nucleus collisions in a detector with both
large acceptance and nearly complete coverage of the various final
states that can result from perturbative QCD processes. 

An important physics topic that can be addressed by proton-nucleus
measurements in ATLAS is the applicability of factorization in hard
processes involving nuclei. No detector that has studied either
nuclear deep inelastic scattering or proton-nucleus collisions has
been able to simultaneously study many different hard processes to
explicitly demonstrate that factorization applies.  In ATLAS, we will
have the opportunity to study single jet and ${\rm jet}-{\rm jet}$ events,
${\rm photon}-{\rm jet}$ events, heavy quark production, $Z^0$ and $W^{\pm}$
production, Drell-Yan dilepton production, etc. and probe the
physics of jet fragmentation in detail.  With the wide variety of
available hard processes that all should be calculable with the same
parton distributions, we should be able to  clearly demonstrate the
success of factorization for the large-$Q^2$ processes.

The ATLAS inner detector has been designed to have good tracking
efficiency in the momentum range of 1~GeV up to 1~TeV over the
pseudorapidity range $| \eta | \le 2.5$. The ability to
reconstruct jets in ATLAS over a large pseudorapidity region (up to
$| \eta | \le 4.9$) and over the full azimuth would  
probe the onset of saturation effects in
high $p_T$ jets.
One aspect of saturation that is not yet well
understood is the effect on hard scattering processes for $Q^2 \sim
Q_s^2$, where $Q_s$ is the scale at which saturation sets in. 
It has been suggested that the gluon $k_T$ distribution could
be significantly modified well above $Q_s$. If so, then ATLAS should
be able to study the $Q^2$ evolution of the saturation effects using
${\rm photon} - {\rm jet}$, $b \bar{b}$ and ${\rm jet}-\rm{jet}$ measurements.

Detailed studies of the behavior of jet fragmentation as a function of
pseudorapidity could be used to determine whether the jet fragmentation is
modified by the presence of the nucleus or its large number of low-$x$
gluons. The observation or lack thereof of modifications to jet
fragmentation could provide sensitive tests of our understanding of
formation time and coherence in the re-dressing of the hard scattered
parton and the longitudinal spatial spread of low-$x$ gluons in a highly
Lorentz-contracted nucleus.



\clearpage


\section{ QCD FACTORIZATION AND RESCATTERING IN $pA$ COLLISIONS}
\label{pASEC_factorization}
{\em Jianwei Qiu}



%
%

Perturbative QCD has been very successful in interpreting and
predicting hadronic scattering processes in high energy collisions, 
even though the physics associated with an individual hadron wave 
function is nonperturbative. It is the QCD factorization theorem 
\cite{Collins:gx} that provides prescriptions to separate long- and
short-distance effects in hadronic cross sections.  The leading power
contributions to a general hadronic cross section involve only one
hard collision between two partons from two incoming hadrons 
of momenta $p_A$ and $p_B$.   The momentum scale of the hard
collision is set by producing either a heavy particle (like $W/Z$ or
virtual photon in Drell-Yan production) or an energetic third parton,
which fragments into a hadron $h$ of momentum $p'$.
The cross section can be factorized as \cite{Collins:gx}
\begin{equation}
E_h {d{\sigma}_{AB\rightarrow h(p')}\over d^3p'}
=
\sum_{ijk}\int dx' f_{j/B}(x')
          \int dx\, f_{i/A}(x)
          \int dz\, D_{h/k}(z)\,
          E_h {d\hat{\sigma}_{ij\rightarrow k}\over d^3p'} 
          (xp_A,x'p_B,\frac{p'}{z}) ,
\label{twist2conv}
\end{equation}
where $\sum_{ijk}$ runs over all parton flavors and all scale
dependence is implicit. The $f_{i/A}$ are twist-2 distributions of
parton type $i$ in hadron $A$, and the $D_{h/k}$ are fragmentation 
functions for a parton of type $k$ to produce a hadron $h$.
For jet production, fragmentation from a parton to a jet, 
suitably defined, is calculable in perturbation theory and 
may be absorbed into the partonic hard  part $\hat{\sigma}$.
For heavy particle production, the fragmentation function is 
replaced by $\delta(1-z)$.

The factorized formula in Eq.~(\ref{twist2conv}) illustrates
the general leading power collinear factorization theorem
\cite{Collins:gx}. It consistently separates perturbatively calculable
short-distance physics into $\hat{\sigma}$, and isolates long-distance
effects into universal nonperturbative matrix elements (or
distributions), such as $f_{i/A}$ or $D_{h/k}$, associated with each
observed hadron.  Quantum interference between  long- and
short-distance physics is power-suppressed by the large energy
exchange of the collisions.  Predictions of pQCD follow when processes
with different hard scatterings but the same set of parton
distributions and/or fragmentation functions are compared
\cite{Qiu:2001hj}.  

With the vast data available, the parton distributions of a free
nucleon are well determined by QCD global analyses 
\cite{Pumplin:2002vw,Martin:2002dr}.  Thanks to recent efforts, 
sets of fragmentation functions for light hadrons are becoming
available though the precision is 
not as good as the
parton distributions due to the limited data 
\cite{Kniehl:2000hk,Kretzer:2001pz}.


Studies of hard processes at the LHC will cover a very large range of
longitudinal momentum fractions $x$ of parton distributions:
$x \ge x_T\, e^y/(2-x_T e^{-y})$ with $x_T = 2p_T/\sqrt{s}$ 
for inclusive jet production in Eq.~(\ref{twist2conv}), where 
$y$ and  $p_T$ are the rapidity and transverse momentum of the
produced jets, respectively.  For the most forward or backward jets,
or low $p_T$ Drell-Yan dileptons,  $x$ can be as small as
10$^{-6}$ at $\sqrt{s}=14$~TeV.  The number of gluons having such 
small $x$ and transverse size 
$\Delta r_\perp \propto 1/p_T$ may be so large that gluons appear more
like a collective wave than individual particles, and a new
nonperturbative regime of QCD, such as gluon saturation or the color
glass condensate \cite{McLerran:1993ni}, might be reached.

\begin{figure}
\begin{minipage}[c]{7.6cm}
\centerline{\includegraphics[width=7.0cm]{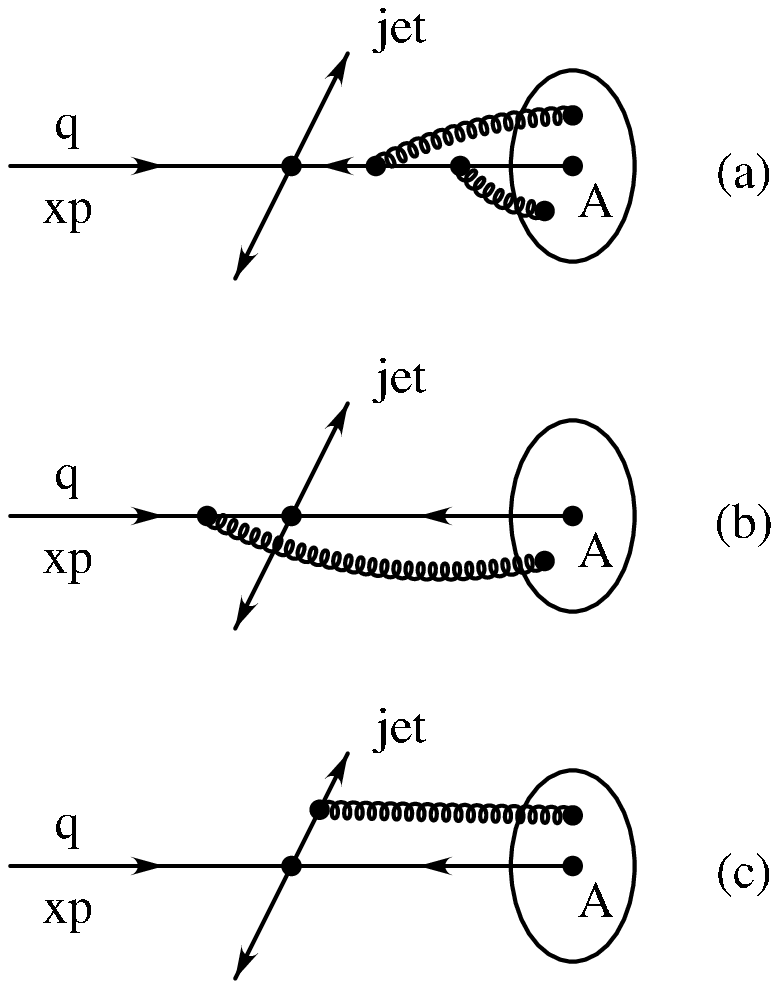}} 
\vspace{-0.1in}
\caption{Classification of parton multiple scattering in nuclear
medium: (a) interactions internal to the nucleus, (b) initial-state
interactions, and (c) final-state interactions.}
\label{qiu:fig1}
\end{minipage}
\hfill
\begin{minipage}[c]{7.6cm}
\centerline{\includegraphics[width=6.0cm]{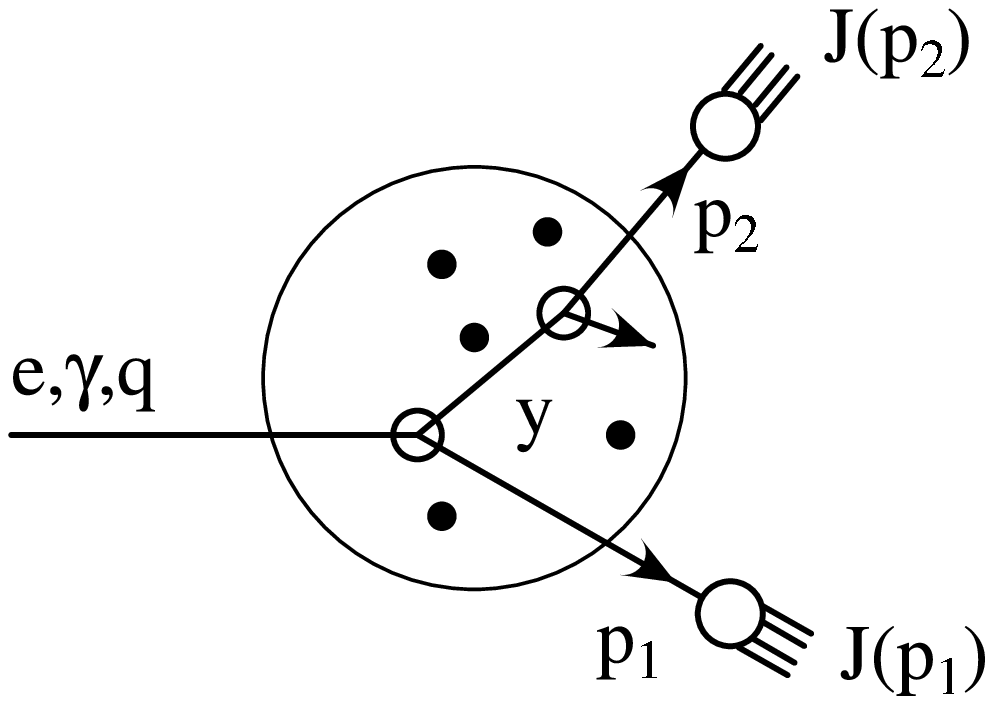}} 
\vspace{-0.1in}
\caption{Sketch for the scattering of an elementary particle or a
parton of momentum $xp$ in a large nucleus.}
\label{qiu:fig2}
\vspace{0.1in}
\centerline{\includegraphics[width=6.5cm]{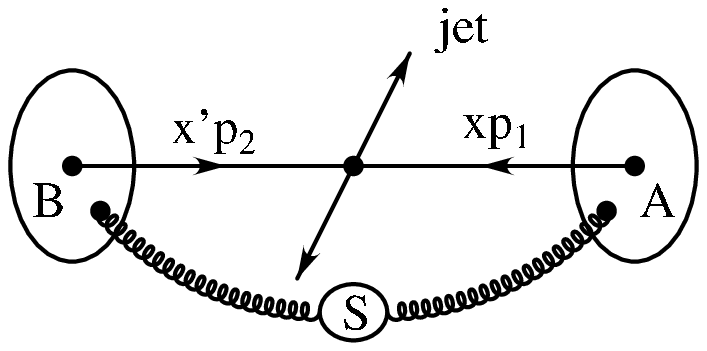}} 
\vspace{-0.1in}
\caption{Soft interactions of long-range fields that might enhance the
  gluon density in nuclear medium and affect the FSI.}
\label{qiu:fig3}
\end{minipage}
\end{figure}

The use of heavy ion beams allows one to enhance the coherent effects
because of the larger number of gluons.  In the small $x$ regime of
hadron-nucleus collisions, a hard collision of a parton of the
projectile nucleon with a parton of the nucleus occurs coherently with
all the nucleons at a given impact parameter.  The coherence length
($\sim 0.1/x$~fm) far exceeds the nuclear size.  To distinguish
parton-nucleus multiple scattering from partonic dynamics internal to
the nucleus, we classify multiple scattering in the following
three categories: (a) interactions internal to the
nucleus, (b) initial-state parton-nucleus interactions (ISI), and (c)
final-state parton-nucleus interactions (FSI), as shown in
Fig.~\ref{qiu:fig1} \cite{Qiu:2001hj}.  

Interactions internal to the nucleus change the
parton distributions of the nucleus, as shown in
Fig.~\ref{qiu:fig1}a.  As a result, the effective
parton distributions of a large nucleus are different from a
simple sum of individual nucleon parton distributions. 
Since only a single parton from the nucleus participates in the hard
collision  to leading power, the effect of the initial-state
interactions internal to the nucleus does not change the hard
collisions between two incoming partons. This preserves the factorized
single scattering formula in Eq.~(\ref{twist2conv}) \cite{Qiu:2002mh} 
except now the twist-2 parton distributions $f_{i/A}$ are replaced by
the corresponding effective nuclear parton distributions, which are
defined in terms of the same operators but on a nuclear state.  Such
effective nuclear parton distributions include the ``EMC'' effect and
other nuclear effects so that they differ from the parton
distributions of a free nucleon.  However, they are still twist-2
distribution functions by definition of the operators of the
matrix elements and are still universal.   

Because of the twist-2 nature of the nuclear parton distributions,
their nuclear dependence can only come from the $A$-dependence of the
nonperturbative input parton distributions at a low momentum scale
$Q_0$ and the $A$-dependent modifications to the DGLAP evolution
equations from $Q_0$ to $Q$ due to the interactions between partons
from different nucleons in the nucleus \cite{Gribov:tu,Mueller:wy}.
The nonperturbative nuclear dependence in the input distributions can
be either parameterized with parameters fixed by fits to 
experimental data \cite{Eskola:1998df,Eskola:1998iy,Hirai:2001np} or
calculated with some theoretical inputs
\cite{Frankfurt:cernrpt,Kovchegov:cernrpt}.

Let hard probes be those whose cross sections are dominated by the
leading power contributions in Eq.~(\ref{twist2conv}).  That is,
nuclear dependence of the hard probes comes entirely from that of
nuclear parton distributions and is universal.  Because of the wide
range of $x$ and $Q$ covered, the hard probes in hadron-nucleus
collisions at the LHC directly detect the partonic dynamics internal
to the nucleus and provide excellent information on nuclear parton
distributions.  The knowledge of these distributions is very useful
for understanding gluon saturation, a new nonperturbative regime
of QCD \cite{Kovchegov:cernrpt,Eskola:2003gc}.    

On the other hand, the hadronic cross sections receive
power-suppressed corrections to Eq.~(\ref{twist2conv})
\cite{Qiu:xx,Brodsky:cernrpt,Qiu:2001zj}.  These
corrections can come from several different sources, including the
effect of partonic non-collinear momentum components and effects of
non-vanishing invariant mass of the fragmenting parton $k$, as well as
the effects of interactions involving more than one parton from each
hadron, as shown in Fig.~\ref{qiu:fig1}b and \ref{qiu:fig1}c.
Although such multiple coherent scattering is formally a 
higher-twist effect and suppressed by powers of the large momentum
scale of the hard collision, the corrections to the leading power
factorized formula in Eq.~(\ref{twist2conv}) are proportional to the 
density of additional scattering centers and can be substantial due to 
a large density of soft gluons in the nucleus-nucleus collisions.  For 
the ISI, the number of gluons available at the same impact
parameter of the hard collision is enhanced due to the large nucleus,
while the densities of soft partons available to the FSI can be even 
larger since they can come either from the initial wave
functions of the colliding nuclei or be produced in the long-range soft
parton interactions between different nucleons at the time when the hard
collision took place.
%


Consider the scattering of an elementary particle or a parton (a quark
or a physically polarized gluon) of momentum $xp$ in nuclear matter,
as shown in Fig.~\ref{qiu:fig2}.  A hard scattering with momentum transfer 
$Q$ can resolve states whose lifetimes are as short as $1/Q$
\cite{Qiu:2001hj}.  The off-shellness of the scattered parton 
increases with the momentum transfer simply because the number of
available states increases with increasing momentum. Typically, 
the scattered parton (say, of momentum $p_1$) is off-shell by 
$m_J\le Q$, where $m_J$  is the invariant mass of the jet into which
parton fragments.  Further interactions of the off-shell parton are
suppressed by an overall factor of $1/m_J^2\sim 1/Q^2$, since the
effective size of the scattered parton decreases with momentum
transfer, and by the strong coupling evaluated at scale $m_J\sim Q$.
Thus rescattering in nuclear collisions is suppressed by
$\alpha_s(Q)/Q^2$ compared to single scattering.   

Counting of available states ensures that $m_J\ge \sum_h \langle
N_h\rangle m_h \gg\Lambda_{\rm QCD}$, where $\sum_h$ runs over all
hadron types in the jet and $\langle N_h\rangle$ is the corresponding
multiplicity.  On the other hand, if we are to recognize the jet, 
we must have $m_J\ll E_J=p_1^0$, where $E_J$ is the jet energy.  
In the rest frame of the nucleus, the scattered parton has a lifetime
$\Delta t \sim ({E_J/ m_J})/m_J$.  Thus, at
high enough jet energy, $\Delta t > R_A$, the lifetime of the
scattered parton will exceed the size of nuclear matter even though
the parton itself is far off mass shell.  Then the
interactions of the scattered off-shell parton with nuclear matter may
be treated by the formalism of pQCD \cite{Qiu:2001hj}.

In order to consistently treat the power-suppressed multiple
scattering, we need a factorization theorem for higher-twist 
(i.e., power suppressed) contributions to hadronic hard scattering.
It was shown in Ref.~\cite{Qiu:xx} that the first power-suppressed
contribution to the hadronic cross section can be factorized into the
form
\begin{eqnarray}
E_h\, {d\sigma^{(4)}_{AB\rightarrow h(p')}\over d^3p'}
&=&
\sum_{(ii')jk}\int dx'\,f_{j/B}(x')\,
              \int dz \,D_{h/k}(z)\,
\label{twist4conv} \\
&\ & \times
\int dx_1 dx_2 dx_3\; T_{(ii')/A}(x_1,x_2,x_3)\,
     E_h {d\hat{\sigma}^{(4)}_{(ii')j\rightarrow k}\over d^3p'} 
         (x_ip_A,x'p_B,\frac{p'}{z})\, ,
\nonumber
\end{eqnarray}
where the partonic hard part 
$\hat{\sigma}^{(4)}_{(ii')j\rightarrow k}$ is infrared safe 
and depends on the identities and momentum fractions of the incoming
partons but is otherwise independent of the structure -- in particular
the size -- of the hadron and/or heavy ion beams.  
The superscript ``(4)'' on the partonic hard part
indicates the dependence on twist-4
operators \cite{Qiu:xx,Luo:ui}. 
The correlation functions $T_{(ii')/A}$ are
defined in terms of matrix elements of twist-4 operators made of
two pairs of parton fields of flavor $i$ and $i'$, respectively. 
For example, a correlation function between quark and gluon in a hadron
$h$ of momentum $p$ can be expressed as
\begin{eqnarray}
   T_{(qG)/A}(x_1,x_2,x_3)
 &=& \int{dy^-_1dy^-_2dy^-_3\over(2\pi)^3}\,   
     {\rm e}^{ip^+(x_1y^-_1+x_2y^-_2+x_3y^-_3)}
 \nonumber \\
 &\ & \times
  \langle p|\bar{\psi}(0)\frac{\gamma^+}{2}
   F^{+\alpha}(y^-_3)F_{\alpha}^{\ +}(y^-_2)\psi(y^-_1) |p\rangle\, ,
 \label{Tdef}  
 \end{eqnarray}
with quark field $\psi$ and gluon field strength $F^{+\alpha}$.

Demonstrating factorization at the next-to-leading power is a 
starting point for 
a unified discussion of the power-suppressed effects in a wide
class of processes.  A systematic treatment of double scattering in a
nuclear medium is an immediate application of the generalized
factorization theorem \cite{Qiu:2001hj}. Because of the infrared safe
nature of the partonic hard part
$\hat{\sigma}^{(4)}_{(ii')j\rightarrow k}$ in Eq.~(\ref{twist4conv}),
the nuclear dependence of the double scattering comes entirely from
the correlation functions of two pairs of parton fields which can be
linearly proportional to $A^{1/3}$ (or nuclear size)
\cite{Luo:ui}.  Therefore, if the scattered off-shell parton has a
lifetime longer than the nuclear size, rescattering is enhanced by
$A^{1/3}$ 
from the medium size and gets an
overall suppression factor,  
${\alpha_s(Q)\, A^{1/3}\, \lambda^2}/{Q^2}$, 
where $\lambda$ has the dimension of mass and represents the
nonperturbative scale of the twist-4 correlation functions
in Eq.~(\ref{Tdef}). A
semiclassical estimate gives $\lambda^2 \sim 
\langle F^{+\alpha}F_{\alpha}^{\ +}\rangle {\rm fm}^2/\pi$ \cite{Luo:ui}.  

For the ISI encountered by the incoming parton, $\lambda^2$ is
proportional to the average squared transverse field strength, 
$\langle F^{+\alpha}F_{\alpha}^{\ +}\rangle$, inside the nucleus, 
which should be more or less universal.
From the data on Drell-Yan transverse momentum broadening, it was
found \cite{Guo:1998rd} that $\lambda^2_{\rm ISI} \approx
\lambda^2_{\rm DY} \sim 0.01$~GeV$^2$.  However, the numerical value
of $\lambda^2$ for the FSI does not have to be equal to the  
$\lambda^2_{\rm ISI}$ due to extra soft gluons produced 
by the instantaneous soft interactions of long-range fields
between the beams at the same time when the jets were produced by two
hard partons, as shown in Fig.~\ref{qiu:fig3}.

It was explicitly shown \cite{Doria:ak,Brandt:xt,Basu:1984ba} that the
corrections to hadronic Drell-Yan cross sections cannot be factorized
beyond the next-to-leading power.  However, it can be shown using
the technique developed in Ref.~\cite{Qiu:xx} that the type of
$A^{1/3}$-enhanced power corrections to
hadronic cross sections in hadron-nucleus collisions can be factorized
to all powers, 
\begin{equation}
E_h\, {d\sigma^{(2n)}_{AB\rightarrow h(p')}\over d^3p'}
=
\sum_{(i_n)jk}\, f_{j/B}(x')\otimes D_{h/k}(z)
                \otimes T^{(2n)}_{(i_n)/A}(x_i)\otimes
     E_h {d\hat{\sigma}^{(2n)}_{(i_n)j\rightarrow k}\over d^3p'} 
         (x_ip_A,x'p_B,\frac{p'}{z}),
\label{twistnconv}
\end{equation}
where $\otimes$ represents convolutions in fractional momenta carried
by the partons and $T^{(2n)}(x_i)$ represent the correlation functions
of $n$ pairs of parton fields with flavors $i_n=1,2,...,n$.  In
Eq.~(\ref{twistnconv}), no power corrections are initiated by the
higher-twist matrix elements of the incoming hadron $B$ because such
contributions do not have the $A^{1/3}$-type enhancement in
hadron-nucleus collisions.  
For example, the recently proposed reaction operator approach to
multiple scattering \cite{Gyulassy:2002yv,Gyulassy:2000er} should fit
into the type of factorization in Eq.~(\ref{twistnconv}) with the
independent-scattering-center approximation to the $T^{(2n)}(x_i)$ and
keeping the lowest order in $\alpha_s$ to the partonic
$\hat{\sigma}^{(2n)}$, which can be obtained using a recursive method.
To the contrary, in nucleus-nucleus collisions, even the nuclear size
enhanced power corrections cannot be formally factorized beyond the
next-to-leading power.

Let semi-hard probes in hadronic collisions be those with a large
momentum exchange as well as large power corrections.  We expect that
pQCD has good predictive power for semi-hard observables in
hadron-nucleus collisions but the pQCD factorization approach does
not work well for semi-hard observables in nucleus-nucleus collisions.


In conclusion, the factorized single scattering formula in
Eq.~(\ref{twist2conv}) remains valid for hard probes in nuclear
collisions, except that the parton distributions are replaced by
corresponding effective nuclear parton distributions which are
independent of the hard scattering and thus universal.  Hard probes in
nuclear collisions at the LHC can provide excellent information on
nuclear parton distributions and detect the partonic dynamics internal
to the nucleus.

However, the power-suppressed corrections to the single scattering
formula can be substantial and come from several different sources.
The effect of the off-shellness of the fragmenting parton $k$ leads to 
a correction of the order $(m_J/p_T)^2$ with $m_J=\sum_h \langle
N_h\rangle m_h$.  Both the ISI and FSI double scatterings give 
power-suppressed corrections proportional to $\alpha_s A^{1/3}
\lambda^2 / p_T^2$, where $\lambda^2_{\rm ISI}$ is relatively small and
almost universal and $\lambda^2_{\rm FSI}$ is sensitive to the number of
soft partons produced by the instantaneous collisions of long-range
fields between nucleons in the incoming beams.

Beyond double scattering (or next-to-leading power corrections), pQCD
calculations might not be reliable due to the lack of factorization
theorems at this level.   However, pQCD factorization for the type of 
$A^{1/3}$-enhanced power corrections in hadron-nucleus collisions is
likely to be valid to all powers.



\clearpage


\section{NUCLEAR PARTON DISTRIBUTION FUNCTIONS (nPDFs)}
\label{pA_SEC:nPDF}
\vspace{-0.25cm}
\vspace{0.25cm}

\subsection{Global DGLAP fit analyses of the nPDFs: EKS98 and HKM}
\label{subsec:eskola}
{\em Kari J. Eskola, Heli Honkanen, Vesa J. Kolhinen, and Carlos A. Salgado}

\def\lsim{~\raise0.3ex\hbox{$<$\kern-0.75em\raise-1.1ex\hbox{$\sim$}}~}
\def\gsim{~\raise0.3ex\hbox{$>$\kern-0.75em\raise-1.1ex\hbox{$\sim$}}~}








\subsubsection{Introduction}

Inclusive cross sections for hard processes $A+B\rightarrow c+X$
involving a sufficiently large scale $Q\gg\Lambda_{\rm QCD}$ are
computable by using collinear factorization. In the leading-twist
approximation power corrections $\sim 1/Q^2$ are neglected and
\begin{eqnarray}
& \displaystyle 
   d\sigma(Q^2,\sqrt s)_{AB\rightarrow c+X} =
   \sum_{i,j=q,\bar q,g} 
   \bigg[ Z_Af_i^{p/A}(x_1,Q^2)+(A-Z_A)f_i^{n/A}(x_1,Q^2)\bigg] 
   \otimes \nonumber  \\
 & \otimes\bigg[ Z_Bf_j^{p/B}(x_2,Q^2)+(B-Z_B)f_j^{n/B}(x_2,Q^2)\bigg]
   \otimes d\hat \sigma(Q^2,x_1,x_2)_{ij\rightarrow c+x}
\label{hardAA}
\end{eqnarray}
where $A,B$ are the colliding hadrons or nuclei containing $Z_A$ and
$Z_B$ protons respectively, $c$ is the produced parton, $x$ and $X$
are anything, $d\hat \sigma(Q^2,x_1,x_2)_{ij\rightarrow c+x}$ is the
perturbatively calculable differential cross section for the
production of $c$ at the scale $Q$, and $x_{1,2}\sim Q/\sqrt s$ are the
fractional momenta of the colliding partons $i$ and $j$. The number
distribution function of the parton flavour $i$ of the protons
(neutrons) in $A$ is denoted as $f_i^{p/A}$ $(f_i^{n/A})$, and
similarly for partons $j$ in $B$.

In the leading-twist approximation, multiple scattering of the bound
nucleons does occur but all collisions are independent, correlations
between partons from the same object $A$ are neglected, and only
one-parton densities are needed. The parton distribution functions
(PDFs) $f_i^{p/A}$ are universal quantities applicable in all
collinearly factorizable processes. The PDFs cannot be computed by
perturbative methods and, so far, it has not been possible to compute
them from first principles, either. Thus, nonperturbative input from
data on various hard processes is needed for the extraction of the
PDFs.  However, once the PDFs are known at some initial (lowest) scale
$Q_0\gg\Lambda_{\rm QCD}$, QCD perturbation theory predicts the
scale evolution of the PDFs to other (higher) values of $Q^2$ in form
of the Dokshitzer-Gribov-Lipatov-Altarelli-Parisi (DGLAP) equations
\cite{Dokshitzer:sg}.

The method of extracting the PDFs from experimental data is well
established in the case of the free proton: the initial
nonperturbative distributions are parametrized at some $Q_0^2$ and
evolved to higher scales according to the DGLAP equations. Comparison
with the data is made in various regions of the $(x,Q^2)$-plane. The
parameters of the initial distributions, $f_i^{p}(x,Q_0^2)$, are fixed
when the best global fit is found. The data from deeply inelastic $lp$
scattering (DIS) are most important for these global DGLAP fits,
especially the HERA data at small values of $x$ and $Q^2$. The sum
rules for momentum, charge and baryon number give further
constraints. In this way, through the global DGLAP fits, groups like
MRST \cite{Martin:2001es}, CTEQ \cite{Pumplin:2002vw} or GRV
\cite{Gluck:1998xa} obtain their well-known sets of the free proton
PDFs.

The nuclear parton distribution functions (nPDFs) differ in magnitude
from the PDFs of the free proton. In the measurements of the structure
function, $F_2^A = Z_AF_2^{p/A}+ (A-Z_A)F_2^{n/A}$, of nuclear targets
in DIS (see e.g. Ref. \cite{Arneodo:1992wf} and references therein), and 
especially of the ratio
\begin{equation}
R_{F_2}^A(x,Q^2) \equiv
\frac{\frac{1}{A}{d\sigma^{lA}}/{dQ^2dx}}{\frac{1}{2}{d\sigma^{lD}}/{dQ^2dx}}
\approx \frac{\frac{1}{A}F_2^A(x,Q^2)}{\frac{1}{2}F_2^D(x,Q^2)},
\end{equation}
the following nuclear effects have been discovered in different Bjorken-$x$
regions:
\begin{itemize}
\item shadowing; a depletion, $R_{F_2}^A<1$,  at $x \lsim 0.1$,
\item antishadowing; an excess, $R_{F_2}^A>1$, at $0.1 \lsim x \lsim 0.3$,
\item EMC effect; a depletion at $0.3 \lsim x\lsim0.7$, and
\item Fermi motion; an excess towards $x\rightarrow1$ and beyond.
\end{itemize}
The $Q^2$ dependence of $R_{F_2}^A$ is weaker and has thus been more
difficult to measure. Data with high enough precision, however, exist.
The NMC collaboration discovered a clear $Q^2$ dependence in the ratio
$d\sigma^{\mu{\rm Sn}}/d\sigma^{\mu{\rm C}}$
\cite{Arneodo:1996ru}, i.e. the scale dependence of the ratio
$F_2^{\rm Sn}/F_2^{\rm C}$, at $x\gsim 0.01$. Since $F_2^{p(n)/A}=
\sum_{q} e_q^2x[f_q^{p(n)/A}+f_{\bar q}^{p(n)/A}]+ {\cal
O}(\alpha_s)$, the nuclear effects in the ratio $R_{F_2}^A$ directly
translate into nuclear effects in the parton distributions:
$f_i^{p/A}\ne f_i^{p}$.

\begin{figure}[htb]
\begin{center}
\vspace{-3cm}
\includegraphics[width=12cm]{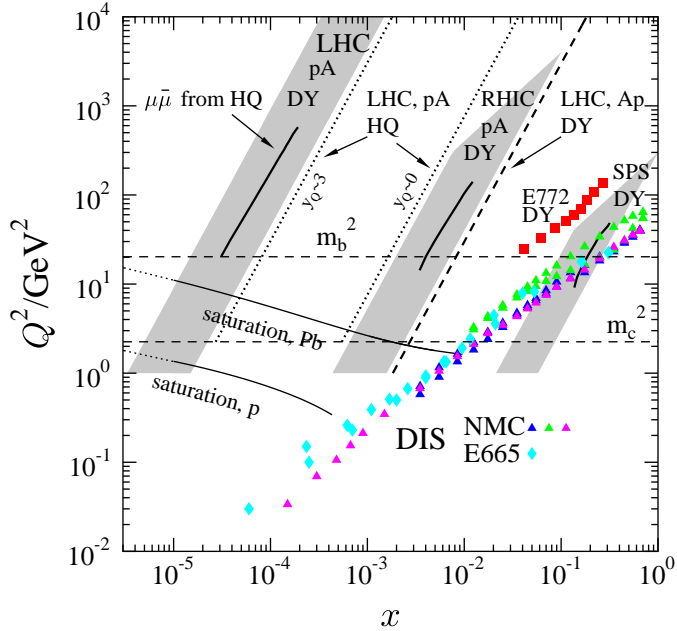} 
\vspace{-3.5cm}
\caption{The average values of $x$ and $Q^2$ of the DIS data from NMC
\cite{Amaudruz:1995tq,Arneodo:1995cs,Arneodo:1996rv} (triangles) and
E665 \cite{Adams:1992nf,Adams:1995is} (diamonds) in $lA$, and of $x_2$
and $M^2$ of the DY dilepton data \cite{Alde:im} (squares) in
$pA$. The heavy quark mass scales are shown by the horizontal dashed
lines.  The initial scale $Q_0^2$ is $m_c^2$ in EKRS and 1 GeV$^2$ in
HKM. For the rest of the figure, see the text in
Sec.~\ref{future}.  }
\label{fig:q2_vs_x}
\end{center}
\end{figure}

The nPDFs, $f_i^{p/A}$, also obey the DGLAP equations in the large-$Q^2$
limit. They can be determined by using a global DGLAP fit
procedure similar to the case of the free proton PDFs.  Pioneering studies
of the DGLAP evolution of the nPDFs are found in
e.g. Ref. \cite{Qiu:wh,Frankfurt:xz,Eskola:1992zb,Kumano:1994pn}. References
for various other studies of perturbative evolution of the nPDFs and
also to simpler $Q^2$-independent parametrizations of the nuclear
effects in the PDFs can be found e.g. in Refs.
\cite{Eskola:2002us,Eskola:2001ms}. The nuclear case is, however, more
complicated because of additional variables, the mass number $A$ and
the charge $Z$, and, because the number of data points available in
the perturbative region is more limited than for the PDFs of the free
proton.  The DIS data play the dominant role in the nuclear case as
well. However, as illustrated by Fig.~\ref{fig:q2_vs_x}, no data are
available from nuclear DIS experiments below $x\lsim 5\cdot 10^{-3}$
at $Q^2\gsim1$~GeV$^2$. This makes the determination of the nuclear
gluon distributions especially difficult. Further constraints on the
global DGLAP fits of the nPDFs can be obtained from e.g. the Drell-Yan
(DY) process measured in fixed-target $pA$ collisions
\cite{Alde:im,Vasilev:1999fa}. Currently, there are two sets of nPDFs
available which are based on the global DGLAP fits to the data: (i)
EKS98 \cite{Eskola:1998df,Eskola:1998iy} (the code in
Refs. \cite{EKS98_code,Plothow-Besch:1992qj}), and (ii) HKM
\cite{Hirai:2001np} (the code in Ref. \cite{HKM_code}). We shall
compare the main features of these two analyses and comment on their
differences below.

\subsubsection{Comparison of EKS98 and HKM}

\paragraph{EKS98 and overview of constraints available from data.}

The parametrization EKS98 \cite{Eskola:1998df} is based on the results
of the DGLAP analysis \cite{Eskola:1998iy} and its follow-up
\cite{Eskola:1998df}. We refer to these together as EKRS here.  In the
EKRS approach, the nPDFs, the parton distributions of the bound {\em
protons}, $f_i^{p/A}$ , are defined through the modifications of the
corresponding distributions in the free proton,
\begin{equation}
R_i^A(x,Q^2)\equiv {f_i^{p/A}(x,Q^2)\over f_i^p(x,Q^2)},
\label{eqratios}
\end{equation}
where the PDFs of the free proton are from a known set such as CTEQ,
MRS or GRV. As in the case of the free nucleons, for isoscalar nuclei
the parton distributions of bound neutrons are obtained through
isospin symmetry, $f_{u(\bar u)}^{n/A}=f_{d(\bar d)}^{p/A}$ and
$f_{d(\bar d)}^{n/A}=f_{u(\bar u)}^{p/A}$.  Although exact only for
isoscalar and mirror nuclei, this is expected to be a good first
approximation for all nuclei.

To simplify the determination of the input nuclear effects for valence
and sea quarks, the following flavour-independent {\em initial}
modifications are assumed: $R_{u_V}^A(x,Q_0^2)\approx
R_{d_V}^A(x,Q_0^2) \approx R_V^A(x,Q_0^2)$, and $R_{\bar
u}^A(x,Q_0^2)\approx R_{\bar d}^A(x,Q_0^2) \approx
R_{s}^A(x,Q_0^2)\approx R_S(x,Q_0^2)$.  Thus only three independent
initial ratios, $R_V^A$, $R_S^A$ and $R_G^A$ are to be determined at
$Q_0^2=m_c^2=2.25$~GeV$^2$. Note also that the approximations above
are needed and used {\em only} at $Q_0^2$ in \cite{Eskola:1998iy}. In
the EKS98-parametrization \cite{Eskola:1998df} of the DGLAP-evolved
results, it was observed that, to a good approximation, $R_{u_V}^A =
R_{d_V}^A$ and $R_{\bar u}^A=R_{\bar d}^A$ for all $Q^2$.  Further
details of the EKRS analysis can be found in
Ref. \cite{Eskola:1998iy}, here we summarize the constraints available
in each $x$-region. Consider first quarks and antiquarks:

\begin{itemize}
\item 
At { $x\gsim 0.3$} the DIS data constrains only the ratio $R_V^A$:
valence quarks dominate $F_2^A$, so that $R_{F_2}^A\approx R_V^A$ but
$R_S^A$ and $R_G^A$ are left practically unconstrained. An EMC effect
is, however, assumed also for the initial $R_S^A$ and $R_G^A$ since in
the scale evolution the EMC effect of valence quarks is transmitted to
gluons and, further on, to sea quarks \cite{Eskola:1992zb}. In this way the
nuclear modifications $R_i^A$ also remain stable against the
evolution. 

\item 
At {$0.04\lsim x \lsim 0.3$} both the DIS and DY data constrain
$R_S^A$ and $R_V^A$ but in different regions of $Q^2$, as shown in
Fig. \ref{fig:q2_vs_x}. In addition, conservation of baryon number
restricts $R_V^A$. The use of DY data \cite{Alde:im} is essential in
order to fix the relative magnitude of $R_V^A$ and $R_S^A$ since the
DIS data alone cannot distinguish between them. As a result, no
antishadowing appears in $R_S^A(x,Q_0^2)$.

\item 
At { $5\cdot10^{-3}\lsim x\lsim0.04$}, only DIS data exist in the
region $Q\gsim1$ GeV where the DGLAP analysis can be expected to
apply. Once $R_V^A$ is fixed by the DIS and DY data at larger $x$, the
magnitude of nuclear valence quark shadowing in the EKRS approach
follows from the conservation of baryon number. As a result of these
contraints, nuclear valence quarks in EKRS are less shadowed than the
sea quarks, $R_V^A>R_S^A$.

\item 
At $x\lsim 5\cdot10^{-3}$, as shown by Fig. \ref{fig:q2_vs_x}, the DIS
data for the ratio $R_{F_2}^A$ lie in the region $Q\lsim1$ GeV where
the DGLAP equations are unlikely to be applicable in the nuclear
case. Indirect constraints, however, {\em can} be obtained by
noting the following: (i) At the smallest values of $x$ (where $Q\ll1$~GeV),
$R_{F_2}^A$ depends only very weakly, if at all, on $x$
\cite{Arneodo:1995cs,Adams:1992nf}. (ii) $\partial (F_2^{\rm
Sn}/F_2^{\rm C})/\partial \log Q^2>0$ at $x\sim0.01$
\cite{Arneodo:1996ru}, indicating that $\partial R_{F_2^A}/\partial
\log Q^2>0$. (iii) Negative $\log Q^2$-slopes of $R_{F_2}^A$ at
$x\lsim 5\cdot10^{-3}$ have not been observed \cite{Arneodo:1995cs}.
Based on these experimental facts, the EKRS approach assumes
that the sign of the $\log Q^2$-slope of $R_{F_2}^A$ remains
non-negative at $x<0.01$ and therefore the DIS data
\cite{Arneodo:1995cs,Adams:1992nf} in the nonperturbative region
gives a {\em lower bound} on $R_{F_2}^A$ at $Q_0^2$ for very small
$x$.  The sea quarks dominate over the valence quarks at small values
of $x$ so that only $R_S^A$ is constrained by the DIS data. $R_V^A$
remains restricted by baryon number conservation.

\end{itemize}

\noindent 
Pinning down the nuclear gluon distributions is difficult in the
absence of stringent direct constraints. What is available in
different regions of $x$ can be summarized as follows:

\begin{itemize}

\item 
At $x\gsim 0.2$, no experimental constraints are currently available on
the gluons. Conservation of momentum is used as an overall constraint
but this alone is not sufficient to determine whether there is an EMC effect
for gluons. Only about 30 \% of the gluon momentum comes
from $x\gsim 0.2$ so that fairly sizable changes in $R_G^A(x,Q_0^2)$ in
this region can be compensated by smaller changes at $x<0.2$ without
violating the constraints discussed below.  As mentioned above, in the
EKRS approach an EMC effect is initially assumed for $R_G$. This
guarantees a stable scale evolution and  the EMC effect remains 
at larger $Q^2$.

\item 
At $0.02 \lsim x \lsim0.2$ the $Q^2$ dependence of the ratio $F_2^{\rm
Sn}/F_2^{\rm C}$ measured by NMC \cite{Arneodo:1996ru} currently sets
the most important constraint on $R_G^A$ \cite{Gousset:1996xt}. In the
small-$x$ region where gluons dominate the DGLAP evolution, the $Q^2$
dependence of $F_2(x,Q^2)$ is dictated by the gluon distribution as
$\partial F_2(x,Q^2)/\partial \log Q^2\sim \alpha_s xg(2x,Q^2)$ 
\cite{Prytz:1993vr}. As discussed in Refs. \cite{Eskola:1998iy,Eskola:2002us}, 
this leads to $\partial R_{F_2}^A(x,Q^2)/\partial
\log Q^2 \sim \alpha_s [R_G^A(2x,Q^2)-R_{F_2}^A(x,Q^2)]
xg(2x,Q^2)/F_2^D(x,Q^2)$.  The
$\log Q^2$ slopes of $F_2^{\rm Sn}/F_2^{\rm C}$ measured by NMC
\cite{Arneodo:1996ru} therefore constrain $R_G^A$. Especially, 
 the positive $\log Q^2$-slope of $F_2^{\rm Sn}/F_2^{\rm C}$ measured
by NMC indicates that $R_G^A(2x,Q_0^2)>R_{F_2}^A(x,Q_0^2)$ at
$x\sim0.01$ \cite{Eskola:1998iy,Eskola:2002us}. Thus, within the DGLAP
framework, much stronger gluon than antiquark shadowing at $x\sim 0.01$,
as suggested e.g. in Ref.~\cite{Li:2001xa}, is not supported by the
NMC data. The antishadowing in the EKRS gluons follows from the
constraint $R_G^A(0.03,Q_0^2)\approx 1$ imposed by the NMC data (see
also Ref.
\cite{Gousset:1996xt}) combined with the requirement of momentum
conservation. The EKRS antishadowing is consistent with the E789 data
\cite{Leitch:1994vc} on $D$-meson production in $pA$ collisions
(note, however, the large error bar on the data). It also seems
to be supported by $J/\psi$ production in DIS, measured by NMC
\cite{Amaudruz:1991sr}.

\item 
At $x\lsim0.02$, stringent experimental constraints do not yet exist for
the nuclear gluons. It should be emphasized, however,
that the initial $R_G^A$ in this region is directly connected to the
initial $R_{F_2}^A$. As discussed above, related to quarks at small
$x$, taking the DIS data on $R_{F_2}^A$ in the nonperturbative region
as a lower limit for $R_{F_2}^A$ at $Q_0^2$ corresponds to $\partial
R_{F_2^A}/\partial \log Q^2\ge0$ so that $R_G^A(x<0.02,Q_0^2)\ge
R_{F_2}^A(x/2,Q_0^2)\ge R_{F_2}^A(x/2,Q^2\ll1\,{\rm GeV}^2)$. Reference
\cite{Eskola:1998iy} observed that setting $R_G^A(x,Q_0^2)\approx 
R_{F_2}^A(x,Q_0^2)$ at $x\lsim 0.01$ fulfills this constraint. This 
approximation remains fairly good even after DGLAP evolution from 
$Q_0\sim 1$ GeV to $Q\sim100$ GeV, see Ref. \cite{Eskola:1998df}.

\end{itemize}

\bigskip

As explained in detail in Ref. \cite{Eskola:1998iy}, in the EKRS
approach the initial ratios $R_V^A(x,Q_0^2)$, $R_S^A(x,Q_0^2)$ and
$R_G^A(x,Q_0^2)$ are constructed piecewise in different regions of
$x$. Initial nPDFs are computed at $Q_0^2$. Leading order (LO) DGLAP
evolution to higher scales is performed and comparisons with the DIS and DY
data are made. The parameters in the input ratios are iteratively
changed until the best global fit to the data is achieved. The
determination of the input parameters in EKRS has so far been done
only manually with the best overall fit determined by eye. For the
quality of the fit, see the detailed comparison with the data
in Figs. 4-10 of Ref.
\cite{Eskola:1998iy}.  The parametrization EKS98 \cite{Eskola:1998df}
of the nuclear modifications $R_i^A(x,Q^2)$ was prepared on the basis
of the results from Ref. \cite{Eskola:1998iy}. It was also shown that when
the free proton PDFs were changed from GRVLO
\cite{Gluck:1991ng} to CTEQ4L \cite{Lai:1996mg} (differing from each
other considerably), the changes induced in $R_i^A(x,Q^2)$ were within
a few percent \cite{Eskola:1998df}.  Therefore, accepting this range
of uncertainty, the EKS98 parametrization can be used together with
any (LO) set of the free proton PDFs.

\paragraph{The HKM analysis.}

In principle, the definition of the nPDFs in the HKM analysis
\cite{Hirai:2001np} differs slightly from that in EKRS: instead of the
PDFs of the bound protons, HKM define the nPDFs as the {\em average}
distributions of each flavour $i$ in a nucleus $A$: $f_i^A(x,Q^2) =
(Z/A)f_i^{p/A}(x,Q^2) + (1-Z/A)f_i^{n/A}(x,Q^2)$. Correspondingly, the
HKM nuclear modifications at the initial scale $Q_0^2=1$~GeV$^2$ are
then defined through
\begin{equation}
f_i^A(x,Q_0^2) = 
w_i(x,A,Z)[(Z/A)f_i^{p}(x,Q_0^2) + (1-Z/A)f_i^{n}(x,Q_0^2)].
\end{equation}
In practice, however, since in the EKS98 parametrization 
$R_{u_V}^A = R_{d_V}^A$ and $R_{\bar u}^A=R_{\bar d}^A$, the EKS98 
modifications also represent average modifications. Flavour-symmetric
sea quark distributions are assumed in HKM whereas the flavour
asymmetry of the sea quarks in EKRS follows from that of the free
proton.

An improvement relative to EKRS is that the HKM method to extract the
initial modifications $w_i(x,A,Z)$ at $Q_0^2$ is more automatic and
more quantitative in the statistical analysis since the HKM analysis is
strictly a minimum-$\chi^2$ fit. Also, with certain assumptions of a
suitable form (see Ref. \cite{Hirai:2001np}) for the initial modifications
$w_i$, the number of parameters has been conveniently reduced to
seven. The form used in the fit is
\begin{equation}
w_i(x,A,Z)=1+\left(1-{1\over A^{1/3}}\right){a_i(A,Z)+H_i(x)\over
(1-x)^{\beta_i}}\quad,
\label{eqSalgado2}
\end{equation}
where $H_i(x)=b_ix+c_ix^2$. An analysis with a cubic polynomial is also
performed, with similar results. Due to the flavour symmetry, the sea
quark parameters are identical for all flavours. For valence quarks,
conservation of charge $Z$ (not used in EKRS) and baryon number $A$
are required, fixing $a_{u_V}$ and $a_{d_V}$. In the case of
non-isoscalar nuclei $w_{u_V}\ne w_{d_V}$. Also, momentum conservation
is imposed fixing $a_g$. Taking $\beta_V$, $b_V$ and $c_V$ to be equal
for $u_V$ and $d_V$, and $\beta_{\bar q}=\beta_g=1$, $b_g=-2c_g$,
the remaining seven parameters $b_V$, $c_V$, $\beta_V$, $a_{\bar q}$
$b_{\bar q}$, $c_{\bar q}$ and $c_g$ are determined by a global DGLAP fit
which minimizes $\chi^2$.

\paragraph{The comparison.}

Irrespective of whether the best fit is found automatically or by eye,
the basic procedure to determine the nPDFs in the EKRS and HKM analyses
is the same.  The results obtained for the nuclear effects are,
however, quite different, as can be seen in the comparison at $Q^2 =
2.25$~GeV$^2$ shown in Fig. \ref{fig:initial}. The main reason for
this is that different data sets are used:

\begin{figure}[htb]
\begin{center}
\vspace{0cm}
\includegraphics[width=13cm]{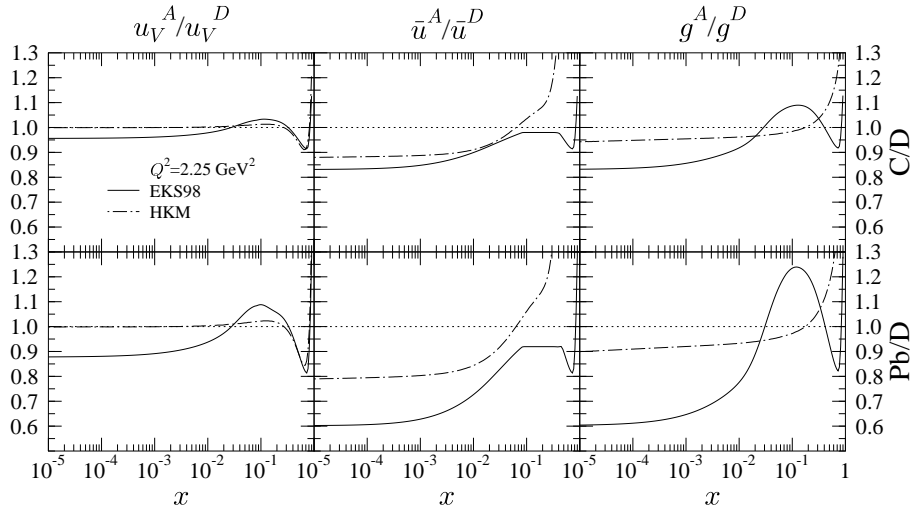} 
\vspace{0cm}
\caption{The nuclear modifications for the $u_V$, $\bar u$ and gluon
distributions in EKS98 (solid) and in HKM (dot-dashed) at $Q^2 =
2.25$~GeV$^2$. Upper panels: C/D. Lower panels: Pb/D with isoscalar
Pb.  }
\label{fig:initial}
\end{center}
\end{figure}

\begin{itemize}

\item The HKM analysis \cite{Hirai:2001np} uses only the DIS data
whereas EKRS include also the DY data from $pA$ collisions. As
explained above, the DY data are very important in the EKRS analysis to
fix the relative modifications of valence and sea quarks at
intermediate $x$. Preliminary results reported in Ref. \cite{Kumano:2002ii}
show that, when the DY data are included in the HKM analysis, the
antiquark modifications become more similar to those of EKRS.

\item The NMC data on $F_2^A/F_2^{\rm C}$ \cite{Arneodo:1996rv},
which imposes quite stringent constraints on the $A$-systematics in
the EKRS analysis, is not used in HKM. As a result, $R_{F_2}^A$ has
less shadowing in HKM than in EKRS (see also Fig. 1 of Ref.
\cite{Eskola:2002us}), especially for heavy nuclei.

\item In addition to the recent DIS data sets, some older ones are
used in the HKM analysis. The older sets are not included in EKRS.
This, however, presumably does not cause any major differences between
EKRS and HKM since the older data sets have larger error bars and will
therefore typically have less statistical weight in the $\chi^2$
analysis.

\item The HKM analysis does not make use of the NMC data
\cite{Arneodo:1996ru} on the $Q^2$ dependence of $F_2^{\rm
Sn}/F_2^{\rm C}$. As explained above, these data are the main
experimental constraint on the nuclear gluons in the EKRS
analysis. Figure~\ref{fig:snc_eks_hkm} shows a comparison of the EKRS
(solid) and the HKM (dot-dashed) results to the NMC data on
$F_2^{\rm Sn}/F_2^{\rm C}$ as a function of $Q^2$.  The HKM results do
not reproduce the measured $Q^2$ dependence of $F_2^{\rm Sn}/F_2^{\rm
C}$ at the smallest values of $x$.  This figure demonstrates
explicitly that the nuclear modifications of gluon distributions at
$0.02\lsim x\lsim 0.1$ {\em can} be pinned down with the help of these NMC
data.

\begin{figure}[htb]
\begin{center}
\vspace{0cm}
\includegraphics[width=12cm]{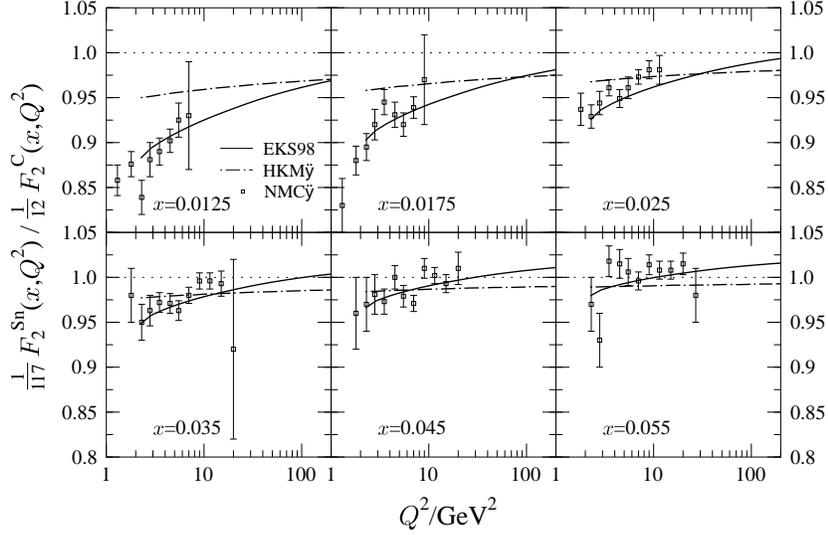} 
\vspace{-1cm}
\caption{ The ratio $F_2^{\rm Sn}/F_2^{\rm C}$ as a function of $Q^2$
at different fixed values of $x$. The data are from NMC
\cite{Arneodo:1996ru} and the curves are EKS98 \cite{Eskola:1998df}
(solid) and HKM \cite{Hirai:2001np} (dot-dashed).  }
\label{fig:snc_eks_hkm}
\end{center}
\end{figure}

\end{itemize}

\subsubsection{Future prospects}

\paragraph{Constraints from $pA$ collisions at LHC, RHIC and SPS.}
\label{future}
The hard probes in $pA$ collisions, especially those at the LHC, will play
a major role in probing the nPDFs at regions not accessible before. 
Figure~\ref{fig:q2_vs_x} shows the $x$ and $Q^2$ regions of the 
phase space
probed by certain hard processes at different center of mass
energies\footnote{No counting rates have been considered for the
figure.}.  Let us consider the figure in more detail:

\textbullet~The measurements of semileptonic decays of $D$ and $\overline
D$ mesons in $pA$ will help to pin down the nuclear gluon
distributions \cite{Lin:1995pk,Eskola:2001gt}.  Borrowed from a LO
analysis \cite{Eskola:2001gt}, the thick solid lines in
Fig.~\ref{fig:q2_vs_x} show how the average scale $Q^2 = \langle
m_T^2\rangle$ of open charm production is correlated with the average
fractional momentum $x=\langle x_2\rangle$ of the incoming nuclear
gluon in dimuon production in correlated $D\overline D$ decays at the LHC
(computed for $p$Pb at $\sqrt s = 5500$~GeV and single muon acceptance
$2.5\le y_{\mu} \le 4.0$, with no rapidity shifts), at RHIC ($p$Au at
$\sqrt s =200$~GeV, $1.15\le y_{\mu} \le 2.44$) and at the SPS ($p$Pb
at $\sqrt s = 17.3$~GeV, $0\le y_{\mu} \le 1$). For each solid curve,
the smallest $Q^2$ shown corresponds to a dimuon invariant mass
$M=1.25$ GeV, and the the largest $Q^2$ to $M=9.5$ GeV (LHC), 5.8 GeV
(RHIC) and 4.75 GeV (SPS) (cf. Fig. 5 in Ref.~\cite{Eskola:2001gt}). We
observe that data from the SPS, RHIC and the LHC will offer
constraints in different regions of $x$. The data from NA60 at the SPS
will probe the antishadowing of nuclear gluons, while the RHIC data probe 
the onset of gluon shadowing, and the LHC data constrain the
nuclear gluons at very small $x$, deep in the shadowing region.

\textbullet~ 
Similarly, provided that the DY dimuons can be distinguished from
those coming from the $D$ and $B$ decays, the DY dimuon $pA$ cross
sections are expected to set further constraints on the nuclear
effects on the sea quark distributions.  In $2\rightarrow 1$
kinematics, $x_2=(M/\sqrt s)e^{-y}$ where $M$ and $y$ are the
invariant mass and rapidity of the lepton pair.  The rapidity of the
pair is always between those of the leptonic constituents of the
pair. The shaded regions in Fig.~\ref{fig:q2_vs_x} illustrate the
regions in $x=x_2^A$ and $Q^2=M^2$ probed by DY dimuons in $pA$
collisions at the LHC, RHIC and SPS.  The parameters used for the LHC
are $\sqrt s=5500$~GeV and $2.5\le y\le 4.0$ with no rapidity shifts
\cite{MORSCH}. Note that from the (phase space) point of view of
$x_2^A$ (but {\em not} of $x_1^p$), Pb+Pb collisions at $\sqrt s_{\rm
PbPb}=5500$~GeV at the LHC are equivalent to $p$Pb collisions at
$\sqrt s_{p\rm Pb}=8800$~GeV: the decrease of $x_2$ due to the
increase in $\sqrt s$ is compensated by the rapidity
shift $y_0$ \cite{MORSCH}: $x_2^{pA} = (M/\sqrt s_{pA})
e^{-(y_{AA}+y_0)} = (M/\sqrt s_{AA})e^{-y_{AA}}=x_2^{AA}$.  For RHIC,
the shaded region corresponds to $\sqrt s = 200$~GeV and $1.2\le y\le
2.2$, the rapidity acceptance of the PHENIX forward muon arm
\cite{Nagle:2002ib}.  For the SPS, we have again used $\sqrt s=17.3$~GeV and
$0\le y\le 1$. Single muon $p_T$-cuts (see Refs.
\cite{BOTJE_proc,Nagle:2002ib}) have not been applied and the shaded
regions all correspond to $M^2\ge1$~GeV$^2$.  At the highest scales
shown, the dimuon regions probed at the SPS and RHIC are limited by
the phase space.  The SPS data may shed more light on the existence of
an EMC effect for the sea quarks \cite{Eskola:2000xv}.
At the LHC the dimuons from decays of $D$ and $B$ mesons are expected 
to dominate the dilepton continuum up to the $Z$ mass 
\cite{Alessandro:bv,Gavin:ma,Lokhtin:2001nh}. If, however, the DY 
contribution can be identified (see Sec.~\ref{section511} for more
discussion), DY dimuons at the LHC will probe the sea quark
distributions deep in the shadowing region and also at high scales,
up to $Q^2=M_Z^2$.  The RHIC region again lies conveniently between the
SPS and the LHC, mainly probing the beginning of sea quark shadowing
region.  In $Ap$ collisions at the LHC, DY dimuons with
$y=(y_{\min}+y_{\rm max})/2$ probe the regions indicated by the dashed
line (the spread due to the $y$ acceptance is not shown).

\textbullet~ The dotted line shows the kinematical region probed by
open heavy flavor production in $pA$ (and also in $AA$) collisions,
when both heavy quarks are at $y=0$ in the detector frame. In
this case, again assuming $2\rightarrow 2$ kinematics, $x_2^{pA} =
x_2^{AA} = 2Q/\sqrt s_{AA}$ with $Q=m_T$. To illustrate the effect of
moving towards forward rapidities, the case $y_Q=y_{\overline Q}=3$ is 
also shown. Only the region where $Q^2\gsim m_b^2$ (above the upper
horizontal line) is probed by $b\bar b$ production. Due to the same
kinematics (in $x_2$ and $Q^2$) as in open heavy quark production, the
dotted lines also correspond to the regions probed by direct photon
production (with $Q=p_T$).

\textbullet~ As demonstrated by Fig.~\ref{fig:q2_vs_x}, hard
probes in $pA$ collisions at RHIC and at the LHC in particular will
provide us with very important constraints on the nPDFs at scales
where the DGLAP evolution is expected to be applicable. Power
corrections to the evolution \cite{Gribov:tu,Mueller:wy} can, however,
be expected to play an increasingly important role towards small
values of $x$ and $Q^2$. The effects of the first of such corrections,
the GLRMQ terms \cite{Gribov:tu,Mueller:wy} leading to nonlinear
DGLAP+GLRMQ evolution equations, have been  studied recently in light
of the HERA data \cite{Eskola:2002yc}.  From the point of view of the
DGLAP evolution of the (n)PDFs, gluon saturation occurs when their
evolution becomes dominated by power corrections \cite{Qiu_proc}.
Figure~\ref{fig:q2_vs_x} shows the saturation limits obtained for the
free proton and for Pb in the DGLAP+GLRMQ approach
\cite{Eskola:2003gc} (solid curves, the dotted curves are
extrapolations to guide the eye).  The saturation limit obtained for
the free proton in the DGLAP+GLRMQ analysis should be taken as an
upper limit in $Q^2$. It is constrained quite well by the HERA data
(see Ref. \cite{Eskola:2002yc}).  The constraints from the $Q^2$
dependence of $F_2^{\rm Sn}/F_2^{\rm C}$, however, were not taken into
account in obtaining the saturation limit shown for the Pb nucleus
\cite{Eskola:2003gc}. For a comparison of the saturation limits
obtained in other models, see Ref. \cite{Eskola:2003gc}.

\textbullet~ Figure~\ref{fig:q2_vs_x} also shows which hard probes of
$pA$ collisions can be expected to probe the nPDFs in the gluon
saturation region. The probes directly sensitive to the gluon
distributions are especially interesting from this point of view. Such
probes include open $c\bar c$ production at small $p_T$ and direct
photon production at $p_T\sim $ few GeV, both as far forward in
rapidity as possible (see the dotted lines). Note that open $b\bar b$
production is already in the region where linear DGLAP evolution is
applicable. In light of the saturation limits shown, the chances of
measuring the effects of nonlinearities in the evolution through open
$c\bar c$ production in $pA$ at RHIC would seem marginal.  At the LHC,
however, measuring saturation effects in the nuclear gluon
distributions through open $c\bar c$ in $pA$ could be possible.

\paragraph{Improvements of the DGLAP analyses.}
On the practical side, the global DGLAP fits of the nPDFs discussed
above can be improved in obvious ways. The EKRS analysis should be
made more automatic and a proper statistical treatment, such as in
HKM, should be added.  The automatization alone is, however, not
expected to change the nuclear modifications of the PDFs significantly
from EKS98 but more quantitative estimates of the uncertainties and of
their propagation would be obtained.  This work is in progress. As
also discussed above, more contraints from data should be added to the
HKM analysis.  More generally, all presently available data from hard
processes in DIS and $pA$ collisions have not yet been exhausted: for
instance, the recent CCFR DIS data for $F_2^{\nu{\rm Fe}}$ and
$F_3^{\nu{\rm Fe}}$ from $\nu$Fe and $\bar \nu$Fe collisions
\cite{Seligman:mc} (not used in EKRS or in HKM), could help to pin
down the valence quark modifications \cite{Botje:1999dj}. In the
future, hard probes of $pA$ collisions at the LHC, RHIC and the SPS
will offer very important constraints on the nPDFs, especially for
gluons and sea quarks. Eventually, the nPDF DGLAP analyses should be
extended to NLO perturbative QCD. As discussed above, the effects of
power corrections \cite{Gribov:tu,Mueller:wy} to the DGLAP equations
and also to the cross sections \cite{Guo:2001tz} should be analysed in
detail in the context of global fits to the nuclear data.


\subsection{Nonlinear corrections to the DGLAP equations; looking for the
saturation limits}
\label{subsec:kolhinen}
{\em Kari J. Eskola, Heli Honkanen, Vesa J. Kolhinen, Jianwei Qiu, 
and Carlos A. Salgado}











%
%


Free proton PDFs, $f_i(x,Q^2)$, are needed for the calculation of hard 
process cross sections in hadronic collisions. Once they are
determined at a certain initial scale $Q_0^2$, the DGLAP equations
\cite{Dokshitzer:sg} well describe their evolution to large scales. 
Based on global fits to the available data, several different sets
of PDFs have been obtained
\cite{Martin:2001es,Martin:2002dr,Lai:1999wy,Pumplin:2002vw}.  The
older PDFs do not adequately describe the recent HERA data
\cite{Adloff:2000qk} on the structure function $F_2$ at the
perturbative scales $Q^2$ at small $x$.  In the analysis of newer PDF
sets, such as CTEQ6 \cite{Pumplin:2002vw} and MRST2001
\cite{Martin:2001es}, these data have been taken into
account. However, difficulties arise when fitting both small and large
scale data simultaneously. In the new MRST set, the entire H1 data set
\cite{Adloff:2000qk} has been used in the analysis, leading to a good
average fit at all scales but at the expense of a negative NLO gluon
distribution at small $x$ and $Q^2\lsim 1$ GeV$^2$.  In the CTEQ6 set
only the large scale ($Q^2>4$ GeV$^2$) data have been included, giving
a good fit at large $Q^2$ but resulting in a poorer fit at small-$x$
and small $Q^2$ ($Q^2<4$ GeV$^2$). Moreover, the gluon distribution at
small $x$ with $Q^2 \lsim 1.69$ GeV$^2$ has been set to zero.

These problems are interesting since they could indicate a new QCD
phenomenon. At small values of momentum fraction $x$ and scales $Q^2$,
gluon recombination terms, which lead to nonlinear corrections to the
evolution equations, can become significant.  The first of these nonlinear
terms have been calculated by Gribov, Levin and Ryskin
\cite{Gribov:tu}, and Mueller and Qiu \cite{Mueller:wy}.  In the
following, these corrections shall be referred to as the GLRMQ terms
for short. With these modifications, the evolution equations become
\cite{Mueller:wy}
\begin{eqnarray}
\frac{\partial xg(x,Q^2) }{\partial \log Q^2} 
  &=&   \frac{\partial xg(x,Q^2) }{\partial \log Q^2}\bigg|_{\rm DGLAP} 
   - \quad \frac{9\pi}{2} \frac{\alpha_s^2}{Q^2} 
    \int_x^1 \frac{dy}{y} y^2 G^{(2)}(y,Q^2), \label{gl-evol} \\
\frac{\partial x\bar{q}(x,Q^2)}{\partial \log Q^2}   & = & 
\frac{\partial x\bar{q}(x,Q^2)}{\partial \log Q^2}\bigg|_{\rm DGLAP}
  -  \quad \frac{3\pi}{20}\frac{\alpha_s^2}{Q^2} 
     x^2 G^{(2)}(x,Q^2)
   + \ldots G_{\rm HT}, \label{sea-evol}
\end{eqnarray} 
where the two-gluon density can be modelled as $ x^2G^{(2)}(x,Q^2)=
\frac{1}{\pi R^2}[xg(x,Q^2)]^2, $ with proton radius $R=1$~fm. The 
higher dimensional gluon term, $G_{\rm HT}$ \cite{Mueller:wy}, is here
assumed to be zero.  The effects of the nonlinear corrections on the
DGLAP evolution of the free proton PDFs were studied in Ref.
\cite{Eskola:2002yc} in view of the recent H1 data. The results are
discussed below.

\subsubsection{Constraints from the HERA data}

\begin{figure}[tbh]
\begin{center}
\includegraphics[width=8cm]{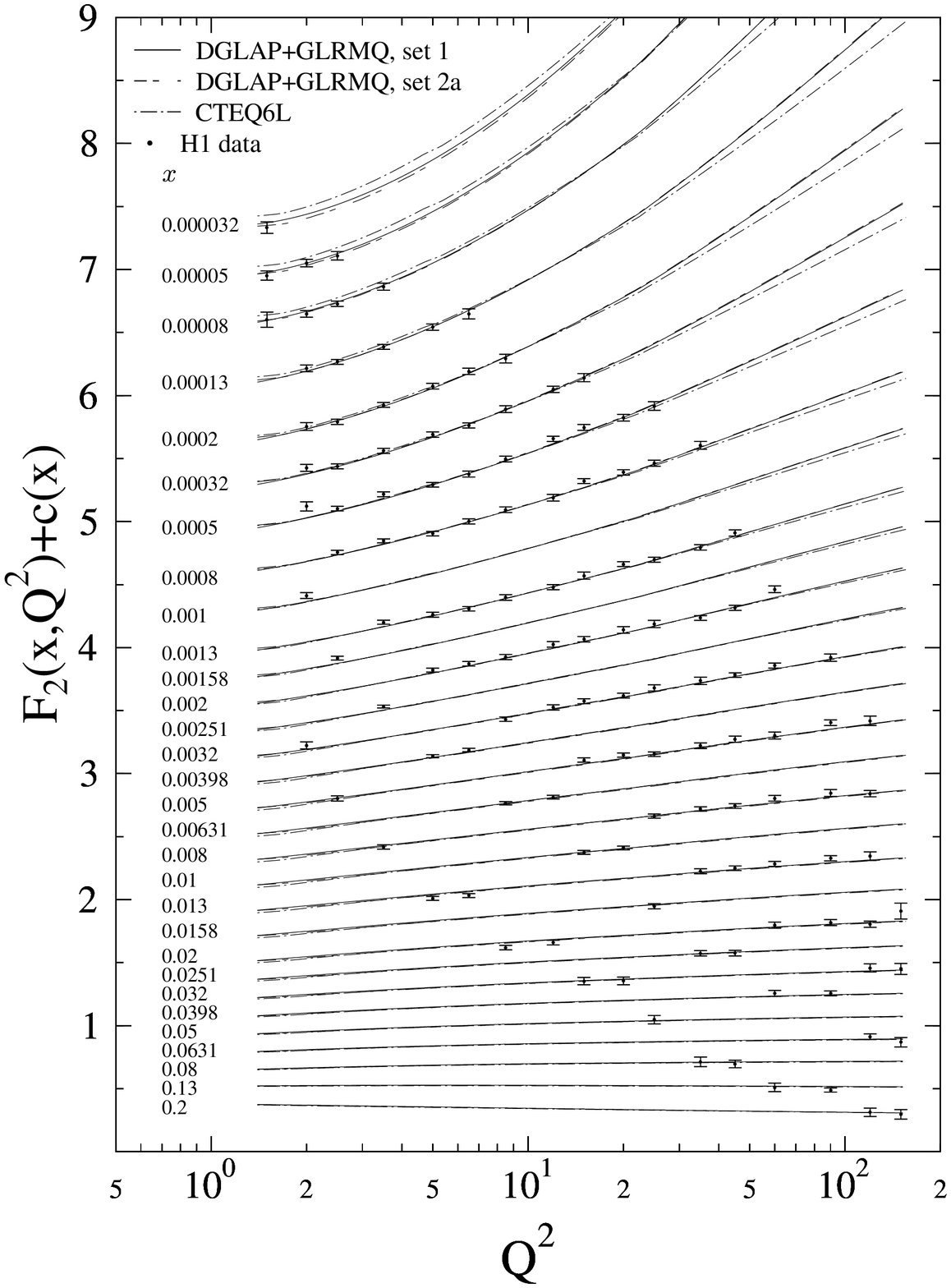}
\includegraphics[width=6.9cm]{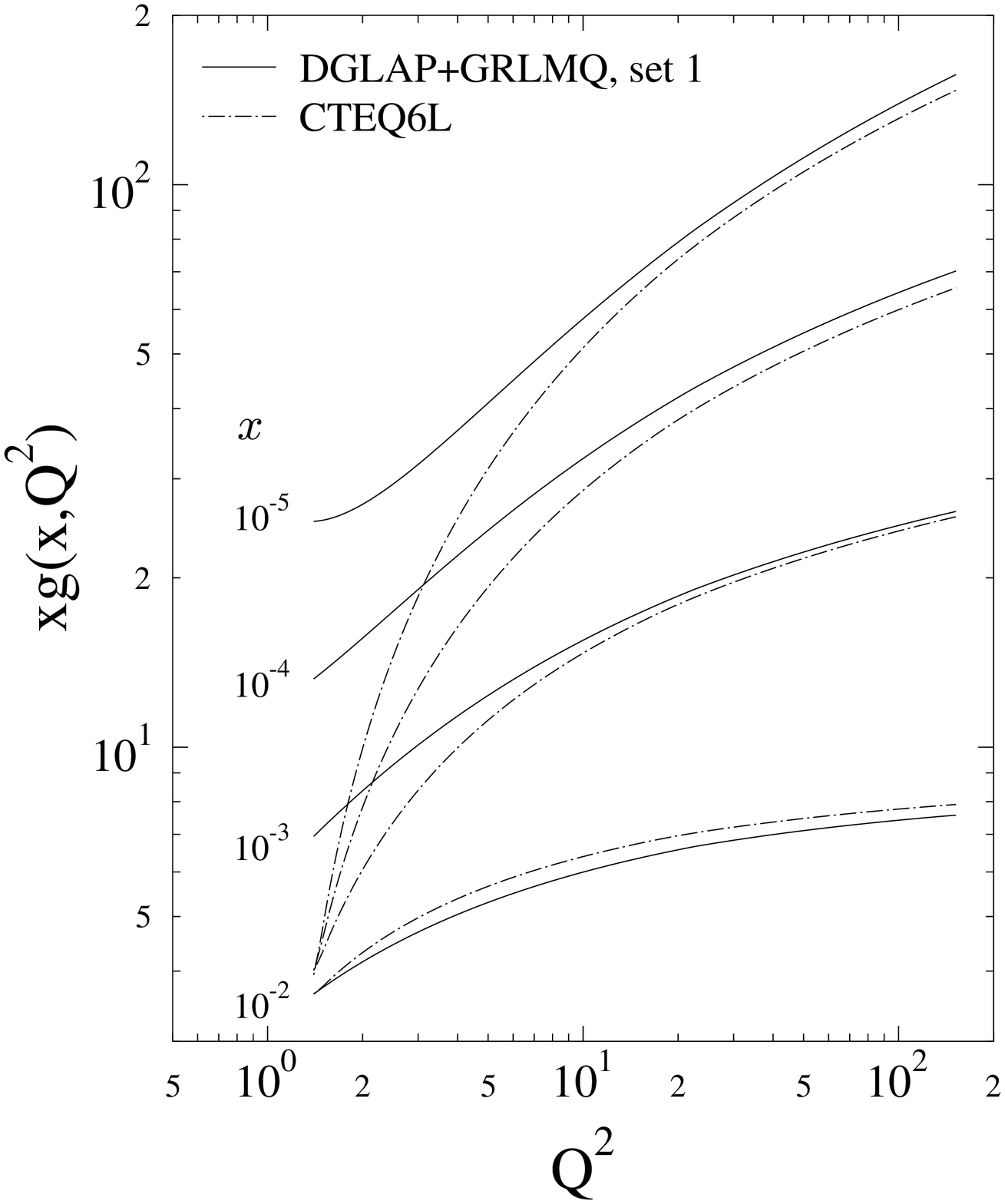} 
\caption[a]{{\small {\bf Left:} $F_2(x,Q^2)$ calculated using
CTEQ6L \cite{Pumplin:2002vw} (dotted-dashed) and the DGLAP+GLRMQ
results with set 1 (solid) and set 2a (double dashed)
\cite{Eskola:2002yc}, compared with the H1 data
\cite{Adloff:2000qk}. {\bf Right:} The $Q^2$ dependence of the gluon
distribution function at fixed $x$, from set 1 evolved with
DGLAP+GLRMQ (solid) and directly from CTEQ6L (dotted-dashed).} }
\label{F2_vs_cteq6}
\end{center}
\end{figure}

The goal of the analysis in Ref. \cite{Eskola:2002yc} was (1) to possibly
improve the (LO) fit of the calculated $F_2(x,Q^2)$ to the H1 data
\cite{Adloff:2000qk} at small $Q^2$ while (2) concurrently
maintaining the good fit at large $Q^2$ and, finally, (3) to study the
interdependence of the initial distributions and the evolution.

In CTEQ6L, a good fit to the H1 data is obtained (see
Fig.~\ref{F2_vs_cteq6}) with a flat small-$x$ gluon distribution at
$Q^2\sim 1.4$~GeV$^2$.  As can be seen from
Eqs.~(\ref{gl-evol})-(\ref{sea-evol}), the GLRMQ corrections slow the
scale evolution. Now one may ask whether the H1 data can be reproduced
equally well with different initial conditions (i.e. assuming larger
initial gluon distributions) and the GLRMQ corrections included in the
evolution.  This question was addressed in Ref. \cite{Eskola:2002yc} by
generating three new sets of initial distributions using DGLAP + GLRMQ
evolved CTEQ5 \cite{Lai:1999wy} and CTEQ6 distributions as guidelines.
The initial scale was chosen to be $Q_0^2=1.4$ GeV$^2$, slightly below
the smallest scale of the data.  The modified distributions at
$Q_0^2$ were constructed piecewise from CTEQ5L and CTEQ6L
distributions evolved down from $Q^2$ = 3 and 10 GeV$^2$ (CTEQ5L) and
$Q^2$ = 5 GeV$^2$ (CTEQ6L). A power law form was used in the small-$x$
region to tune the initial distributions until good agreement with
the H1 data was found.

The difference between the three sets in Ref. \cite{Eskola:2002yc} is
that in set 1 there is still a nonzero charm distribution at
$Q_0^2=1.4$ GeV$^2$, slightly below the charm mass threshold
with $m_c=1.3$ GeV in CTEQ6.  In set 2a and set 2b the charm
distribution has been removed at the initial scale and the resulting
deficit in $F_2$ has been compensated by slightly increasing the other
sea quarks at small $x$. Moreover, the effect of charm was studied
by using different mass thresholds: $m_{\rm c}=1.3$ GeV in set 2a and
$m_{\rm c}=\sqrt{1.4}$ GeV in set 2b, i.e. charm begins to evolve
immediately above the initial scale.

The results of the DGLAP+GLRMQ evolution with the new initial
distributions are shown in Fig.~\ref{F2_vs_cteq6}. The left panel
shows the comparison between the H1 data and the (LO) structure
function $F_2(x,Q^2)=\sum_i e_i^2 x[q_i(x,Q^2)+\bar q_i(x,Q^2)]$
calculated from set 1 (solid lines), set 2a (double dashed) and the
CTEQ6L parametrization (dotted-dashed lines). As can be seen, the
results are very similar, showing that with modified initial
conditions and DGLAP+GLRMQ evolution, one obtains an as good or even a
better fit to the HERA data ($\chi^2/N = 1.13$, 1.17, 0.88 for the sets
1, 2a, 2b, respectively) as with the CTEQ6L distributions ($\chi^2/N
= 1.32$).

The evolution of the gluon distribution functions in the DGLAP+GLRMQ
and DGLAP cases is illustrated more explicitly in the right panel of
Fig. \ref{F2_vs_cteq6}, where the absolute distributions for fixed $x$
are plotted as a function of $Q^2$ for set~1 and for CTEQ6L. The
figure shows how large differences at the initial scale vanish during
the evolution due to the GLRMQ effects.  At scales $Q^2
\gsim 4$~GeV$^2$, the GLRMQ corrections fade out rapidly and the DGLAP
terms dominate the evolution.

\subsubsection{Saturation limits: $p$ and Pb}

The DGLAP+GLRMQ approach also offers a way to study the gluon
saturation limits. For each $x$ in the small-$x$ region, the
saturation scale $Q_{\rm sat}^2$ can be defined as the value of the
scale $Q^2$ where the DGLAP and GLRMQ terms in the nonlinear evolution
equation become equal, $\frac{\partial xg(x,Q^2)}{\partial \log
Q^2}|_{Q^2=Q_{\rm sat}^2(x)}=0$.  The region of applicability of the
DGLAP+GLRMQ is $Q^2>Q^2_{\rm sat}(x)$ where the linear DGLAP part
dominates the evolution.  In the saturation region, at $Q^2<Q^2_{\rm
sat}(x)$, the GLRMQ terms dominate, and all nonlinear terms become
important.

\begin{figure}[hbt]
\vspace{-1,0cm}
\begin{center}
\includegraphics[width=6.5cm]{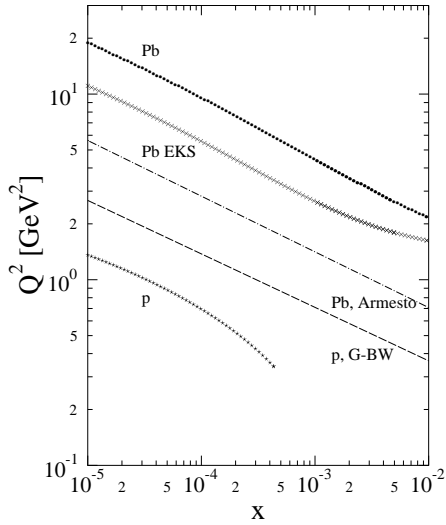} 
\vspace{-0.5cm}
\caption{{\small The gluon saturation limits in the DGLAP+GLRMQ approach 
for proton (asterisks) and Pb ($A=208$), with (crosses) and 
without (dots) nuclear modifications \cite{Eskola:2002yc}.  The
saturation line for the proton from the geometric saturation model 
\cite{Golec-Biernat:1998js} (dashed line), and for Pb from \cite{Armesto:2002ny} (dotted-dashed) are also plotted.
 }}
\label{satur}
\end{center}
\end{figure}

In order to find the saturation scale $Q^2_{\rm sat}(x)$, for the free
proton, the initial distributions (set 1) at $Q_0^2=1.4$~GeV$^2$ have
to be evolved downwards in scale using the DGLAP+GLRMQ equations.  As
discussed in Ref. \cite{Eskola:2002yc}, since only the first
correction term has been taken into account, the gluon distribution
near the saturation region should be considered as an upper
limit. Consequently, the saturation scale obtained is an upper limit
as well. The result is shown by the asterisks in Fig. \ref{satur}.  The
saturation line for the free proton from the geometric saturation
model by Golec-Biernat and W\"usthoff (G-BW)
\cite{Golec-Biernat:1998js} is also plotted (dashed line) for
comparison. It is interesting to note that although the DGLAP+GLRMQ
and G-BW approaches are very different, the slopes are very similar at
the smallest values of $x$.

Saturation scales for nuclei can be determined similarly.  For a
nucleus $A$, the two-gluon density can be modelled as
$x^2G^{(2)}(x,Q^2)= \frac{A}{\pi R_A^2}[xg_A(x,Q^2)]^2$, i.e. the
effect of the correction is enhanced by a factor of $A^{1/3}$. An
initial estimate of the saturation limit can be obtained by starting
the downwards evolution at high enough scales, $Q^2=100 \ldots 200$
GeV$^2$, for the GLRMQ terms to be negligible. 
Like in the proton case, the result is an upper limit. It is shown for Pb in
Fig.~\ref{satur} (dots).  The effect of the nuclear modifications was
also studied by applying the EKS98 \cite{Eskola:1998df}
parametrization at the high starting scale.  As a result, the
saturation scales $Q_{\rm sat}^2(x)$ are somewhat reduced, as shown in
Fig.~\ref{satur} (crosses).  The saturation limit obtained for a Pb
nucleus by Armesto in a Glauberized geometric saturation model
\cite{Armesto:2002ny} is shown (dotted-dashed) for comparison. Again,
despite the differences between the approaches, the slopes are
strikingly similar.

For further studies and for more accurate estimates of $Q_{\rm
sat}^2(x)$ in the DGLAP+GLRMQ approach, a full global analysis of
the nuclear parton distribution functions should be performed, along
the same lines as in EKRS \cite{Eskola:1998iy,Eskola:1998df} and 
HKM \cite{Hirai:2001np}.


\subsection{Gluon distributions in nuclei at small $x$: 
guidance from different models}
\label{subsec:armesto_salgado}
{\em N\'estor Armesto and Carlos A. Salgado}

The difference between the structure functions measured in nucleons
and nuclei \cite{Arneodo:1992wf} is a very important and well known
feature of nuclear structure and nuclear
collisions. At small values of $x$ ($< 0.01$,
shadowing region), the structure function $F_2$ per nucleon turns out
to be smaller in nuclei than in a free nucleon. This
corresponds to shadowing of parton densities in nuclei.  While at
small $x$ valence quarks are of little importance and the behaviour of
the sea is expected to follow that of $F_2^A$, the gluon distribution,
which is not an observable quantity, is badly determined and
represents one of the largest uncertainties for cross
sections both at moderate and large $Q^2$ in collinear
factorization \cite{Collins:gx}. For example, the
uncertainty in the determination of the gluon distributions
for Pb at $Q^2\sim 5$ GeV$^2$ at
LHC for $y=0$, $x\sim m_T/\sqrt{s}\sim 10^{-4}\dots 10^{-3}$, is a factor
$\sim 3$ (see Fig.~\ref{compglue}),
which may result in  an uncertainty of up to a factor of $\sim 9$
for the corresponding cross sections in PbPb collisions.

In this situation, while waiting for new data
from lepton-ion \cite{Arneodo:1996qa,Abramowicz:2001qt,eacoll} or $pA$
colliders, guidance from different theoretical models is of
utmost importance for controlled extrapolations from the region
where experimental data exist to those interesting for the LHC.
Two different approaches to the problem have been suggested: On one
hand, some models try to explain the origin of shadowing,
usually in terms of multiple scattering (in the frame where the
nucleus is at rest) or parton interactions in the nuclear wave
function (in the frame in which the nucleus is moving fast). On the
other hand, other models parameterize parton densities inside the
nucleus at some scale $Q_0^2$ large enough for perturbative QCD to be
applied reliably and then evolve these parton densities using the
DGLAP \cite{Dokshitzer:sg} evolution
equations. Then, the origin of the differences between parton
densities in nucleons and nuclei is not addressed, but
contained in the parameterization at $Q_0^2$, obtained from a
fit to data.

\subsubsection{Multiple scattering and saturation models}

The nature of shadowing is well understood qualitatively. In the rest
frame of the nucleus, the incoming photon splits, at small enough $x$,
into a $q\bar q$ pair long before reaching the nucleus, with a
coherence length $l_c \propto 1/(m_Nx)$ with $m_N$ the nucleon mass.
At small enough $x$, $l_c$ becomes on the order of or greater than the
nuclear size.  Thus the $q\bar q$ pair interacts coherently with the
nucleus with a typical hadronic cross section, resulting in
absorption
\cite{Brodsky:1989qz,Barone:ej,Kopeliovich:1995yr,Armesto:1996id,Nikolaev:1990ja,Frankfurt:nt,Frankfurt:1988zg,Kopeliovich:2000ra}.
(See Ref. \cite{Armesto:2000zh} for a simple geometrical approach in this
framework.)  In this way nuclear shadowing is a consequence of
multiple scattering and is thus related to diffraction (see e.g. Refs.
\cite{Armesto:2003fi,Frankfurt:2002kd}).

Multiple scattering is usually formulated in the dipole model
\cite{Nikolaev:1990ja,Mueller:1994jq}, equivalent to $k_T$-factorization
\cite{Catani:1990eg} at
leading order. In this framework, the $\gamma^*$-nucleus cross section is
expressed as the convolution of the probability of the transverse
or longitudinal $\gamma^*$ to split into a $q\bar q$ pair of transverse
dimension $r$ times the cross section for dipole-nucleus scattering.
It is this dipole-nucleus cross section at fixed impact parameter which
saturates, i.e. reaches the maximum value allowed by unitarity.
Most often this multiple scattering problem is modelled through 
the Glauber-Gribov approach
\cite{Armesto:2002ny,Armesto:2001vm,Huang:1997ii,Frankfurt:2002kd}. 
The dipole-target cross section is related
through a Bessel-Fourier transform to the so-called unintegrated gluon
distribution $\varphi_A(x,k_T)$ in $k_T$-factorization
\cite{Andersson:2002cf,Armesto:2002ny,Armesto:2001vm}, which in turn
can be related to the usual collinear gluon density by
\begin{equation}
xG_A(x,Q^2)=\int_{\Lambda^2}^{Q^2} dk_T^2\  \varphi_A(x,k_T)
\label{gluglu}
\end{equation}
where $\Lambda^2$ is some infrared cut-off, if required.
This identification is only true
for $Q^2\gg Q_{\rm sat}^2$ \cite{Kovchegov:1998bi,Mueller:1999wm},
where $Q_{\rm sat}^2$
corresponds to the transverse momentum scale at which the saturation of the
dipole-target cross section occurs.

Other formulations of multiple scattering do not use the dipole
formalism but relate shadowing to
diffraction by Gribov
theory. See recent applications in 
\cite{Armesto:2003fi,Frankfurt:2002kd} and the discussion 
in Sec.~\ref{subsec:frankfurt_etal}.

An  explanation equivalent to multiple scattering, in the frame in which
the nucleus is moving fast, is gluon recombination due to the
overlap of the gluon clouds from different nucleons. This makes the gluon
density in a nucleus with mass number
$A$ smaller than $A$ times that of a free nucleon \cite{Gribov:tu,Mueller:wy}. At small
$x$, the interaction develops over longitudinal distances $z\sim 1/(m_Nx)$
which become of the order of or larger than the nuclear size, leading to
the overlap of gluon clouds from different nucleons located in a
transverse area $\sim 1/Q^2$.
Much recent work in this field stems from the development of the semiclassical
approximations
in QCD and the appearance of non-linear evolution equations
in $x$
(see Refs.
\cite{Mueller:2002kw,Venugopalan:1999wu}, the contributions by Mueller,
Iancu {\it et al.}, Venugopalan, Kaidalov and Kharzeev in
\cite{cargese}, and references therein),
although saturation appears to be
different from shadowing \cite{Kovchegov:1998bi,Mueller:1999wm} (i.e. the reduction in the
number of gluons as defined in DIS is not apparent). In
the semiclassical framework, the gluon field at saturation reaches a
maximum value and becomes proportional to the inverse QCD coupling
constant, gluon correlations are absent and the dipole-target
cross section has a form which leads to geometrical scaling of the
$\gamma^*$-target
cross section. Such scaling has been found in small $x$ nucleon data
\cite{Stasto:2000er} and also in analytical and numerical solutions of the
non-linear equations
\cite{Lublinsky:2001bc,Golec-Biernat:2001if,Armesto:2001fa,Iancu:2002tr}
but the data indicate that
apparently the region where this scaling should be seen in
lepton-nucleus collisions has not yet been reached
\cite{Freund:2002ux}.  Apart from proposed $eA$ colliders
\cite{Arneodo:1996qa,Abramowicz:2001qt,eacoll}, the LHC will be the place
to look for non-linear effects.  However, the consequences of such
high density configurations may be masked in $AB$ collisions by other
effects, such as final-state interactions. Thus $pA$ collisions would
be essential for these studies, see
e.g. Ref. \cite{Dumitru:2002qt}. Moreover, they would be required to fix
the baseline for other studies in $AB$ collisions such as the search for and
characterization of the QGP.
As a final comment, we note that the non-linear terms in Refs.
\cite{Gribov:tu,Mueller:wy} are of a higher-twist nature. Then, in
the low density limit, the DGLAP equations
\cite{Dokshitzer:sg} are recovered.
However, the non-linear evolution equations
in $x$
\cite{Mueller:2002kw,Venugopalan:1999wu,cargese} do not correspond to
any definite twist. Their linear limit corresponds to the BFKL equation
\cite{Fadin:cb,Balitsky:ic}.

We now comment further on the difference between gluon shadowing and
saturation. In saturation models,
see Refs. \cite{Mueller:2002kw,Venugopalan:1999wu,cargese},
saturation, defined as either
a maximum value of the gluon field or by the scattering
becoming black,
and gluon shadowing, defined as $xG_A/(xG_N)<1$,
are apparently different phenomena, i.e. saturation does not necessarily
lead to gluon shadowing \cite{Kovchegov:1998bi,Mueller:1999wm}.
Indeed, in
numerical studies of the
non-linear small $x$ evolution
equations, the unintegrated
gluon distribution turns out to be a universal function of just one variable
$\tau=k_T^2/Q_{\rm sat}^2$
\cite{Lublinsky:2001bc,Golec-Biernat:2001if,Armesto:2001fa}, vanishing quickly
for $k^2_T>Q_{\rm sat}^2$. (In Ref.
\cite{Iancu:2002tr} this universality is analytically shown to be
fulfilled up to $k_T^2$ much larger than $Q_{\rm sat}^2$).
This scaling also arises in the analysis of DIS
data on nucleons \cite{Stasto:2000er}. It has also been
searched for in nuclear data \cite{Freund:2002ux}.  In nuclei, $Q_{\rm sat}^2$
increases with increasing nuclear size, centrality and energy. Thus,
through the relation to the collinear gluon density given in
Eq. (\ref{gluglu}), this scaling implies that the integral gives the
same value (up to logarithmic corrections if a perturbative tail
$\propto 1/k_T^2$ exists for $k^2_T\gg Q_{\rm sat}^2$), provided that
the upper limit of the integration domain $Q^2\gg Q_{\rm sat}^2$. Thus the
ratio $xG_A/(xG_N)$ remains equal to 1 (or approaching 1 as a ratio of
logs if the perturbative tail exists) for $Q^2\gg Q_{\rm sat}^2$. So
saturation or non-linear evolution do not automatically guarantee the
existence of gluon shadowing.  Since shadowing is
described in terms of multiple scattering in many models, this result
\cite{Kovchegov:1998bi,Mueller:1999wm} seems somewhat surprising.

\subsubsection{DGLAP evolution models}

A different approach is taken in Refs.
\cite{Eskola:1998iy,Eskola:1998df,Hirai:2001np,Indumathi:1996pb}. Parton
densities inside the nucleus are parameterized at some scale
$Q_0^2\sim 1\dots 5$ GeV$^2$ and then evolved using the DGLAP 
\cite{Dokshitzer:sg}
equations.
Now all nuclear effects on the parton densities are included in the
parameterization at $Q_0^2$, obtained from a fit to experimental data.
The differences between the approaches mainly come from the data used
to constrain the parton distributions (e.g. including or excluding
Drell-Yan data). In these calculations lack of data leaves the gluon
badly constrained at very small $x$. Data on the $Q^2$-evolution of
$F_2^A$ give direct constraints \cite{Eskola:2002us} through DGLAP
evolution but only for $x\geq 0.01$.
See Sec.~\ref{subsec:eskola} for more discussion of this framework. 

\subsubsection{Comparison between different approaches}
Model predictions usually depend on additional
semi-phenomenological assumptions. Since these assumptions are typically
different, the predictions often contradict each other. For
example, in Refs.
\cite{Brodsky:1989qz,Barone:ej,Kopeliovich:1995yr,Armesto:1996id,Nikolaev:1990ja}
it is argued that large size $q\bar q$ configurations give
the dominant contribution to the absorption, 
which does not correspond to any definite twist but to an admixture 
of all twists, resulting in essentially $Q^2$-independent shadowing.
On the other hand, in the gluon recombination approach
\cite{Gribov:tu,Mueller:wy} the absorption is clearly a
higher-twist effect dying out at large $Q^2$.  Finally, in the DGLAP-based
models
\cite{Eskola:1998iy,Eskola:1998df,Hirai:2001np,Indumathi:1996pb},
all $Q^2$-dependence comes from QCD evolution and is thus
of a logarithmic, leading-twist nature.
Thus these approaches all lead to very different predictions at small $x$,
particularly for the gluon density.

\begin{figure}[bht]
\begin{center}
\vspace{-0.5cm}
\includegraphics[width=9.5cm]{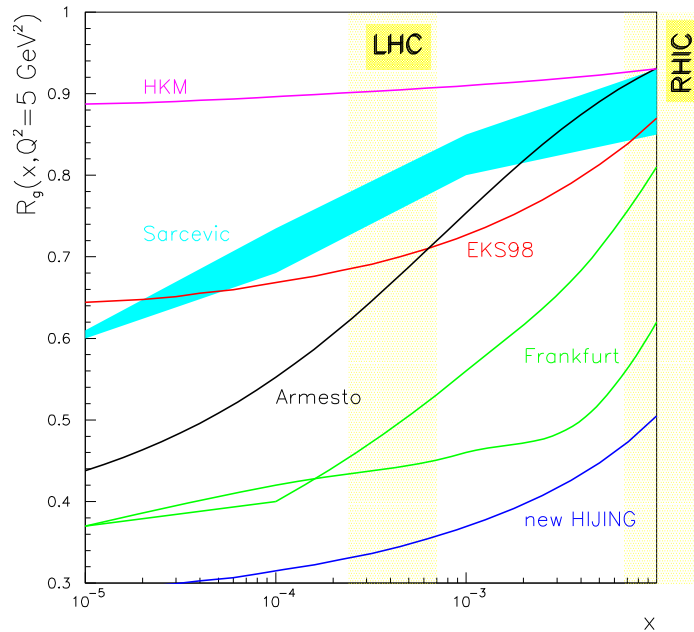}
\vspace{-0.5cm}
\caption{Ratios of gluon distribution functions from different models at $Q^2=5$
GeV$^2$; HKM refers to the results from Ref. \protect{\cite{Hirai:2001np}},
Sarcevic, Ref. \protect{\cite{Huang:1997ii}}, EKS98, Refs.
\protect{\cite{Eskola:1998iy,Eskola:1998df}}, Frankfurt, Ref.
\protect{\cite{Frankfurt:2002kd}}, Armesto, Refs.
\protect{\cite{Armesto:2002ny,Armesto:2001vm}} and new HIJING, Ref.
\protect{\cite{Li:2001xa}}. The bands represent the ranges of $x=(Q/\sqrt s)e^y$
for processes with $|y|\le0.5$, $Q^2=5$ GeV$^2$ at RHIC 
($\sqrt s =200$ GeV) and LHC ($\sqrt s=5.5$~TeV).}
\label{compglue}
\end{center}
\end{figure}

We now compare available numerical results from different approaches.
Recall that in dipole approaches such as that of Ref.
\cite{Armesto:2002ny}, there are difficulties in
identifying the unintegrated gluon distribution with the ordinary gluon density
at small and moderate $Q^2$, 
see the discussion of Eq. (\ref{gluglu}).  A comparison at
$Q^2=5$ GeV$^2$ for the Pb/$p$ gluon ratio
can be found in Fig.~\ref{compglue}. In the RHIC region, $x\simeq 10^{-2}$,
the results from Refs.
\cite{Armesto:2002ny,Eskola:1998iy,Eskola:1998df,Hirai:2001np,Huang:1997ii}
roughly coincide but lie above
those from Refs.
\cite{Frankfurt:2002kd,Li:2001xa}. At $x\simeq 10^{-4}\dots 10^{-5}$
(accessible at the LHC)
the results of Ref.
\cite{Armesto:2002ny} are smaller
than those of Refs.
\cite{Eskola:1998iy,Eskola:1998df,Hirai:2001np,Huang:1997ii}, approaching
those of Ref. \cite{Frankfurt:2002kd} but remaining larger than those
of Ref.~\cite{Li:2001xa}. Apart from the loose small $x$ gluon constraints
from DIS data on nuclei, in
Refs. \cite{Eskola:1998iy,Eskola:1998df,Hirai:2001np} saturation is
imposed on the initial conditions for DGLAP evolution, while
Refs. \cite{Armesto:2002ny,Frankfurt:2002kd,Huang:1997ii} are multiple
scattering models.  In Ref. \cite{Li:2001xa}, the $Q^2$-independent
gluon density is tuned at $x\sim 10^{-2}$ to reproduce charged
particle multiplicities in Au+Au collisions at RHIC.  The strongest
gluon shadowing is thus obtained. However, it seems to be in
disagreement with existing DIS data \cite{Eskola:2002us}.  Additional
caution has to be taken when comparing results from multiple
scattering models with those from DGLAP analysis
\cite{Eskola:1998iy,Eskola:1998df,Hirai:2001np}. The gluon ratios at
some moderate, fixed $Q^2$ and very small $x$ may become smaller
(e.g. in Ref. \cite{Armesto:2002ny}) than the $F_2$ ratios at the same
$x$ and $Q^2$, which might lead to problems with leading-twist DGLAP
evolution \cite{Eskola:2002us}.



\subsection{Predictions for the leading-twist shadowing at the LHC}
\label{subsec:frankfurt_etal}
{\em Leonid~Frankfurt, Vadim~Guzey, Martin ~McDermott, and Mark~Strikman}

\newcommand \Pomeron {I\!\!P}

The basis of our present understanding of nuclear shadowing in
high-energy scattering on nuclei was formulated in the seminal work by
V.~Gribov~\cite{Gribov:1968ia}.  The key observation was that, within
the approximation that the range of the strong interactions is much
smaller than the average inter-nucleon distances in nuclei, there is a
direct relationship between nuclear shadowing in the total hadron-{\it
nucleus} cross section and the cross section for diffractive
hadron-{\it nucleon} scattering.  While the original derivation was
presented for hadron-deuterium scattering, it can be straightforwardly
generalized to lepton-nucleus DIS.

\begin{figure}[bh]
\begin{center}
\includegraphics[width=12.5cm]{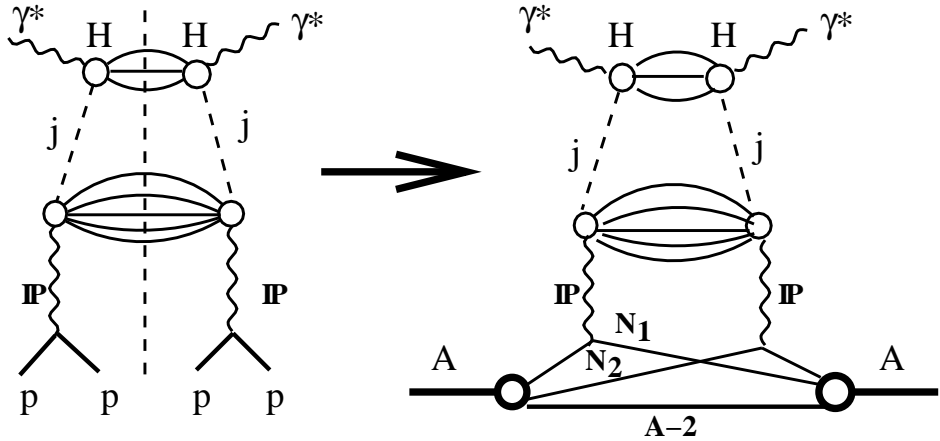}
\caption{
Diagrams for hard diffraction in $ep$ scattering
and for the leading-twist nuclear shadowing.
}
\label{fig:nuc}
 \end{center}
\end{figure}

Reference~\cite{Frankfurt:1998ym} demonstrated how to use the
Collins factorization theorem for hard diffraction in DIS
\cite{Collins:1997sr}  to generalize Gribov theory to
calculate the leading-twist component of nuclear shadowing {\it separately} 
for each
nuclear parton distribution. The correspondence
between the two phenomena is illustrated in Fig.~\ref{fig:nuc}.

As a result, the double scattering term of the shadowing correction
to the nuclear parton distributions, $\delta f_{j/A}^{(2)}(x,Q^2)$,
takes the form
\begin{eqnarray}
&&\delta f_{j/A}^{(2)}(x,Q^2)=\frac{A(A-1)}{2} 16 \pi \, {\rm Re}
\Bigg[\frac{(1-i\eta)^2}{1+\eta^2}\int d^2b \int_{-\infty}^{\infty}
dz_1 \int_{z_{1}}^{\infty} dz_2 \int_x^{x_{\Pomeron, 0}} dx_{\Pomeron}
\times \nonumber \\ &&f_{j/N}^{D}(\beta, Q^2,x_{\Pomeron},0)
~\rho_A(b,z_1)~\rho_A(b,z_2)~e^{ix_{\Pomeron}m_N(z_1-z_2)} \Bigg] \, .
\label{deltaf}
\end{eqnarray}
Here $j$ is a generic parton label (i.e. a gluon or a quark of
a particular flavor), $f_{j/N}^{D}$ is the diffractive parton
distribution function (DPDF) of the nucleon for parton $j$,
$\rho_A(b,z)$ is the nucleon density normalized to unity, $\eta$ is
the ratio of the real to imaginary parts of the nucleon diffractive
amplitude, while $x_{\Pomeron, 0}=0.1$ for quarks and $x_{\Pomeron, 0}=0.03$
for gluons.  Since $f_{j/N}^{D}$ obeys the leading-twist DGLAP
evolution equation, the $Q^2$-evolution of $\delta f_{j/A}^{(2)}$ is
also governed by DGLAP, i.e. it is {\it by definition} a leading-twist
contribution.  This explains why the approach of
Ref.~\cite{Frankfurt:1998ym} can be legitimately called a
leading-twist approach. However, in this approach, one can also take
into account higher-twist effects in diffraction.

\begin{figure}[htb]
  \begin{center}
\includegraphics[width=10cm]{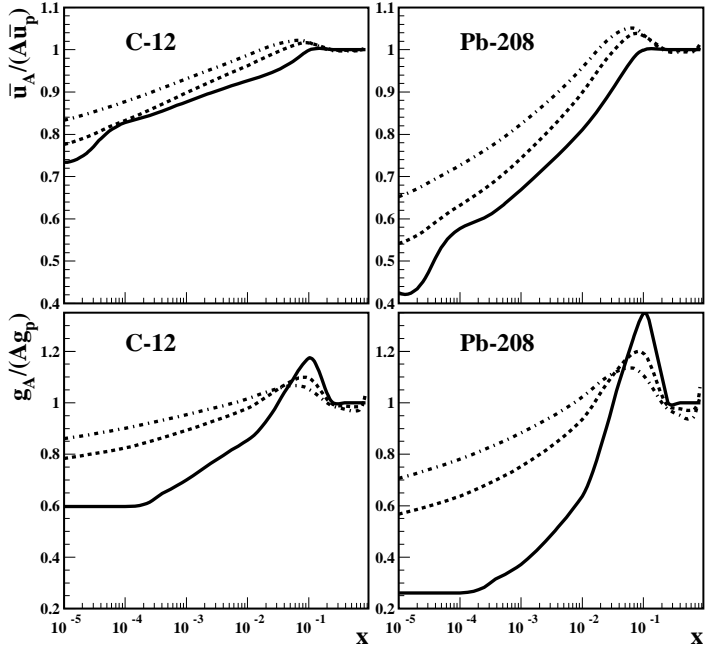}
\caption{
Predictions of quark and gluon shadowing for $Q=2$ GeV (solid curves),
 $Q=10$ GeV (dashed curves) and $Q=100$
GeV (dot-dashed curves).
}
\label{fig:q}
\end{center}
\end{figure}

The HERA diffractive data are consistent with the dominance of 
leading-twist diffraction in the process $\gamma^{\ast} +p \to M_X +p$
for $Q^2\ge 4$ GeV$^2$. These studies have determined NLO quark DPDFs
directly and the gluon diffractive DPDF indirectly through
scaling violations.  More recent studies of  charm production and
diffractive dijet production  in the direct photon kinematics
measured  the gluon DPDF directly. It was  found to be consistent (within
a 30\% uncertainty primarily
determined by the 
 treatment of the NLO contribution to two jet production) with the
inclusive diffraction in DIS. 
 As a result, we were able to determine
the strength of interaction in the quark and gluon channels,
$\sigma_{{\rm eff}}^j$.  This strength turns out to be large at $Q_0^2
\sim 4$ GeV$^2$, where one can set up the boundary
condition\footnote{For $Q^2 \le 4$ GeV$^2$, higher-twist effects in
$ep$ diffraction appear to be significant. This leads to $30\dots50$\%
contributions of higher-twist effects in nuclear shadowing at $Q^2
\sim 1\dots2$ GeV$^2$, which may make the use of the small-$x$ NMC data
for the extraction of nuclear PDFs problematic.}
for the subsequent QCD evolution.  The interaction strength is
typically of the order of 20 mb for the quark channel and 30-40
 mb
for the gluon channel at $Q^2 \sim 4$ GeV$^2$, $x\le
10^{-3}$. Consequently, it is necessary to include the interactions
with $N\ge 3 $ nucleons in the calculations. These interactions were
modeled using the quasi-eikonal approximation.  Corrections due to
fluctuations in the cross section were also analyzed and found to be
rather small for reasonable models of these effects
\cite{Alvero:1998bz}.  Therefore, the master equation for the
evaluation of nuclear shadowing takes the following form
\begin{eqnarray}
&&\delta f_{j/A}(x,Q^2)=\frac{A(A-1)}{2} 16
\pi \, {\rm Re} \Bigg[\frac{(1-i\eta)^2}{1+\eta^2}\int d^2b 
\int_{-\infty}^{\infty} dz_1 \int_{z_{1}}^{\infty} dz_2 
\int_x^{x_{\Pomeron, 0}}
dx_{\Pomeron} \times \nonumber \\ &&f_{j/N}^{D}(\beta,
Q^2,x_{\Pomeron},0)
~\rho_A(b,z_1)~\rho_A(b,z_2)~e^{ix_{\Pomeron}m_N(z_1-z_2)}
e^{-(A/2)(1-i\eta)\sigma_{{\rm eff}}^j \int_{z_1}^{z_2} dz
\rho_A(b,z)}\Bigg] \, .
\label{deltaf2}
\end{eqnarray}
It is important to note that Eq.~(\ref{deltaf2}) defines the
input nuclear PDFs at the initial, low scale $Q=Q_0=2$ GeV for
subsequent QCD evolution to higher scales $Q^2$.  In order to have
consistent results after evolution, Eq.~(\ref{deltaf2}), 
applicable for $10^{-5} \leq x \leq 0.05$, should be supplemented by
a model of nuclear PDFs for larger $x$, $x > 0.05$.  In our
analysis, this is done by assuming no enhancement for quarks and some
enhancement (antishadowing) for gluons (see Fig.~\ref{fig:q}).

We have recently performed a detailed analysis of nuclear shadowing
within the leading-twist approach using the current HERA diffractive
data~\cite{Frankfurt:2002kd}.´ We have analyzed the uncertainties
originating from  the input diffractive parton
distribution functions, related to the uncertainties in the data, and
found that they are less than  $\le 20\dots30\%$ for $x\le 10^{-3}$.
The biggest uncertainty in the gluon channel originates from the
$t$-dependence of the gluon DPDF, not directly measured.  An
example of our calculations~\cite{Frankfurt:2003zd} is presented in
Fig.~\ref{fig:q}.
%

We also calculated the higher-twist shadowing, which we estimated in
the eikonal approximation, and demonstrated that the leading-twist
contribution dominates down to very small $x$ at $Q^2
\geq 4$ GeV$^2$. (Numerical results are available from V.~Guzey upon
request.) Also, we found large differences between the LO and NLO
calculations of the nuclear structure function $F_2^A$.

One can see from Fig.~\ref{fig:q} that the leading-twist shadowing
effects are expected to be large for small $x$  for both the quark and
gluon channels at  $Q\sim$~few~GeV. These effects are expected to
survive up to scales characteristic of $Z$ and $W$ production.

\begin{figure}[htb]
  \begin{center} 
\includegraphics[width=10cm]{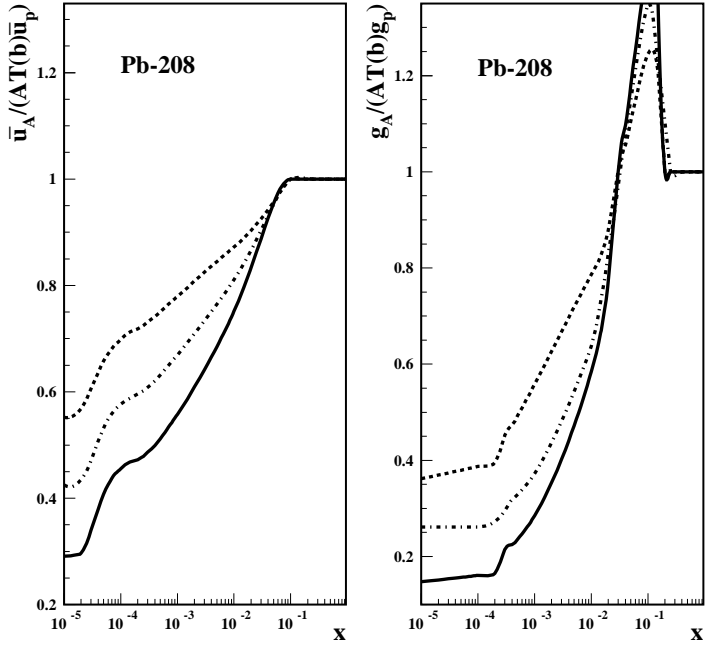}
\caption{ Predictions of quark and gluon shadowing as a function
of the impact parameter: $b=0$ (solid curves); $b=6$ rm (dashed
curves); $b$-averaged ratios (dot-dashed curves). For all curves, $Q=2$
GeV.}
\label{fig:g}
\end{center}
\end{figure}

Our formalism can be readily applied to evaluate nuclear shadowing in
nPDFs at all impact parameters. The nuclear shadowing correction to
the impact-parameter-dependent nPDFs, $\delta f_{j/A}(x,Q^2,b)$, can
be found from Eq.~(\ref{deltaf2}) by simply differentiating with
respect to the impact parameter $b$ \cite{Frankfurt:2002kd}.
%
The effect of nuclear shadowing increases with decreasing impact parameter.
This is illustrated in Fig.~\ref{fig:g}, where we plot the
ratios $f_{j/A}(x,Q^2,b)/(A T_A(b)f_{j/N}(x,Q^2))$,
where $T_A(b)=\int dz \rho_A(b,z)$, for quarks and
gluons at $b=0$ (solid curves) and $b=6$ fm (dashed curves). For
comparison, the impact-parameter-averaged ratios $f_{j/A}(x,Q^2)/(A
f_{j/N}(x,Q^2))$ are given by the dot-dashed curves.
Figure~\ref{fig:g} demonstrates that nuclear shadowing significantly
depends on the impact parameter $b$. Hence, comparing peripheral
and central $pA$ collisions will provide an additional measure of 
the nuclear shadowing.

It is worth emphasizing that the diagrams corresponding to 
leading-twist shadowing are usually neglected in the BFKL-type
approaches to nuclear scattering. For example, these diagrams
are neglected in the model of Balitsky~\cite{Balitsky:1995ub} and
Kovchegov~\cite{Kovchegov:1999ua}. Note also that  higher-twist
effects become important at sufficiently small $x$ and $Q\le
Q_{{\rm min}}(x)$. Rough estimates indicate that, in the gluon sector with
$A\sim 200$, $Q_{{\rm min}}(10^{-3})\sim 2\dots5$ GeV, depending on 
how the NLO effects are handled.
Also, $ Q_{{\rm min}}(x)$
should increase with decreasing $x$.

In conclusion, leading-twist shadowing effects are expected to lead to
significant modifications of the hard process cross sections at the
LHC in $pA$ and $AA$ collisions.  Further progress in the diffraction
studies at HERA will significantly reduce the uncertainties in the
predictions.

\subsection{Structure functions are not parton distributions}
\label{subsec:brodsky}
{\em Stanley J. Brodsky}

\newcommand{\qu}{{\rm q}}
\newcommand{\qb}{${\rm\bar q}$}
\newcommand{\pvec}{\vec p}
\newcommand{\kvec}{\vec k}
\newcommand{\rvec}{\vec r}
\newcommand{\Rvec}{\vec R}
\newcommand{\ieps}{i\varepsilon}
\newcommand{\pl}{{||}}

\newcommand{\order}[1]{${ O}\left(#1 \right)$}
\newcommand{\eq}[1]{(\ref{#1})}

\newcommand{\beq}{\begin{equation}}
\newcommand{\eeq}{\end{equation}}





\begin{figure}[htb]
\begin{center}
\includegraphics[width=5in]{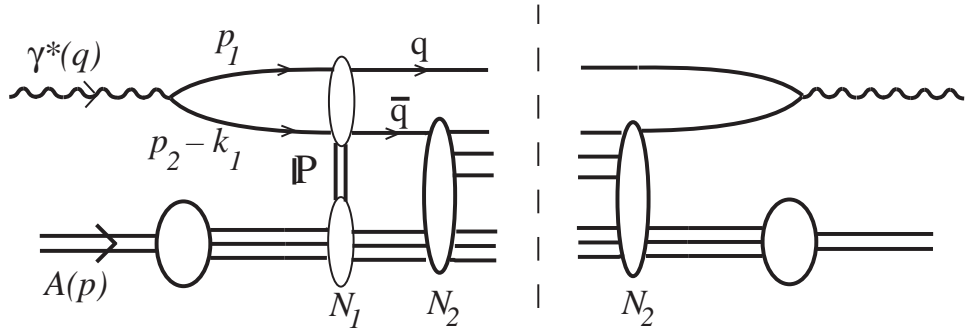}
\caption[*]{Glauber-Gribov shadowing involves interference between
rescattering amplitudes. 
\label{brodsky2}}
\end{center}
\end{figure}

Ever since the earliest days of the parton model, it has been
assumed that the leading-twist structure functions $F_i(x,Q^2)$
measured in deep inelastic lepton scattering are determined by the
{\it probability} distributions of quarks and gluons as determined
by the light-cone (LC) wavefunctions of the target.  For example, the
quark distribution is
$$
{ P}_{\qu/N}(x_B,Q^2)= \sum_n \int^{k_{iT}^2<Q^2}\left[
\prod_i\, dx_i\, d^2k_{T i}\right] |\psi_n(x_i,k_{T i})|^2
\sum_{j=q} \delta(x_B-x_j).
$$
The identification of structure functions with the square of
light-cone wavefunctions is usually made in the LC gauge, $n\cdot A =
A^+=0$, where the path-ordered exponential in the operator product
for the forward virtual Compton amplitude apparently reduces to
unity. Thus the deep inelastic lepton scattering cross section
(DIS) appears to be fully determined by the probability
distribution of partons in the target. However, Paul Hoyer, Nils
Marchal, Stephane Peigne, Francesco Sannino, and I have recently
shown that the leading-twist contribution to DIS is affected by
diffractive rescattering of a quark in the target, a coherent
effect which is not included in the light-cone wavefunctions, even
in light-cone gauge
\cite{Brodsky:2002ue,Belitsky:2002sm,Collins:2003fm}
.  The distinction between structure functions
and parton probabilities is already implied by the Glauber-Gribov
picture of nuclear
shadowing~\cite{Gribov:1968jf,Brodsky:1969iz,Brodsky:1990qz,Piller:2000wx}.
In this framework shadowing arises from interference between
complex rescattering amplitudes involving on-shell intermediate
states, as in Fig.\ref{brodsky2}.  In contrast, the wavefunction of a stable
target is strictly real since it does not have on energy-shell
configurations.  A probabilistic interpretation of the DIS cross
section is thus precluded.

\begin{figure}[htb]
\begin{center}
\includegraphics[width=5in]{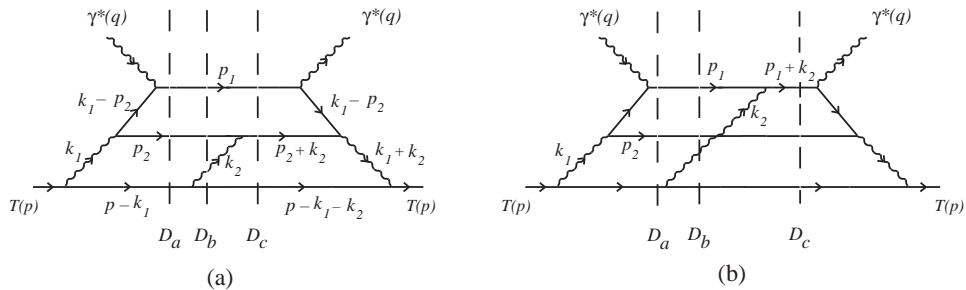}
\caption[*]{Two types of final state interactions.  (a) Scattering
of the antiquark ($p_2$ line), which in the aligned jet kinematics
is part of the target dynamics.  (b) Scattering of the current
quark ($p_1$ line).  For each LC time-ordered diagram, the
potentially on-shell intermediate states -- corresponding to the
zeroes of the denominators $D_a, D_b, D_c$ -- are denoted by
dashed lines.} 
\label{brodsky1}
\end{center}
\end{figure}

It is well-known that in the Feynman and other covariant gauges one
has to evaluate the corrections to the ``handbag" diagram due to
the final-state interactions of the struck quark (the line
carrying momentum $p_1$ in Fig.~\ref{brodsky1}) with the gauge field of the
target.  In light-cone gauge, this effect also involves
rescattering of a spectator quark, the $p_2$ line in Fig.~\ref{brodsky1}.  The
light-cone gauge is singular -- in particular, the gluon
propagator $ d_{LC}^{\mu\nu}(k) =
\frac{i}{k^2+\ieps}\left[-g^{\mu\nu}+\frac{n^\mu k^\nu+ k^\mu
n^\nu}{n\cdot k}\right] \label{lcprop} $ has a pole at $k^+ = 0$
which requires an analytic prescription.  In final-state
scattering involving on-shell intermediate states, the exchanged
momentum $k^+$ is of \order{1/\nu} in the target rest frame, which
enhances the second term in the propagator.  This enhancement
allows rescattering to contribute at leading twist even in LC
gauge.

The issues involving final-state interactions even occur in the simple
framework of abelian gauge theory with scalar quarks.  Consider a
frame with $q^+ < 0$. We can then distinguish FSI from ISI using LC
time-ordered perturbation theory, LCPTH
\cite{Lepage:1980fj}. Figure~\ref{brodsky1} illustrates two LCPTH
diagrams which contribute to the forward $\gamma^* T \to \gamma^* T$
amplitude, where the target $T$ is taken to be a single quark.  In the
aligned jet kinematics the virtual photon fluctuates into a
\qu\qb\ pair with limited transverse momentum, and the (struck)
quark takes nearly all the longitudinal momentum of the photon.
The initial \qu\ and \qb\ momenta are denoted $p_1$ and $p_2-k_1$,
respectively.

The calculation of the rescattering effects on DIS in Feynman and
light-cone gauge through three loops is given in detail in
Ref.~\cite{Brodsky:2002ue}.  The result can be resummed and is
most easily expressed in eikonal form in terms of transverse
distances $r_T, R_T$ conjugate to $p_{2T}, k_T$.
The DIS cross section can be expressed as \beq
Q^4\frac{d\sigma}{dQ^2\, dx_B} =
\frac{\alpha_{\rm em}}{16\pi^2}\frac{1-y}{y^2} \frac{1}{2M\nu} \int
\frac{dp_2^-}{p_2^-}\,d^2\rvec_T\, d^2\Rvec_T\, |\tilde
M|^2 \label{transcross} \eeq where \beq |\tilde{
M}(p_2^-,\rvec_T, \Rvec_T)| = \left|\frac{\sin \left[g^2\,
W(\rvec_T, \Rvec_T)/2\right]}{g^2\, W(\rvec_T,
\Rvec_T)/2} \tilde{A}(p_2^-,\rvec_T, \Rvec_T)\right|
\label{Interference} \eeq is the resummed result.  The Born
amplitude is \beq \tilde A(p_2^-,\rvec_T, \Rvec_T) = 2eg^2
M Q p_2^-\, V(m_\pl r_T) W(\rvec_T, \Rvec_T)
\label{Atildeexpr} \eeq where $ m_\pl^2 = p_2^-Mx_B + m^2
\label{mplus}$ and \beq V(m\, r_T) \equiv \int
\frac{d^2\pvec_T}{(2\pi)^2}
\frac{e^{i\rvec_T\cdot\pvec_{T}}}{p_T^2+m^2} =
\frac{1}{2\pi}K_0(m\,r_T). \label{Vexpr} \eeq The rescattering
effect of the dipole of the \qu\qb~ is controlled by \beq
W(\rvec_T, \Rvec_T) \equiv \int
\frac{d^2\kvec_T}{(2\pi)^2}
\frac{1-e^{i\rvec_T\cdot\kvec_{T}}}{k_T^2}
e^{i\Rvec_T\cdot\kvec_{T}} = \frac{1}{2\pi}
\log\left(\frac{|\Rvec_T+\rvec_T|}{R_T} \right).
\label{Wexpr} \eeq The fact that the coefficient of $\tilde A$ in
Eq.~\eq{Interference} is less than unity for all $\rvec_T,
\Rvec_T$ shows that the rescattering corrections reduce the
cross section.  It is the analog of nuclear shadowing in our
model.

We have also found the same result for the DIS cross
sections in light-cone gauge.  Three prescriptions for defining
the propagator pole at $k^+ =0$ have been used in the literature:
\beq \label{prescriptions} \frac{1}{k_i^+} \rightarrow
\left[\frac{1}{k_i^+} \right]_{\eta_i} = \left\{
\begin{array}{cc}
k_i^+\left[(k_i^+ -i\eta_i)(k_i^+ +i\eta_i)\right]^{-1} & ({\rm PV}) \\
\left[k_i^+ -i\eta_i\right]^{-1} & ({\rm K}) \\
\left[k_i^+ -i\eta_i \epsilon(k_i^-)\right]^{-1} & ({\rm ML})
\end{array} \right.
\eeq the principal-value (PV), Kovchegov (K)~\cite{Kovchegov:1997pc}, and
Mandelstam-Leibbrandt (ML)~\cite{Leibbrandt:1987qv} prescriptions. The
`sign function' is denoted $\epsilon(x)=\Theta(x)-\Theta(-x)$.
With the PV prescription we have $ I_{\eta} = \int dk_2^+
\left[k_2^+\right]_{\eta_2}^{-1} = 0. $ Since an individual
diagram may contain pole terms $\sim 1/k_i^+$, its value can
depend on the prescription used for light-cone gauge. However, the
$k_i^+=0$ poles cancel when all diagrams are added.  The net is
thus prescription-independent and agrees with the Feynman
gauge result. It is interesting to note that the diagrams
involving rescattering of the struck quark $p_1$ do not contribute
to the leading-twist structure functions if we use the Kovchegov
prescription to define the light-cone gauge.  In other
prescriptions for light-cone gauge the rescattering of the struck
quark line $p_1$ leads to an infrared divergent phase factor $\exp
(i\phi)$, where \beq \phi = g^2 \, \frac{I_{\eta}-1}{4 \pi} \, K_0(\lambda
R_{T}) + {{O}}(g^6) \eeq where $\lambda$ is an infrared
regulator, and $I_{\eta}= 1$ in the K prescription. The phase is
exactly compensated by an equal and opposite phase from
FSI of line $p_2$. This irrelevant change of
phase can be understood by the fact that the different
prescriptions are related by a residual gauge transformation
proportional to $\delta(k^+)$ which leaves the light-cone gauge
$A^+ = 0$ condition unaffected.

Diffractive contributions which leave the target intact thus
contribute at leading twist to deep inelastic scattering.  These
contributions do not resolve the quark structure of the target,
and thus they are contributions to structure functions which are
not parton probabilities. More generally, the rescattering
contributions shadow and modify the observed inelastic
contributions to DIS.

Our analysis in the K prescription for light-cone gauge
resembles the ``covariant parton model" of Landshoff, Polkinghorne
and Short~\cite{Landshoff:1971ff,Brodsky:1973hm} when interpreted
in the target rest frame.  In this description of small $x$ DIS,
the virtual photon with positive $q^+$ first splits into the pair
$p_1$ and $p_2$.  The aligned quark $p_1$ has no final state
interactions.  However, the antiquark line $p_2$ can interact in
the target with an effective energy $\hat s \propto {k_T^2/x}$
while staying close to mass shell.  Thus at small $x$ and
large $\hat s$, the antiquark $p_2$ line can first multiply
scatter in the target via pomeron and Reggeon exchange, and then
it can finally scatter inelastically or be annihilated. The DIS
cross section can thus be written as an integral of the
$\sigma_{\bar q p \to X}$ cross section over the $p_2$ virtuality.
In this way, the shadowing of the antiquark in the nucleus
$\sigma_{\bar q A \to X}$ cross section yields the nuclear
shadowing of DIS~\cite{Brodsky:1990qz}.  Our analysis, when
interpreted in frames with $q^+ > 0,$ also supports the color
dipole description of deep inelastic lepton scattering at small
$x$.  Even in the case of the aligned jet configurations, one can
understand DIS as due to the coherent color gauge interactions of
the incoming quark-pair state of the photon interacting first
coherently and finally incoherently in the target.
For further discussion see 
Refs.~\cite{Brodsky:2001rz,Hoyer:2002fc,Belitsky:2002sm}. 
The same final-state interactions which
produce leading-twist diffraction and shadowing in DIS 
also lead to Bjorken-scaling single-spin asymmetries in
semi-inclusive deep inelastic, see 
Refs.~\cite{Brodsky:2002cx,Collins:2002kn,Efremov:2003tf}.

The analysis presented here has important implications for the
interpretation of the nuclear structure functions measured in deep
inelastic lepton scattering. Since leading-twist nuclear shadowing
is due to the destructive interference of diffractive processes
arising from final-state interactions (in the $q^+ \le 0$ frame),
the physics of shadowing is not contained in the wavefunctions of
the target alone. For example, the light-front wavefunctions of
stable states computed in light-cone gauge (PV prescription) are
real, and they only sum the interactions within the bound-state
which occur up to the light-front time $\tau=0$ when the current
interacts. Thus the shadowing of nuclear structure functions is
due to the mutual interactions of the virtual photon and the
target nucleon, not the nucleus in isolation~\cite{Brodsky:2002ue}.


\subsection{Nuclear parton distribution functions in the saturation regime}
\label{subsec:kovchegov}
{\em Yuri V. Kovchegov}


\subsubsection{Introduction}

At the high collision energies 
reached at the LHC, parton
densities in the nuclear 
wavefunctions will become very large, resulting
in the onset of 
parton recombination and {\it
saturation} of parton distribution functions
\cite{Gribov:tu,Mueller:wy,McLerran:1993ni}. In the
saturation regime, the growth of partonic structure functions with
energy is expected to slow down sufficiently to unitarize the total
nuclear cross sections. Gluonic fields in the saturated nuclear 
wavefunction would become very strong, reaching the maximum possible
magnitude in QCD of $A_\mu \sim 1/g$
\cite{Kovchegov:1996ty,Kovchegov:1997pc,Jalilian-Marian:1996xn}.
The transition to the saturation region can be characterized by the {\it
saturation scale} $Q_s^2 (s, A)$, which is related to the typical two
dimensional density of the 
parton color charge in the infinite
momentum frame of the hadronic or nuclear 
wavefunctions \cite{McLerran:1993ni,Kovchegov:1996ty,Kovchegov:1997pc,Jalilian-Marian:1996xn}. The
saturation scale 
is an increasing function of energy $\sqrt s$ and of the atomic 
number $A$
\cite{Mueller:st,Kovchegov:1999yj,Kovchegov:1999ua,Levin:1999mw,Braun:2000wr,Golec-Biernat:2001if,Iancu:2002tr,Mueller:2002zm}. 
It is expected that it (roughly) scales as
\begin{equation}
Q_s^2 (s, A) \, \sim \, A^\alpha \, s^\delta
\end{equation}
where $\delta \, \approx \, 0.2 - 0.3$ based on HERA data \cite{Golec-Biernat:1998js,Golec-Biernat:1999qd}
and $\alpha \, \ge \, 1/3$ \cite{Levin:1999mw,Braun:2000wr,Golec-Biernat:2001if,Iancu:2002tr,Mueller:2002zm}. Various estimates predict the
saturation scale for lead 
nuclei at $14$~TeV to be $Q_s^2 \, \approx
\, 10-20$~GeV$^2$. This high value of the saturation scale would ensure that the 
strong coupling constant $\alpha_s (Q_s^2)$ is small, making the vast
majority of partons in the nuclear wave functions at the LHC {\it
perturbative} \cite{Mueller:1999fp,Mueller:1999wm,Mueller:2001fv,Kovchegov:1998bi,Jalilian-Marian:1996xn}! The presence of 
an 
intrinsic large
momentum scale $Q_s$ justifies the use of 
a
perturbative QCD expansion
even for such traditionally nonperturbative observables as total
hadronic cross sections \cite{McLerran:1993ni,Mueller:st}. 

Below we 
first discuss the predictions for gluon and quark
distribution functions of a large nucleus, given by the quasi-classical
model of McLerran and Venugopalan
\cite{McLerran:1993ni,Mueller:st,Kovchegov:1996ty,Kovchegov:1997pc,Kovchegov:1998bi}. We
then discuss how quantum corrections should be included to obtain
an equation for the realistic $F_2$ structure function of the nucleus
\cite{Kovchegov:1999yj,Kovchegov:1999ua,Balitsky:1997mk,Balitsky:1995ub,Balitsky:1998ya}. We will conclude by deriving the gluon production 
cross section for {\it pA} scattering in the quasi-classical
approximation \cite{Kovchegov:1998bi} and generalizing it to include quantum
corrections \cite{Kovchegov:2001sc}.

\subsubsection{Structure functions in the quasi-classical approximation}

\subsubsection*{The gluon distribution}

\begin{figure}
\begin{center}
\includegraphics[width=8cm]{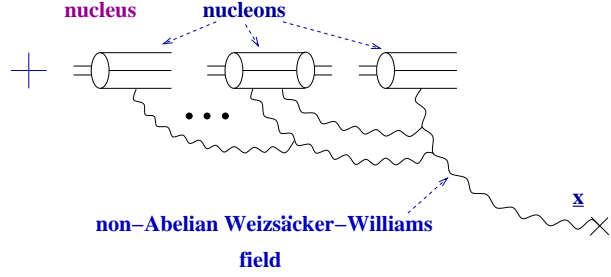} 
\caption{The non-Abelian Weizs\"{a}cker-Williams field of a nucleus.}
\label{wwf}
\end{center}
\end{figure}

As suggested originally by McLerran and Venugopalan \cite{McLerran:1993ni}, the
gluon field of a large ultrarelativistic nucleus is classical and is
given by the solution of the Yang-Mills equations of motion with the
nucleus providing the source current. The classical solution in the
light cone gauge of the nucleus, known as the non-Abelian
Weizs\"{a}cker-Williams field (WW), has been found in Refs. 
\cite{Kovchegov:1996ty,Kovchegov:1997pc,Jalilian-Marian:1996xn}. 
It includes the effects of all multiple rescatterings 
of the small-$x$ gluon on the nucleons in the nucleus located at the
same impact parameter as the gluon. This resummation parametrically
corresponds to resumming powers of $Q_s^2/q^2$ (higher twists), where
$q$ is the gluon transverse momentum. Diagrammatically, the field is
shown in Fig. \ref{wwf}.
The correlator of two WW fields gives rise to the unintegrated nuclear
gluon distribution given by \cite{Jalilian-Marian:1996xn,Kovchegov:1998bi}
\begin{eqnarray}
\frac{dxG_A(x,\underline q^2)}{d\underline q^2} &=&
\frac{2}{(2\pi)^2}\,\int\, d^2\underline z \,
e^{-i\underline z\cdot\underline q}\, \int\, d^2\underline b
\,\mathrm{Tr}\,\langle \underline A^{WW}(\underline 0)\,\underline
A^{WW}(\underline z)\rangle\nonumber\\ &=& \frac{2}{\pi
(2\pi)^2}\,\int\, d^2\underline z\, e^{-i\underline z\cdot\underline
q}\,
\frac{S_\bot C_F}{\alpha_s \,\underline z^2}\left(
1-e^{-\frac{1}{4}\underline z^2 Q_s^2}\right),\label{WW}
\end{eqnarray}
where $\underline b$ is the gluon impact parameter and
\begin{equation}\label{SATSCALE}
Q_s^2(\underline z)\,=\, \frac{4\pi^2\, \alpha_s N_c}{N_c^2 - 1}\,\rho\,
xG_N (x,1/\underline z^2)\, T_A(\underline b)
\end{equation}
is the saturation scale with 
atomic number density $\rho\,=\, A/[(4/3) \pi R_A^3]$ 
and 
$T_A(\underline b)=2 \sqrt{R_A^2 - b^2}$ for spherical nuclei
$A$ considered here.
The gluon distribution in a single nucleon, $xG_N (x, Q^2)$, is
assumed to be a slowly varying function of $Q^2$. In the two-gluon
exchange approximation
\cite{Kovchegov:1998bi} $xG_N (x, Q^2) \, = \, \frac{\alpha_s C_F}{\pi} \ln Q^2/\Lambda^2$. 

The gluon distribution function of Eq. (\ref{WW}) falls off
perturbatively as $xG_A \, \sim \, Q_s^2/{\underline q}^2$ for large
transverse momenta $q$, but is much less singular in the infrared
region, scaling as $xG_A \, \sim \, \ln Q_s/q$ for small $q$
\cite{Jalilian-Marian:1996xn}. We conclude that multiple rescattering 
softens the 
infrared 
singular behavior of the traditional perturbative distribution. Note
also that the number of gluons given by Eq. (\ref{WW}) is of the order
of $1/\alpha_s$, corresponding to strong gluonic fields of the order
of $A_\mu \sim 1/g$.

\subsubsection*{The quark distribution}

\begin{figure}[t]
\begin{center}
\includegraphics[width=8cm]{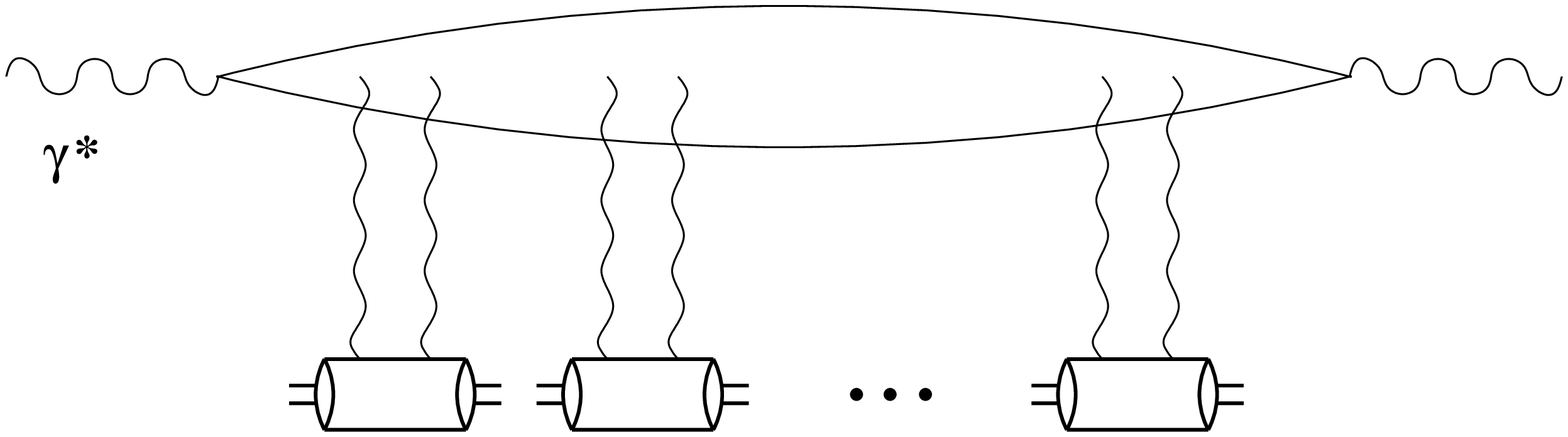} 
\caption{DIS on a nucleus in the quasi-classical approximation.}
\label{f2cl}
\end{center}
\end{figure}

To calculate the quark distribution function in the quasi-classical
approximation, one has to consider deep inelastic scattering in the
rest frame of the nucleus including all multiple rescatterings, as
shown in Fig. \ref{f2cl}. 
The $F_2$ structure function (and the quark distribution) was first calculated
by Mueller \cite{Mueller:st} in this approximation, yielding
\begin{equation}\label{f2}
F_2 (x, Q^2) = \sum_q e_q^2[xq (x, Q^2) + x{\overline q} (x, Q^2)] 
=  \frac{Q^2}{4 \pi^2 \alpha_{\rm em}} \sum_q e_q^2
\int\frac{d^2 r d z }{2 \pi} \, \Phi^{\gamma^* \rightarrow q{\bar q}}
({\underline r},z) \ d^2 b \ N({\underline r},{\underline b} , Y) ,
\end{equation}
where $\Phi^{\gamma^* \rightarrow q{\bar q}} ({\underline r},z)$ is the
wavefunction of a virtual photon in 
DIS,
splitting into a $q\bar q$ pair with transverse separation ${\underline r}$.
The fraction of photon longitudinal momentum carried by the
quark is $z$. The quantity $N({\underline r},{\underline b} , Y)$ 
is the
forward scattering amplitude of a dipole with transverse
size $\underline r$ at impact parameter $\underline b$ with rapidity
$Y$ on a target nucleus. In the quasi-classical approximation of
Refs. 
\cite{Mueller:st,McLerran:1993ni},
$N({\underline r},{\underline b} , Y=0)$ is given by
\begin{equation}\label{init}
N({\underline r},{\underline b} , Y=0) \, = \, 1 - e^{ - r^2
Q_{s,{\rm quark}}^2 / 4}
\end{equation}
where 
$Q_{s,{\rm quark}}^2$
is the quark saturation scale 
which can be obtained
from the gluon saturation scale in Eq. (\ref{SATSCALE}) by replacing $N_c$ by
$C_F$ in the numerator \cite{Mueller:st,Kovchegov:2001sc}.

\subsubsection{Inclusion of quantum corrections}

At high energy, the quasi-classical approximation previously
considered 
ceases to be valid. The energy dependence of the cross sections comes in
through the quantum corrections 
which resum
logarithms of energy
\cite{Kuraev:fs,Balitsky:ic,Jalilian-Marian:1998cb,Kovner:2000pt,Weigert:2000gi,Ferreiro:2001qy}. 
The evolution equation for $N$, resumming logarithms of energy, 
closes only in the large-$N_c$ limit of QCD and reads
\cite{Kovchegov:1999yj,Kovchegov:1999ua,Balitsky:1997mk,Balitsky:1995ub,Balitsky:1998ya,Mueller:1993rr,Mueller:1994jq,Mueller:gb,Chen:1995pa}
\begin{eqnarray*}
  N({\underline x}_{01},{\underline b}, Y) = N ({\underline
  x}_{01},{\underline b}, 0) \, e^{ - \frac{4 \alpha C_F}{\pi} \ln
  \left( \frac{x_{01}}{\rho} \right) Y } + \frac{\alpha C_F}{\pi^2}
  \int_0^Y d y \, e^{ - \frac{4 \alpha C_F}{\pi} \ln \left(
  \frac{x_{01}}{\rho} \right) (Y - y) }
\end{eqnarray*}
\begin{eqnarray}\label{eqN}
\times \int_\rho d^2 x_2 \frac{x_{01}^2}{x_{02}^2 x_{12}^2} \, [ 2
  \, N({\underline x}_{02},{\underline b} + \frac{1}{2} {\underline
  x}_{12}, y) - N({\underline x}_{02},{\underline b} + \frac{1}{2}
  {\underline x}_{12}, y) \, N({\underline x}_{12},{\underline b} +
  \frac{1}{2} {\underline x}_{02}, y) ], 
\end{eqnarray}
with 
initial condition in
Eq. (\ref{init}) and ${\underline
x}_{ij} = {\underline x}_{i} - {\underline x}_{j}$. In the usual
Feynman diagram formalism, it resums multiple pomeron exchanges (fan
diagrams). Together with Eq. (\ref{f2}), it provides the quark
distribution function in
a nucleus at high energies. The equation has
been solved 
both
by approximate analytical methods and numerically in Refs.
\cite{Levin:1999mw,Braun:2000wr,Golec-Biernat:2001if,Iancu:2002tr,Mueller:2002zm}, providing unitarization of the total DIS cross section and
saturation of the quark distribution functions in the nucleus.

Using Eq.~(\ref{eqN}) together with Eq.~(\ref{WW}), one may conjecture
the following expression for the unintegrated gluon distribution function
including quantum corrections \cite{Kovchegov:2001sc}
\begin{equation}\label{gluev}
\frac{dxG_A(x,\underline q^2)}{d\underline q^2} \, = \, \frac{2}{\pi (2\pi)^2}\,
\int\, d^2 b \, d^2\underline z\, e^{-i\underline z\cdot\underline q}\, 
\frac{C_F}{\alpha_s\,\underline z^2} 
\, N_G (\underline z, \underline b, Y = \ln 1/x)
\end{equation}
with $N_G$ the forward amplitude of a gluon dipole scattering on the
nucleus which, in the large-$N_c$ limit, 
is $N_G = 2 N - N^2$ \cite{Kovchegov:2001sc}, where $N$ is the forward
amplitude of the $q\bar q$ dipole scattering on the nucleus used in
Eq. (\ref{eqN}).  Equation (\ref{gluev})
must still be proven. 

\subsubsection{Gluon production in $pA$ collisions}

The $pA$ program at the LHC will 
test the above predictions for
parton distribution functions by measuring particle production cross
sections which depend on these quantities. Below we 
discuss gluon
production in $pA$.

\subsubsection*{Quasi-classical result}

\begin{figure}[t]
\begin{center}
\includegraphics[width=8cm]{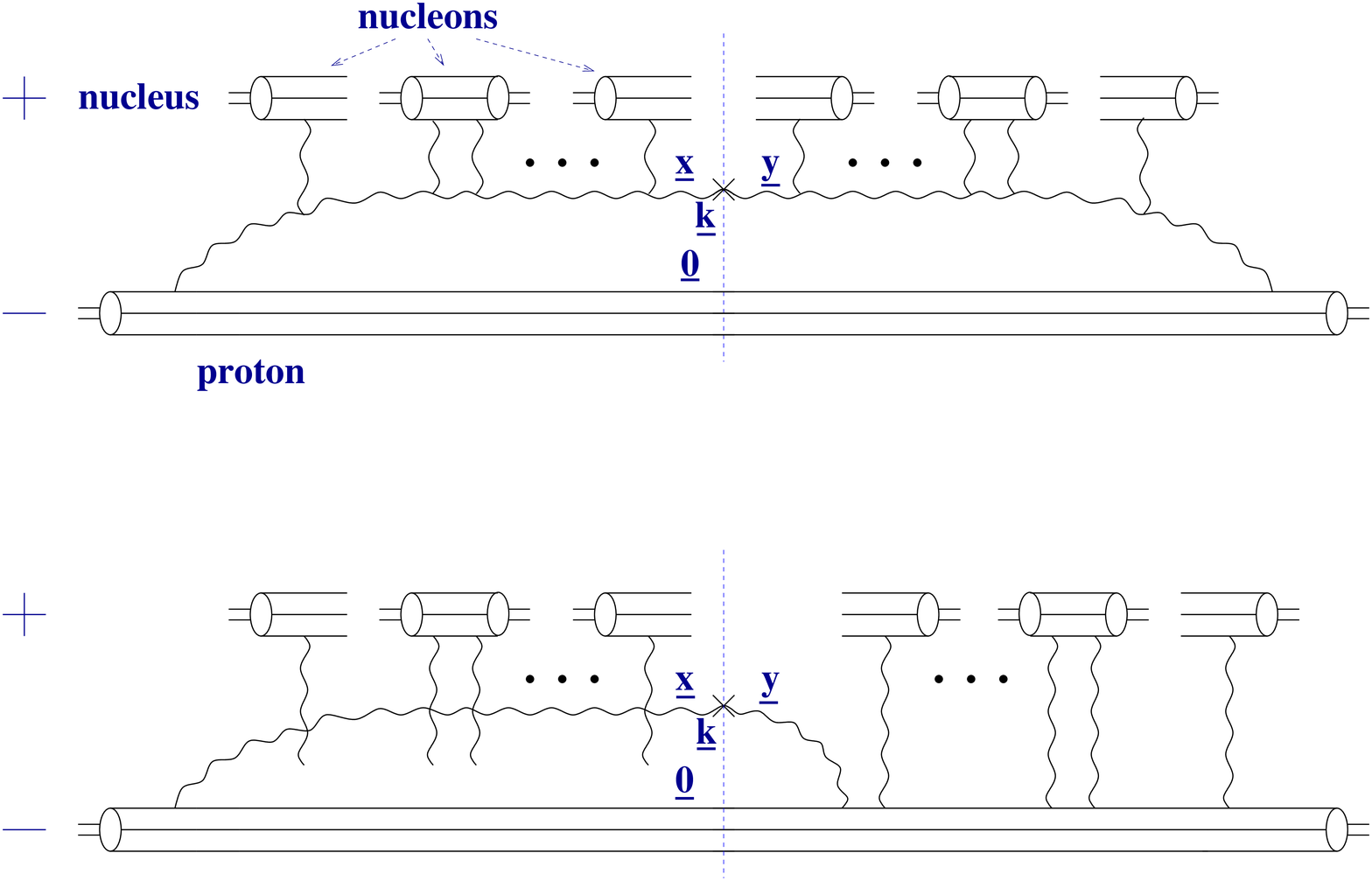} 
\caption{Gluon production in {\it pA} collisions in the quasi-classical 
approximation.}
\label{pa}
\end{center}
\end{figure}

Gluon production in $pA$ 
interactions 
has been calculated in the
quasi-classical approximation in Ref. \cite{Kovchegov:1998bi}. 
The relevant diagrams in
the light cone gauge of the proton are shown in Fig. \ref{pa}. 
The cross section is \cite{Kovchegov:1998bi}
\begin{equation}\label{pax}
\frac{d \sigma^{pA}}{d^2 k \ dy} \ = \ \int \ d^2 b \, d^2 x 
\, d^2 y \, \frac{\alpha_s C_F}{\pi^2 \, (2 \pi)^2} \frac{{\underline x} 
\cdot {\underline y}}{{\underline x}^2 {\underline y}^2} \, e^{i {\underline k} 
\cdot ({\underline x} - {\underline y})} \, \left( 1 -  e^{- {\underline x}^2 \ 
Q_s^2 /4 } - e^{- {\underline y}^2 \ Q_s^2 /4 } + e^{- ({\underline x}
- {\underline y})^2 \ Q_s^2 /4 } \right).
\end{equation}

\subsubsection*{Gluon production including quantum corrections}

To include quantum corrections in Eq. (\ref{pax}), 
it must be rewritten
in the factorized form \cite{Kovchegov:2001sc} and 
generalized to include
the linear evolution of the proton structure function and nonlinear
evolution of the nuclear structure function. 
The result, see also Ref. \cite{Kovchegov:2001sc}, is
\begin{equation}\label{paxe}
\frac{d \sigma^{pA}}{d^2 k \ dy} \ = \ \frac{C_F}{\alpha_s \pi (2 \pi)^3} \, 
\int d^2 z \ n_G (\underline z, Y-y) \, \frac{\stackrel{\textstyle\leftarrow}
{\nabla}^2_z \, e^{i {\underline k} \cdot {\underline z}} \,
\stackrel{\textstyle\rightarrow}{\nabla}^2_z}{\underline k^2} \, N_G 
(\underline z, y),
\end{equation}
where $Y$ is the total rapidity interval between the proton and the
nucleus, $N_G (\underline z, y) = \int d^2 b \, N_G (\underline z,
\underline b, y)$ and $\stackrel{\textstyle\rightarrow}{\nabla}^2_z$
is the transverse gradient squared. 
The function
$n_G (\underline z, Y-y)$ in
Eq. (\ref{paxe}) is the solution of the linear part of Eq. (\ref{eqN})
(BFKL) integrated over all $b$ with the two-gluon exchange amplitude
between the proton and a gluon dipole as initial conditions. In the
no-evolution limit, Eq. (\ref{paxe}) reduces to
Eq. (\ref{pax}). Equation (\ref{paxe}) gives the saturation physics
prediction for particle production at midrapidity in $pA$ collisions
at the LHC.




\clearpage


\section{THE BENCHMARK TESTS IN $pA$}
\label{pA_SEC:benchmarks}
\vspace{-0.25cm}
\vspace{0.5cm}

\subsection{$W$, $Z$, and Drell-Yan cross sections}
\label{section411}
{\em Ramona Vogt}


Studies of $W^\pm$ and $Z^0$ gauge bosons as well as high mass Drell-Yan 
production in $pA$ interactions
at the LHC will provide information on the nuclear parton densities at much
higher $Q^2$ than previously available.  The high $Q^2$ measurements
will probe the $x$ range around 0.015 at $y=0$.  This range of $x$ has been 
probed before by fixed target nuclear deep-inelastic scattering so that a
measurement at the LHC will constrain the
evolution of the nuclear quark distributions over a large lever arm in $Q^2$.
Knowledge of the nuclear quark distributions in this region will aid in studies
of jet quenching with a $Z^0$ opposite a jet in $AA$ collisions.  At least one
of the initial partons will be a quark or antiquark in the reactions $q
\overline q \rightarrow Z^0 g$ and $q(\overline q) g \rightarrow q(\overline q)
Z^0$. 

Since the $W$ and $Z^0$ studies will open a new window in the
$(x,Q^2)$ plane for nuclear parton distribution functions, these
measurements are obvious benchmarks.  The $Z^0$ measurement probes
symmetric $q \overline q$ annihilation while measurements such as the
$W^+ - W^-$ asymmetry via high $p_T$ decay leptons are better probes
of the nuclear valence quark distributions.  Drell-Yan production at
high $Q^2$ is a somewhat different story because, to be an effective
benchmark, the Drell-Yan signal should dominate the dilepton mass
spectrum.  In $AA$ collisions, the signal dilepton mass distribution
is dominated by $b \overline b$ decays until $M>M_Z$ \cite{cmsnote}.
Thus only in this large mass region is Drell-Yan a useful benchmark.

We very briefly outline the next-to-leading order, NLO, calculation
here.  The NNLO cross sections are available \cite{Hamberg:1990np} but
are only a small addition (a few percent) compared to the NLO
correction.  The NLO cross section for
production of a vector boson, $V$, with mass $M$ at scale $Q$ in a
$pp$ interaction is
\begin{eqnarray} \frac{d\sigma^V_{pp}}{dy} & = & 
H_{ij}^V \int dx_1\, dx_2 \,dx \, \delta
\bigg(\frac{M_V^2}{s} - x x_1 x_2 \bigg) 
\delta \bigg( y - \frac{1}{2} \ln \bigg( \frac{x_1}{x_2} \bigg) \bigg) 
\label{sigmajpsi} \\
&   & \mbox{} \times
\bigg\{ \sum_{i,j \in Q,\overline Q} C^{\rm ii}(q_i, \overline q_j)
\Delta_{q \overline q}(x,Q^2) f_{q_i}^p(x_1,Q^2) f_{\overline 
q_j}^p(x_2,Q^2) \nonumber \\
&   & \mbox{} +
\sum_{i,k \in Q, \overline Q} C^{\rm if}(q_i, q_k) \Delta_{qg}(x,Q^2) \bigg[
f_{q_i}^p(x_1,Q^2) f_g^p(x_2,Q^2) 
+ f_g^p(x_1,Q^2) 
f_{q_j}^p(x_2,Q^2) \bigg] \bigg\} \, \,  , \nonumber
\end{eqnarray}
where $H_{ij}^V$ is proportional to the LO 
partonic cross section, $ij \rightarrow V$, and the sum over $Q$ includes $u$,
$d$, $s$ and $c$.  The matrices $C^{\rm ii}$ and
$C^{\rm if}$ contain information on the coupling of the various quark flavors
to boson $V$.  The parton densities in the proton, $f_i^p$,
are evaluated at momentum fraction $x$ and scale $Q^2$.  The convolutions of
the parton densities, including the proton and neutron (isospin) content of
the nucleus, with shadowing parameterizations formulated assuming $f_i^A(x,Q^2)
= f_i^p(x,Q^2) R_i^A(x,Q^2)/A$ are given in the appendix of
Ref.~\cite{Vogt:2000hp}\footnote{In Ref.~\cite{Vogt:2000hp}, $S^i(x,Q^2,A)
\equiv R_i^A(x,Q^2)$.}.   All the results are calculated with the MRST HO 
densities \cite{Martin:1998sq} at $Q = M_V$ and the EKS98 shadowing
parameterization \cite{Eskola:1998iy,Eskola:1998df}. 
In the Drell-Yan process, the
mass distribution can also be calculated by adding a $dM$ in the denominator of
the left-hand side of Eq.~(\ref{sigmajpsi}) and the delta function $\delta(M -
M_V)$ on the right-hand side.
The prefactors $H_{ij}^V$ are rather simple \cite{Hamberg:1990np},
\begin{eqnarray}
H_{ij}^{Z^0} = \frac{8\pi}{3} \frac{G_F}{\sqrt{2}} [(g_V^i)^2 
+ (g_A^i)^2] \frac{M_Z^2}{s} \, \, , \,\,\,\,\,\,\,\,\,\,\,\,\,\,\,
H_{ij}^{W^\pm} = \frac{2\pi}{3} \frac{G_F}{\sqrt{2}} \frac{M_W^2}{s} \, \, ,
\label{part} 
\end{eqnarray}
where $G_F = 1.16639 \times 10^{-5}$ GeV$^2$, $M_Z = 91.187$ GeV, 
and $M_W = 80.41$ GeV.  
In the case of the Drell-Yan process,
there are three contributions to $H_{ij}^V$: virtual photon exchange, 
$Z^0$ exchange, and $\gamma^* -Z^0$ interference so that
\begin{eqnarray}
H_{ij}^M & = & H_{ij}^{\gamma^*} + H_{ij}^{\gamma^*-Z^0} + H_{ij}^{Z^0} : 
\label{dypart} \\
H_{ij}^{\gamma^*} & = & \frac{4 \pi \alpha^2}{9 M^2 s} |e_i|^2 \, \, ,
\nonumber \\
H_{ij}^{\gamma^* -Z^0} 
& = & \frac{\alpha}{9}\frac{G_F}{\sqrt{2}} \frac{M_Z^2}{s}
\frac{(1 - 4\sin^2\theta_W)(M^2 - M_Z^2)}{(M^2 - M_Z^2)^2 + M_Z^2 \Gamma_Z^2}
|e_i|(1 - 4|e_i|\sin^2\theta_W) \, \, , \nonumber \\
H_{ij}^{Z^0} & = & \frac{1}{3} \frac{G_F}{\sqrt{2}} \frac{M^2}{s}
\frac{M_Z \Gamma_{Z \rightarrow l^+ l^-}}{(M^2 - M_Z^2)^2 + M_Z^2 \Gamma_Z^2}
(1 + (1 - 4|e_i| \sin^2\theta_W)^2) \, \, ,
\nonumber 
\end{eqnarray}
where $\sin^2 \theta_W = 1 - M_W^2/M_Z^2$, $\Gamma_Z = 
2.495$ GeV, and $\Gamma_{Z \rightarrow l^+ l^-} = 3.367$\%.
The functions $\Delta_{ij}(x,Q^2)$ in Eq.~(\ref{sigmajpsi}) are 
universal for gauge bosons and Drell-Yan production \cite{Hamberg:1990np}.  
We work in the
${\overline {\rm MS}}$ scheme.  The NLO correction to the $q \overline q$
channel includes contributions from soft and virtual gluons as well as hard
gluons from the process $q \overline q \rightarrow V g$.  The quark-gluon 
contribution only appears at NLO through the real correction 
$qg \rightarrow qV$.  At NLO, $\alpha_s(Q^2)$ is calculated to two loops
with $n_f = 5$.

\begin{figure}[h]
\setlength{\epsfxsize=0.95\textwidth}
\setlength{\epsfysize=0.4\textheight}
\centerline{\epsffile{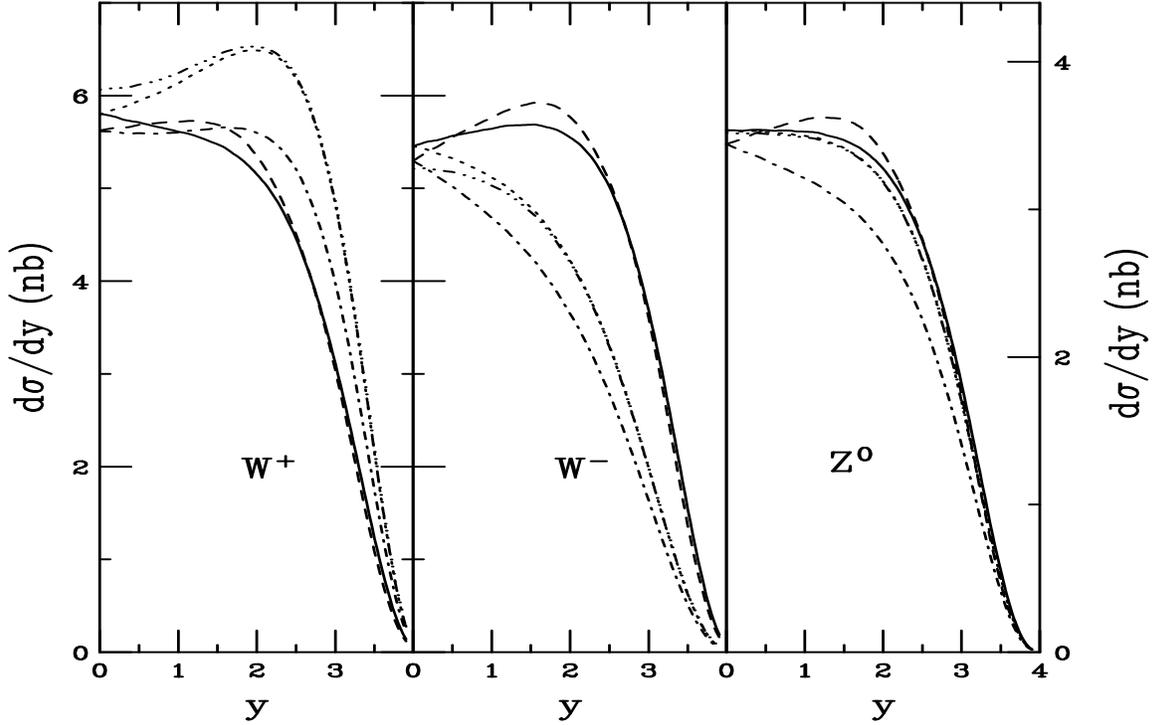}}
\caption{The $W^+$, $W^-$ and $Z^0$ rapidity distributions in $pp$, $p$Pb and
Pb$p$ collisions at 5.5 TeV/nucleon evaluated at $Q = M_V$.  
The solid and dashed curves show the results without and with shadowing 
respectively in Pb$p$ collisions while the dotted and dot-dashed
curves give the results without and with shadowing for $p$Pb collisions.
The dot-dot-dot-dashed curve is the $pp$ result.
}
\label{wpdist}
\end{figure}

As we show in Fig.~\ref{wpdist}, the isospin of the nucleus is most important
for $q \overline q$-dominated processes.  Gauge boson and Drell-Yan production
are the best examples of this domination in hadroproduction.  
We compare the $p$Pb and Pb$p$ $W^+$, $W^-$ and $Z^0$
distributions per nucleon with and without the EKS98
shadowing parameterization to $pp$ production at 5.5 TeV. 
The $pp$ distributions closely follow the $p$Pb distributions without shadowing
except for $y<2$.  Since the $W^+$ distribution is
most sensitive to the $f_u^p(x_1) f_{\overline d}^p(x_2)$ contribution, the
$pp$ and $p$Pb distributions peak away from $y = 0$.  The Pb$p$ distribution,
on the other hand, decreases with increasing $y$ because the neutron content
of lead washes out the peak.  The opposite effect is seen for $W^-$ production
where, in Pb$p$, the neutrons produce a peak at forward rapidity since
$f_d^n = f_u^p$.  The $Z^0$ distribution is almost independent of isospin, as
are the high mass Drell-Yan distributions.  We see antishadowing in
Pb$p$ interactions with large nuclear $x_1$ while we have small nuclear $x_2$ 
in the $p$Pb calculations, reducing 
the $y$ distributions.  The $W^+$ $pA$ distributions
are all reduced compared to the $pp$ distributions, the $W^-$ $pA$ results are
typically enhanced relative to $pp$ while the $Z^0$ $pA/pp$ ratios do not
change much with $y$.

The gauge boson cross sections in $pp$, Pb$p$, and $p$Pb collisions at 5.5
TeV/nucleon 
as well as Pb$p$ and $p$Pb collisions at 8.8 TeV/nucleon are given in
Table~\ref{gbsigs}. 
\begin{table}[htb]
\caption{Gauge boson cross sections per nucleon. No decay channel is 
specified. 
}
\label{gbsigs}
\begin{center}
\begin{tabular}{ccccccc}
$\Delta y$ & $\sigma_{\rm NS}$ (nb) & $\sigma_{\rm EKS98}$ (nb) 
& $\sigma_{\rm NS}$ (nb) & $\sigma_{\rm EKS98}$ (nb) 
& $\sigma_{\rm NS}$ (nb) & $\sigma_{\rm EKS98}$ (nb)
\\ \hline
& \multicolumn{2}{c}{$Z^0$} & \multicolumn{2}{c}{$W^+$} &
\multicolumn{2}{c}{$W^-$} \\ \hline
\multicolumn{7}{c}{$pp$ 5.5 TeV} \\
$0< y < 2.4$ &  8.10 & - & 15.13 & - & 11.41 & - \\
$0< y < 1$   &  3.51 & - &  6.13 & - &  5.16 & - \\
$2.4<y<4$ &  2.01 & - &  5.68 & - &  2.42 & - \\ \hline
\multicolumn{7}{c}{Pb$p$ 5.5 TeV} \\
$0< y < 2.4$ &  8.23 & 8.37 & 13.12 & 13.34 & 13.37 & 13.63 \\
$0< y < 1$   &  3.53 & 3.53 &  5.71 &  5.67 &  5.56 &  5.54 \\
$2.4<y<4$ &  2.14 & 2.06 &  3.72 &  3.62 &  4.36 &  4.24 \\ \hline
\multicolumn{7}{c}{$p$Pb 5.5 TeV} \\
$0< y < 2.4$ &  8.12 & 7.39 & 14.87 & 13.43 & 11.62 & 10.57 \\
$0< y < 1$   &  3.52 & 3.33 &  5.95 &  5.60 &  5.31 &  5.01 \\
$2.4<y<4$ &  2.01 & 1.68 &  5.67 &  4.69 &  2.43 &  2.02 \\ \hline
\multicolumn{7}{c}{Pb$p$ 8.8 TeV} \\
$0< y < 2.4$ & 12.37 & 12.47 & 19.62 & 19.71 & 19.87 & 20.01 \\
$0< y < 1$   &  5.22 &  5.11 &  8.37 &  8.16 &  8.22 &  8.02 \\
$2.4<y<4$ &  5.07 &  5.03 &  8.30 &  8.31 &  9.46 &  9.47 \\ \hline
\multicolumn{7}{c}{$p$Pb 8.8 TeV} \\
$0< y < 2.4$ & 12.26 & 10.93 & 21.41 & 18.96 & 18.08 & 16.09 \\
$0< y < 1$   &  5.20 &  4.82 &  8.60 &  7.92 &  7.98 &  7.35 \\
$2.4<y<4$ &  4.83 &  3.95 & 11.80 &  9.58 &  5.98 &  4.87 \\ \hline
\end{tabular}
\end{center}
\end{table}
The Drell-Yan cross sections in the mass bins $60<M<80$ GeV,
$80<M<100$ GeV, and $100<M<120$ GeV are given in Table~\ref{dysigs} for $pp$,
Pb$p$, and $p$Pb collisions at 5.5 TeV/nucleon.  
\begin{table}[htb]
\caption{Drell-Yan cross sections in the indicated mass intervals at 5.5 
TeV/nucleon in $pp$, Pb$p$ and $p$Pb collisions.  
}
\label{dysigs}
\begin{center}
\begin{tabular}{cccccc}
$\Delta y$ & $\sigma$ (nb) 
& $\sigma_{\rm NS}$ (nb) & $\sigma_{\rm EKS98}$ (nb) 
& $\sigma_{\rm NS}$ (nb) & $\sigma_{\rm EKS98}$ (nb)
\\ \hline
& $pp$ & \multicolumn{2}{c}{Pb$p$} &
\multicolumn{2}{c}{$p$Pb} \\ \hline
\multicolumn{6}{c}{$60<M<80$ GeV} \\
$0<y<2.4$ & 0.012  & 0.012 & 0.012 & 0.012 & 0.011 \\
$0<y<1$   & 0.0048 & 0.0048 & 0.0047 & 0.0048 & 0.0044 \\ 
$2.4<y<4$ & 0.0038 & 0.0036 & 0.0035 & 0.0038 & 0.0031 \\ \hline 
\multicolumn{6}{c}{$80<M<100$ GeV} \\
$0<y<2.4$ & 0.26 & 0.27 & 0.27 & 0.26 & 0.24 \\
$0<y<1$   & 0.11 & 0.11 & 0.11 & 0.11 & 0.10 \\ 
$2.4<y<4$ & 0.055 & 0.059 & 0.057 & 0.055 & 0.046 \\ \hline 
\multicolumn{6}{c}{$100<M<120$ GeV} \\
$0<y<2.4$ & 0.0088 & 0.0088 & 0.0090 & 0.0088 & 0.0080 \\
$0<y<1$   & 0.0037 & 0.0037 & 0.0037 & 0.0037 & 0.0035 \\ 
$2.4<y<4$ & 0.0015 & 0.0015 & 0.0014 & 0.0015 & 0.0012 \\ \hline 
\end{tabular}
\end{center}
\end{table}
The cross sections are presented in three rapidity intervals, $0<y< 2.4$, 
corresponding to the CMS barrel+endcaps, $0<y<1$, the
central part of ALICE, and $2.4<y<4$, the ALICE muon arm.  The difference in
the Pb$p$ and $p$Pb cross sections without shadowing is entirely due to
isospin.   Thus the $pA$ cross section in
$|y|<2.4$ is not a simple factor of two larger than the $0<y<2.4$ cross section
since the results are not symmetric
around $y=0$, as can be seen in Fig.~\ref{wpdist}.  

We have checked the dependence of the gauge boson results on scale, parton
density, and shadowing parameterization.  The scale 
dependence enters in the PDFs, $\alpha_s$, and the 
scale-dependent logarithms $\ln (Q^2/M_V^2)$ \cite{Hamberg:1990np}.  
At these high $Q^2$ values, $M_V/2 < Q < 2M_V$, the change in $\alpha_s$ 
is $\approx \pm
10$\%, small compared to the variation at lower scales. 
When $Q^2 = M_V^2$, the scale dependent
terms drop out.  If $Q^2 > M_V^2$, the logs are
positive, enhancing the ${\cal O}(\alpha_s)$ component, but if $Q^2 < M_V^2$,
the scale-dependent logs change sign, reducing the NLO contribution.
Evolution with $Q^2$ increases the low $x$ density of the sea
quarks and gluons while depleting the high $x$ component and generally
reducing the valence distributions.  Since the $x$ values do not change 
when $Q^2$ is varied, the higher scales also tend to enhance the 
cross sections.  The gauge boson cross sections are 
decreased $\sim 30$\% when $Q^2$ is decreased and
increased by 20\% with increased $Q^2$.
The scale change is even smaller for high mass Drell-Yan production, a few
percent in the 60-80 GeV bin and 1\% or less for higher masses.

The variation with parton density can be as large as the scale dependence.
The MRST and CTEQ \cite{Lai:1999wy} rapidity distributions are similar but the 
CTEQ cross sections are 7\%
higher.  The GRV 98 \cite{Gluck:1998xa} results are stronger functions of
rapidity than the MRST.  The total GRV 98 cross sections, however,
are about 13\% smaller.

We have also compared the $pA/pp$ ratios to the HPC
\cite{HPC} and HKM \cite{Hirai:2001np} parameterizations.  The HPC
parameterization does not distinguish between the sea and valence effects and
includes no $Q^2$ evolution.  HKM actually  parameterizes the
nuclear parton distributions themselves and focuses on the distinction between
the up and down nuclear distributions.  It is only useful to LO.
However, shadowing at LO and NLO is nearly identical \cite{Emel'yanov:1999bn}.

In Pb$p$ collisions, the
EKS98 and HKM ratios are similar for $y\leq 1.5$.  At higher
rapidities, the HKM ratio is larger because its sea distribution increases
more with $x_1$ but its valence shadowing is weaker.  The HPC
ratio is lower than the EKS98 ratio at $y\sim 0$, $\sim 0.8$ relative to $\sim
0.97$.  However the two ratios
are very similar for $y>2$.  In $p$Pb collisions,
$x_2$ is in the shadowing region and all three ratios decrease with rapidity. 
Again, near $y\sim 0$, the HKM and EKS98 ratios are similar but the HKM
sea quark shadowing is nearly independent of $x_2$
while the EKS98 ratio drops from 0.97 to 0.8 between $y=0$ and 4.  The HPC
ratio is parallel to but lower than the EKS98 ratio.

Finally, we note that, without keeping the
energy fixed, extracting the nuclear quark distributions from $pp$ and $pA$
and applying them to $AA$ is difficult\footnote{For a discussion of shadowing
effects on gauge boson production in Pb+Pb collisions at 5.5 TeV/nucleon and 
the possible
impact parameter dependence of the effect, see Ref.~\cite{Vogt:2000hp}.} 
since the isospin is
an additional, nonnegligible uncertainty.  Therefore, it is desirable 
to make $pp$ and $p$Pb runs at 5.5 TeV for a more precise
correlation with the Pb+Pb data at 5.5 TeV/nucleon.  
Indeed, for best results, both $p$Pb and Pb$p$ runs should be done
since the ALICE muon detector is not symmetric around $y=0$.  Its large
rapidity coverage would extend the nuclear $x$ 
coverage to much higher (Pb$p$) and 
lower ($p$Pb) values since  $y\sim 4$ is the maximum rapidity for
$Z^0$ production at 5.5 TeV/nucleon, 
the upper limit of the muon
coverage, where $x_1 \sim 1$ and $x_2 \sim 0.0003$. 

Systematic studies with nuclear beams could be very useful for fully 
mapping out the $A$ dependence of the nuclear parton densities.  
The LHC Ar beam would
link to the NMC Ca data while the LHC O beam could connect back to previous
C data \cite{Amaudruz:1995tq}.  
However, these $A$ systematics are best tuned to the
$AA$ collision energies, different for O, Ar, and Pb.
Comparing $pA$ and $pp$ at the same energies does not shift $x$ while
a comparison to $pp$ at 14 TeV rather than 5.5 TeV 
would shift the $x$ values by a factor of $\sim 3$ between Pb+Pb and $pp$.  
Energy systematics of gauge boson production
between RHIC and LHC are more difficult because $W$ and $Z^0$ production 
in $pA$ at RHIC is quite close to threshold.  However, at RHIC, larger
$x$ values can be probed at $y \sim 0$, $x \sim 0.45$, where the cross
sections are quite sensitive to the valence distributions.  High mass Drell-Yan
is likely to be beyond the reach of RHIC.

\vspace{0.3cm}
\subsection{$Z^0$ and $W$ transverse momentum spectra }
\label{subsec:zhang_fei}
{\em Xiaofei Zhang and George Fai}

\def\tb{\tilde{b}}




At LHC energies, perturbative QCD (pQCD) provides a powerful
calculational tool.  For $Z^{0}$ and $W$ transverse momentum spectra,
pQCD theory agrees with the CDF \cite{Affolder:1999jh} and D0
\cite{Abbott:2000xv} data very well at Tevatron energies \cite{Qiu:2000ga}.
The LHC $pp$ program will test pQCD at an unprecedented energy.  The
heavy-ion program at the LHC will make it possible for the first time
to observe the full $p_T$ spectra of heavy vector bosons in nuclear
collisions and will provide a testing ground for pQCD resummation
theory \cite{Collins:1984kg}.

In nuclear collisions, power corrections will be enhanced by
initial and final state multiple scattering.  As we will show, 
higher-twist 
effects are small at the LHC for heavy boson production.  The
only important nuclear effect is the nuclear modification of the
PDFs (shadowing).  Because of their large
masses, the $W$ and $Z^0$ will tell us about the 
nPDFs at large scales.  Since 
contributions from different scales need to be
resummed, the $p_T$ spectra of heavy vector bosons will provide
information about the evolution of nPDFs from small scales to large
scales.  Since it is more difficult to detect $W^\pm$ than $Z^0$, we
will concentrate on 
the 
$Z^0$ here (the results for $W^{\pm}$
production are very similar \cite{Zhang:2002jf}).

Resummation of large logarithms in QCD can be carried out either
directly in $p_T$-space, or in the so-called 
``impact parameter'', $\tb$-space, 
the
Fourier conjugate of the $p_T$-space.
Using the renormalization group
equation technique, 
Collins, Soper, and Sterman (CSS) derived a formalism
for the transverse momentum distribution of vector boson production
in hadronic collisions \cite{Collins:1984kg}. In the CSS formalism, 
nonperturbative
input is needed for the large $\tb$ region.  The dependence of the pQCD
results on the nonperturbative input is not weak if the original
extrapolation proposed by CSS is used. Recently, a new extrapolation
scheme was introduced, based on solving the renormalization group equation
including power corrections \cite{Qiu:2000ga}. Using
the new extrapolation formula, the dependence of the pQCD results
on the nonperturbative input was significantly reduced.

For vector boson ($V$) production in a hadron collision, 
the CSS resummation formalism yields \cite{Collins:1984kg}
\begin{eqnarray}
\frac{d\sigma(h_A+h_B\rightarrow V+X)}{dM^2\, dy\, dp_T^2} =
\frac{1}{(2\pi)^2}\int d^2 \tb\, e^{i\vec{p}_T\cdot \vec{\tb}}\,
\tilde{W}(\tb,M,x_A,x_B) + Y(p_T,M,x_A,x_B) \,\,\, ,
\label{css-gen}
\end{eqnarray}
where $x_A= e^y\, M/\sqrt{s}$ and $x_B= e^{-y}\, M/\sqrt{s}$, with
rapidity $y$ and collision energy $\sqrt{s}$.
In Eq.~(\ref{css-gen}), the 
$\tilde{W}$ term dominates the $p_T$ distributions
when $p_T \ll M$, and the $Y$ term gives corrections 
that are negligible
for small $p_T$ but become important when $p_T\sim M$. 

The function $\tilde{W}(\tb,M,x_A,x_B)$ can be  
calculated perturbatively for small $\tb$, but an
extrapolation to the large $\tb$ region, requiring nonperturbative input,
is necessary in order to complete the Fourier transform in Eq.~(\ref{css-gen}).
In order to improve the situation, a new form was proposed \cite{Qiu:2000ga}
by solving the renormalization equation including power
corrections. In the new formalism, 
$\tilde{W}(\tb,M,x_A,x_B)=\tilde{W}^{\rm pert}(\tb,M,x_A,x_B)$, when 
$\tb \leq \tb_{\rm max}$, with 
\begin{equation}
\tilde{W}^{\rm pert}(\tb,M,x_A,x_B) =
{\rm e}^{S(\tb,M)}\, \tilde{W}(\tb,c/\tb,x_A,x_B) \,\,\, ,
\label{Collins:1984kg-sol}
\end{equation}
where all large logarithms from $\ln(1/\tb^2)$ to $\ln(M^2)$ have
been completely resummed into the exponential factor
$S(\tb,M)$, and $c$ is a constant of order
unity \cite{Collins:1984kg}. For $\tb>\tb_{\rm max}$,
\begin{eqnarray}
\tilde{W}(\tb,M,x_A,x_B)
=\tilde{W}^{\rm pert}(\tb_{\rm max}, M, x_A, x_B)
F^{\rm NP}(\tb;\tb_{\rm max})  \,\,\, ,
\label{qz-W-sol-m}
\end{eqnarray}
where the
nonperturbative function $F^{\rm NP}$ is given by
\begin{eqnarray}
F^{\rm NP}
=\exp\bigl\{ -\ln(M^2 \tb_{\rm max}^2/c^2) 
\left[ g_1 \left( (\tb^2)^\alpha - (\tb_{\rm max}^2)^\alpha\right) \right.
 \left.   +g_2 \left(\tb^2 - \tb_{\rm max}^2\right) \right]
-\bar{g}_2 \left(\tb^2 - \tb_{\rm max}^2\right) \bigr\}.
\label{qz-fnp-m}
\end{eqnarray}
Here, $\tb_{\rm max}$ is a parameter that separates 
the perturbatively calculated part from the nonperturbative
input. Unlike in the original CSS formalism,
$\tilde{W}(\tb,M,x_A,x_B)$ is 
independent of 
the nonperturbative parameters  when $\tb < \tb_{\rm max}$. 
In addition, the $\tb$-dependence in 
Eq.~(\ref{qz-fnp-m}) is separated according to different physics
origins. The $(\tb^2)^\alpha$-dependence mimics the
summation of the perturbatively calculable leading-power
contributions to the  renormalization group equations
to all orders in the running
coupling constant $\alpha_s(\mu)$. The $\tb^2$-dependence of the
$g_2$ term is a direct consequence of dynamical power corrections to the
renormalization group equations 
and has an explicit dependence on $M$. 
The ${\bar g}_{2}$ term represents the effect 
of the non-vanishing intrinsic parton transverse momentum.

A remarkable feature of the $\tb$-space resummation formalism is that
the resummed exponential factor $\exp[S(\tb,M)]$ suppresses the
$\tb$-integral when $\tb$ is larger than $1/M$.  It can be shown using
the saddle point method that, for a large enough $M$, QCD perturbation
theory is valid even at $p_T=0$ \cite{Collins:1984kg}.  As discussed
in Refs. \cite{Qiu:2000ga,Zhang:2002jf}, the value of the saddle point
strongly depends on the collision energy $\sqrt{s}$, in addition to
its well-known $M^2$ dependence.
The predictive power of the $\tb$-space resummation formalism 
should be even better at  the LHC than at the Tevatron.

In $Z^0$ production, since final-state interactions are negligible,
power corrections can arise only from initial state multiple scattering.
Equations (\ref{qz-W-sol-m}) and (\ref{qz-fnp-m}) represent 
the most general form of $\tilde{W}$ and thus, apart from isospin and 
shadowing effects, which will be discussed later. The only way nuclear
modifications associated with scale evolution
enter the $\tilde{W}$ term is through the coefficient 
${g}_{2}$. 

The parameters $g_1$ and $\alpha$ of Eq.~(\ref{qz-fnp-m})
are fixed by the requirement of continuity of the function
$\tilde{W}(\tb)$  and its derivative at $\tb=\tb_{\rm max}$.
(The results are insensitive to changes of $\tb_{\rm max}$ $\in [0.3$ GeV$^{-1}$,\,0.7~GeV$^{-1}]$.
We use $\tb_{\rm max}=$ 0.5 GeV$^{-1}$.) 
A fit to low energy Drell-Yan data gives 
$\bar g_2 = 0.25 \pm 0.05$~GeV$^2$ and $g_2 = 0.01 \pm 0.005$~GeV$^2$.
As the $\tb$ dependence of the  $g_2$ and ${\bar g}_2$ 
terms in Eq.~(\ref{qz-fnp-m}) is identical, it is convenient to combine
these terms and define
$G_{2}= \ln({M^2 \tb_{\rm max}^2/  {c^2}})g_2 + \bar{g}_2 \,\,\, .$
Using the values of the parameters listed above, we get 
$G_2 = 0.33 \pm 0.07$ GeV$^2$ 
for $Z^0$ production in $pp$ collisions. 
The parameter $G_2$ can be considered 
the only free parameter of the nonperturbative input in Eq. (\ref{qz-fnp-m}),
arising from the power corrections in the renormalization group equations.
An impression of the importance of the power corrections can be obtained by
comparing results with the above value of $G_2$ 
to those with power corrections turned off ($G_2=0$).
We therefore define the ratio
\begin{equation}
R_{G_2}(p_T) \equiv \left.
\frac{d\sigma^{(G_2)}(p_T)}{dp_T} \right/
\frac{d\sigma(p_T)}{dp_T} \,\,\, .
\label{Sigma-g2}
\end{equation}
The cross sections in the above equation and in the results presented
here have been integrated over rapidity ($-2.4 \leq y \leq 2.4$) 
and 
$M^2$ in the  narrow width approximation.
For the PDFs, we use the CTEQ5M set \cite{Lai:1999wy}.

\begin{figure}
\begin{center}
\hspace{-3cm}
\includegraphics[width=8.0cm]{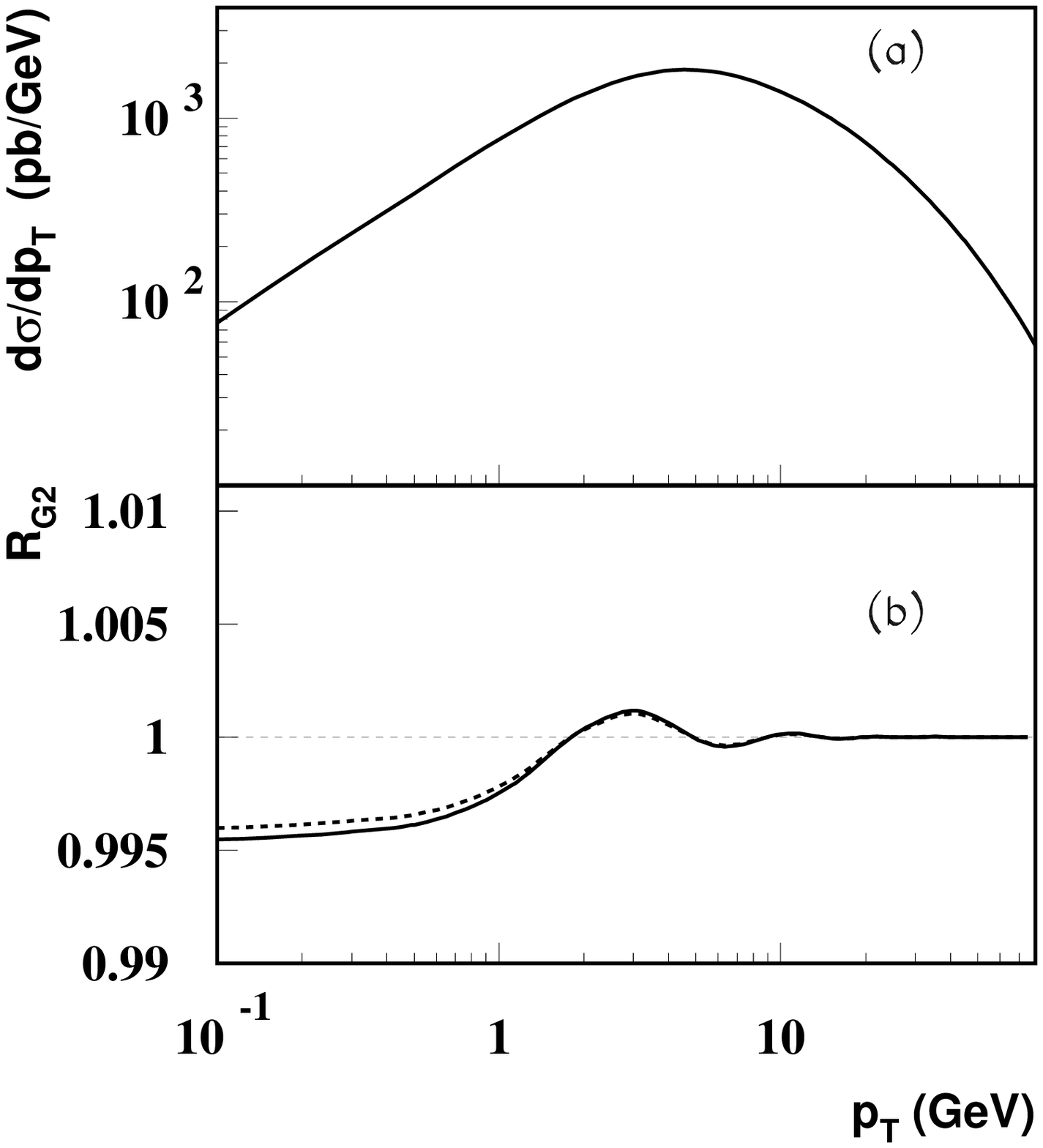}
\vspace{1.0cm}
\includegraphics[width=8.0cm, bbllx=55pt,bblly=123pt,bburx=557,bbury=652pt]{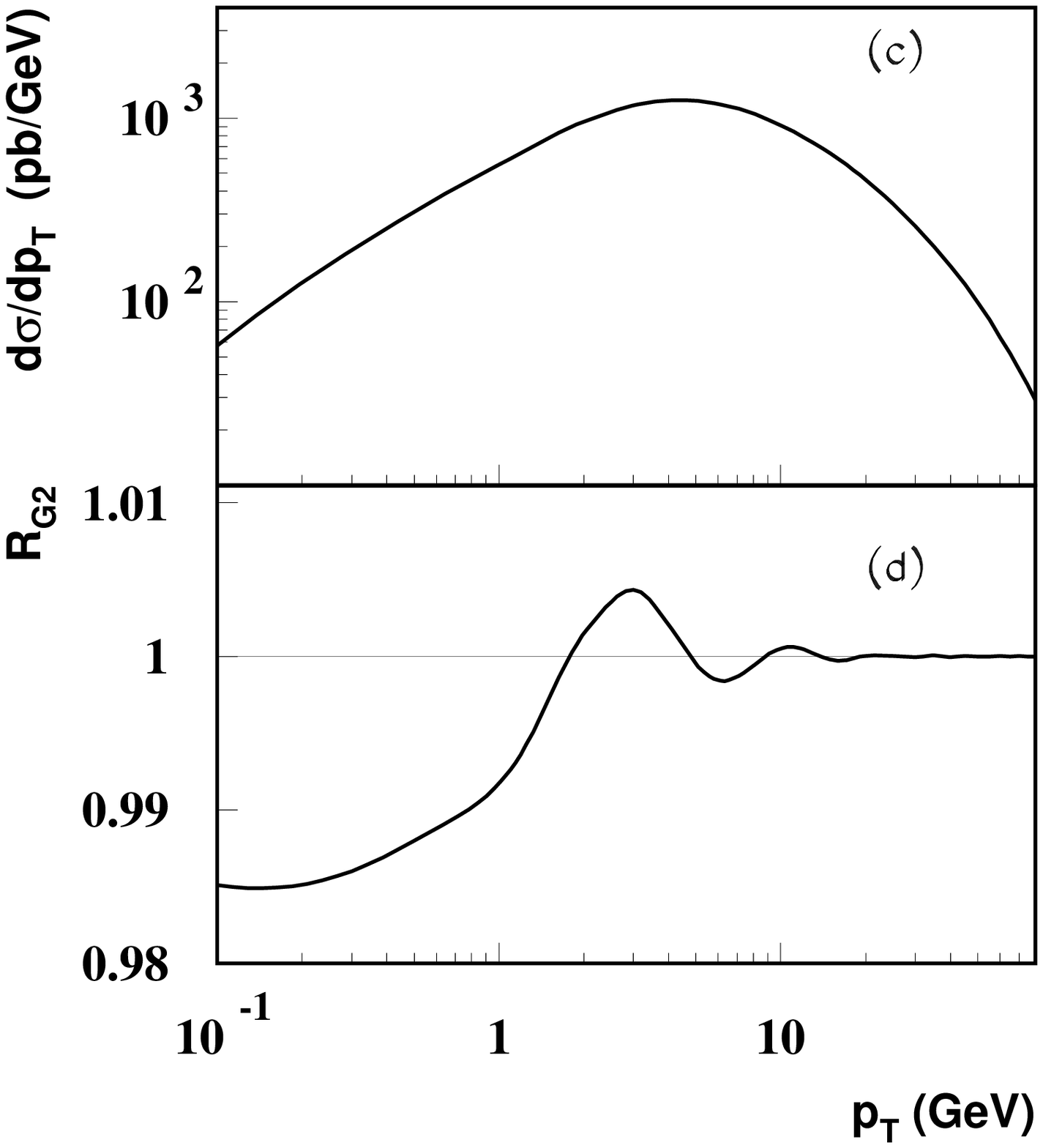} \hspace{-3cm} 
\vspace{-1.8cm}
\caption{(a) Cross section ${d\sigma / dp_T}$ for $Z^0$ production 
in $pp$ collisions at the LHC with
$\sqrt{s}=14$ TeV; (b) $R_{G_2}$ defined in Eq.~(\protect\ref{Sigma-g2})
with $G_2=$ 0.26 GeV$^2$  (dashed) and 0.40 GeV$^2$  (solid).
(c) Cross section in $p$Pb collisions at $\sqrt s=8.8$ TeV;
(d) $R_{G_2}$  in $p$Pb with $G_2=0.8$GeV$^2$.
}
\vspace{-0.3in}
\label{zfig1}
\end{center}
\end{figure}

Figure \ref{zfig1} displays the differential cross sections 
and the corresponding $R_{G_2}$ ratio 
(with the limiting values of $G_2=0.26$ GeV$^2$
(dashed) and $G_2=0.40$  GeV$^2$ (solid))
for $Z^0$ production as functions of $p_T$ in $pp$ collisions 
at $\sqrt s= 14$ TeV.   
The deviation of $R_{G_2}$ from unity
decreases rapidly as $p_T$ increases.
It 
is smaller than one percent for both $\sqrt{s}=5.5$~TeV (not shown)
and $\sqrt{s}=14$ TeV in $pp$ collisions, even when $p_T=0$. 
In other words, the effect of power corrections
is very small at the LHC over the whole $p_T$  region.

Without nuclear effects on the hard collision, the production of heavy
vector bosons in nucleus-nucleus ($AB$) collisions should scale 
with the number of binary collisions, and
$\sigma_{AB}\propto AB$.
However, there are several
additional nuclear effects on the hard collision in a heavy-ion
reaction.  First of all, the isospin effects, which come from the
difference between the neutron PDFs and the proton PDFs, is about 2\%
at LHC.  This is expected, since at the LHC $x \sim 0.02$ where the
$u-d$ asymmetry is very small. 

The dynamical power corrections entering the parameter $g_2$ 
should be enhanced by the nuclear
size, i.e. proportional to $A^{1/3}$. Taking into account the $A$-dependence,
we obtain $G_2 = 1.15 \pm 0.35$ GeV$^2$ for PbPb reactions.
We find that with this larger value of $G_2$, 
the effects of power corrections appear to be enhanced by a factor of  
about three from $pp$ to PbPb collisions at the LHC. 
Thus, even the enhanced power corrections remain
under 1\% when 3 GeV $< p_T < $ 80 GeV. This 
small effect is taken into account in the following nuclear calculations.

Next we turn to the phenomenon of shadowing, 
expected to be a function of $x$,
the scale $\mu$, and of the position in the nucleus. The latter
dependence means that in heavy-ion collisions, shadowing
should be impact parameter ($b$) dependent. 
Here we concentrate on impact-parameter integrated 
results, where the effect of the $b$-dependence of shadowing is relatively 
unimportant \cite{Zhang:2001ce}, and 
focus
on the scale dependence. 
We therefore use 
the EKS98 parametrization \cite{Eskola:1998df}.
We define
\begin{equation}
R_{\rm sh}(p_T) \equiv \left.
\frac{d\sigma^{\rm (sh)}(p_T,Z_A/A,Z_B/B)}{dp_T} \right/
\frac{d\sigma(p_T)}{dp_T} \,\,\, ,
\label{Sigma-sh}
\end{equation}
where $Z_A$ and $Z_B$ are the atomic numbers and $A$ and $B$ are 
the mass numbers of the colliding nuclei. The cross section
$d\sigma^{\rm (sh)}(p_T,Z_A/A,Z_B/B)/dp_T$ has been averaged over $AB$,
while $d\sigma(p_T)/dp_T$ is the $pp$ cross section.
We have seen above that 
shadowing remains  the only significant
effect responsible for nuclear modifications.  
Figure~\ref{zfig1}(c) shows $d\sigma^{\rm (sh)}/dp_T$ and 
(d)  the corresponding $R_{G_2}$ 
for $p$Pb collisions at $\sqrt s=8.8$~TeV.

\begin{figure}
\begin{center}
\includegraphics[width=8.0cm]{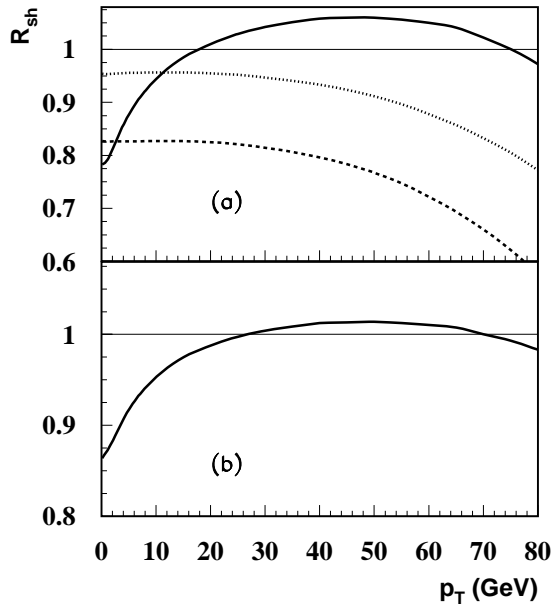}
\vspace{-0.7cm}
\caption{
  (a)  Cross section ratios for $Z^0$ production in PbPb collisions at 
  $\sqrt{s}=5.5$ TeV: $R_{\rm sh}$ of Eq.~(\protect\ref{Sigma-sh}) 
  (solid line), and $R_{\rm sh}$ with the scale fixed at 
  5 GeV (dashed) and 90 GeV (dotted); (b) $R_{\rm sh}$ in $p$Pb at 
  $\sqrt s=8.8$~TeV.}
\label{zfig3}
\vspace{-0.3in}

\end{center}
\end{figure}

In Fig.~\ref{zfig3}(a), 
the result for 
$R_{\rm sh}$ (solid line) is surprising
because even at $p_T=90$ GeV, $x\sim 0.05$, and we are still in the
``strict shadowing'' region. Therefore, the fact that $R_{\rm sh} > 1$
for 20 GeV $< p_T <$ 70 GeV is not ``anti-shadowing''.  To better
understand the shape of the ratio as a function of $p_T$, we also show
$R_{\rm sh}$ with the scale fixed 
to 
5 GeV (dashed line)
and 90 GeV (dotted), respectively, in Fig.~\ref{zfig3}(a).  The
nuclear modification to the PDFs is only a function of $x$ and flavor
in the calculations represented by the dashed and dotted lines. These
two curves are similar in shape, but rather different from the solid
line. In $\tb$ space, $\tilde{W}(\tb,M,x_A,x_B)$ is almost equally
suppressed in the whole $\tb$ region if the fixed scale shadowing is
used.  However, with scale-dependent shadowing, the suppression
increases as $\tb$ increases
since 
the scale $\mu\sim 1/\tb$
in the nPDFs.  
Thus the scale dependence redistributes the
shadowing effect.  
Here the redistribution brings
$R_{\rm sh}$ above unity for 20 GeV $< p_T <$ 70 GeV. When $p_T$
increases further, the contribution from the $Y$ term starts to be
important and $R_{\rm sh}$ dips back below one to match the fixed
order pQCD result. 
Figure \ref{zfig3}(b) shows $R_{\rm sh}$ in $p$Pb at $\sqrt s=8.8$~TeV.

We see from Fig.~\ref{zfig3} that the shadowing effects on the $p_T$
distribution of $Z^0$ bosons at the LHC are intimately related to the
scale dependence of the nPDFs 
where the data are very limited
\cite{Eskola:1998df}. Theoretical studies such as EKS98 are based on
the assumption that the nPDFs differ from the parton distributions in
the free proton but obey the same DGLAP evolution
\cite{Eskola:1998df}. Therefore, the transverse momentum distribution
of heavy bosons at the LHC in PbPb 
and $p$Pb
collisions can further
test our understanding of the nPDFs.

In summary, higher-twist nuclear effects appear to be negligible in
$Z^{0}$ production at LHC energies.  We have demonstrated that the
scale dependence of shadowing 
may lead to unexpected
phenomenology of shadowing at these energies. Overall, the $Z^0$
transverse momentum distributions can be used as a precision test for
leading-twist pQCD in the TeV energy region for both proton-proton
and nuclear collisions. We propose that measurements of the $Z^{0}$
spectra be of very high priority at the LHC.

\vspace{0.1in}





\vspace{0.3cm}
\subsection{Jet and dijet rates in $pA$ collisions}
\label{subsec:accardi_armesto}
{\em Alberto Accardi and N\'estor Armesto}


The study of jet production has become one of the precision tests of
QCD. Next-to-leading order (NLO) computations have been successfully
confronted with experimental
data. For example, jet production at the Tevatron \cite{Affolder:2001fa}
provides a very stringent test of available QCD Monte Carlo
codes and gives very
valuable information on the behaviour of the gluon distribution in the
proton at large $x$.

On the other hand, few studies on jets in a nuclear environment have been
performed in fixed target experiments \cite{Bromberg:fn,Stewart:1990wa}
and at colliders.
There are prospects to measure high-$p_T$ particles at RHIC but
the opportunity of detecting and studying jets with
$E_T\sim 100$ GeV, where $E_T$ is the total transverse energy of all jets, will appear only
at the LHC. Concretely, such measurements in $pA$
collisions would settle some uncertainties such as the validity
of collinear factorization
\cite{Collins:1985ue,Collins:ig,Collins:gx}, the modifications of
nPDFs inside nuclei
\cite{Arneodo:1992wf,Geesaman:1995yd}, and the existence and amount of
energy loss of fast partons inside cold nuclear matter
\cite{Baier:2000mf}.  Such issues have to be well understood
before any claim, based on jet studies, of the existence of new
physics in nucleus-nucleus collisions (hopefully QGP formation), can
be considered conclusive.  Furthermore, the study of high-$p_T$
particles and jets can help not only to disentangle the existence of
such new state of matter but also to characterize its properties, see
e.g. \cite{Gyulassy:2001nm,Wang:2002ri,Salgado:2002cd}.

For our computations we will use a NLO Monte Carlo code
\cite{Frixione:1995ms,Frixione:1997np,Frixione:1997ks}, adapted to include isospin effects and
modifications of the PDFs inside nuclei.  This code is based on the
subtraction method to cancel the infrared singularities between real
and virtual contributions.  A full explanation, a list of available
codes and detailed discussions on theoretical uncertainties in
nucleon-nucleon collisions can be found in
Ref.~\cite{Catani:2000jh}. The accuracy of our computations, limited
by CPU time, is estimated to be:
\begin{itemize}
\item 2~\% for the lowest
and 15~\% for the highest $E_T$-bins of the transverse energy distributions.
\item 3~\% for the pseudorapidity distributions.
\item
20~\% for the least populated and 3~\% for the most populated
bins of the dijet distributions of the angle between the two hardest
jets.
\end{itemize}
The results will be presented in the LHC lab frame, so that
for $p$Pb collisions we have a 7 TeV proton
beam on a 2.75 TeV Pb beam, see 
Sec. \ref{subsec:morsch} on the experimental parameters in $pA$ collisions.
All the energies will be given per
nucleon and, in order to compare with the $pp$ case, all cross sections
will be presented per nucleon-nucleon pair, i.e. divided by $AB$.

Unless explicitly stated, we will use the MRST98 central
gluon distribution \cite{Martin:1998sq} with the EKS98
parameterization \cite{Eskola:1998iy,Eskola:1998df} and identical
factorization and
renormalization scales $\mu=\mu_F=\mu_R=E_T/2$.
We will employ 
the $k_T$-clustering algorithm for jet reconstruction
\cite{Catani:1993hr,Ellis:tq} with $D=1$ which is
more sound on theoretical grounds than the cone algorithm
\cite{Ellis:1991xa,Sterman:1977wj,Bethke:1988zc}.
The kinematical regions we
consider are:
\begin{itemize}
\item $|\eta_i|<2.5$, where $\eta_i$ is the pseudorapidity of the jet,
corresponding to
the acceptance of the central part of the CMS detector.
\item $E_{Ti}>20$ GeV for a single jet in the pseudorapidity distributions,
to ensure the validity
of perturbative QCD.
\item $E_{T1}>20$ GeV and $E_{T2}>15$ GeV for the hardest
and next-to-hardest jets entering the CMS acceptance in the $\phi$-distributions,
where $\phi$ is the angle
between these two jets.
\end{itemize}
Please note that, even in the absence of nuclear effects, the
$\eta_i$-distributions may be asymmetric with respect to $\eta_i=0$ since
we present our results in the lab frame, not the
center-of-mass frame.

Asymmetric $E_{Ti}$-cuts have been imposed on the $\phi$ distributions
to avoid the generation of large
logarithms at certain points
in phase space, see Ref. \cite{Frixione:1997ks}. Also, the results near $\phi=\pi$ are
unreliable \cite{Frixione:1997ks}, because they require all-order resummation,
not available in the code we are using.

We caution that a comparison of our results with data is not straightforward.
Comparison with $AB$ data will be even more problematic, see the jet Chapter
for details.
Several physical effects are not included in our
computations. For example, the so-called underlying event (i.e. the soft
particle production that coexists with the hard process) is not
considered here. It may cause difficulties with jet reconstruction and
increase the uncertainties in jet-definition algorithms. This effect
has been considered in antiproton-proton collisions
at the Tevatron \cite{Huston:zr,Field:2000dy} but its estimate is model
dependent, relying on our limited
knowledge of soft multiparticle production. In
collisions involving nuclei, the situation is even worse since our
knowledge is more limited. The only way to take this into account
would be to use available Monte Carlo simulations for the full $pA$ event
\cite{Armesto:2000xh,aliceppr}, modified to include the NLO
contributions.
Other processes which might need to be taken into account more
carefully
in our computations are multiple hard parton collisions
\cite{Accardi:ur,Accardi:2001ih,Ametller:1985tp}. They may be divided
into two classes:
\begin{itemize}
\item[1)] Disconnected collisions: 
more than one independent parton-parton collisions in the same event, each
one
producing a pair of high-$p_T$ jets.
This has been studied at the Tevatron
\cite{Abe:1997xk,Abe:1997bp} for single $p\bar p$
collisions,
but its quantitative explanation and the
extension to nuclear collisions is model dependent. Simple
estimates of the influence of disconnected collisions on jet production in
$pA$ show it to be
negligible\footnote{We make a simple estimate of the average number
of binary nucleon-nucleon collisions $\langle n \rangle$ associated with jet production for
$E_{Ti}>E_{T0}$.
In the Glauber model \cite{Capella:1981ju,Braun:me}, one obtains:
\[ \displaystyle
  \langle n \rangle (b,E_{T0})\,\frac{d\sigma_{pA}}{d^2b}(E_{T0})
    = AT_A(b) \, \sigma_{pp}(E_{T0}) \ ,
\]
where $b$ is the impact
parameter and the nuclear profile function $T_A(b)$ is normalized to unity.
The cross sections
are $\sigma(E_{T0})\equiv \sigma_{pp}(E_{T0})=\int_{E_{T0}}^\infty dE_{Ti}
\ d\sigma_{pp}(E_{T})/dE_{Ti}$, the jet cross section for jets
with $E_{Ti} > E_{T0}$ in $pp$ collisions and $\sigma_{pA}(b,E_{T0})\equiv
d\sigma_{pA}(E_{T0})/d^2b=
1-[1-T_A(b)\sigma_{pp}(E_{T0})]^A$, the inelastic jet production cross section
for
$E_{Ti} > E_{T0}$ in $pA$ collisions as a function of $b$.
In both cases the cross sections are integrated over the pseudorapidity
acceptance.
Taking $\sigma(E_{T0})=527$, 1.13 and 0.046 $\mu$b as representative values
in $p$Pb collisions at 8.8 TeV for $E_{T0}=20$, 100 and 200 GeV respectively,
see Fig. \ref{pafig5-7},
$\langle n \rangle =1.0$ for all $E_{T0}$, both for minimum
bias collisions (i.e. integrating $T_A(b)$ and $d\sigma_{pA}(b,E_{T0})/d^2b$
over $b$ between $b=0$ and $\infty$) and for central
collisions (e.g. integrating $0<b<1$ fm). Thus
the contribution of multiple hard scattering coming from
different nucleon-nucleon collisions seems to be negligible.}.
This is not the case for $AB$ collisions, see the jet Chapter
of this report.
\item[2)] Rescatterings:
a high-$p_T$ parton may undergo several hard collisions before
hadronizing into a jet, mimicking a single higher-order process.
Quantitative descriptions of the
modification of the jet $E_{Ti}$ spectrum are model
dependent. However, at the LHC the effect is expected to be very small,
almost negligible at the transverse energies considered here
(see Sec. \ref{subsec:accardi} on the Cronin effect in $pA$ collisions).
%
\end{itemize}
Finally, no centrality dependence has been studied. It is not
clear how to implement such dependence in theoretical computations
because both the number of nucleon-nucleon collisions
and the modification of the nPDF
may be centrality dependent.

\begin{figure}[!]
\begin{center}
\includegraphics[width=17cm,bbllx=0pt,bblly=30pt,bburx=560,bbury=545pt]
{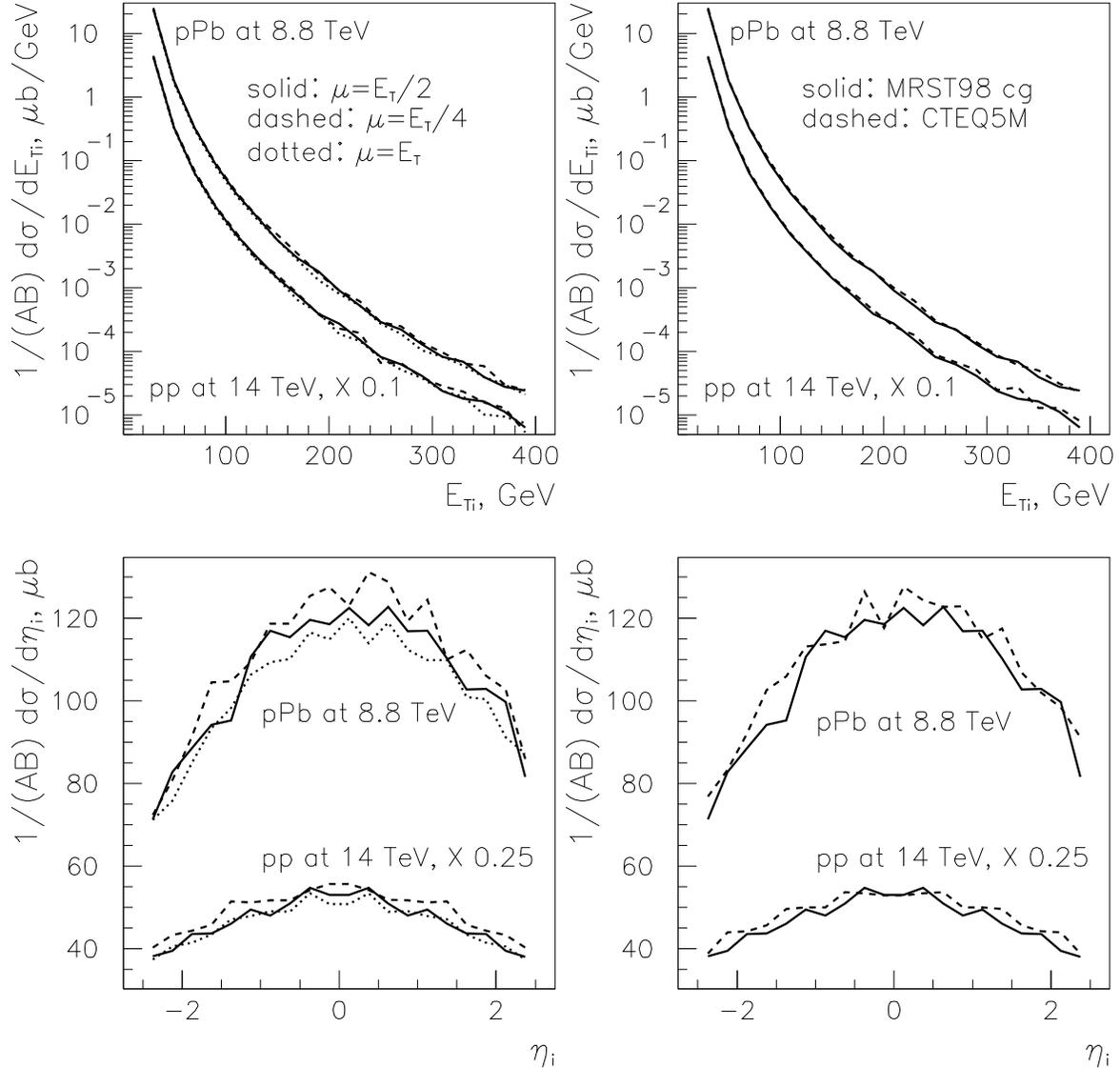}
\caption{{\it Left-hand side}:
Scale dependence of jet cross sections [we use $\mu=E_T/2$ (solid),
$\mu=E_T/4$ (dashed) and $\mu=E_T$ (dotted) lines] as a function of $E_{Ti}$
for $|\eta_i|<2.5$ (upper plot) and  $\eta_i$
for $E_{Ti}> 20$ GeV (lower plot).  {\it Right-hand side}:
PDF dependence of jet cross sections, MRST98 central
gluon (solid) and CTEQ5M (dashed), as a function of $E_{Ti}$
for $|\eta_i|<2.5$ (upper plot) and  $\eta_i$
for $E_{Ti}> 20$ GeV (lower plot).
In each plot, results for $p$Pb
at 8.8 TeV (upper lines) and $pp$ at 14 TeV (lower lines, scaled
by 0.1 in the  upper plots and by 0.25 in the lower plots) are
shown. Unless otherwise stated, default options are used, see text.}
\label{pafig1-2}
\end{center}
\end{figure}

\subsubsection{$d\sigma/dE_T$ and $d\sigma/d\eta$ for jets at large $E_T$}

\subsubsection*{Uncertainties}

\begin{figure}[!]
\begin{center}
\includegraphics[width=17cm,bbllx=0pt,bblly=30pt,bburx=560,bbury=545pt]
{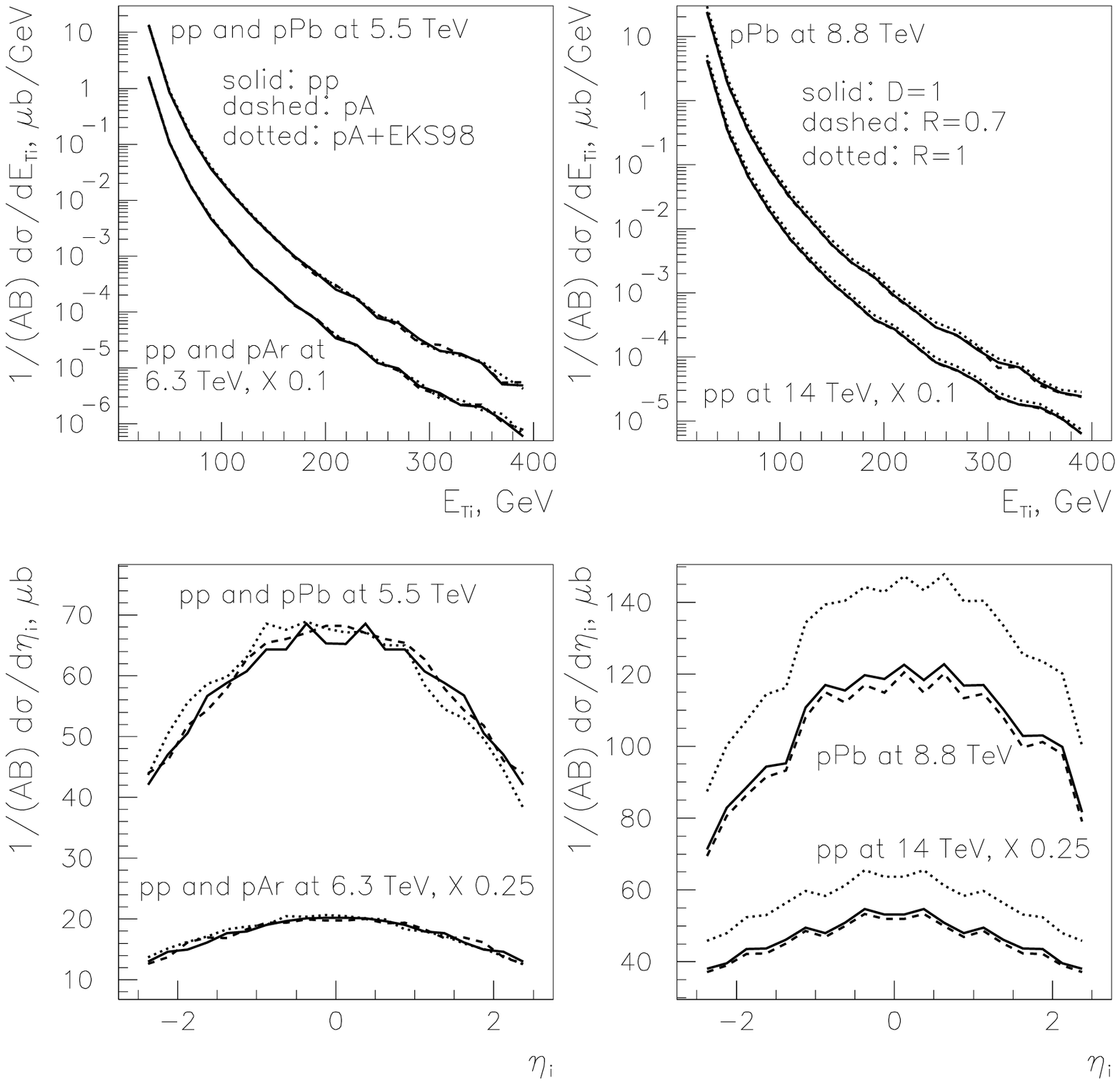}
\caption{{\it Left-hand side}:
Isospin and nuclear PDF dependence of jet cross sections [we give the $pp$
results
(solid), $pA$ results without
(dashed) and $pA$ results with EKS98
(dotted)] as a function of $E_{Ti}$
for $|\eta_i|<2.5$ (upper plot) and
$\eta_i$ for $E_{Ti}> 20$ GeV (lower plot). In each
plot,
results for  $pp$ and $p$Pb at 5.5 TeV (upper lines) and $pp$ and $p$Ar at
6.3 TeV (lower
lines, scaled by 0.1 in the upper plot and by 0.25 in the
lower plot) are shown.
{\it Right-hand side}: Dependence of jet cross sections
on the jet reconstruction algorithm
[we show the $k_T$-algorithm
with $D=1$ (solid), the cone algorithm with $R=0.7$ (dashed)  and the cone
algorithm with $R=1$ (dotted)]
as a function of $E_{Ti}$
for $|\eta_i|<2.5$ (upper plot) and
$\eta_i$ for $E_{Ti}> 20$ GeV (lower plot). In each
plot,
results for $p$Pb at 8.8 TeV (upper lines) and $pp$ at 14 TeV (lower
lines, scaled by 0.1 in the  upper plot and by 0.25 in the
lower plot) are shown. Unless otherwise stated, default options are
used, see text.}
\label{pafig3-4}
\end{center}
\end{figure}

A number of theoretical uncertainties in $pp$ and $p$Pb collisions are
examined in Figs. \ref{pafig1-2} and \ref{pafig3-4}.
Varying the scale between $E_T/4$ and $E_T$
gives differences of order $\pm 10$~\%, the smaller scale giving the
larger result. We have estimated the uncertainty due to the PDF by
using CTEQ5M
\cite{Lai:1999wy}. It gives a $\sim 5$~\% larger results than the default
MRST98 central gluon \cite{Martin:1998sq}.
The effect of isospin alone (obtained from
the comparison of $pp$ and $p$Pb without any modification of nPDF
at the same energy per nucleon) is very small, as one
would expect from $gg$ dominance.
The effect of nPDFs, estimated
using EKS98 \cite{Eskola:1998iy,Eskola:1998df}, is
also small. However it produces an extra asymmetry (an excess in the region
$\eta_i<0$) in $pA$ collisions at
reduced energy (e.g. $p$Pb at 5.5 TeV),
which
disappears at maximum energy (e.g. $p$Pb at 8.8 TeV), see Fig.
\ref{pafig3-4} (bottom-left) and Fig. \ref{pafig5-7} (bottom-right),
due to the asymmetry of the projectile and target momentum in the
LHC lab frame.
Finally, there is some uncertainty due to the choice of jet-finding algorithm.
The cone algorithm
\cite{Ellis:1991xa,Sterman:1977wj,Bethke:1988zc} with $R=0.7$ 
slightly reduces the cross section $\sim 1\dots2$~\% compared to the
$k_T$-clustering algorithm
\cite{Catani:1993hr,Ellis:tq} with $D=1$. The cone with $R=1$ gives 
results $\sim 15$~\% larger than our default choice.

\begin{figure}[!]
\begin{center}
\includegraphics[width=16cm,bbllx=0pt,bblly=30pt,bburx=560,bbury=545pt]
{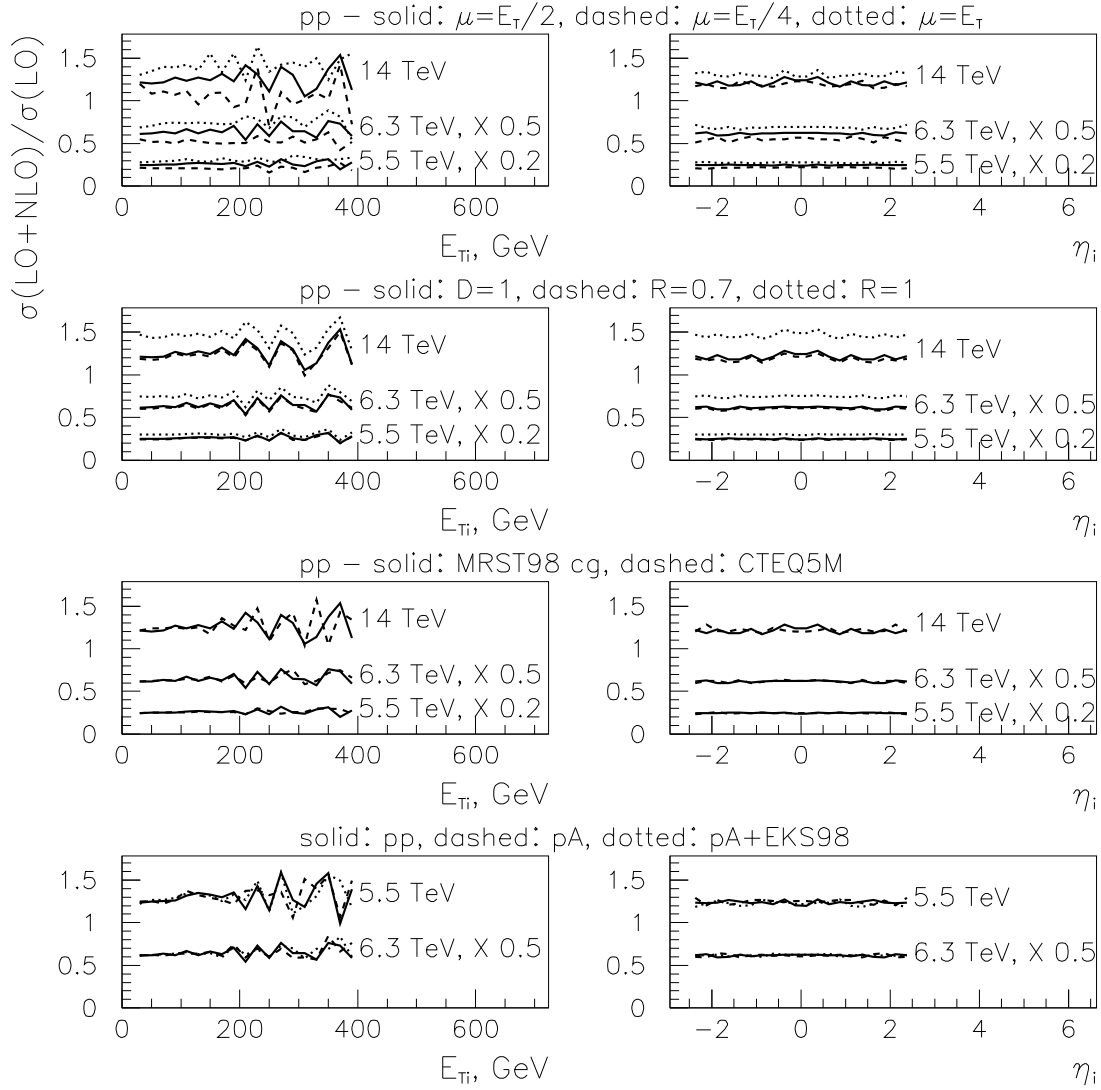}
\caption{The $K$-factors
as a function of $E_{Ti}$ for $|\eta_i|<2.5$ (left-hand side)
and $\eta_i$ for $E_{Ti}> 20$ GeV (right-hand side).
From top to bottom:
scale dependence
($\mu=E_T/2$: solid; $\mu=E_T/4$: dashed;
$\mu=E_T$: dotted); dependence
on the jet reconstruction algorithm
($k_T$-algorithm
with $D=1$: solid; cone algorithm with $R=0.7$: dashed; cone
algorithm with $R=1$: dotted); nucleon PDF dependence (MRST98
central gluon: solid; CTEQ5M: dashed lines); and isospin and nuclear
PDF dependence ($pp$:
solid; $pA$ without:
dashed; $pA$ with EKS98:
dotted).
{\it Six upper plots}: $pp$ at 14, 6.3
(scaled by 0.5) and 5.5 (scaled by 0.2) TeV. {\it
Two bottom plots}:
$pp$ and $p$Ar
at 6.3 TeV (scaled
by 0.5), and $pp$ and $p$Pb at 5.5 TeV.
Unless otherwise stated, default options are
used, see text.}
\label{kfact}
\end{center}
\end{figure}

Finally, we discuss the ratio of the NLO to LO cross sections,
the
$K$-factor. In Fig. \ref{kfact} the $K$-factor is calculated
for different collision systems and energies, varying the inputs.
For our default options, the ratio 
is  
quite constant, $\sim 1.2$, over all $E_{Ti}$ and $\eta_i$ for
all energies. The
$\mu$-dependence results in
$K \sim 1.35$ for $\mu=E_T$ and $\sim 1.15$ for $\mu=E_T/4$. The
cone algorithm with $R=0.7$ gives results very similar to those
obtained with the $k_T$-clustering algorithm with $D=1$, while a cone
with $R=1$ produces $K \sim 1.45$. (Note that at LO the choice of
jet-finding algorithm has no influence on the results.)  Finally,
the choice of PDF,
isospin and
modifications of the nPDFs do not sizably influence the $K$-factor.

\begin{table}[thb]
\begin{center}
\caption{Luminosities ${\cal L}$ (cm$^{-2}$s$^{-1}$) and 
${\cal L}\times10^6$~s ($\mu{\rm b}/AB$) for different
collision systems at the LHC.  The numbers of expected jets and dijets in a
certain kinematical range are obtained by multiplying the last
column by the cross sections given in Figs. \ref{pafig1-2},
\ref{pafig3-4}, \ref{pafig5-7} (jets) and
\ref{pafig8}, \ref{pafig9} (dijets).}
\vskip0.2cm
\begin{tabular}{|c|c|c|c|}
\hline
System & $\sqrt{s_{NN}}$ (TeV) & $\cal{L}$ (cm$^{-2}$s$^{-1}$)
& ${\cal L}\times10^6$~s ($\mu$b/($AB$)) \\
\hline
$pp$ & 14 & $10^{34}$ & $10^{10}$ \\
\hline
$pp$ & 14 & $3\cdot 10^{30}$ & $3\cdot 10^6$ \\
\hline
$p$Ar & 9.4 & $4\cdot 10^{30}$ & $1.6\cdot 10^8$ \\
\hline
$p$Ar & 9.4 & $1.2\cdot 10^{29}$ & $4.8\cdot 10^6$ \\
\hline
$p$Pb & 8.8 & $10^{29}$ & $2.1\cdot 10^7$ \\
\hline
\end{tabular}
\label{armesto:table1}
\end{center}
\end{table}

\subsubsection*{Results}

\begin{figure}[bth]
\vspace{-0.5cm}
\begin{center}
\includegraphics[width=12.5cm,bbllx=0pt,bblly=30pt,bburx=545,bbury=535pt]
{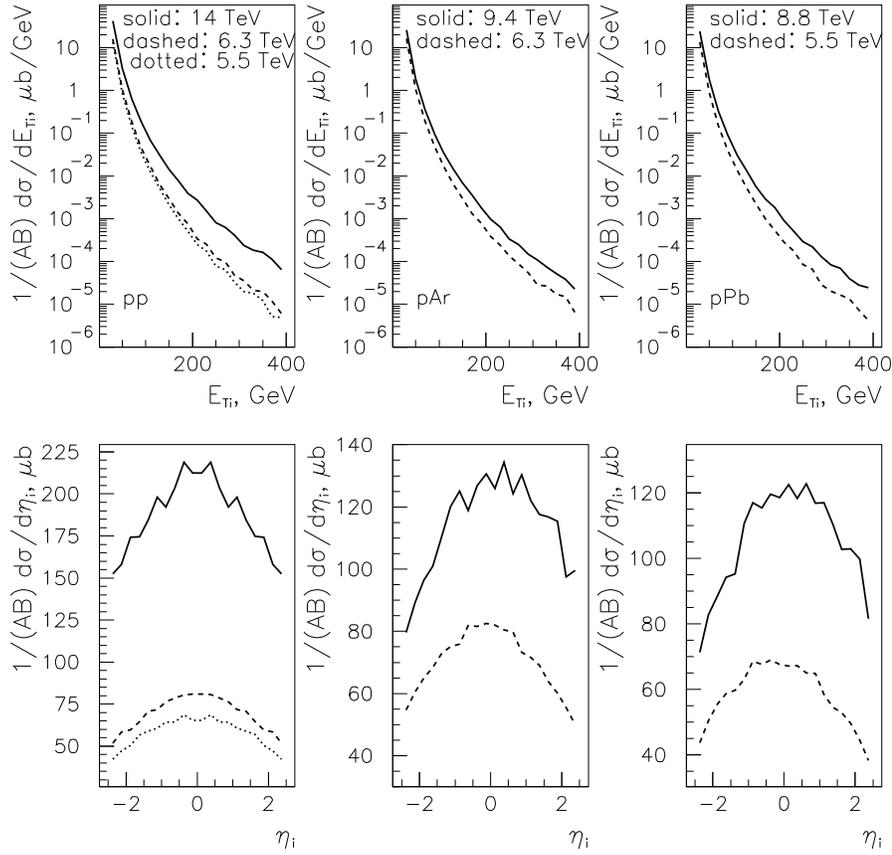}
\caption{Energy dependence of jet cross sections
as a function of $E_{Ti}$ for $|\eta_i|<2.5$ (upper plots)
and $\eta_i$ for $E_{Ti}> 20$ GeV (lower
plots). {\it Left-hand side}: $pp$ collisions at 14 (solid),
6.3 (dashed) and 5.5 TeV (dotted). {\it Center}:
$p$Ar collisions at 9.4 (solid) and 6.3 TeV (dashed).
{\it Right-hand side}: $p$Pb collisions at 8.8 (solid)
and 5.5 TeV (dashed). Default options are used, see text.}
\label{pafig5-7}
\end{center}
\end{figure}
\begin{figure}[!]
\begin{center}
\vspace{-0.5cm}
\includegraphics[width=10.5cm,bbllx=0pt,bblly=30pt,bburx=560,bbury=545pt]
{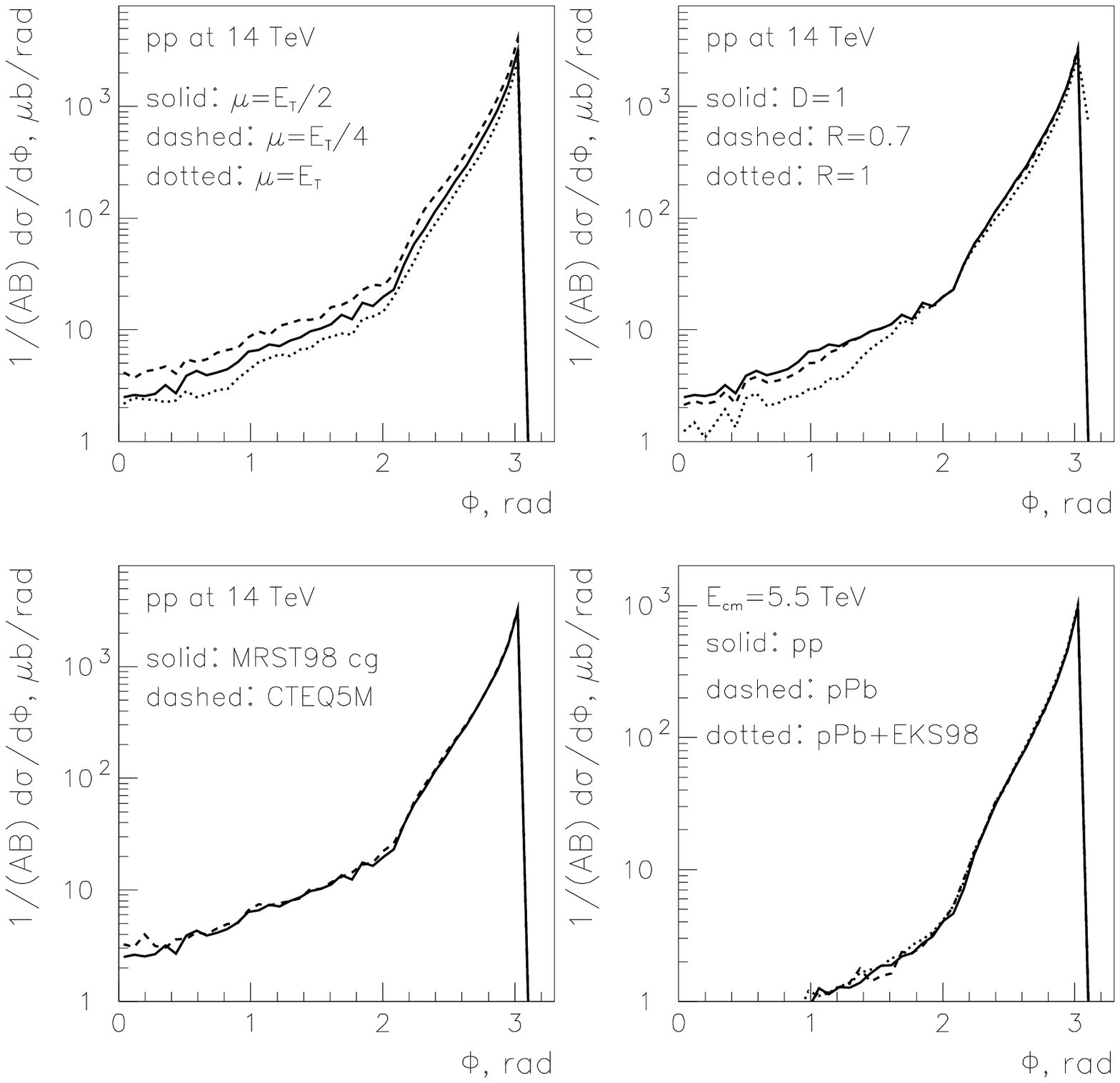}
\vspace{-0.3cm}
\caption{Uncertainties in dijet cross sections
as a function of $\phi$ for $E_{T1}>20$ GeV,
$E_{T2}>15$ GeV and $|\eta_1|,|\eta_2|<2.5$ in $pp$ collisions at 14 TeV.
{\it Upper left}: Scale
dependence;
{\it Upper right plot}:
dependence on jet
reconstruction algorithm;
{\it Lower left}:
PDF dependence
in $pp$ collisions at 14 TeV.
{\it Lower right plot} shows the
isospin and nuclear
PDF dependence in $pp$ and $p$Pb collisions at 5.5 TeV.
Unless otherwise stated, default
options are
used, see text.}
\label{pafig8}
\end{center}
%
\begin{center}
\vspace{-0.7cm}
\includegraphics[width=11.0cm,bbllx=0pt,bblly=30pt,bburx=560,bbury=545pt]
{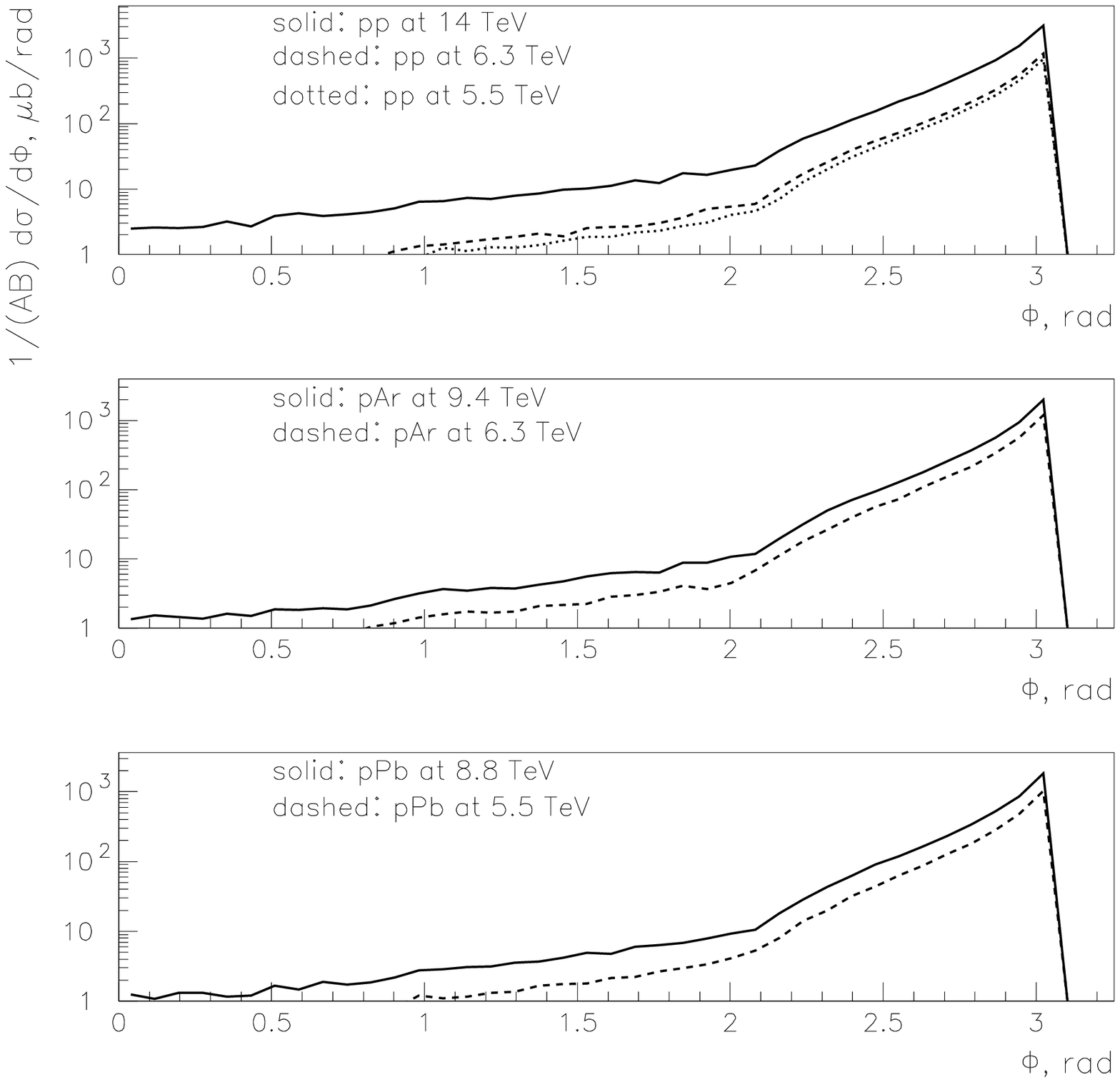}
\vspace{-0.3cm}
\caption{Energy dependence of dijet cross sections
as a function of $\phi$ for $E_{T1}>20$ GeV,
$E_{T2}>15$ GeV and $|\eta_1|,|\eta_2|<2.5$. {\it Upper}: $pp$
collisions;
{\it Center}: $p$Ar collision;
{\it Lower}: $p$Pb collisions.
Default options
are
used, see text.}
\label{pafig9}
\end{center}
\end{figure}

The expected LHC luminosities for different collisions are shown
in Table~\ref{armesto:table1}, along with the one month ($10^6$ s)
integrated luminosity expressed
in units of $\mu{\rm b}/(AB)$.  Using this table and the
cross sections for inclusive single jet and dijet production
in the figures, it is possible to extract the number of expected jets
(Figs.~\ref{pafig1-2}, \ref{pafig3-4} and \ref{pafig5-7}) or dijets
(Figs.~\ref{pafig8} and \ref{pafig9}).
For example,
from the solid curve on the upper left-hand side of Fig.~\ref{pafig5-7} and
${\cal L}=10^{34}$ cm$^{-2}$s$^{-1}$, we expect
$10^{10}$ jets with $E_{Ti}\sim 70$ GeV from a cross section of
1 $\mu$b/($AB$), and $10^6$ jets
with $E_{Ti}\sim 380$ GeV from a cross section of
$10^{-4}$ $\mu$b/($AB$), in our $\eta_i$ range.

Looking at the results of
Fig. \ref{pafig5-7}, it is evident that samples of
$\sim 10^3$ jets are feasible,
up to $E_{Ti}\sim 325$ GeV in one month run
at the given luminosity. Indeed,  from
Table~\ref{armesto:table1}, $10^3$ jets in $p$Pb at 8.8 TeV would
correspond to a cross section of $4.8\cdot 10^{-5}$ $\mu$b/($AB$) which
intersects the upper right solid curve in Fig. \ref{pafig5-7} at
$E_{Ti}\sim 325$ GeV.

Some conclusions can now be drawn.
First, since nuclear effects on the PDFs are
not large (at most 10~\% at $E_T>20$ GeV for Pb according to
EKS98), an extensive systematic study of the $A$-dependence of the
large-$E_T$ jet cross sections does not seem to be the most urgent need.
If, however, the nuclear effects in these cross sections turn out to be
significant, the measurements of the $A$ systematics, especially towards
the smallest values of $E_T$, would be necessary in order to understand
the new underlying QCD collision dynamics.
Second, noting the
asymmetry in the pseudorapidity distributions in $p$Pb collisions at
the reduced energy of 5.5 TeV, and to a lesser extent in $p$Ar at 6.3
TeV, when EKS98 corrections are implemented, a $pA$ run at
reduced proton energies might test the nuclear PDF effects;
a study of
$E_{Ti}$-distributions in different $\eta_i$-slices going from the
backward ($A$) to the forward ($p$) region would scan different regions in
$x$ of the ratio of the nuclear gluon distribution over that
of a free nucleon, which is so far not well constrained.
Nevertheless, this would require a detailed comparison with $pp$
results at the same energy in order to disentangle possible biases
(e.g. detector effects). If these $pp$ results were not available
experimentally, the uncertainty due to the scale dependence, which in
our computations at reduced energies is as large as the asymmetry, in
the extrapolation from 14 TeV to the $pA$ energy
should be reduced as much as possible for this study to be useful.
%

\subsubsection{High-$E_T$ dijet momentum imbalance}

Dijet distributions as a function of $\phi$, the angle between the two hardest
jets,  can test
the perturbative QCD expansion. At LO
in collinear factorization
\cite{Collins:1985ue,Collins:ig,Collins:gx}, the two jets are produced
back-to-back, so that any deviation from a peak at $\phi=\pi$
is a signal of NLO corrections. However, recall that the
results near $\phi=\pi$ are unreliable \cite{Frixione:1997ks} since
the NLO corrections become large and negative,
requiring an all-order resummation in this region.

\subsubsection*{Uncertainties}

In Fig. \ref{pafig8}, the same uncertainties examined for $E_{Ti}$ and $\eta_i$
are studied in the dijet cross
sections as a function of $\phi$. First, the smaller
the scale, the larger the results, so that the results for $\mu=E_T$ are
up to $\sim 50$~\% smaller than for $\mu=E_T/4$. Second, the PDF choice
has less than a 10~\% effect
when comparing CTEQ5M with MRST98 central gluon. Third, the
variation due to isospin, obtained from the comparison of $pp$
and $p$Pb without nuclear modifications of PDFs,
is again negligible, and the effect of including
EKS98
\cite{Eskola:1998iy,Eskola:1998df} is also small,
$\sim 3$~\%. Finally, the jet-finding algorithm gives an uncertainty
of order 20~\% for a
cone with $R=1$, $R=0.7$ and the $k_T$-clustering
algorithm with $D=1$.

\subsubsection*{Results}

Our results for $pp$ and $pA$ collisions at different energies are presented in
Fig. \ref{pafig9}.
It is clear that the dijet momentum imbalance should
be measurable provided jet reconstruction is possible in the
nuclear environment and no other physics contribution, such the underlying
event or multiple hard parton scattering, interferes, spoiling the
comparison with experimental data.
Only extensive studies using Monte
Carlo simulations including full event reconstruction
will be able to clarify whether such measurements
are feasible, see the Section on Jet Detection at CMS
in the jet Chapter.

%

\vspace{0.2cm}
\subsection{Direct photons}
For direct photons, we refer to the photon Chapter by P. Aurenche 
{\em et al} in this report.

\eject

\subsection{Quarkonium transverse momentum distributions}
\label{section441}
{\em Ramona Vogt}

Since a major experimental focus at the LHC will be quarkonium suppression in
heavy ion collisions, an important plasma probe \cite{Matsui:1986dk} 
already at the SPS
\cite{Abreu:wb}, it is important that systematic studies also be undertaken at
the LHC.  These systematics are especially valuable for the $\Upsilon$ family 
where high statistics measurements of $\Upsilon$
production will be available in $pp$, $pA$, and $AA$ interactions for the
first time at the LHC.

Studies of quarkonium production in $pA$ interactions are particularly relevant
for LHC because the energy dependence of cold nuclear matter effects such as 
nucleon absorption are not fully understood and need to be checked with
measurements. The existence of nuclear effects beyond the modifications of the
parton densities in the nucleus argues that quarkonium production is not a
good reaction by which to study the nuclear gluon distribution.  However,
the nuclear gluon distribution can be obtained through careful 
measurements of heavy flavor production in $pp$ and $pA$ interactions and 
used as an input to quarkonium production.  Thus 
quarkonium is still a benchmark process to test cold matter effects in $pA$
interactions at the LHC.
It is important to study the shape of the quarkonium $p_T$ distributions in
$pA$ interactions because plasma effects on the $p_T$ distributions 
are expected to persist to high $p_T$ \cite{Gunion:1996qc,Vogt:cu}.  Detailed
comparisons of the $p_T$ dependencies of different quarkonium states in $AA$
collisions will
reveal whether or not the ratios of {\it e.g.}\ $\psi'/J/\psi$ and
$\Upsilon'/\Upsilon$ and $\Upsilon''/\Upsilon$ are independent of $p_T$, as
predicted by the color evaporation model 
\cite{Barger:1979js,Barger:1980mg,Gavai:1994in}
or vary over the $p_T$ range, as expected from nonrelativistic QCD
\cite{Bodwin:1994jh}.  Indeed quarkonium studies in $pA$ may provide a 
crucial test
of the quarkonium production process because it is likely to be possible to
separate the $P$ and $S$ states only in $pA$ interactions where the
accompanying decay photons can be found.  

Two approaches have been successful in describing quarkonium
production phenomenologically---the color evaporation model (CEM) 
\cite{Barger:1979js,Barger:1980mg}
and nonrelativistic QCD (NRQCD) \cite{Bodwin:1994jh}.  
Both are briefly discussed.

In the CEM, any quarkonium production cross section is some fraction $F_C$ of 
all $Q \overline Q$ pairs below the $H \overline H$ threshold where $H$ is
the lowest mass heavy hadron containing $Q$.  Thus the CEM cross section is
simply the $Q \overline Q$ production cross section with a mass cut imposed but
without any contraints on the 
color or spin of the final state.  The produced $Q
\overline Q$ pair then neutralizes its color by
interaction with the collision-induced color field---``color evaporation''.
The $Q$ and the $\overline Q$ either combine with light
quarks to produce heavy-flavored hadrons or bind with each other 
in a quarkonium state.  The additional energy needed to produce
heavy-flavored hadrons is obtained nonperturbatively from the
color field in the interaction region.
The yield of all quarkonium states
may be only a small fraction of the total $Q\overline 
Q$ cross section below the heavy hadron threshold, $2m_H$.
At leading order, the production cross section of quarkonium state $C$ is
\begin{eqnarray}
\sigma_C^{\rm CEM} = F_C \sum_{i,j} \int_{4m_Q^2}^{4m_H^2} d\hat s \int dx_1 
dx_2~f_i^p(x_1,Q^2)~f_j^p(x_2,Q^2)~ \hat\sigma_{ij}(\hat s)~\delta(\hat 
s-x_1x_2s)\, \, , \label{sigtil}
\end{eqnarray} 
where $ij = q \overline q$ or $gg$ and $\hat\sigma_{ij}(\hat s)$ is the
$ij\rightarrow Q\overline Q$ subprocess cross section.
The fraction $F_C$ must be universal so that, once it is fixed by data, the
quarkonium production ratios should be constant as a function of $\sqrt{s}$,
$x_F$ and $p_T$.  The actual value of $F_C$ depends on the heavy quark mass, 
$m_Q$, the scale, $Q^2$, the parton densities and the order of the calculation.

Of course the leading order calculation in Eq.~(\ref{sigtil}) is insufficient
to describe high $p_T$ quarkonium production since the $Q \overline Q$ pair
$p_T$ is zero at LO.  Therefore, the CEM was 
taken to NLO \cite{Gavai:1994in,Schuler:1996ku} using the exclusive 
$Q \overline Q$ hadroproduction code of Ref.~\cite{Mangano:jk}.  
At NLO in the CEM, the process
$gg \rightarrow g Q \overline Q$
is included, providing a good
description of the quarkonium $p_T$ distributions at the Tevatron
\cite{Schuler:1996ku}. 
The values of $F_C$ for the individual charmonium and bottomonium states 
have been calculated from
fits to the total $J/\psi$ and inclusive $\Upsilon$ 
data combined with relative cross sections
and branching ratios, see Refs.~\cite{Gunion:1996qc,Digal:2001ue} 
for details.  
The $pp$ total cross sections at 5.5 TeV are shown in 
Tables~\ref{psitab} and
\ref{upstab} in Section~\ref{section531}, 
along with the fit parameters for the parton densities, quark
masses and scales given for $Q \overline Q$ production in Table~\ref{qqbtab}
in Section~\ref{section431}.   

NRQCD was motivated by the success of potential models in determining the
quarkonium mass spectra.  It was first applied to high $p_T$ quarkonium
production when it was clear that the leading color singlet contribution
severely underestimated direct $J/\psi$ and $\psi'$ production at the Tevatron
\cite{Abe:1997jz,Abe:1997yz}.
NRQCD describes quarkonium production as an expansion in powers of $v$,
the relative $Q$-$\overline Q$ velocity.  Thus
NRQCD goes beyond the leading color singlet state to include color
octet production.  Now the initial $Q \overline Q$ quantum numbers do not have
to be the same as in the final quarkonium state because an arbitrary number of
soft gluons with $E<m_Q v$ can be emitted before bound state formation and
change the color and spin of the $Q \overline Q$.  These soft gluon emissions
are included in the hadronization process.

The cross section of quarkonium state $C$ in NRQCD is
\begin{eqnarray}  
\sigma_C^{\rm NRQCD} = \sum_{i,j}  \sum_n
\int_0^1 dx_1 dx_2 f_i^p(x_1,Q^2)f_j^A(x_2,Q^2)
C^{ij}_{Q \overline Q \, [n]}(Q^2)
\langle {\cal O}_n^C \rangle \, \, , \label{signrqcd} 
\end{eqnarray}
where the partonic cross section is the product of perturbative
expansion coefficients, $C^{ij}_{Q \overline Q \, [n]}(Q^2)$, and
nonperturbative parameters describing the hadronization, $\langle
{\cal O}_n^C \rangle$.  The expansion coeffients can be factorized
into the perturbative part and the nonperturbative parameters because
the production time, $\propto 1/m_Q$, is well separated from the
timescale of the transition from the $Q \overline Q$ to the bound
state, $\propto 1/m_Q v$.  If $1/m_Q v \gg 1/m_Q$, the bound state
formation is insensitive to the creation process.  In principle, this
argument should work even better for bottomonium because $m_Q$ is
larger.  The color singlet model result is recovered when $v
\rightarrow 0$ \cite{Kramer:2001hh}.  The NRQCD parameters were
obtained from comparison
\cite{Cho:1995ce,Beneke:1996yw,Kniehl:1998qy,Braaten:1999qk} 
to unpolarized high 
$p_T$ quarkonium production at the Tevatron \cite{Abe:1997jz,Abe:1997yz}.  
The calculations 
describe this data rather well but fail to explain the polarization 
data \cite{Affolder:2000nn}.

Both the CEM and NRQCD can explain the Tevatron quarkonium data
because both are dominated by $gg \rightarrow g g^*$ and $gg
\rightarrow g Q \overline Q$ at high $p_T$.  
In the CEM $g^*\rightarrow Q \overline Q$ while in NRQCD,
$g^*\rightarrow Q \overline Q [ ^3S_1^{(8)}]$ but
$Q \overline Q \rightarrow Q \overline Q [ ^1S_0^{(8)},
{}^3P_J^{(8)}]$  \cite{Kramer:2001hh}.  Direct $J/\psi$
and $\psi'$ production at the Tevatron are predominantly by $
^3S_1^{(8)}$ states followed by $ ^1S_0^{(8)} + {}^3P_J^{(8)}$ states
and, while the $\chi_c$ singlet contribution does not underpredict the
yield as badly, a $ ^3S_1^{(8)}$ component is needed for the
calculations to agree with the data.  The CEM includes the octet
components automatically but the relative strengths of the octet
contributions are all determined perturbatively.  The two approaches
differ in their predictions of quarkonium polarization.  NRQCD
predicts transverse polarization at high $p_T$ \cite{Kramer:2001hh}
while the CEM predicts unpolarized quarkonium production. Neither
approach can successfully explain the polarization data
\cite{Affolder:2000nn}.

The bottomonium results at the Tevatron \cite{Abe:1995an} extend to
low $p_T$ where the comparison to NRQCD is somewhat inconclusive.
Some intrinsic $k_T$ is needed to describe the data below $p_T \sim 5$
GeV.

\begin{figure}[!p]
\setlength{\epsfxsize=0.95\textwidth}
\setlength{\epsfysize=0.4\textheight}
\centerline{\epsffile{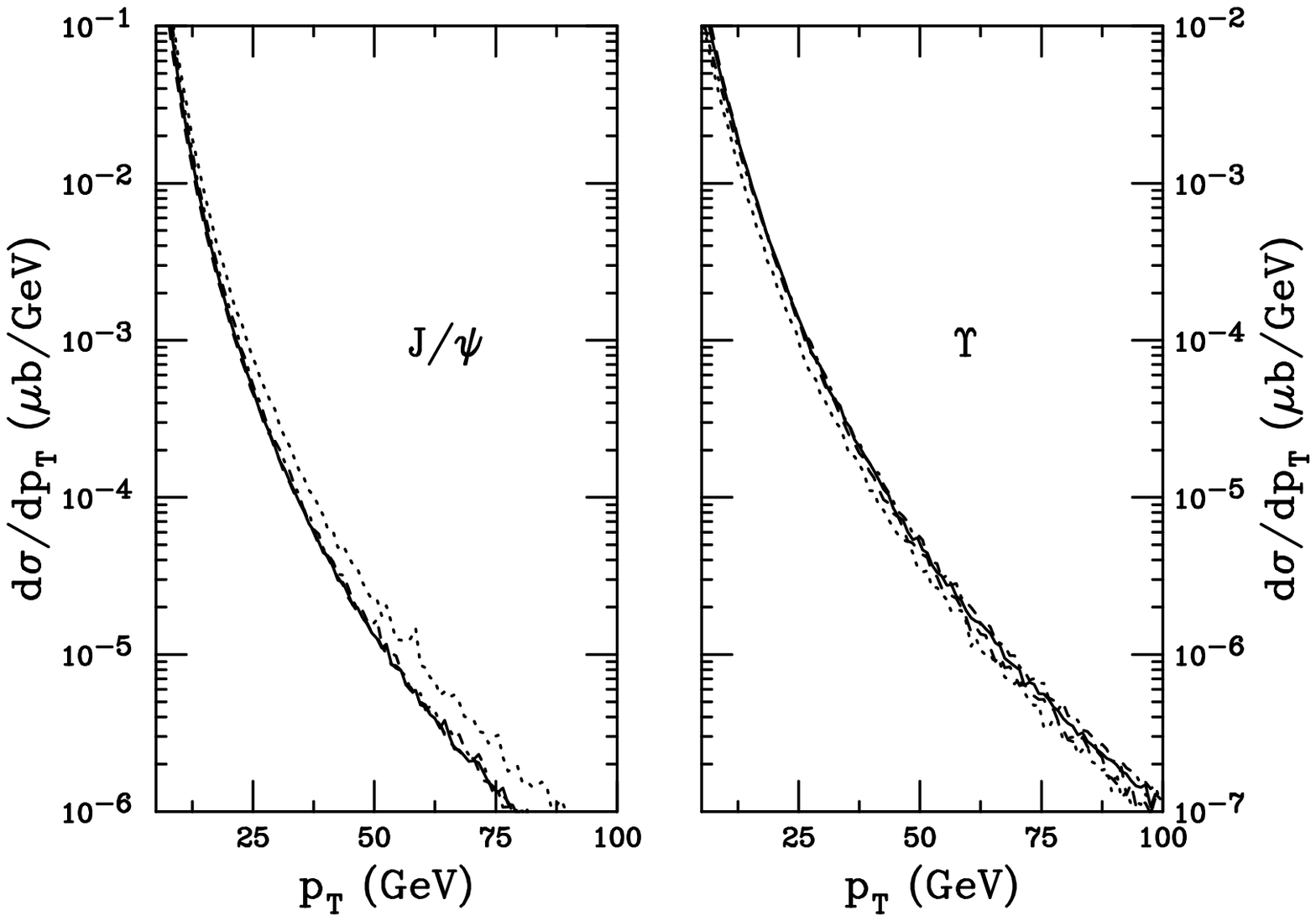}}
\caption{The $p_T$ distributions for $J/\psi$ (left)
and $\Upsilon$ (right) production assuming $\langle k_T^2 \rangle = 1$ GeV$^2$.
On the left hand side, the
solid curve employs the MRST HO distributions with $m_c = Q/2 = 1.2$ GeV,
the dashed, MRST HO with $m_c = Q = 1.4$ GeV, the dot-dashed, CTEQ 5M with
$m_c = Q/2 = 1.2$ GeV, and the dotted, GRV 98 HO with $m_c = Q = 1.3$ GeV.
On the right hand side, the
solid curve employs the MRST HO distributions with $m_b = Q = 4.75$ GeV,
the dashed, $m_b = Q/2 = 4.5$ GeV, the dot-dashed, 
$m_b = 2Q = 5$ GeV, and the dotted, GRV 98 HO with $m_b = Q = 4.75$ GeV.}
\label{ptdistpp}
\vspace{0.5cm}
\setlength{\epsfxsize=0.95\textwidth}
\setlength{\epsfysize=0.4\textheight}
\centerline{\epsffile{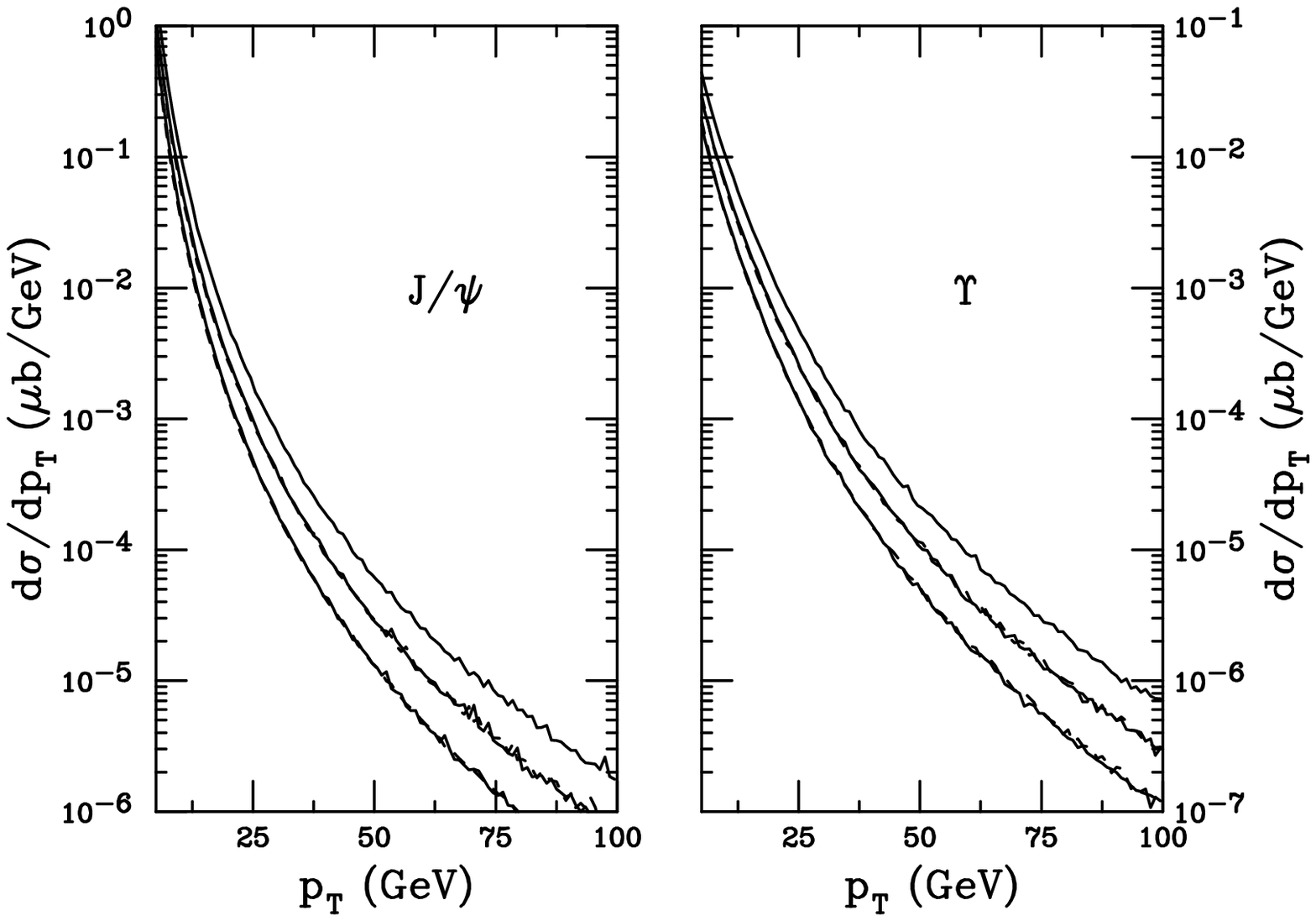}}
\caption{The $p_T$ distributions for $J/\psi$ (left)
and $\Upsilon$ (right) production assuming $\langle k_T^2 \rangle_{pp} = 1$ 
GeV$^2$
for $pp$ interactions at 5.5 (lower), 8.8 (center) and 14 (upper) TeV.
On the left hand side, we use the MRST HO distributions with 
$m_c = Q/2 = 1.2$ GeV while on the right hand side, we
use the MRST HO distributions with $m_b = Q = 4.75$ GeV.}
\label{ptdistpa}
\end{figure}

In this section, we only present the CEM $p_T$ distributions.
The NRQCD approach has not yet been fully calculated beyond LO
\cite{Kramer:2001hh}.  We include intrinsic transverse momentum
broadening, described further in Section~\ref{section532}, although it
is a rather small effect at high $p_T$.  The direct $J/\psi$ and
$\Upsilon$ $p_T$ distributions above 5 GeV in $pp$ interactions at 5.5
TeV are shown in Fig.~\ref{ptdistpp} for all the fits to the
quarkonium data, described in Section~\ref{section531}.
They all agree rather well with each other.  However, the GRV98 result is
slightly larger than the other $J/\psi$ curves at high $p_T$.
Thus changing the quark mass and scale does not have a large effect on
the $p_T$ dependence.  In general, the $J/\psi$ $p_T$ distributions
are more steeply falling than the $\Upsilon$ distributions.

Figure~\ref{ptdistpa} shows the $p_T$
distributions in $pp$ and $p$Pb interactions at 5.5 and 8.8 TeV along with
$pp$ interactions at 14 TeV.
The shadowing effect is negligible 
at high $p_T$ so that the two distributions are indistinguishable.

\clearpage


\section{PROCESSES WITH POTENTIALLY LARGE NUCLEAR EFFECTS}
\label{pA_SEC:Aeffects?}
\vspace{0.25cm}

\subsection{Drell-Yan cross sections at low $M^2$: rapidity distributions}
\label{section511}
{\em Ramona Vogt}

Low mass Drell-Yan production could provide important information on the
quark and antiquark distributions at low $x$ but still rather high $Q^2$
relative to the deep-inelastic scattering measurements.  When $M = 4, 11, 20$
and 40 GeV respectively, $x \sim 0.00073, 0.002, 0.0036$ and 0.0073 at $y=0$
and $\sqrt{s} = 5.5$ TeV.  Such low values of $x$ have been measured before
in nuclear deep-inelastic scattering but always at $Q^2 < 1$ GeV$^2$, out of
reach of perturbative analyses.  Thus, these measurements could be a benchmark
for analysis of the nuclear quark and antiquark distributions.  

As previously explained in Sec.~\ref{section411}, in $AA$
collisions, the inclusive Drell-Yan contribution is unlikely to be
extracted from the dilepton continuum which is dominated by correlated
and uncorrelated $Q\overline Q$ ($D\overline D $, $B\overline B$) pair decays
\cite{Gavin:ma,Lokhtin:2001nh}. 
However, in $pA$, the low mass Drell-Yan may be competitive with
the $Q \overline Q$ decay contributions to the dilepton spectra
\cite{Gavin:ma}.  The number of uncorrelated $Q \overline Q$ pairs is
also substantially reduced from the $AA$ case \cite{Eskola:2001gt}.
Further calculations of the heavy quark contribution to dileptons are
needed at higher masses to determine whether the Drell-Yan signal ever
dominates the continuum or not.  Perhaps the Drell-Yan and $Q \overline Q$ 
contributions can be separated using their different angular dependencies, 
isotropic for $Q \overline Q$ and $\propto (1+\cos^2 \theta)$ for 
Drell-Yan, as proposed by NA60 \cite{NA60} and discussed in 
Sec.~\ref{subsec:fai_qiu_zhang}. When $M$ is small and
$p_T$ is large relative to $M$, the $Q \overline Q$ background may be
reduced and the DY signal observable. This is discussed in the next
section.

The calculation of the Drell-Yan cross section is described in
Section~\ref{section411}.  Since the nuclear effects are larger at low
$Q^2$, we will focus on the variation of the results with scale and shadowing
parameterization.  We again make our calculations with the MRST HO
\cite{Martin:1998sq} parton densities with $Q^2 = M^2$ and the EKS98
parameterization \cite{Eskola:1998iy,Eskola:1998df} in the $\overline{\rm MS}$
scheme.  The Drell-Yan convolutions with nuclear parton densities including
isospin and shadowing effects are the same as for the $Z^0$ in the appendix
of Ref.~\cite{Vogt:2000hp}.  Only the prefactor changes from $H_{ij}^{Z^0}$ to
$H_{ij}^M$.  We present results for $pp$, Pb$p$ and $p$Pb collisions at 5.5
TeV/nucleon, the same energy as the Pb+Pb center of mass for better
comparison without changing $x$.  This is most desirable from the theoretical
point of view because the nuclear parton densities are most easily extracted
from the data.  However, as was not touched upon in Section~\ref{section411},
the desirability of this approach ultimately depends upon the machine
parameters.  The highest $p$Pb and Pb$p$ luminosities can be achieved at 8.8
TeV/nucleon.  Running the machine at lower energies will reduce the maximum
luminosity, up to a factor of 20 if only injection optics are used
\cite{Brandtpriv}, which could reduce the statistical significance of the
measurement.  This issue remains to be settled.

\begin{figure}[!p]
\setlength{\epsfxsize=0.7\textwidth}
\setlength{\epsfysize=0.4\textheight}
\centerline{\epsffile{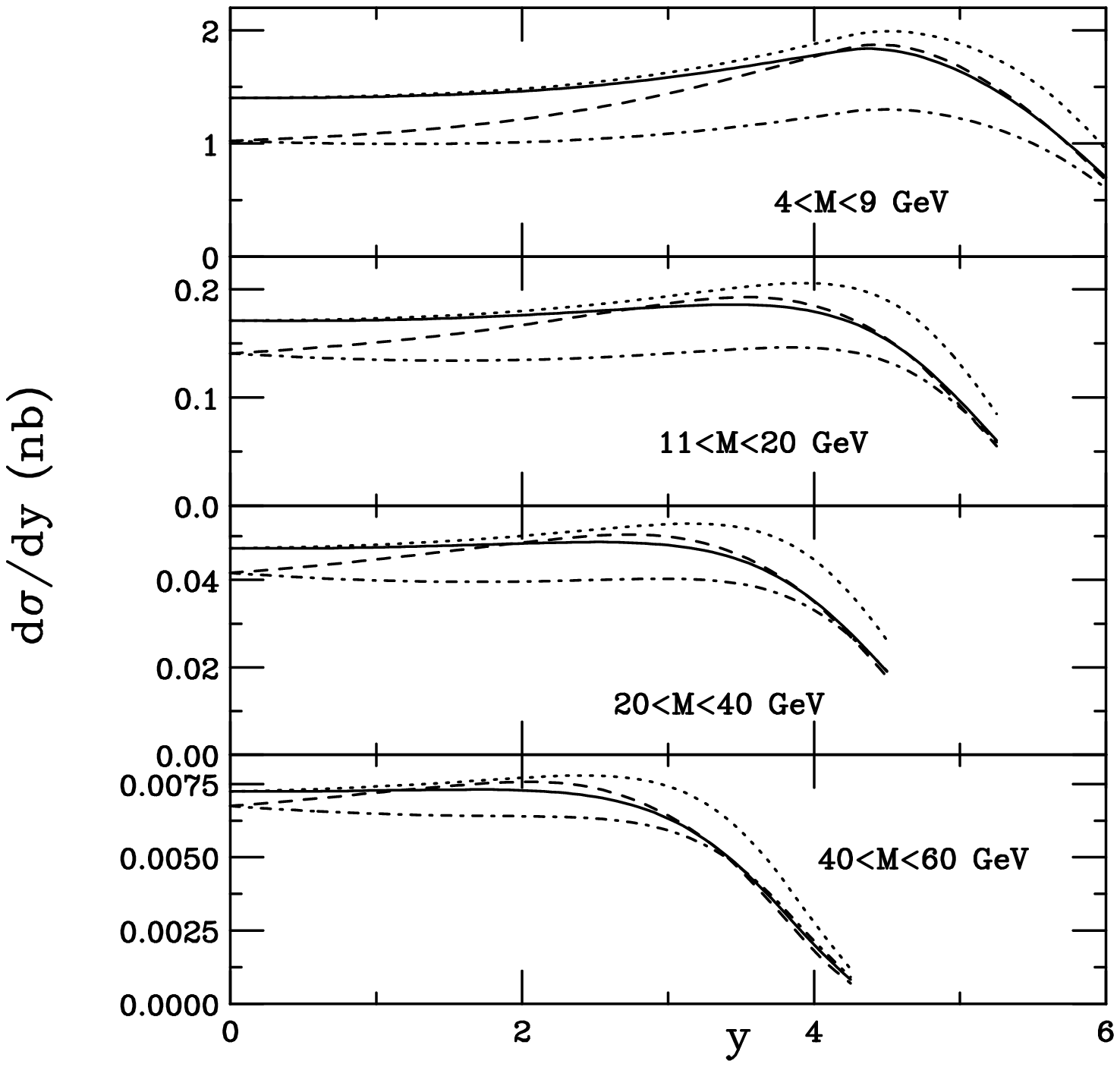}}
\caption{The Drell-Yan rapidity distributions in $p$Pb and
Pb$p$ collisions at 5.5 TeV evaluated at $Q = M$ for $4<M<9$ GeV,
$11<M<20$ GeV, $20<M<40$ GeV and $40<M<60$ GeV. 
The solid and dashed curves show the results without and with shadowing 
respectively in Pb$p$ collisions while the dotted and dot-dashed
curves give the results without and with shadowing for $p$Pb collisions.
}
\label{dyydist}
\vspace{0.5cm}
\setlength{\epsfxsize=0.7\textwidth}
\setlength{\epsfysize=0.4\textheight}
\centerline{\epsffile{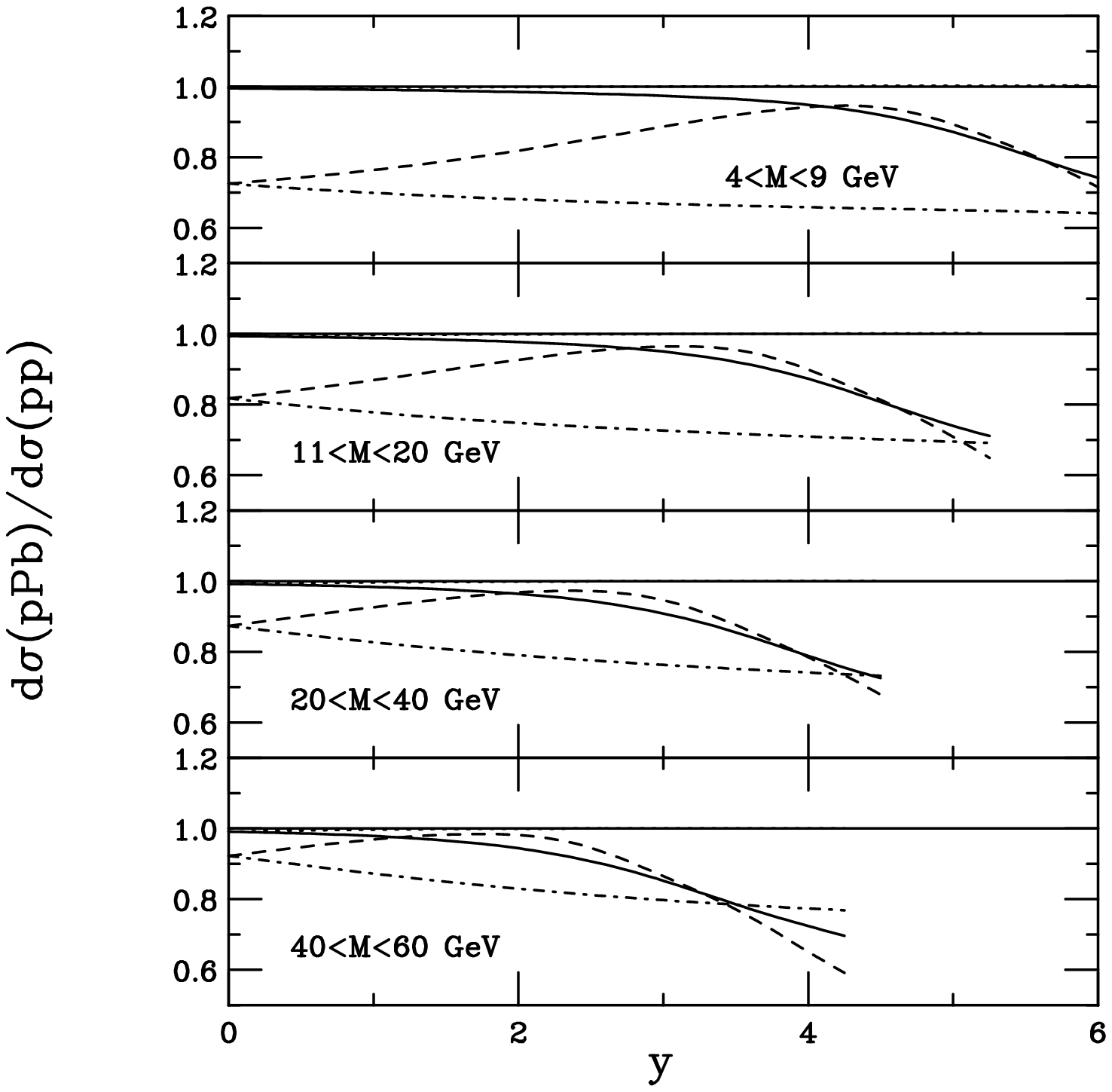}}
\caption{The ratios of $p$Pb and
Pb$p$ collisions to $pp$ collisions at 5.5 TeV evaluated at 
$Q = M$ for $4<M<9$ GeV,
$11<M<20$ GeV, $20<M<40$ GeV and $40<M<60$ GeV, as a function of rapidity. 
The solid and dashed curves show the Pb$p$/$pp$ ratios without and with 
shadowing respectively while the dotted and dot-dashed
curves are the $p$Pb/$pp$ ratios without and with shadowing.
}
\label{dypbpp}
\end{figure}

Figure~\ref{dyydist} shows the rapidity distributions in $p$Pb and Pb$p$
collisions at 5.5 TeV/nucleon in four mass bins:  $4<M<9$ GeV, $11<M<20$ GeV,
$20<M<40$ GeV and $40<M<60$ GeV.  
The shadowing
effect decreases as the mass increases, from a $\sim 25$\% effect at $y=0$
in the lowest mass bin to a $\sim 10$\% effect in the highest bin.  Isospin
is a smaller effect on low mass Drell-Yan, as is clear from Fig.~\ref{dypbpp},
where the $p$Pb/$pp$ and Pb$p$/$pp$ ratios are given. 

The $p$Pb/$pp$ ratio without shadowing is indistinguishable from unity
over the whole rapidity range.
The Pb$p$/$pp$ ratio without shadowing only differs from unity
for $y<2$.
The dashed curves in Fig.~\ref{dypbpp} represent
the Pb$p$/$pp$ ratios with EKS98 shadowing and shows the pattern of
increasing $x_1$ through the `antishadowing' region into the EMC
region.  The ratio does not increase above unity because we present
the $Ap/pp$ ratio rather than a ratio of $Ap$ with shadowing to $Ap$
without shadowing.  This difference can be easily seen by comparison
to Fig.~\ref{dyydist} where the dotted curves are equivalent to the
$pp$ distributions.  The dashed curves in Fig.~\ref{dypbpp} are the
ratios of the dashed to solid curves of Fig.~\ref{dyydist}.  Clearly,
at the peak of the rapidity distributions, the dashed curves are
higher than the solid curves, signifying antishadowing, yet do not
cross the dotted curves.  The isospin of the Pb nucleus thus causes
the Pb$p$/$pp$ ratio to remain less than unity over all rapidity.
Note also that the peaks of the ratios occur at smaller values of
$x_1$, from $x_1 \sim 0.11$ in the lowest bin to $\sim 0.062$ in the
highest.  The evolution of the antishadowing region broadens and the
average $x$ decreases somewhat with increasing $Q^2$.  On the other
hand, the $p$Pb/$pp$ ratios are rather constant, decreasing only
slowly with rapidity although the slope of the ratio increases with
$M$.  The $p$Pb calculations are always at low $x_2$ where the
shadowing parameterization does not change much with $x_2$.  These
results are reflected in the integrated cross sections in
Table~\ref{dysigsloq}.

\begin{table}[htb]
\caption{Rapidity integrated Drell-Yan cross sections at 5.5 TeV/nucleon
in the indicated mass intervals.  
The calculations are to NLO with the MRST HO parton densities and $Q^2 =
M^2$. 
}
\label{dysigsloq}
\begin{center}
\begin{tabular}{cccccc}
$\Delta y$ & $\sigma$ (nb) 
& $\sigma_{\rm NS}$ (nb) & $\sigma_{\rm EKS98}$ (nb) 
& $\sigma_{\rm NS}$ (nb) & $\sigma_{\rm EKS98}$ (nb)
\\ \hline
& $pp$ & \multicolumn{2}{c}{Pb$p$} &
\multicolumn{2}{c}{$p$Pb} \\ \hline
\multicolumn{6}{c}{$4<M<9$ GeV} \\
$0<y<2.4$ & 3.61 & 3.57 & 2.82 & 3.60 & 2.51 \\
$0<y<1$   & 1.41 & 1.40 & 1.05 & 1.40 & 1.00 \\ 
$2.4<y<4$ & 2.52 & 2.44 & 2.28 & 2.52 & 1.68 \\ \hline 
\multicolumn{6}{c}{$11<M<20$ GeV} \\
$0<y<2.4$ & 0.44 & 0.43 & 0.39 & 0.44 & 0.34 \\
$0<y<1$   & 0.17 & 0.17 & 0.14 & 0.17 & 0.14 \\ 
$2.4<y<4$ & 0.30 & 0.27 & 0.28 & 0.30 & 0.21 \\ \hline 
\multicolumn{6}{c}{$20<M<40$ GeV} \\
$0<y<2.4$ & 0.12 & 0.12 & 0.11 & 0.12 & 0.10 \\
$0<y<1$   & 0.048 & 0.047 & 0.043 & 0.048 & 0.041 \\ 
$2.4<y<4$ & 0.077 & 0.068 & 0.069 & 0.077 & 0.058 \\ \hline 
\multicolumn{6}{c}{$40<M<60$ GeV} \\
$0<y<2.4$ & 0.019 & 0.018 & 0.018 & 0.019 & 0.016 \\
$0<y<1$   & 0.0074 & 0.0072 & 0.0070 & 0.0073 & 0.0066 \\ 
$2.4<y<4$ & 0.0094 & 0.0078 & 0.0078 & 0.0094 & 0.0075 \\ \hline 
\end{tabular}
\vspace{0.5cm}
\end{center}
\end{table}

\begin{figure}[htb]
\setlength{\epsfxsize=0.7\textwidth}
\setlength{\epsfysize=0.4\textheight}
\centerline{\epsffile{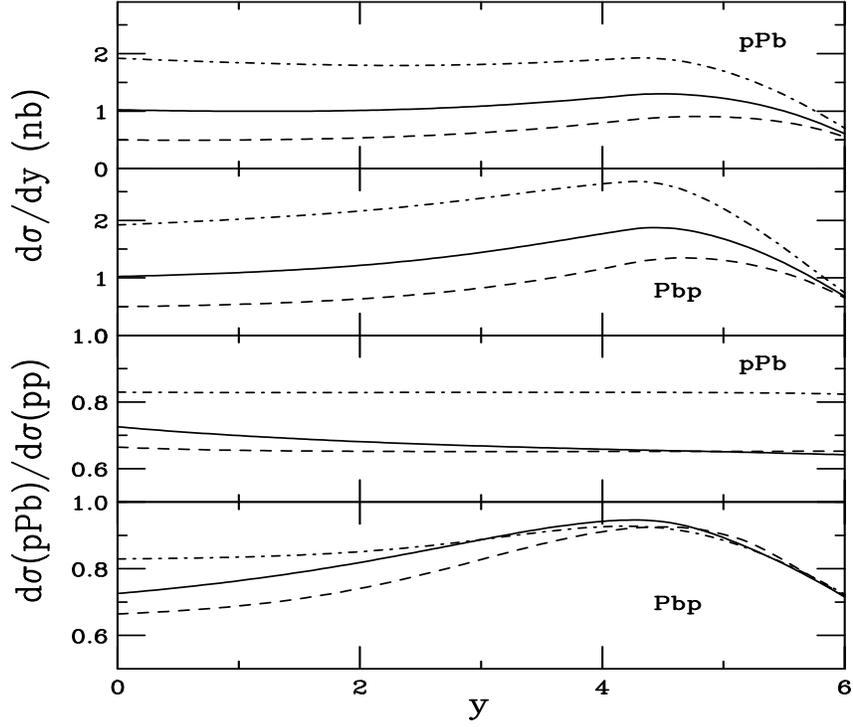}}
\caption{The upper two plots present the scale dependence of the Drell-Yan
cross section in the mass interval $4<M<9$ GeV
in $p$Pb and Pb$p$ collisions at 5.5 TeV respectively.  The EKS98
parameterization is used in all cases.  The solid curve is calculated with
$Q=M$, the dashed curve with $Q = M/2$ and the dot-dashed with $Q=2M$. The
lower plots present the dependence on the shadowing parameterization in the
mass range $4<M<9$ GeV for the ratios $p$Pb/$pp$ and Pb$p$/$pp$ at 5.5 TeV.
The solid curve is the EKS98 parameterization, the dashed curve is the HPC
parameterization and the dot-dashed curve is the HKM parameterization.
}
\label{dydeps}
\end{figure}

In Fig.~\ref{dydeps} we show the effects of scale (upper two plots) and
shadowing parameterization (lower two plots) on the results.

We choose the lowest mass bin to emphasize the overall effect.  The scale
variation, discussed in Section~\ref{section411}, shows that the cross sections
are larger for scale $2M$ and smallest for scale $M/2$.  This scale dependence
is opposite to typical expectations since lower scales generally lead to
larger cross sections.  At lower energies and other processes, this is true
but at high energies, the low $x$ evolution exceeds the effect of the scale
dependent logarithms \cite{Hamberg:1990np}.  The scale variation is a 50\%
effect in this bin, decreasing to 10\% for $40<M<60$ GeV.  The HPC \cite{HPC}
and HKM \cite{Hirai:2001np} shadowing calculations have the same general trends
as the EKS98 parameterization.  The HKM model has the weakest effect on sea
quarks at low $x$ and produces the largest ratios for $p$Pb/$pp$ and in
Pb$p$/$pp$ for $y<3$.  The HPC parameterization has the strongest shadowing
without evolution and is generally lower than the other two.

If the low $Q^2$ Drell-Yan cross section can be extracted from the other
contributions to the continuum in $pA$, it could provide useful information on
the low $x$, moderate $Q^2$ regime not covered in previous experiments.
Comparison could be made to the only previous available $pA$ Drell-Yan
measurements as a function of $x_F$ by the E772 \cite{Alde:im} and E866 
\cite{Vasilev:1999fa} collaborations at much higher $x$.

RHIC $pA$ data should also be available by the time the LHC starts up which
could increase the $\sqrt{s}$ systematics, providing, with both colliders, a
collective measurement over a wide range of $x$ values at similar $Q^2$.

\vfill\eject

\subsection{ Low mass Drell-Yan  production at LHC energies}
\label{subsec:fai_qiu_zhang}
{\em George Fai,  Jianwei Qiu, and Xiaofei Zhang }

\def\tb{\tilde{b}}




\subsubsection{Introduction}

Massive  
dilepton production in hadronic collisions is an excellent 
laboratory for theoretical and experimental investigations of strong 
interaction dynamics, and it is a channel for discovery of quarkonium 
states and a clean process for the study of the PDF.
In the Drell-Yan process, the massive lepton pair is produced via 
the decay of an intermediate virtual photon $\gamma^*$.  
If both the physically measured dilepton mass $M$ and the
transverse momentum $p_T$ are large, the cross section in a collision 
between hadrons (or nuclei) $A$ and $B$,
$A(P_A)+B(P_B)\rightarrow \gamma^*(\rightarrow l\bar{l})+X$, 
can be factorized systematically in QCD perturbation theory and 
expressed as
\cite{Collins:gx}
\begin{equation}
\frac{d\sigma_{AB\rightarrow l\bar{l}(M)X}}{dM^2\, dy\, dp_T^2}
=\left(\frac{\alpha_{\rm em}}{3\pi M^2}\right)
 \sum_{a,b}\int dx_1 f_{a/A}(x_1,\mu) 
           \int dx_2 f_{b/B}(x_2,\mu)\,
 \frac{d\hat{\sigma}_{ab\rightarrow \gamma^* X}}{dp_T^2\, dy}
 (x_1,x_2,M,p_T,y;\mu) .  
\label{Vph-fac}
\end{equation}
The sum $\sum_{a,b}$ runs over all parton flavors, $f_{a/A}$
and $f_{b/B}$ are normal parton distributions and $\mu$ represents
the renormalization and factorization scales.  The
partonic cross section
$d\hat{\sigma}_{ab\rightarrow \gamma^* X}/dp_T^2 dy$ in 
Eq.~(\ref{Vph-fac}), the short-distance probability for partons of 
flavors $a$ and $b$ to produce a virtual photon of invariant mass $M$,
is perturbatively calculable  in terms of a power
series in $\alpha_s(\mu)$.  The scale $\mu$ is 
of the order of the energy exchange in the reaction, 
$\mu \sim \sqrt{M^2 + p_T^2}$.  

When $M\sim M_Z$, high mass dilepton production in heavy ion 
collisions at LHC energies is dominated by the $Z^0$ channel and 
is an excellent hard probe of  
QCD dynamics \cite{Zhang:2002yz}.  
In this contribution, we demonstrate that the transverse momentum 
distribution of  low mass ($\Lambda_{\rm QCD}\ll M\ll M_Z$) 
dilepton production at LHC energies is a reliable probe of 
both hard and semihard physics at LHC energies and 
is an advantageous source of constraints on gluon distributions in 
the proton and in nuclei \cite{Fai:2003qz}.  
In addition, it provides a significant contribution to the total
dilepton spectra at the LHC, also an important channel
for quarkonium and heavy flavour decays.

The transverse momentum ($p_T$) distribution of the dileptons 
can be divided into three regions: low $p_T$ ($\ll M$), intermediate 
$p_T$ ($\sim M$), and high $p_T$ ($\gg M$).
When both the physically measured $M$ and $p_T$ are large and are of the 
same order, the short-distance partonic part  
$d\hat{\sigma}_{ab\rightarrow \gamma^* X}/dp_T^2 dy$ in 
Eq.~(\ref{Vph-fac}) can be calculated reliably in conventional
fixed-order QCD perturbation theory in terms of a power series in
$\alpha_s(\mu)$.  
However, when $p_T$ is very different from $M$, 
the calculation of Drell-Yan production in both low and high $p_T$ 
regions becomes a two-scale problem in perturbative QCD and 
the calculated partonic parts include potentially 
large logarithmic terms proportional to a 
power of $\ln(M/p_T)$.  As a result, the higher-order corrections in powers 
of $\alpha_{s}$ are not necessary small.  
The ratio $\sigma^{\rm NLO}/\sigma^{\rm LO}$ $\propto \alpha_s \times$
(large logarithms) can be of order 1 
and convergence of the conventional perturbative expansion 
in powers of $\alpha_{s}$ is possibly impaired.  

In the low $p_T$ region, there are two powers of $\ln(M^2/p_T^2)$ 
for each additional power of $\alpha_s$
and the Drell-Yan $p_T$ distribution calculated in fixed-order QCD 
perturbation theory is known 
to be unreliable.  
Only after all-order
resummation of the large $\alpha_s^n\,\ln^{2n+1}(M^2/p_T^2)$ 
logarithms
do predictions for the $p_T$ 
distributions become consistent with data 
\cite{Qiu:2000ga,Landry:1999an}. 
We demonstrate in Sec.~\ref{zhang2:subsec2} that low mass Drell-Yan
production at $p_{T}$ as low as $\Lambda_{\rm QCD}$ at LHC energies
can be calculated reliably in perturbative QCD with all order
resummation.

When $p_T \ge M/2$, the lowest-order virtual photon
``Compton'' subprocess $g+q \rightarrow \gamma^* + q$ dominates the 
$p_{T}$ distribution, and the high-order contributions including
all-order resummation of $\alpha_s^n\, \ln^{n-1}(M^2/p_T^2)$ preserve the fact
that the $p_{T}$ distributions of low mass Drell-Yan  pairs 
are dominated by gluon initiated partonic subprocesses 
\cite{Berger:2001wr}.  We show in Sec.~\ref{zhang2:subsec3} that the $p_{T}$ distribution of 
low mass Drell-Yan pairs can be a good probe of the gluon distribution 
and its nuclear dependence.  We give our conclusions in Sec.~\ref{zhang2:subsec4}.

\subsubsection{Low mass Drell-Yan production at low transverse momentum}
\label{zhang2:subsec2}

Resummation of large logarithmic terms at low $p_T$ can be carried 
out in either $p_T$ or impact parameter ($\tilde{b}$) space, 
the Fourier conjugate of $p_T$ space.  All else being 
equal, the $\tilde{b}$ space approach has the advantage 
of explicit transverse momentum conservation.
Using renormalization
group techniques, Collins, Soper, and Sterman (CSS)~\cite{Collins:1984kg}
devised a $\tilde{b}$ space resummation formalism that resums all 
logarithmic terms as singular as $(1/p_T^2)\ln^m(M^2/p_T^2)$ 
when $p_T \rightarrow 0$.  This formalism has been widely used for 
computations of 
vector boson $p_T$ distributions
in hadron reactions \cite{Berger:2002ut}.

At low mass $M$ and $p_T$, Drell-Yan transverse momentum distributions
calculated in the CSS $\tilde{b}$-space resummation formalism  
strongly depend on the nonperturbative parameters 
determined
at fixed target
energies.   However, it was pointed out recently that the predictive
power of 
pQCD
resummation improves with 
$\sqrt s$ and, when the energy is high enough,
pQCD should have good predictive power even for low mass Drell-Yan
production \cite{Qiu:2000ga}. The LHC will provide us a chance to 
study low mass Drell-Yan production at unprecedented energies.

In the CSS resummation formalism, the differential cross section 
for Drell-Yan production in Eq.~(\ref{Vph-fac}) is reorganized 
as the sum
\begin{equation}
\frac{d\sigma_{AB\rightarrow l\bar{l}(M)X}}
     {dM^2\, dy\, dp_T^2}
=
\frac{d\sigma_{AB\rightarrow l\bar{l}(M)X}^{\rm (resum)}}
     {dM^2\, dy\, dp_T^2}
+
\frac{d\sigma_{AB\rightarrow l\bar{l}(M)X}^{\rm (Y)}}
{dM^2\, dy\, dp_T^2}\, .
\label{css-gen_dy}
\end{equation}
The all-orders resummed term is a Fourier transform from 
the $\tb$-space,
\begin{equation} 
\frac{d\sigma_{AB\rightarrow l\bar{l}(M)X}^{\rm (resum)}}
     {dM^2\, dy\, dp_T^2}
= \frac{1}{(2\pi)^2}\int d^2\tb\, 
e^{i\vec{p}_T\cdot \vec{\tb}}\, W(\tb,M,x_A,x_B)
= \frac{1}{2\pi}\int d\tb\, J_0(p_T \tb)\, \tb\, W(\tb,M,x_A,x_B) \,\, ,
\label{css-resum}
\end{equation}
where $J_0$ is a Bessel function, $x_A= e^y\, M/\sqrt{s}$ and 
$x_B= e^{-y}\, M/\sqrt{s}$, with rapidity $y$ and collision energy 
$\sqrt{s}$.  In Eq.~(\ref{css-gen_dy}), 
the $\sigma^{\rm (resum)}$ term dominates the $p_T$ distributions when
$p_T\ll M$, 
while the
$\sigma^{(Y)}$ term gives 
negligible corrections
for small $p_T$ but becomes important when $p_T\sim M$.

The function $W(\tb,M,x_A,x_B)$ resums to all orders in QCD perturbation
theory the singular terms that would otherwise behave as $\delta^2(p_T)$ 
and $(1/p_T^2)\ln^m(M^2/p_T^2)$ in transverse momentum space
for all $m\ge 0$.
It
can be calculated perturbatively 
for small \ $\tb$, 
\begin{equation}
W(\tb,M,x_A,x_B) =
{\rm e}^{-S(\tb,M)}\, W(\tb,c/\tb,x_A,x_B) 
\equiv W^{\rm pert}(\tb,M,x_A,x_B)
\,\,\, .
\label{css-W-sol_dy}
\end{equation}
All large logarithms from $\ln(c^2/\tb^2)$ to $\ln(M^2)$ have
been completely resummed into the exponential factor
$S(\tb,M)=\int_{c^2/\tb^2}^{M^2} d\mu^2/\mu^2\left[\ln(M^2/\mu^2)
{\cal A}(\alpha_s(\mu)) + {\cal B}(\alpha_s(\mu))\right]$ 
where the functions
$\cal A$ and $\cal B$ 
are
given in Ref.~\cite{Collins:1984kg} and
$c=2e^{-\gamma_E}$ with Euler's constant $\gamma_E\approx 0.577$.
With only one large momentum scale $1/\tb$, the function 
$W(\tb,c/\tb,x_A,x_B)$ in Eq.~(\ref{css-W-sol_dy}) 
is perturbatively calculable and is factorized as
\begin{equation}
W(\tb,c/\tb,x_A,x_B) = \sum_{a,b}
\left[f_{a/A}(\xi_1,1/\tb)\otimes{\cal C}_{a}
                                (\frac{x_A}{\xi_1})\right]
\left[f_{b/B}(\xi_2,1/\tb)\otimes{\cal C}_{b}
                                (\frac{x_B}{\xi_2})\right]
\sigma_{0},
\label{w-perp-fix}
\end{equation}
where the sum $\sum_{a,b}$ runs over all parton flavors, the
$\otimes$ represents the convolution over parton momentum 
fraction, and $\sigma_0$ is the lowest order partonic 
cross section for the Drell-Yan process~\cite{Collins:1984kg}.  
The functions ${\cal C}= \sum_n {\cal C}^{(n)} (\alpha_s/\pi)^n$ 
in Eq.~(\ref{w-perp-fix}) 
are perturbatively calculable 
and given in Ref.~\cite{Collins:1984kg}.  
Since the perturbatively 
resummed $W^{\rm pert}(\tb,M,x_A,x_B)$ in
Eq.~(\ref{css-W-sol_dy}) is only reliable for the small $\tb$ region, an
extrapolation to the nonperturbative large $\tb$ region is necessary 
in order to complete the Fourier transform in Eq.~(\ref{css-resum}).  

In the original CSS formalism, a variable $\tb_*$
and a nonperturbative function $F_{\rm CSS}^{\rm NP}(\tb,M,x_A,x_B)$
were introduced to extrapolate the perturbatively calculated
$W^{\rm pert}$ into the large $\tb$ region.
The full $\tb$-space distribution was of the form 
\begin{equation}
W^{\rm CSS}(\tb,M,x_A,x_B) \equiv
W^{\rm pert}(\tb_*,M,x_A,x_B)\,
F_{\rm CSS}^{\rm NP}(\tb,M,x_A,x_B)\, ,
\label{css-W-b}
\end{equation}
where $\tb_*=\tb/\sqrt{1+(\tb/\tb_{\rm max})^2}$, with 
$\tb_{\rm max} = 0.5$ GeV$^{-1}$.  This construction ensures that 
$\tb_* \leq \tb_{\rm max}$ for all values of $\tb$.

In terms of the $\tb_*$ formalism, a number of functional forms for
the $F_{\rm CSS}^{\rm NP}$ have been proposed.  A simple Gaussian form in 
$\tb$ was first proposed by Davies, Webber and Stirling (DWS) 
\cite{Davies:1984sp},
\begin{equation}
F_{\rm DWS}^{\rm NP}(\tb,M,x_A,x_B)
=\exp\left\{-(g_1+g_2\ln(M/2M_0))\tb^2\right\},
\label{DS}
\end{equation}
with the parameters $M_0=2$~GeV, $g_1=0.15$~GeV$^2$, and 
$g_2=0.4$~GeV$^2$.
In order to take into account the apparent dependence on collision 
energies, Ladinsky and Yuan (LY) introduced a new functional form 
\cite{Ladinsky:1993zn}, 
\begin{equation}
F_{\rm LY}^{\rm NP}(\tb,M,x_A,x_B)
=\exp\left\{-(g_1+g_2\ln(M/2M_0))\tb^2-g_1\, g_3
\ln(100x_Ax_B)\tb\right\},
\label{LY}
\end{equation}
with $M_0=1.6$~GeV, $g_1=0.11^{+0.04}_{-0.03}$~GeV$^2$, 
$g_2=0.58^{+0.1}_{-0.2}$~GeV$^2$, and $g_3=-1.5^{+0.1}_{-0.1}$~GeV$^{-1}$.  
Recently, Landry, Brook, Nadolsky, and Yuan proposed a modified
Gaussian form \cite{Landry:2002ix},
\begin{equation}
F_{\rm BLNY}^{\rm NP}(\tb,M,x_A,x_B)
=\exp\left\{-\left[g_1+g_2\ln(M/2M_0)+g_1\, g_3 \ln(100x_Ax_B)
             \right]\tb^2\right\},
\label{BLNY}
\end{equation}
with $M_0=1.6$~GeV, $g_1=0.21^{+0.01}_{-0.01}$~GeV$^2$, 
$g_2=0.68^{+0.01}_{-0.02}$~GeV$^2$, and 
$g_3=-0.6^{+0.05}_{-0.04}$.  
All these parameters were obtained 
by fitting low energy Drell-Yan and 
high energy $W$ and $Z$ data.  Note,  however that
the $\tb_*$ formalism introduces a modification to the perturbative 
calculation.
The
size of the modifications strongly depends on 
the nonperturbative parameters in $F^{\rm NP}(\tb,M,x_A,x_B)$, 
$M$, and $\sqrt{s}$ 
\cite{Berger:2002ut}.  

A remarkable feature of the $\tb$-space resummation formalism is
that the resummed exponential factor $\exp[-S(\tb,M)]$
suppresses the $\tb$-integral when $\tb$ is larger than $1/M$. 
It can be shown using the saddle point method that, for a large
enough $M$, QCD perturbation theory is valid even at $p_T=0$
\cite{Collins:1984kg}.  For high energy heavy boson ($W$, $Z$, and 
Higgs) production, the integrand of $\tb$-integration in 
Eq.~(\ref{css-resum}) at $p_T=0$ is proportional to 
$\tb W(\tb,Q,x_A,x_B)$, 
with 
a saddle point $\tb_{\rm sp}$ 
well within the perturbative region ($\tb_{\rm sp}<\tb_{\rm max}$).
Therefore, the $\tb$-integration in Eq.~(\ref{css-resum}) is 
dominated by the perturbatively resummed calculation.  The 
uncertainties from the large-$\tb$ region have very little effect
on the calculated $p_T$ distributions and the resummation formalism 
has 
good predictive power.

On the other hand, in low energy Drell-Yan production, 
there is no saddle point in the perturbative region for 
the integrand in Eq. (\ref{css-resum}).
Therefore the dependence of the 
final result on the nonperturbative input is strong 
\cite{Qiu:2000ga,Landry:1999an}.
However, as discussed in Refs. 
\cite{Qiu:2000ga,Berger:2002ut,Zhang:2002jf}, the value of the 
saddle point strongly depends on 
$\sqrt{s}$ in addition to its well-known $M^2$ dependence.  

\begin{figure}
\begin{minipage}[c]{7.6cm}
\centerline{\includegraphics[width=6.5cm]{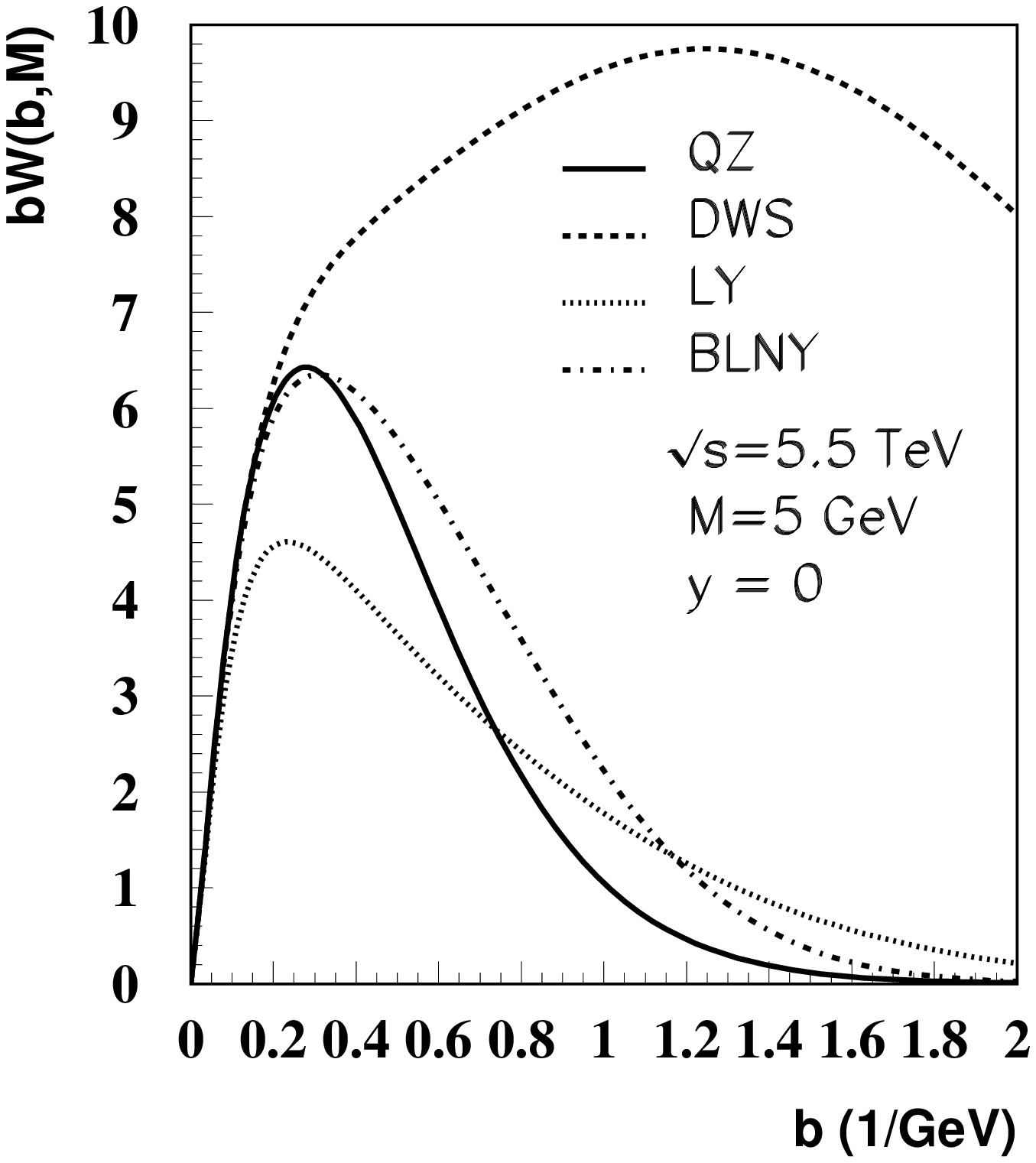}} 
\vspace{-0.1in}
\caption{The $\tb$-space resummed functions $\tb W(\tb)$ in
Eq.~(\protect\ref{css-resum})
for Drell-Yan production of dilepton mass $M=5$~GeV 
at $\sqrt s=5.5$~TeV with the QZ (solid), DWS (dashed), LY (dotted),
and BNLY (dot-dashed) formalism of nonperturbative extrapolation. }
\label{fig1}
\end{minipage}
\hfill
\begin{minipage}[c]{7.6cm}
\vspace{-0.5cm}
\centerline{\includegraphics[width=6.5cm]{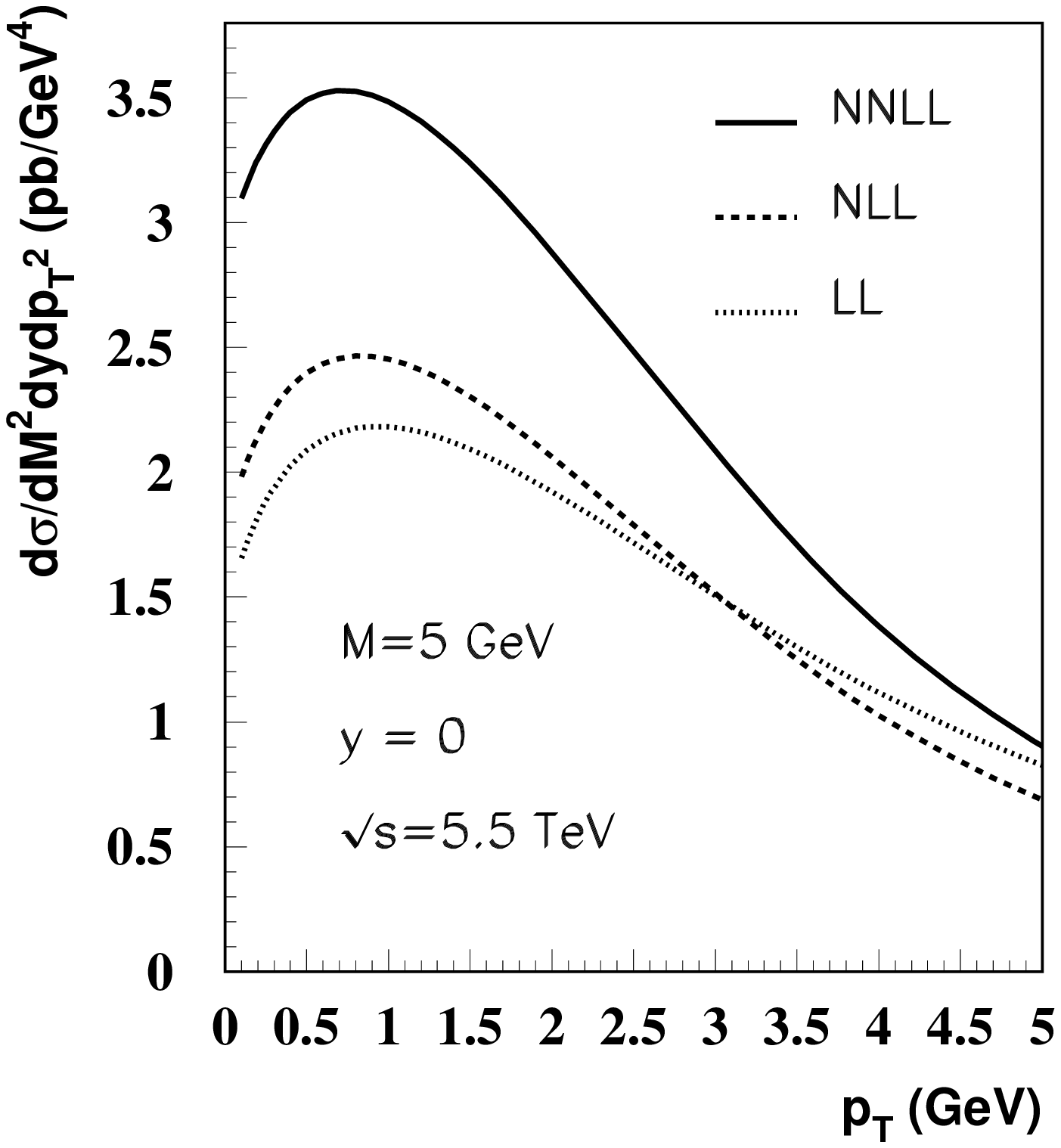}} 
\vspace{-0.1in}
\caption{Differential cross section ${d\sigma /dM^2 dy dp_T^2}$ for 
production of Drell-Yan pairs of $M=5$~GeV in $pp$ collisions 
at the LHC with $y=0$, and $\sqrt{s}=5.5$~TeV with the NNLL (solid),
NLL (dashed), and LL (dotted) accuracy.}
\label{fig2}
\end{minipage}
\end{figure}

Figure~\ref{fig1} shows the integrand of the $\tb$-integration in 
Eq.~(\ref{css-resum}) at $p_T=0$ for production of Drell-Yan pairs 
of mass $M=5$~GeV in $pp$ collisions at $\sqrt{s}=5.5$~TeV 
and $b_{\rm max}=0.5$~GeV$^{-1}$.  
The curves represent different extrapolations to the large-$\tb$
region.  The three curves (dashed, dotted, and dot-dashed) are
evaluated using the $\tb_*$ formalism with the DWS, LY, and BLNY
nonperturbative functions, respectively.  Although these three
nonperturbative functions give similar $\tb$-space distributions for
heavy boson production at Tevatron energies, they predict very
different $\tb$-space distributions for low mass Drell-Yan production
at LHC energies even within the perturbative small-$\tb$ region.
Since the $\tb$-distribution in Fig.~\ref{fig1} completely determines
the resummed $p_T$ distribution through the $\tb$-integration weighted
by the Bessel function $J_0(p_T \tb)$, we need to 
worry about
the uncertainties of the resummed low mass $p_T$ distributions
calculated with different nonperturbative functions.

In order to improve the situation, a new formalism 
of extrapolation (QZ) was proposed \cite{Qiu:2000ga},
\begin{equation}
W(\tb,M,x_A,x_B) = \left\{
\begin{array}{ll}
W^{\rm pert}(\tb,M,x_A,x_B) & \quad \mbox{$\tb\leq \tb_{\rm max}$} \\
W^{\rm pert}(\tb_{\rm max},M,x_A,x_B)\,
F^{\rm NP}(\tb,M,x_A,x_B;\tb_{\rm max},\alpha)
& \quad \mbox{$\tb > \tb_{\rm max}$}
\end{array} \right. \,\, ,
\label{qz-W-sol-m_dy}
\end{equation}
where the nonperturbative function $F^{\rm NP}$ is given by
\begin{eqnarray}
F^{\rm NP}
=\exp\left\{ -\ln(M^2 \tb_{\rm max}^2/c^2) 
\left[ g_1 \left( (\tb^2)^\alpha - (\tb_{\rm max}^2)^\alpha\right) \right.
 \left.   +g_2 \left(\tb^2 - \tb_{\rm max}^2\right) \right]
-\bar{g}_2 \left(\tb^2 - \tb_{\rm max}^2\right) \right\}.
\label{qz-fnp-m_dy}
\end{eqnarray}
Here, $\tb_{\rm max}$ is a parameter to separate the perturbatively 
calculated part from the nonperturbative input, and its role is 
similar to the $\tb_{\rm max}$ in the $\tb_*$ formalism.  The term 
proportional to $g_1$ in Eq.~(\ref{qz-fnp-m_dy}) represents a direct 
extrapolation of the resummed leading power contribution to the large
$\tb$ region.
The parameters $g_1$  and $\alpha$ 
are determined by the continuity of the function  $W(\tb,M,x_A,x_B)$ 
at $\tb_{\rm max}$.  On the other hand, the values of $g_2$ and 
${\bar g}_2$ represent the size of nonperturbative power corrections. 
Therefore, sensitivity to $g_2$ and ${\bar g}_2$ in this formalism 
clearly indicates the precision of the calculated $p_T$ distributions.

The solid line in Fig.~\ref{fig1} is the result of the QZ parameterization with 
$\tb_{\rm max}=0.5$~GeV$^{-1}$ and $g_2={\bar g}_2=0$.  Unlike in the 
$\tb_*$ formalism, the solid line represents the full perturbative 
calculation and is independent of the nonperturbative parameters for 
$\tb < \tb_{\rm max}$.  The difference between the solid line and the other 
curves in the small $\tb$ region, which can be as large as 40\%, indicates 
the uncertainties introduced by the $\tb_*$ formalism.  

It is clear from the solid line in Fig.~\ref{fig1} that 
there is a saddle point in the perturbative 
region even for dilepton masses as low as $M=5$~GeV in Drell-Yan 
production at $\sqrt{s}=5.5$ TeV.  At that energy, $x_A, x_B \sim 0.0045$.  
For such small values of $x$, the PDFs have very strong scaling 
violations, leading to large parton showers.  It is the large 
parton shower at the small $x$ that strongly suppresses the function  
$W(\tb,c/\tb,x_A,x_B)$ in Eq.~(\ref{css-W-sol_dy}) as $\tb$ increases.
Therefore, for  $\tb_{\rm max}\sim$(a few GeV)$^{-1}$, the  predictive power
of the $\tb$-space resummation formalism depends on the relative size
of contributions from the small-$\tb$ ($\tb<\tb_{\rm max}$) and 
large-$\tb$ ($\tb>\tb_{\rm max}$) regions of the $\tb$-integration in
Eq.~(\ref{css-resum}).  With a narrow $\tb$ distribution peaked within 
the perturbative region for the integrand, the $\tb$-integration in 
Eq.~(\ref{css-W-sol_dy}) is dominated by the small-$\tb$ region and,
therefore, we expect pQCD to have good predictive power even for low 
$M$ Drell-Yan production at LHC energies.  

Figure~\ref{fig2} presents our prediction of the fully differential cross 
section $d\sigma/dM^2 dy dp_T^2$ for Drell-Yan production 
in $pp$ collisions at $\sqrt{s}=5.5$ TeV 
and $y=0$
\cite{Fai:2003qz}.
Three curves represent the different order of contributions in 
$\alpha_s$ to the perturbatively calculated functions 
${\cal A}(\alpha_s)$, ${\cal B}(\alpha_s)$, 
and ${\cal C}(\alpha_s)$ in the resummation formalism
\cite{Giele:2002hx}.
The solid line represents a next-to-next-to-leading-logarithmic (NNLL) 
accuracy corresponding to keeping the functions, 
${\cal A}(\alpha_s)$, ${\cal B}(\alpha_s)$, 
and ${\cal C}(\alpha_s)$ 
to order
$\alpha_s^3$, 
$\alpha_s^2$, and $\alpha_s^1$, respectively.  
The dashed line has next-to-leading-logarithmic (NLL) accuracy with 
the functions, ${\cal A}(\alpha_s)$, ${\cal B}(\alpha_s)$, 
and ${\cal C}(\alpha_s)$ at
 $\alpha_s^2$, $\alpha_s$, and $\alpha_s^0$, respectively, while
the dotted line has the lowest leading-logarithmic (LL) accuracy 
with the functions, ${\cal A}(\alpha_s)$, ${\cal B}(\alpha_s)$, 
and ${\cal C}(\alpha_s)$ at
the $\alpha_s$, $\alpha_s^0$, and $\alpha_s^0$, respectively.
Similar to what was seen in the fixed order calculation, the resummed
$p_T$ distribution has a $K$-factor about 1.4--1.6 around the peak
due to the inclusion of the coefficient ${\cal C}^{(1)}$.

\begin{figure}
\begin{minipage}[c]{7.6cm}
\centerline{\includegraphics[width=8cm]{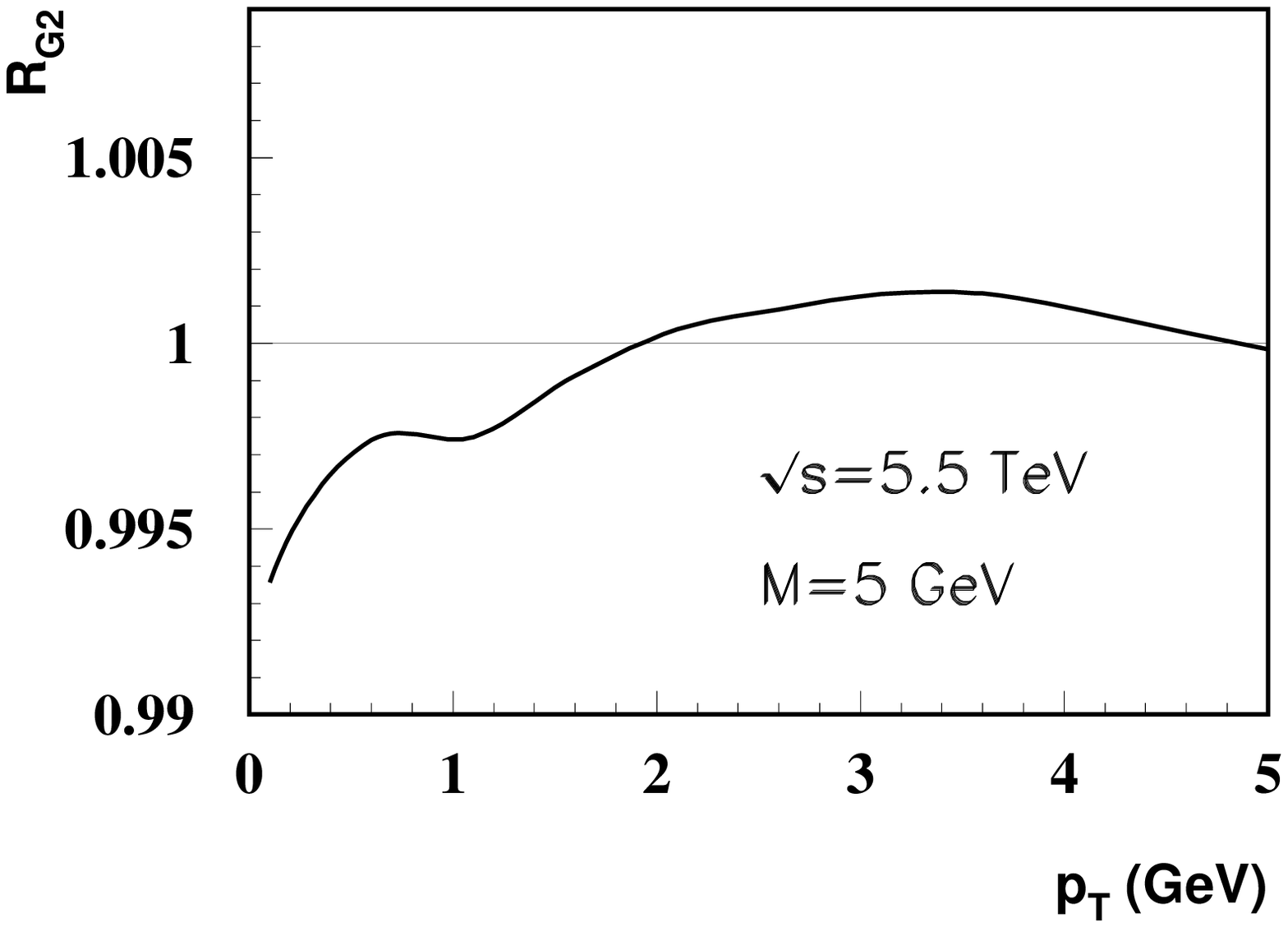}} 
\vspace{-0.1in}
\caption{The ratio $R_{G_2}$ defined in Eq.~(\protect\ref{Sigma-g2_dy})
with $G_2=$ 0.25 GeV$^2$ for the differential cross section shown in
Fig.~\ref{fig2}.}
\label{fig3}
\end{minipage}
\hfill
\begin{minipage}[c]{7.6cm}
\vspace{0.0in}
\includegraphics[width=8cm]{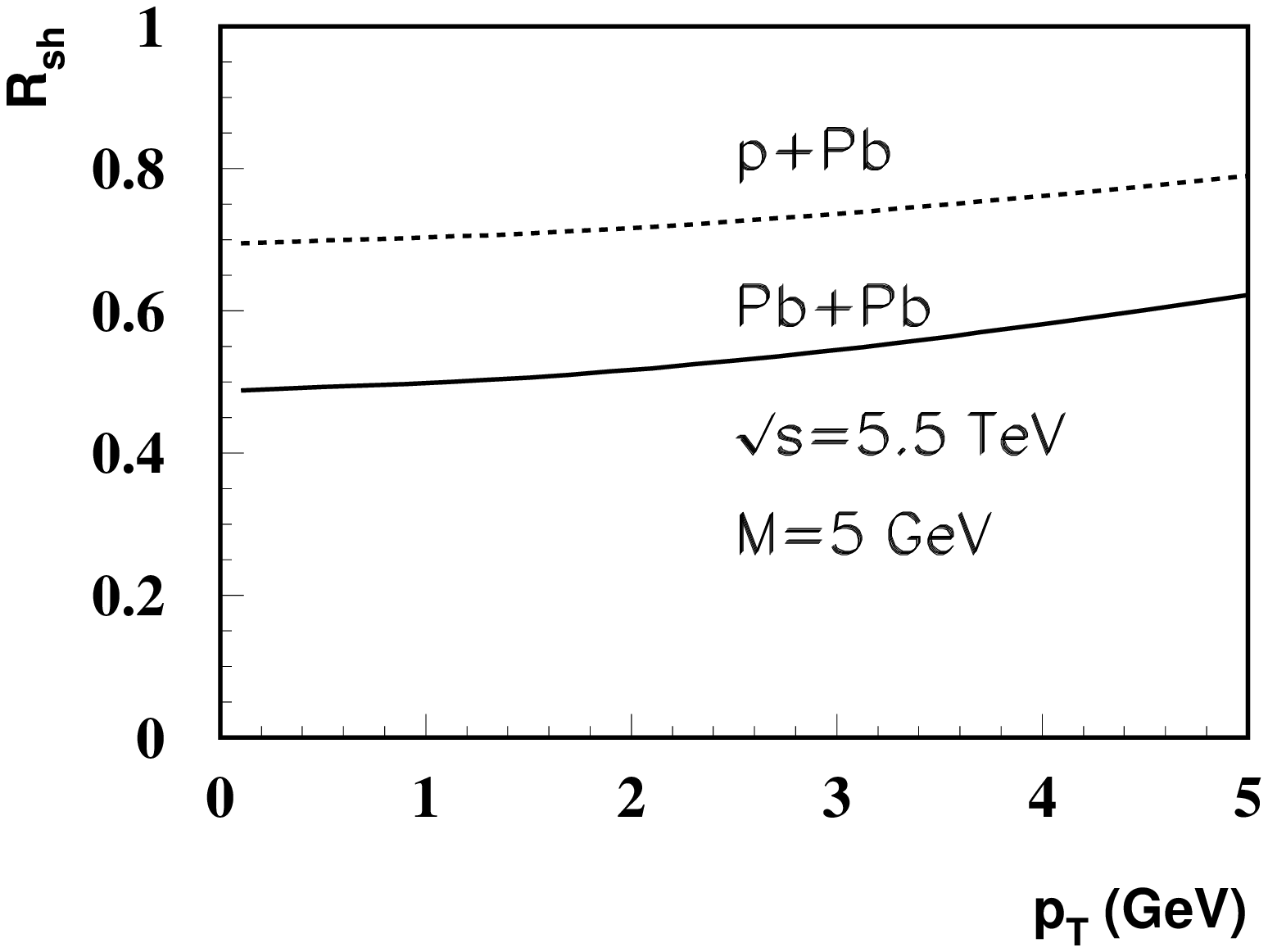} 
\vspace{-0.3in}
\caption{$R_{\rm sh}$  as a function of $p_T$ using EKS shadowing 
in $p$Pb and PbPb collisions at $\sqrt s=5.5$ TeV.}
\label{fig4}
\end{minipage}
\end{figure}

In Eq.~(\ref{qz-fnp-m_dy}), in addition to the $g_1$ term from 
the leading power contribution of soft gluon showers, 
the $g_2$ term corresponds to the first power correction from 
soft gluon showers and the $\bar g_2 $ term is from the intrinsic
transverse momentun of the incident parton.  The numerical values of 
$g_2$ and $\bar g_2$ have to be obtained 
from fits to 
the data.  
From 
fits
of low energy Drell-Yan data and  heavy gauge 
boson data at the Tevatron, we found that the intrinsic transverse 
momentum term dominates the power corrections.
It has a weak energy dependence.  For  convenience,
we combine the parameters of the $\tb^2$ term as
$G_2= g_2\ln({M^2 \tb_{\rm max}^2/{c^2}}) + \bar{g}_2$.
For $M=5$ GeV and $y=0$, we use $G_2\sim 0.25$ in the discussion
here \cite{Qiu:2000ga}. 
To test the $G_2$ dependence of our calculation, we define 
\begin{equation}
R_{G_2}(p_T) \equiv \left.
\frac{d\sigma^{(G_2)}_{AB\rightarrow l\bar{l}(M)X}(p_T)}
     {dM^2\, dy\, dp_T^2} \right/
\frac{d\sigma_{AB\rightarrow l\bar{l}(M)X}(p_T)}
     {dM^2\, dy\, dp_T^2} \,\,\, ,
\label{Sigma-g2_dy}
\end{equation}
where the numerator represents the result with finite $G_2$,
and the denominator contains no power corrections, $G_2 = 0$.

The result for $R_{G_2}$ is shown in Fig.~\ref{fig3}. 
It 
deviates from unity by less than $1\%$. The dependence of our result on 
the nonperturbative input is indeed very weak. 

Since the $G_2$ terms represent the power corrections from soft gluon
showers and 
parton
intrinsic transverse momenta, the smallness of 
the deviation of $R_{G_2}$ from 
unity also means that leading 
power contributions from gluon showers dominate the dynamics of 
low mass Drell-Yan production at LHC energies. 
Even though the power corrections
will be enhanced in nuclear collisions, we expect 
them to be  still less than several percent \cite{Zhang:2002jf}.  
The isospin effects are  also small here because 
$x_A$ and $x_B$ are  small.

Since the leading power contributions from initial-state parton showers 
dominate the production dynamics, the important nuclear effect is 
the modification of the parton distributions.  Because 
$x_A$ and $x_B$ are small for low mass Drell-Yan 
production at LHC energies, shadowing is the only dominant 
nuclear effect.  In order to study the shadowing effects, we define
\cite{Zhang:2002jf}
\begin{equation}
R_{\rm sh}(p_T) \equiv \left.
\frac{d\sigma^{\rm (sh)}_{AB\rightarrow l\bar{l}(M)X}(p_T,Z_A/A,Z_B/B)}
     {dM^2\, dy\, dp_T^2} \right/
\frac{d\sigma_{AB\rightarrow l\bar{l}(M)X}(p_T)}
     {dM^2\, dy\, dp_T^2} \,\,\, .
\label{Sigma-sh_dy}
\end{equation}
We plot the ratio $R_{\rm sh}$ as a function of $p_T$ in PbPb
collisions at $\sqrt s=5.5$~TeV for $M=5$~GeV and $y=0$ in
Fig.~\ref{fig4}.  The EKS parameterization of nuclear parton
distributions
\cite{Eskola:1998iy,Eskola:1998df} was used to evaluate the cross sections
in Eq.~(\ref{Sigma-sh_dy}).  It is clear from Fig.~\ref{fig4} that low 
mass Drell-Yan production can be a good probe of nuclear shadowing 
\cite{Fai:2003qz}, both in PbPb and in $p$Pb collisions at the LHC.  

\subsubsection{Low mass Drell-Yan production at high transverse momentum}
\label{zhang2:subsec3}

The gluon distribution plays a central role in calculating many 
important signatures at hadron colliders because of the dominance of 
gluon-initiated subprocesses.  A precise knowledge of the gluon distribution
as well as its nuclear dependence is absolutely vital for understanding
both hard and semihard probes at LHC energies.  

It was pointed out recently that the transverse momentum distribution 
of massive lepton pairs produced in hadronic collisions is an 
advantageous source of constraints on the gluon distribution
\cite{Berger:1998ev}, free from the experimental and 
theoretical complications of photon isolation that beset studies of 
prompt photon production~\cite{Berger:1990es,Berger:1995cc}.
Other than the difference between a virtual and a real photon, 
the Drell-Yan process and prompt photon production 
share the same partonic subprocesses.  Similar to prompt photon production, 
the lowest-order virtual photon ``Compton'' subprocess 
$g+q\rightarrow \gamma^*+q$ dominates the $p_T$ distribution 
when $p_T > M/2$.
The next-to-leading order contributions 
preserve the fact that the $p_T$ distributions are dominated by 
gluon-initiated partonic subprocesses~\cite{Berger:1998ev}. 

There is a phase space penalty associated with the finite mass of 
the virtual photon, and the Drell-Yan factor 
$\alpha_{\rm em}/(3\pi M^2)< 10^{-3}/M^2$ in Eq.~(\ref{Vph-fac}) 
renders the production rates for massive lepton pairs small 
at large values of $M$ and $p_T$.  In order to enhance the
Drell-Yan cross section while keeping the dominance of the 
gluon-initiated subprocesses, it is useful to study lepton pairs with
low invariant mass and relatively large transverse momenta
\cite{Berger:2001wr}.  
With the large 
$p_T$ setting the hard scale of the 
collision, the invariant mass of the virtual photon $M$ can be small,
as long as the process can be identified experimentally
and 
$M\gg\Lambda_{\rm QCD}$.  
For example, the cross section for Drell-Yan
production was measured by the CERN UA1 Collaboration~\cite{Albajar:1988iq}
for virtual photon mass $M\in [2m_\mu, 2.5]$~GeV.  

When $p_T^2\gg M^2$, the perturbatively calculated short-distance
partonic parts, $d\hat{\sigma}_{ab\rightarrow \gamma^* X}/dp_T^2 dy$
in Eq.~(\ref{Vph-fac}), receive one power of the logarithm
$\ln(p_T^2/M^2)$ at every order of $\alpha_s$ beyond leading
order.  At sufficiently large $p_T$, the coefficients of
the perturbative expansion in $\alpha_s$ will have large 
logarithmic terms, 
so that 
these higher-order corrections may not be small.  
In order to derive reliable QCD predictions, resummation of the 
logarithmic terms $\ln^m(p_T^2/M^2)$ must be considered.  It was 
recently shown \cite{Berger:2001wr} that the large
$\ln^m(p_T^2/M^2)$ terms in the low mass Drell-Yan cross sections
can be systematically resummed into 
a set of perturbatively calculable 
virtual photon fragmentation functions \cite{Qiu:2001nr}.
Similar to Eq.~(\ref{css-gen_dy}), the differential cross section
for low mass Drell-Yan production at large $p_T$ 
can be reorganized as 
\begin{equation}
\frac{d\sigma_{AB\rightarrow l\bar{l}(M)X}}
     {dM^2\, dy\, dp_T^2}
=
\frac{d\sigma_{AB\rightarrow l\bar{l}(M)X}^{\rm (resum)}}
     {dM^2\, dy\, dp_T^2}
+
\frac{d\sigma_{AB\rightarrow l\bar{l}(M)X}^{\rm (Dir)}}
     {dM^2\, dy\, dp_T^2}\, ,
\label{bqz-gen}
\end{equation}
where $\sigma^{\rm (resum)}$ includes the large logarithms 
and can be factorized as \cite{Berger:2001wr}
\begin{eqnarray}
\frac{d\sigma_{AB\rightarrow l\bar{l}(M)X}^{\rm (resum)}}
     {dM^2\, dy\, dp_T^2}
&=&\left(\frac{\alpha_{\rm em}}{3\pi M^2}\right)
\sum_{a,b,c}
\int dx_1 f_{a/A}(x_1,\mu)\, 
\int dx_2 f_{b/B}(x_2,\mu)\,
\nonumber \\
&\ & \times
\int \frac{dz}{z^2}\,
\frac{d\hat{\sigma}_{ab\rightarrow c X}}
     {dp_{c_T}^2\,dy}(p_{c_T}=p_T/z)\
  D_{c\rightarrow \gamma^* X}(z,\mu_F^2;M^2) \,\,  ,
\label{DY-F-fac}
\end{eqnarray}
with the factorization scale $\mu$ and fragmentation scale $\mu_F$,
and virtual photon fragmentation functions
$D_{c\rightarrow \gamma^*}(z,\mu_F^2;Q^2)$.
The $\sigma^{\rm (Dir)}$ term plays the same role as 
$\sigma^{({\rm Y})}$ term in Eq.~(\ref{css-gen_dy}), 
dominating the cross section 
when $p_T\rightarrow M$.

\begin{figure}
\begin{minipage}[c]{7.6cm}
\centerline{\includegraphics[width=8.0cm]{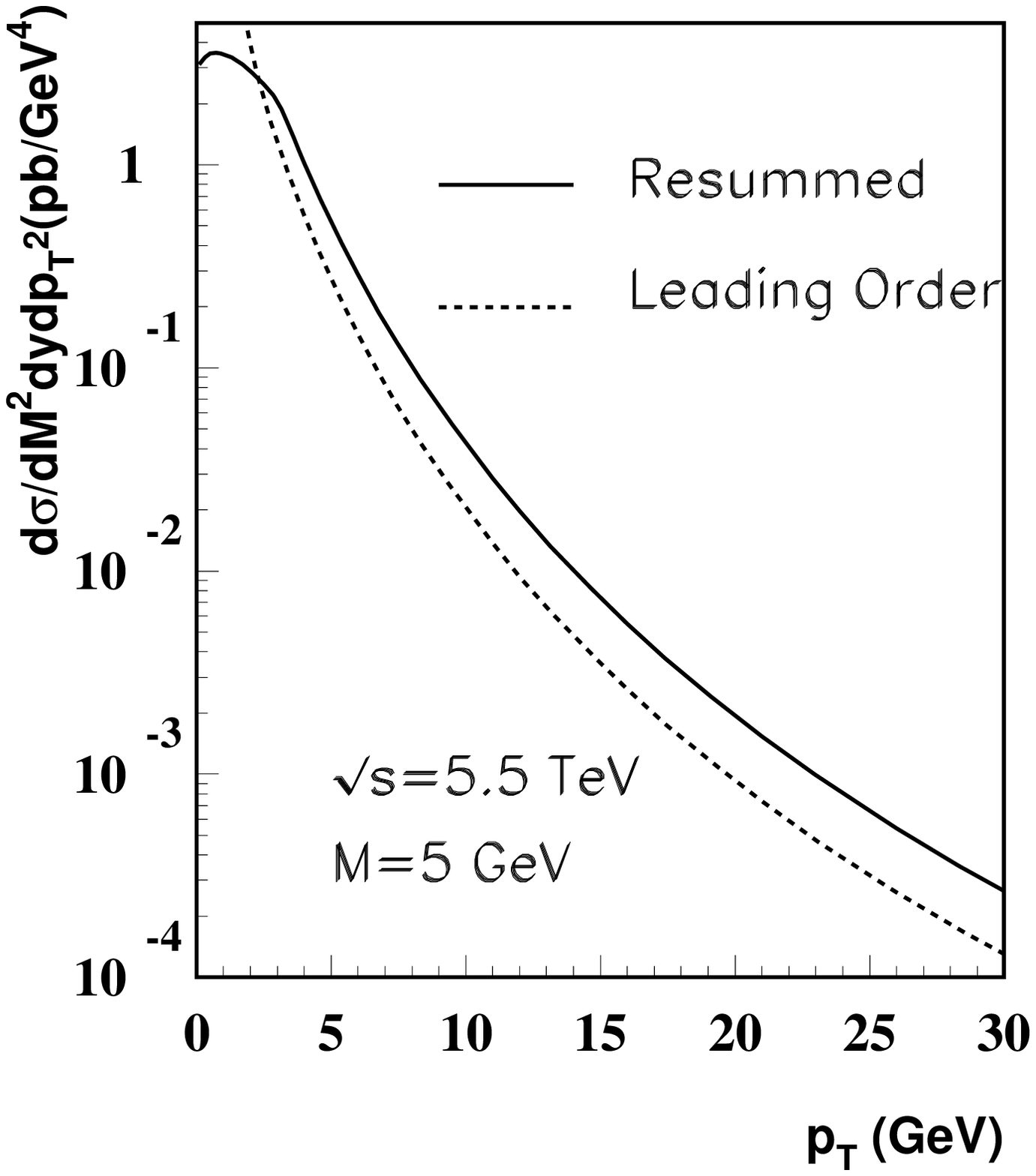}} 
\vspace{-0.1in}
\caption{Differential cross section ${d\sigma / dM^2 dy dp_T^2}$ 
for production of Drell-Yan pairs of $M=5$~GeV in $pp$ collisions 
at $\sqrt s=5.5$~TeV with low and high $p_T$ resummation (solid),
in comparison to  conventional lowest result (dashed).  } 
\label{fig5}
\end{minipage}
\hfill
\begin{minipage}[c]{7.6cm}
\vspace{-0.5cm}
\centerline{\includegraphics[width=8.0cm]{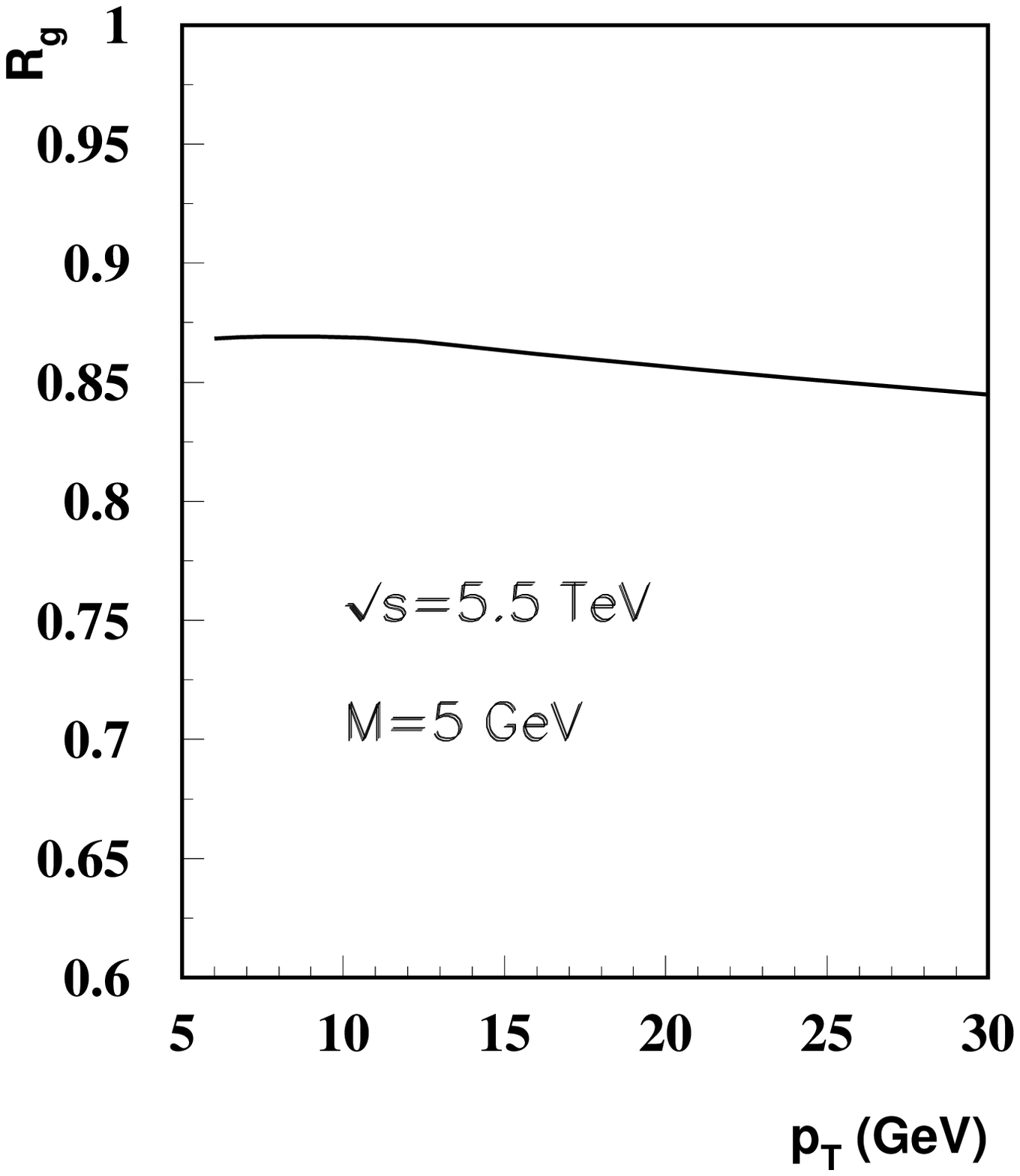}} 
\vspace{-0.1in}
\caption{Ratio of gluonic over total contributions to Drell-Yan
production at the LHC, $R_{g}$, defined in 
Eq.~(\protect\ref{Rg}) with $M=5$~GeV at $\sqrt s=5.5$~TeV.} 
\label{fig6}
\end{minipage}
\end{figure}

Figure~\ref{fig5} presents the fully resummed transverse momentum spectra
of low mass Drell-Yan production in $pp$ collisions with $M=5$~GeV 
at $y=0$ and $\sqrt{s}=5.5$~TeV (solid).  For comparison, we also plot
the leading order spectra calculated in  conventional fixed order pQCD.
The fully resummed distribution is larger in the large $p_T$ region and 
smoothly convergent as $p_T\rightarrow 0$.  In addition, as discussed 
in Ref.~\cite{Berger:2001wr}, the
resummed differential cross section is much less sensitive to the 
changes of renormalization, factorization, and fragmentation scales
and should 
thus
be more reliable than the fixed order calculations.

To demonstrate the relative size of gluon initiated contributions, we 
define the ratio
\begin{equation}
R_g = \left.
\frac{d{\sigma}_{AB\rightarrow \gamma^*(M) X}(\mbox{gluon-initiated})}
     {dp_T^2\, dy} \right/
\frac{d{\sigma}_{AB\rightarrow \gamma^*(M) X}}{dp_T^2\,dy}\, .
\label{Rg}
\end{equation}
The numerator includes the contributions from all partonic subprocesses 
with at least one initial-state gluon
while
the denominator includes all subprocesses.

In Fig.~\ref{fig6}, we show $R_g$ as a function of $p_T$ in $pp$
collisions at $y=0$ and $\sqrt{s}=5.5$~TeV with $M=5$~GeV.  It is
clear from Fig.~\ref{fig6} that gluon-initiated subprocesses dominate
the low mass Drell-Yan cross section and that low mass Drell-Yan
lepton-pair production at large transverse momentum is an excellent
source of information about the gluon distribution~\cite{Berger:2001wr}.
The slow falloff of $R_g$ at large $p_T$ is related to the reduction
of phase space and the fact that cross sections are evaluated at
larger values of 
$x$.

\subsubsection{Conclusions}
\label{zhang2:subsec4}

In summary, we present the fully differential cross section of low mass 
Drell-Yan production calculated in QCD perturbation theory 
with all-order resummation.  For $p_T\ll M$, we use CSS $\tb$-space
resummation formalism to resum large logarithmic contributions
as singular as $(1/p_T^2) \ln^m(M^2/p_T^2)$
to all orders in $\alpha_s$. We show that the resummed $p_T$ distribution
of low mass Drell-Yan pairs at LHC energies is 
dominated by the perturbatively calculable small $\tb$-region and thus 
reliable for $p_T$ as small as $\Lambda_{\rm QCD}$.  Because of the 
dominance of small $x$ PDFs, the low mass Drell-Yan 
cross section is a good probe of the nuclear dependence of 
parton distributions.  For $p_T\gg M$, we use a newly derived 
QCD factorization formalism \cite{Berger:2001wr} to resum 
all orders of $\ln^m(p_T^2/M^2)$ type logarithms.  We show that almost
90\% of the low mass Drell-Yan cross sections at LHC energies is 
from gluon initiated partonic subprocesses.  Therefore, the low mass 
Drell-Yan cross section at $p_T>M$ is an advantageous source of information
on the gluon distribution and its nuclear dependence --- shadowing.  
Unlike other probes of gluon distributions, low mass Drell-Yan 
does not have the problem of isolation cuts associated with direct photon 
production at collider energies and does not have the hadronization 
uncertainties of $J/\psi$ and charm production. Moreover, 
the precise information on dilepton production in the
Drell-Yan channel is critical for studying charm production at 
LHC energies.





\vfill\eject
\subsection{Angular distribution of Drell-Yan pairs in $pA$ at the LHC}
\label{subsec:fries}
{\em Rainer~J.~Fries}










\vspace{-0.2cm}
\subsubsection{Introduction}

The Drell-Yan cross section can be described by a contraction of the lepton 
and hadron tensors
\begin{equation}
  \frac{d\sigma}{d M^2 \, d p_T^2 \, d y \, d \Omega} = 
  \frac{\alpha_{\rm em}^2}{64 \pi^3 sM^4}
  L_{\mu\nu}W^{\mu\nu}.
\end{equation}
Here we parametrize the lepton pair by
the invariant mass $M$ of the virtual photon and its transverse momentum 
$p_T$ and rapidity $y$ in the center of mass frame of the colliding
hadrons. In addition we give the direction of one 
of the leptons, say the positively charged one, in 
the 
photon rest frame 
using polar and azimuthal angles $\phi$ and $\theta$: $d\Omega=d\phi \, d\cos
\theta$.
Now the cross section can be understood in terms of four helicity amplitudes
\cite{Collins:iv,Lam:pu,Mirkes:1992hu}
\begin{equation}
  \begin{split}
  \frac{d\sigma}{d M^2 \, d p_T^2 \, d y \, d \Omega} = 
  \frac{\alpha_{\rm em}^2}{64 \pi^3 sM^2} \Big[ W_{\rm TL} (1+ \cos^2 \theta) +
  W_{\rm L} (1/2 - 3/2 \cos^2 \theta)  \\ + W_{\Delta} \sin 2\theta \cos\phi
  + W_{\Delta\Delta} \sin^2 \theta \cos 2\phi \Big].
  \end {split}
  \label{eq:dy}
\end{equation}
These are defined as contractions $W_{\sigma,\sigma'} = \epsilon_\mu(\sigma)
W^{\mu\nu} \epsilon^*_\nu(\sigma')$
of the hadron tensor with polarization 
vectors of the virtual photon for polarizations $\sigma=0,\pm 1$. Only four
out of all possible contractions are independent, the others can be related 
by symmetries of the hadron tensor. The usual choice is to 
pick the longitudinal $W_{\rm L}= W_{0,0}$, the 
helicity flip $W_{\Delta} = (W_{1,0} + W_{0,1})/\sqrt{2}$ and the double 
helicity-flip
amplitude $W_{\Delta\Delta} = W_{1,-1}$ together with the trace
$W_{\rm TL} = W_{\rm T}+W_{\rm L}/2=-W^\mu_\mu /2$ as a basis.
Note that integration over the angles $\theta$ and $\phi$ leaves only 
contributions from the trace
\begin{equation}
  \frac{d\sigma}{d M^2 \, d p_T^2 \, d y} = \frac{\alpha_{\rm em}^2}{64 \pi^3 sM^2}
  \frac{16 \pi}{3} W_{\rm TL} =
  \frac{\alpha_{\rm em}^2}{24 \pi^2 sM^2} (-g_{\mu\nu}) W^{\mu\nu}.
\end{equation}

On the other hand, if we are only interested in relative angular 
distributions --- i.e.\ in the ratio 
\begin{equation}
  \frac{16\pi}{3} \> \frac{d\sigma}{dM^2 \, 
  dp_T^2 \, dy \, d\Omega}\> \Big/ \>
  \frac{d\sigma}{dM^2 \, dp_T^2 \, dy }
\end{equation}
--- we can make use of angular coefficients. Two different sets can 
be found in the literature \cite{Fries:2000da}. One set consists of the 
coefficients
\begin{equation}
  \label{eq:angcoeff}
  A_0 = \frac{W_{\rm L}}{W_{\rm TL}}, \quad A_1 = 
  \frac{W_{\Delta}}{W_{\rm TL}}, \quad
  A_2 = \frac{2 W_{\Delta\Delta}}{W_{\rm TL}},
\end{equation}
the other one is defined by
\begin{equation}
  \lambda = \frac{2-3A_0}{2+A_0}, \quad \mu = \frac{2A_1}{2+A_0}, 
  \quad \nu = \frac{2A_2}{2+A_0}.
  \label{eq:angcoeff2}
\end{equation}

\begin{figure}
\begin{center}
\includegraphics[width=8cm]{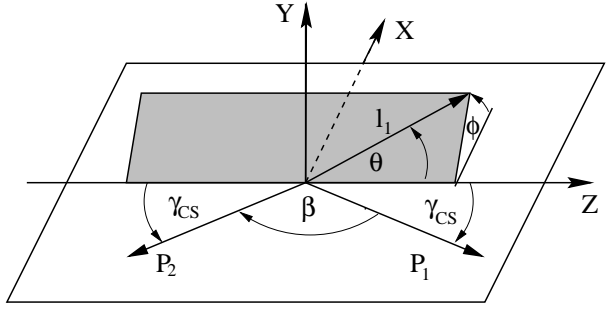} 
\caption{The Collins-Soper frame: the $z$-axis cuts the angle between
 $\mathbf{P}_1$ and $-\mathbf{P}_2$ into halves (the half angle is called the 
 Collins-Soper angle $\gamma_{\rm CS}$) while the $x$-axis is 
 perpendicular to $\mathbf{P}_1$ and $\mathbf{P}_2$. The direction of one
 lepton momentum $\mathbf{l}_1$ can then be given by the angles $\theta$ and 
 $\phi$. }
\label{fig:csframe}
\end{center}
\end{figure}

The helicity amplitudes are frame dependent. In principle we allow all frames
where the photon is at rest, i.e.\ $q^\mu = (M,0,0,0)$. However
there are some frames with particular properties studied in \cite{Lam:pu}. 
Here we only use the Collins-Soper (CS) frame. 
It is characterized by two properties. First, the $y$-axis is perpendicular to 
the plane spanned by the two hadron momenta $\mathbf{P}_1$ and $\mathbf{P}_2$ 
(which are no longer collinear in the photon rest frame as long as $p_T 
\not= 0$, 
as
in our kinematic domain) and second, the $z$-axis cuts 
the angle between $\mathbf{P}_1$ and $-\mathbf{P}_2$ into two equal halves, see
Fig.~\ref{fig:csframe}.

\subsubsection{Leading twist}

\begin{figure}[b]
\begin{center}
\includegraphics[width=5cm]{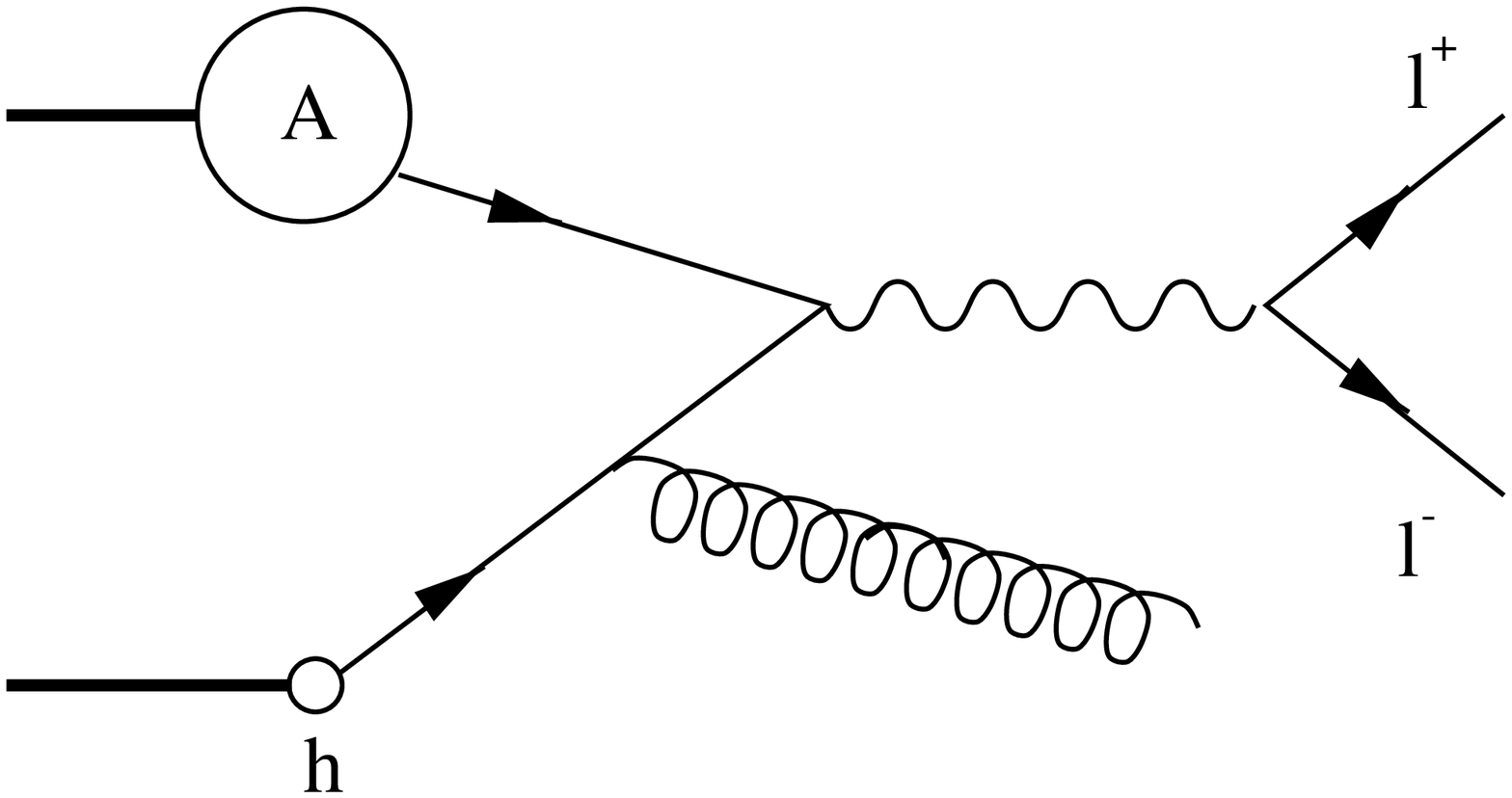} \hspace{4em}
\includegraphics[width=5cm]{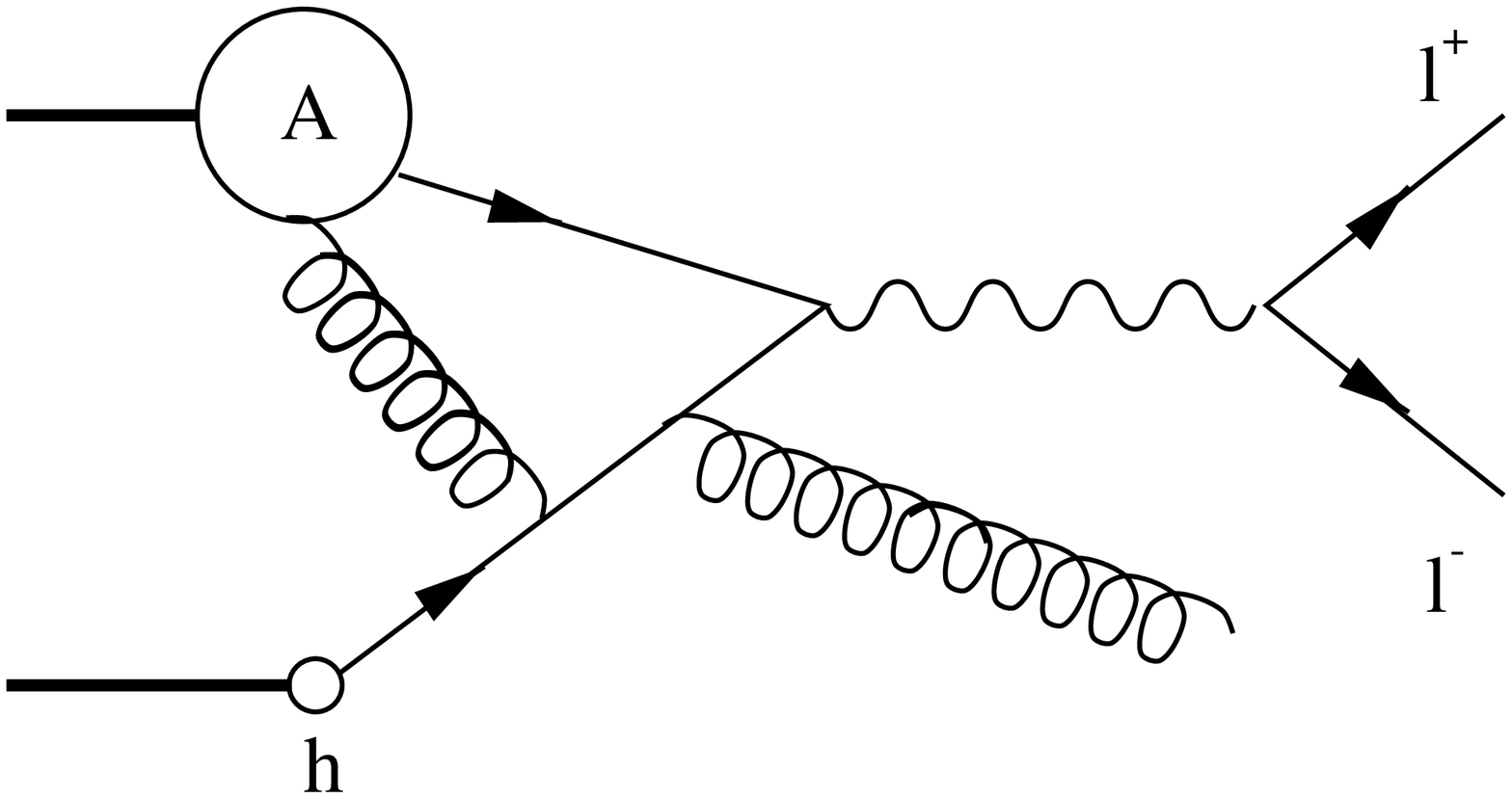}
\caption{Examples for diagrams contributing to the dilepton production in
hadron (h) nucleus (A) scattering at 
twist-2 (left) and twist-4 (right) level.}
\vspace{-0.2cm}
\label{fig:diag}
\end{center}
\end{figure}

The hadron tensor for the Drell-Yan process in a leading-twist (twist-2)
calculation is given by the well known factorization formula as a convolution
of two parton distributions with a perturbative parton cross section.
We are interested here in the kinematic region characterized by intermediate
photon mass $M$ 
of a few GeV
and intermediate transverse momentum $p_T \sim M$.
For these values, the dominant contribution to twist-2 is given
by the 
NLO
perturbative diagrams like the one in 
Fig.~\ref{fig:diag} (left) and we can safely omit logarithmic corrections of 
type $\ln^2(M^2/p_T^2)$. It has been shown that a leading-twist
calculation up to NLO respects the so-called Lam-Tung sum rule, 
$W_{\rm L} = 2 W_{\Delta\Delta}$ \cite{Lam:1980uc}.
In terms of angular coefficients this can be rewritten as $A_0=A_2$ or
$2\nu=1-\lambda$. Furthermore, the spin-flip amplitude $W_\Delta$ has to vanish
for a symmetric colliding system like $pp$. 
For $pA$ we expect small contributions for $W_\Delta$ 
due to lost isospin symmetry and nuclear corrections to the parton 
distributions. Results for $pp$ at $\sqrt{s}=5.5$ TeV have already been 
presented elsewhere \cite{Gavin:1995ch}.

\subsubsection{Nuclear-enhanced twist-4}

For large nuclei, corrections to the leading-twist calculation, induced by 
multiple scattering, play an important role.
The formalism 
for including
these nuclear-enhanced
higher-twist contributions was worked out by Luo, Qiu and 
Sterman \cite{Luo:fz,Luo:ui,Luo:np}. The leading nuclear corrections 
(twist-4 or double scattering) have already been calculated for some 
observables. For Drell-Yan this was first done by Guo \cite{Guo:1998rd} and 
later generalized to the Drell-Yan angular distribution 
\cite{Fries:2000da,Fries:1999jj}.

Figure~\ref{fig:diag} (right) shows an example for a diagram contributing at 
twist-4 level. Now two partons $a$, $b$ from the nucleus and one ($c$) from 
the single hadron are involved. As long as $p_T \sim M$, twist-4 is 
dominated by two different contributions. The double-hard (DH) process where 
each parton from the nucleus has a finite momentum fraction and the soft-hard 
(SH) process where one parton has vanishing momentum fraction. The 
factorization formulas are given by
\begin{eqnarray}
  \label{eq:twist-4}
  W^{\mu\nu} &=& \sum_{a,b,c} \int \frac{d x_c}{x_c} \> T^{\rm DH}_{ab}
  (x_a,x_b) \,
  H_{ab+c}^{\mu\nu} (q,x_a,x_b,x_c) \, f_c(x_c), \\
  W^{\mu\nu} &=& \sum_{a,b,c} \int \frac{d x_c}{x_c} \> D_{x_a,x_b}
  (q,x_c) T^{\rm SH}_{ab} (x_a) \,
  H_{ab+c}^{\mu\nu} (q,x_a,x_b=0,x_c) \, f_c(x_c)
\end{eqnarray}
for double-hard and soft-hard scattering respectively 
\cite{Fries:2000da,Guo:1998rd,Fries:1999jj}. 
The function
$D_{x_a,x_b}(q,x_c)$ is a second order differential operator in $x_a$ and 
$x_b$
while
$T^{\rm DH}_{ab}$ and 
$T^{\rm SH}_{ab}$ are new 
twist-4 matrix elements
which encode nonperturbative correlations between the partons $a$ and $b$. 
Since we are still
missing solid experimental information about these new quantities, they are 
usually modeled in a simple way through parton distributions.
We use $T^{\rm DH}_{ab}(x_a,x_b)= C A^{4/3}f_a(x_a) f_b(x_b)$ and
$T^{\rm SH}_{ab}(x_a)= \lambda^2 A^{4/3}f_a(x_a)$ where $C$ and $\lambda^2$
are normalization constants. The key feature,
the
nuclear enhancement, is their scaling with the nuclear size.

\begin{figure}[t]
\begin{center}
\hspace{-1.5cm}
\includegraphics[width=9cm]{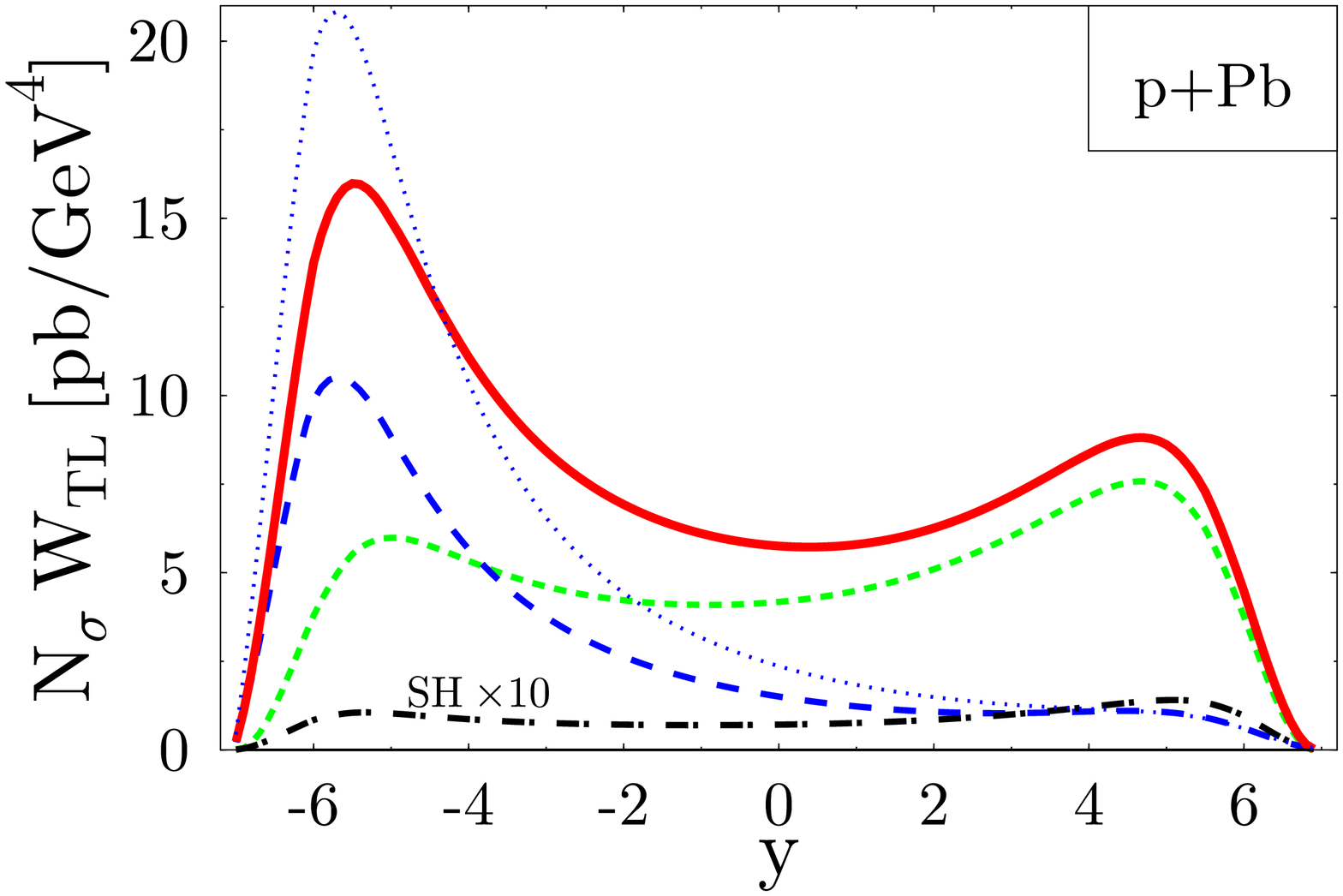} \hspace{-1.0cm}
\includegraphics[width=9cm]{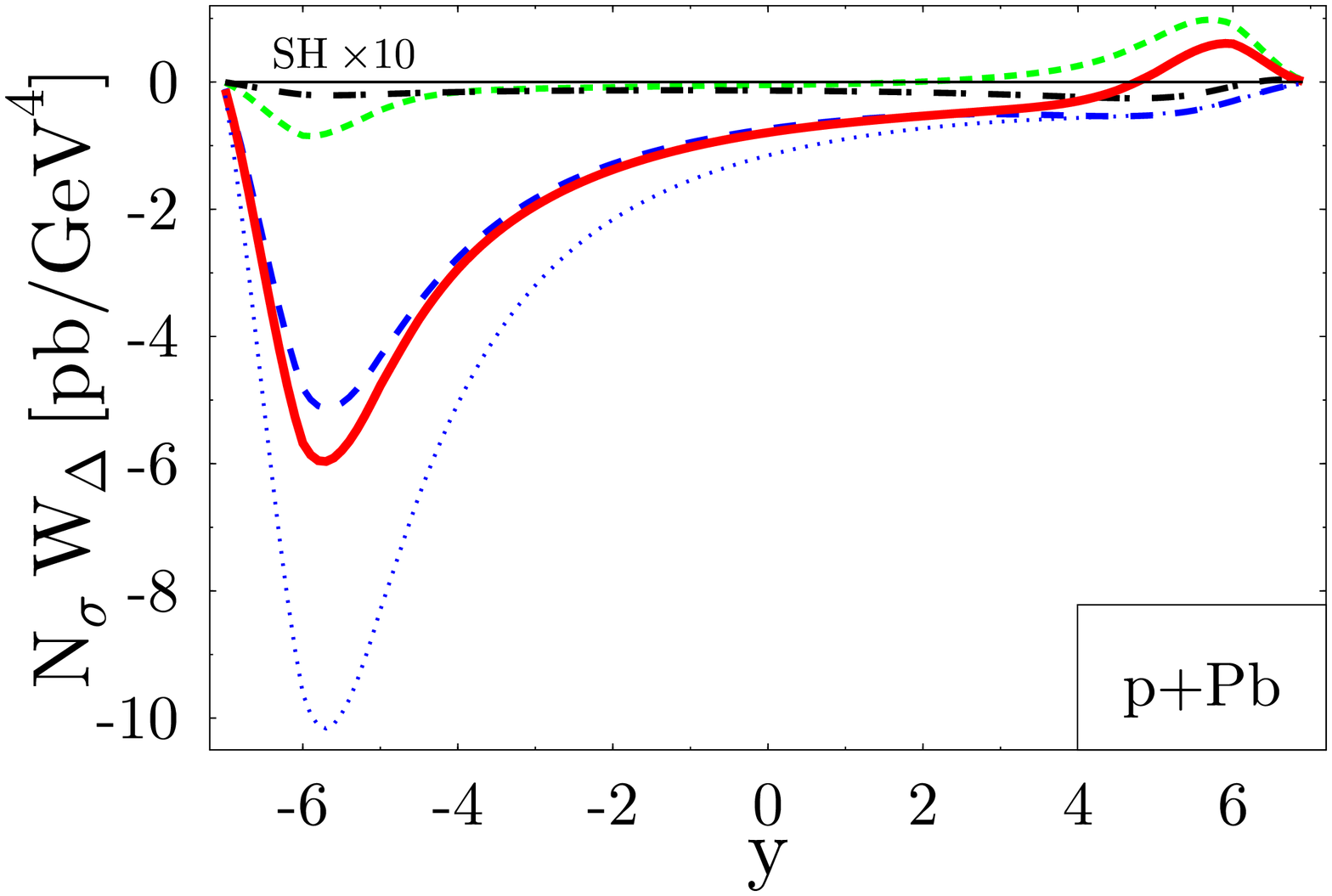}
\caption{Rapidity dependence of $N_\sigma W_{\rm TL}$ (left) and $N_\sigma 
 W_{\Delta}$ (right) for $p$Pb at $\sqrt{s}=8.8$ TeV, $M=5$ GeV and 
 $p_T=4$ GeV: twist-2 NLO (short 
 dashed), double hard with EKS98 modifications (long dashed), soft hard
 (dot-dashed, scaled up by a factor of 10) and the sum of all contributions
 (bold solid line). The double hard contribution 
 calculated without EKS98 modifications is also shown (dotted line). 
 Note that the incoming nucleus has positive rapidity.}
 \label{fig:wy}
\vspace{-0.2cm}
\end{center}
\end{figure}

It has been shown in Refs. \cite{Fries:2000da,Fries:1999jj} that the
DH process shows a trivial angular pattern in the sense that it is
similar to the 
LO
annihilation of on-shell quarks.
The only difference is that one of the quarks now carries finite
$p_T$. In this spirit, it is no surprise that the
DH contribution respects the Lam-Tung relation. On the other hand, SH
scattering is more complicated and violates the Lam-Tung sum rule.
We also
expect that the spin-flip amplitude, $W_\Delta$, can receive
large contributions from the twist-4 calculation.

\subsubsection{Numerical results}

In this section, we present some numerical results obtained for $p$Pb
collisions at the LHC energy $\sqrt{s}= 8.8$ TeV. Results for RHIC energies
can be found elsewhere \cite{Fries:2000da}.
We use the CTEQ5L parton distributions \cite{Lai:1999wy} combined with EKS98
\cite{Eskola:1998df} 
nuclear modifications both for the nuclear parton distributions and
for the models of the twist-4 matrix elements. In some plots we also
give results for double-hard contributions 
without nuclear modification.
Since we do not know anything about the
correct $x$-dependence of the higher-twist matrix elements, this gives
an impression of the theoretical uncertainty we may 
assume for the higher-twist calculation.  The normalization constants 
for the twist-4 matrix
elements are chosen to be $\lambda^2 = 0.01$ GeV$^2$ and $C=0.005$
GeV$^2$.  In order to enable convenient comparison with cross sections,
we show all helicity amplitudes multiplied by the prefactor
$N_\sigma=\alpha_{\rm em}^2 / (64 \pi^3 s M^2)$ of Eq.~(\ref{eq:dy}).

In Fig.~\ref{fig:wy} we give results for the helicity amplitudes $W_{\rm TL}$
and $W_\Delta$ as functions of rapidity. We observe that DH scattering
gives a large contribution at negative rapidities (the direction of the 
proton) which can easily balance 
the suppression of the twist-2 contribution by shadowing.
Soft-hard scattering is strongly suppressed at these energies. 
$W_{\rm L}$ (not shown) has qualitatively the same behavior as $W_{\rm TL}$. 
The helicity-flip amplitude $W_{\Delta}$ picks up only a small contribution 
from twist-2 and is entirely dominated by double-hard scattering, as already 
expected. This would be a good observable to pin down nuclear effects in
$pA$ collisions.
\begin{figure}[t]
\begin{center}
\hspace{-1.5cm}
\includegraphics[width=9.0cm]{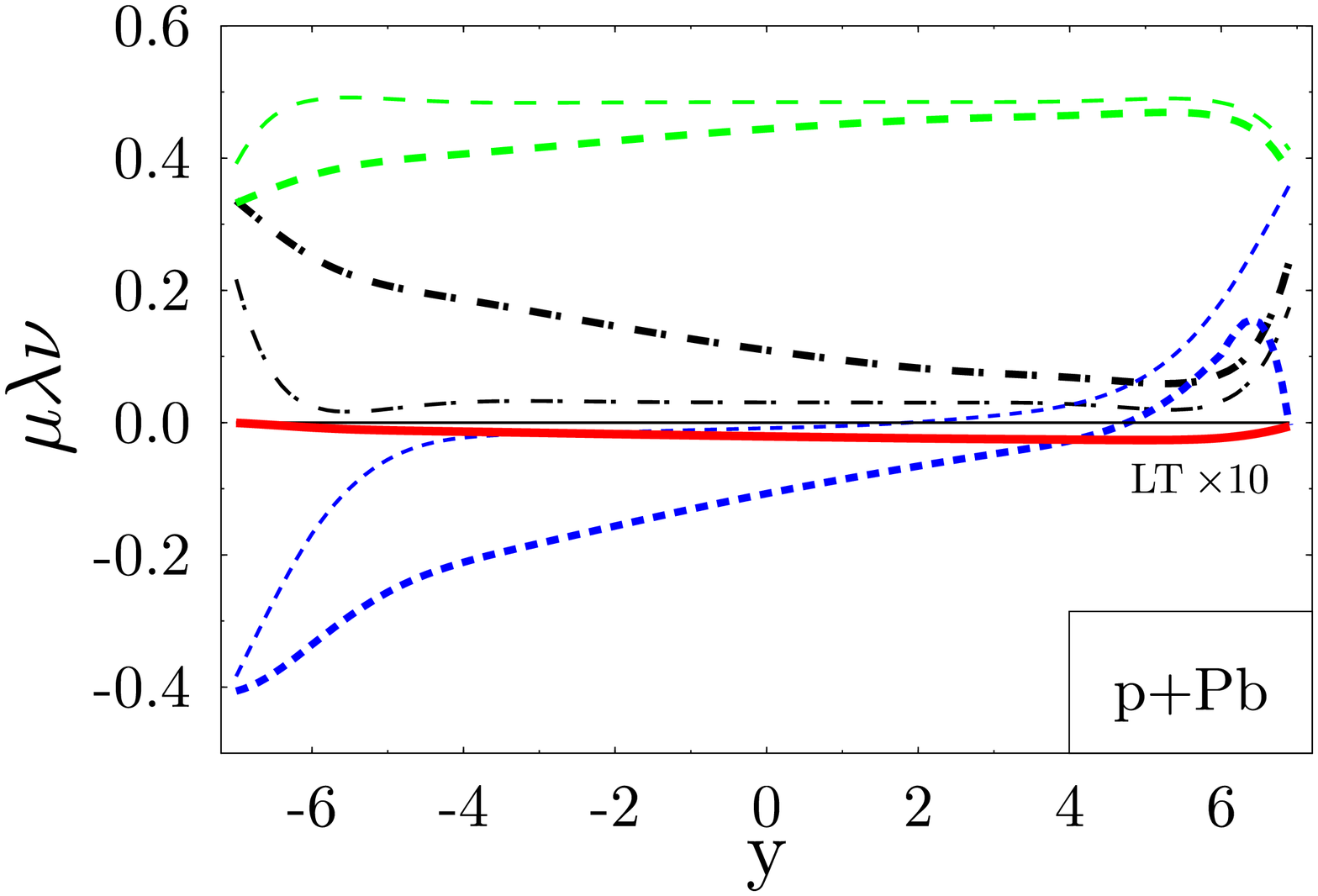} \hspace{-1.0cm}
\includegraphics[width=9.0cm]{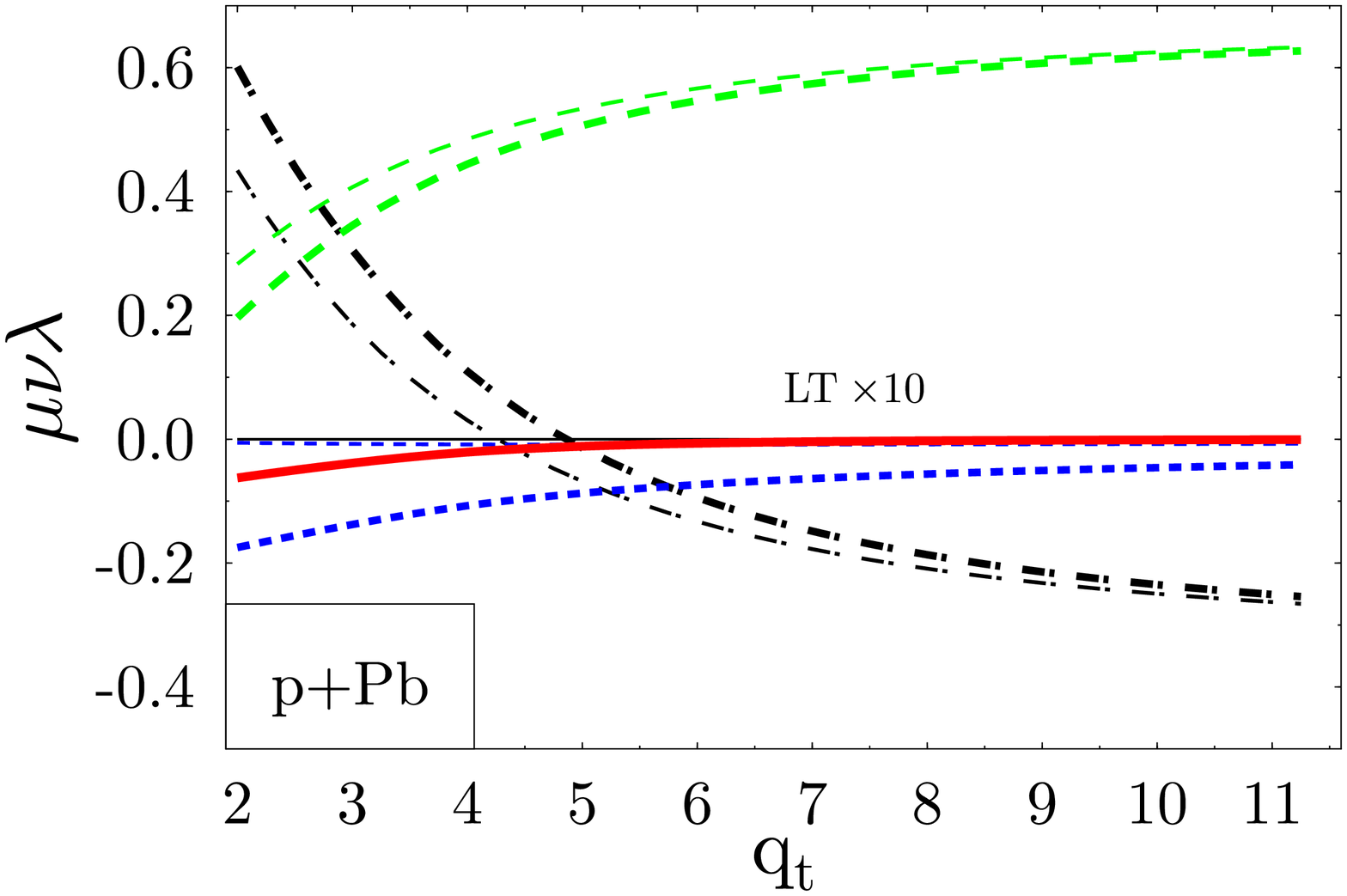}
\caption{The angular coefficients $\lambda$ (long dashed), $\mu$ 
 (short dashed) and $\nu$ (dot dashed) for $p$Pb at $\sqrt{s}=8.8$ TeV
 and $M=5$ GeV as functions of rapidity (left, at $p_T = 4$ GeV) and
 transverse momentum (right, at $y=0$). Thin lines represent pure
 twist-2 calculations, thick lines show the results including
 twist-4. The solid line gives the violation of the Lam-Tung relation
 $2\nu-(1-\lambda)$ scaled up by a factor of 10.}  \label{fig:lmn}
\end{center}
\vspace{-0.5cm}
\end{figure}
Figure~\ref{fig:lmn} gives the full set of angular coefficients $\lambda$, 
$\mu$ and $\nu$ as functions of $y$ and $p_T$. Both the twist-2 
results and the modifications by twist-4 are given. The violation of the 
Lam-Tung sum rule is numerically almost negligible since the soft-hard 
contribution is so small. Note here that earlier experiments have discovered 
a large violation of the Lam-Tung relation in $\pi A$ collisions 
\cite{Conway:fs}. This issue is still not fully resolved. 

Thus
$pA$ collisions offer the unique opportunity to study nuclear effects directly
via the rapidity dependence of observables. Helicity amplitudes and angular 
coefficients can help to pin down the role of nuclear-enhanced higher twist.
The helicity-flip amplitude $W_{\Delta}$ and the 
coefficient $\mu$, which both vanish for
$pp$, 
are particularly promising. 
Several different species of nuclei are advisable 
to address the important question of $A$ scaling.

In addition, the angular distribution of Drell-Yan lepton pairs might
play a crucial role in the extraction of the Drell-Yan process from
dilepton data. Dilepton spectra at the LHC have a large contribution
from the decay of correlated and uncorrelated heavy quark pairs
\cite{Gavin:ma,Lokhtin:2001nh}.
However the angular pattern of these dileptons is isotropic rather
than modulated as in the Drell-Yan case.  This provides additional
means to discriminate between Drell-Yan dileptons and those from heavy
quark decays.

\vspace{0.5cm}
\subsection{Heavy flavor total cross sections} 
\label{section431}
{\em Ramona Vogt}

Heavy quark production will be an important part of the LHC program, with both
open heavy flavors and quarkonium studied extensively.  Heavy quark
production will probe the small $x$ region of the gluon distribution in the
proton and in the nucleus: at $y=0$, $x \sim 4 \times 10^{-4}$ for charm and
$x \sim 0.0011$ for bottom.  Direct information on the gluon distribution in
the proton is still lacking, let alone the nucleus.  Open heavy flavor 
production is a better direct test of the nuclear gluon distribution than
quarkonium which is subject to other cold matter effects such as
nuclear absorption.  

Heavy quark studies in $pA$ interactions provide important benchmarks for 
further measurements in $AA$. 
More than one $Q \overline Q$ pair (bottom as well as charm) will be produced 
in each central $AA$ collision.  Therefore heavy quark decays will dominate
the lepton pair continuum between the $J/\psi$ and $Z^0$ peaks \cite{cmsnote}.
Although heavy quark energy loss in the medium is currently expected 
to be small \cite{Dokshitzer:2001zm}, 
the shape of the dilepton continuum could be significantly altered
 due to modifications of
the heavy quark distributions by energy loss \cite{Lokhtin:2001nh,Lin:1998bd}.
An understanding of heavy quark production in $pA$ is
also a necessity for the study of quarkonium production in $AA$ collisions.
Models predicting a quarkonium enhancement relative to heavy flavor 
production in nuclear collisions require a reliable measure of the heavy flavor
production rates in $AA$ collisions with shadowing included to either verify
or disprove these predictions.

In this section, we first discuss our estimates of the $Q \overline Q$ total
cross sections and some of their associated uncertainties.  We then describe a
more detailed measurement of the nuclear gluon distribution using dileptons
from heavy quark decays.

At LO heavy quarks are produced by $gg$ fusion and
$q \overline q$ annihilation while at NLO $qg$ and
$\overline q g$ scattering is also included.  To any order, the partonic 
cross section may be expressed in terms of dimensionless scaling functions
$f^{(k,l)}_{ij}$ that depend only on the variable $\eta$ 
\cite{Kidonakis:2001nj},
\begin{eqnarray}
\label{scalingfunctions}
\hat \sigma_{ij}(\hat s,m_Q^2,Q^2) = \frac{\alpha^2_s(Q^2)}{m^2}
\sum\limits_{k=0}^{\infty} \,\, \left( 4 \pi \alpha_s(Q^2) \right)^k
\sum\limits_{l=0}^k \,\, f^{(k,l)}_{ij}(\eta) \,\,
\ln^l\left(\frac{Q^2}{m_Q^2}\right) \, , 
\end{eqnarray} 
where $\hat s$ is the partonic center of mass energy squared, 
$m_Q$ is the heavy quark mass,
$Q$ is the scale and $\eta = \hat s/4 m_Q^2 - 1$.  
The cross section is calculated as an expansion in powers of $\alpha_s$
with $k=0$ corresponding to the Born cross section at order ${\cal
O}(\alpha_s^2)$.  The first correction, $k=1$, corresponds to the NLO cross
section at ${\cal O}(\alpha_s^3)$.  It is only at this order and above that
the dependence on renormalization scale, $Q_R$, enters the calculation
since when $k=1$
and $l=1$ the logarithm $\ln(Q^2/m_Q^2)$ 
appears, as in gauge boson and Drell-Yan production at ${\cal O}(\alpha_s)$,
described in Section~\ref{section411}.  
The dependence on the factorization scale, $Q_F$, in the argument of
$\alpha_s$, appears already at LO.  We assume that $Q_R = Q_F = Q$.  
The next-to-next-to-leading
order (NNLO) corrections to next-to-next-to-leading logarithm (NNLL)
have been calculated near threshold \cite{Kidonakis:2001nj} but
the complete calculation only exists to NLO.

The total hadronic cross section is obtained by convoluting the total partonic
cross section with the parton distribution functions (PDFs)
of the initial hadrons,
\begin{eqnarray}
\label{totalhadroncrs}
\sigma_{pp}(s,m_Q^2) = \sum_{i,j = q,{\overline q},g} 
\int_{\frac{4m_Q^2}{s}}^{1} \frac{d\tau}{\tau}\, \delta(x_1 x_2 - \tau) \,
\frac{f_i^p(x_1,Q^2)}{x_1} \frac{f_j^p(x_2,Q^2)}{x_2} \, 
\hat \sigma_{ij}(\tau ,m_Q^2,Q^2)\, , 
\end{eqnarray}
where the sum $i$ is over all massless partons and
$x_1$ and $x_2$ are fractional momenta.
The parton densities are evaluated at
scale $Q$.  All our calculations are fully NLO, applying NLO parton
distribution functions and the two-loop $\alpha_s$ to both the ${\cal
O}(\alpha_s^2)$ and ${\cal O}(\alpha_s^3)$ contributions, as is typically done
\cite{Kidonakis:2001nj,Mangano:jk}.

To obtain the $pp$ cross sections at LHC, we compare the NLO
cross sections to the available $c \overline c$ and $b \overline b$ production
data by varying the mass, $m_Q$, and scale, $Q$, to obtain the `best'
agreement with the data for several combinations of $m_Q$, $Q$, and PDF.
We use the recent MRST HO central gluon \cite{Martin:1998sq}, CTEQ 5M 
\cite{Lai:1999wy}, and
GRV 98 HO \cite{Gluck:1998xa} distributions.   The charm mass is
varied between 1.2 and 1.8 GeV with scales between $m_c$ and $2m_c$.  The
bottom mass is varied over 4.25 to 5 GeV with $m_b/2 \leq Q \leq 2m_b$.
The best agreement for $Q =
m_c$ is with $m_c = 1.4$ GeV and $m_c = 1.2$ GeV is the best choice using $Q =
2m_c$ for the MRST HO and CTEQ 5M distributions.  The best agreement with GRV
98 HO is $Q = m_c = 1.3$ GeV while the results with $Q = 2m_c$ lie below
the data for all $m_c$.  All five results agree very well with each other for
$pp \rightarrow c \overline c$.
We find reasonable agreement with all three PDFs for $m_b = Q = 4.75$ GeV, 
$m_b = Q/2 = 4.5$ GeV and $m_b = 2Q = 5$ GeV.

Before calculating the $Q \overline Q$
cross sections at nuclear colliders, some comments need
to be made about the validity of the procedure.  
Since the $c \overline c$ 
calculations can only be made to agree with the data when somewhat lower than
average quark masses are used and the $pp \rightarrow b \overline b$ data
also need smaller $m_b$ to agree at NLO, it is reasonable to expect that 
corrections beyond NLO could be large.  Indeed, the preliminary HERA-B 
cross section agrees with the NNLO-NNLL cross section in 
Ref.~\cite{Kidonakis:2001nj}, 
suggesting that the next-order correction could be nearly a
factor of two.  Thus the NNLO correction could be nearly as large as the NLO
cross section.

Unfortunately, the NNLO-NNLL calculation is not a full result
and is valid only near threshold.  The $p \overline p$ data at higher energies,
while not total cross sections, also show a large discrepancy between the
perturbative NLO result and the data, nearly a factor of three 
\cite{Affolder:1999hm}.  
This difference could be accounted for using unintegrated parton densities
\cite{Lipatov:2001ny} 
although the unintegrated distributions vary widely.  
The problem remains how to connect the regimes where 
near-threshold corrections are applicable and where high-energy, small-$x$ 
physics dominates.  The problem is increased for charm where, even at low
energies, we are far away from threshold.  

We take the most straightforward approach--comparing the NLO calculations to 
the $pp$ and $\pi^- p$ data without higher-order corrections.  The 
$c \overline c$ data are 
extensive while the $b \overline b$ data are sparse.  The 
$c \overline
c$ data tend to favor lighter charm masses but the $b
\overline b$ data are less clear.  A value of $m_b = 4.75$ GeV
underpredicts the Tevatron results \cite{Affolder:1999hm} at NLO but agrees 
reasonably with the average of the $\pi^- p$ data.  However, for the HERA-B
measurement \cite{Abt:2002rd}
to be compatible with a NLO evaluation, the $b$ quark mass would
have to be reduced to 4.25 GeV, perhaps more compatible with
the Tevatron results.  Therefore, if the NNLO cross section could be calculated
fully, it would likely be more compatible with a larger quark mass.
A quantitative statement is not possible, particularly for charm
which is far enough above threshold at fixed-target energies for the
approximation to be questionable \cite{Smith:1996sb}, see however 
Ref. \cite{Kidonakis:2002vj}.

If we assume that the energy dependence of the NNLO and higher order
cross sections is not substantially different from that at LO and NLO,
then we should be in the right range for collider energies.  The
theoretical $K$ factors have a relatively weak $\sqrt{s}$ dependence,
$\leq 50$\% for $20 \, {\rm GeV} \leq \sqrt{s} \leq 14 \, {\rm TeV}$
\cite{Vogt:2002eu}.
The heavy quark distributions might be slightly affected since the
shapes are somewhat sensitive to the quark mass but the LHC energy is
far enough above threshold for the effect to be small.

\begin{table}[htb]
\begin{center}
\caption{Charm and bottom total cross sections per nucleon at 5.5, 8.8 and
14 TeV.}
\label{qqbtab}
\begin{tabular}{cccccccc}
\multicolumn{3}{c}{} & \multicolumn{2}{c}{5.5 TeV} 
& \multicolumn{2}{c}{8.8 TeV} & 
14 TeV \\ 
\multicolumn{3}{c}{} & $pp$ & $p$Pb & $pp$ & $p$Pb & $pp$ \\ \hline 
\multicolumn{8}{c}{$c \overline c$} \\
PDF & $m_c$ (GeV) & $Q/m_c$ & $\sigma$ (mb) &  $\sigma$ (mb) & 
$\sigma$ (mb) & $\sigma$ (mb) & $\sigma$ (mb)
\\ \hline
MRST HO & 1.4 & 1   & 3.54 & 3.00 & 4.54 & 3.76 & 5.71 \\
MRST HO & 1.2 & 2   & 6.26 & 5.14 & 8.39 & 6.76 & 11.0 \\
CTEQ 5M & 1.4 & 1   & 4.52 & 3.73 & 5.72 & 4.61 & 7.04 \\
CTEQ 5M & 1.2 & 2   & 7.39 & 5.98 & 9.57 & 7.56 & 12.0 \\
GRV 98 HO & 1.3 & 1 & 4.59 & 3.70 & 6.20 & 4.89 & 8.21 \\ \hline
\multicolumn{8}{c}{$b \overline b$} \\
PDF & $m_b$ (GeV) & $Q/m_b$ & $\sigma$ ($\mu$b) & $\sigma$ ($\mu$b) 
& $\sigma$ ($\mu$b) & $\sigma$ ($\mu$b) & $\sigma$ ($\mu$b) \\ \hline
MRST HO & 4.75 & 1     & 195.4 & 180.1 & 309.8 & 279.7 & 475.8 \\
MRST HO & 4.5  & 2     & 201.5 & 186.6 & 326.1 & 296.0 & 510.2 \\
MRST HO & 5.0  & 0.5   & 199.0 & 183.9 & 302.2 & 272.6 & 445.1 \\
GRV 98 HO & 4.75 & 1   & 177.6 & 162.7 & 289.7 & 259.7 & 458.3 \\
GRV 98 HO & 4.5  & 2   & 199.0 & 183.9 & 329.9 & 298.7 & 530.7 \\
GRV 98 HO & 5.0  & 0.5 & 166.0 & 151.2 & 262.8 & 233.6 & 403.2 \\ \hline
\end{tabular}
\end{center}
\end{table}

The total $pp$ and $pA$ cross sections at 5.5, 8.8 and 14 TeV,
calculated with a code originally from J. Smith,
are given in  Table~\ref{qqbtab}.  
Shadowing is the only nuclear effect included in the calculation.
The cross sections  do not reflect any
possible difference between the $pA$ and $Ap$ rates because they are integrated
over all $x_1$ and $x_2$.  However, the different $x$ regions probed
by flipping the beams could be effectively studied through the dilepton decay
channel.  Thus we
now turn to a calculation of the nuclear gluon distribution in $pA$
interactions \cite{Eskola:2001gt}.  
We show that the dilepton continuum can be used to
study nuclear shadowing and reproduces the input shadowing function well, in
this case, the EKS98 parameterization \cite{Eskola:1998iy,Eskola:1998df}.  
To simplify
notation, we refer to generic heavy quarks, $Q$, and heavy-flavored mesons,
$H$.  The lepton pair production cross section in $pA$ interactions is
\begin{eqnarray}
  \frac{d \sigma^{pA \rightarrow l \overline l +X}}
  {dM_{l \overline l} dy_{l \overline l}} & = &
   \int d^3\vec p_{l} d^3\vec p_{\overline l} 
                 \int d^3\vec p_{H} d^3\vec p_{\overline H}
    \,\delta(M_{l \overline l}-M(p_l,p_{\overline l})) 
    \,\delta(y_{l \overline l}-y(p_l,p_{\overline l}))  \nonumber \\
  & & \times
    \frac{d\Gamma^{H \rightarrow l+X}(\vec p_H)}{d^3 \vec p_l}
    \,\, 
    \frac{d\Gamma^{\overline H \rightarrow \overline l+X}(\vec p_{\overline 
     H})}
         {d^3 \vec p_{\overline l}}
    \,\, 
    \frac{d\sigma^{{\rm p}A \rightarrow H \overline H+X}}
         {d^3 \vec p_{H} d^3 \vec p_{\overline H }} \qquad \qquad
    \nonumber \\
  & & \times 
    \theta(y_{\rm min}<y_{l},y_{\overline l}<y_{\rm max})
    \theta(\phi_{\rm min}<\phi_{l},\phi_{\overline l}<\phi_{\rm max})  
   \qquad \label{dspAll}
\end{eqnarray}
where $M  (p_l,p_{\overline l})$ and $ y(p_l,p_{\overline l})$ are the
invariant mass and rapidity of the lepton pair.
The decay rate, $d\Gamma^{H \rightarrow l+X}(\vec p_H)/d^3 \vec p_l$, is 
the probability that meson $H$ with momentum $\vec p_H$ decays
to a lepton $l$ with momentum $\vec p_l$.  The $\theta$ functions define
single lepton rapidity and azimuthal angle cuts used to simulate
detector acceptances.

Using a fragmentation function $D^{H}_{Q}$ to describe quark
fragmentation to mesons, the $H \overline H$ production cross section 
can be written as
\begin{eqnarray}
  \frac{d \sigma^{pA \rightarrow H \overline H+X}}
       {d^{3} \vec p_H d^3 \vec p_{\overline H}}  &= & 
    \int \frac{d^3\vec p_{Q}}{E_{Q}} 
         \frac{d^3\vec p_{\overline Q}}{E_{\overline Q}}
   \, E_{Q}E_{\overline Q}\frac{d\sigma^{pA \rightarrow Q 
    \overline Q+X}}
        {d^{3} \vec p_Q d^3 \vec p_{\overline Q}}  
    \int dz_1 dz_2 D^{H}_{Q}(z_1)
     D^{\overline H}_{\overline Q}(z_2) \nonumber \\
   & \times & \, \delta^{(3)}(\vec p_H - z_1 \vec p_Q)
             \, \delta^{(3)}(\vec p_{\overline H} - z_2 \vec p_{\overline Q})
\, \, . \label{dspADD}
\end{eqnarray}
Our calculations proved to be
independent of $D^H_Q$.
The hadronic heavy quark production cross section per nucleon in
Eq.~(\ref{dspADD}) is
\begin{eqnarray}
\frac{1}{A} E_{Q}E_{\overline Q} \frac{d\sigma^{pA \rightarrow
   Q \overline Q + X}}
    {d^{3} \vec p_Q d^3 \vec p_{\overline Q}} = \sum_{i,j}
    \int dx_1 dx_2 \, f_i^p(x_1,Q^2) f_j^A(x_2,Q^2) 
     E_{Q}E_{\overline Q}
    \frac{d\hat \sigma^{ij \rightarrow Q \overline Q}}
                       {d^3 \vec p_{Q} d^3 \vec p_{\overline Q}}
\label{dspnuc}
\end{eqnarray}
where the parton densities in the nucleus are related to those in the 
proton by $f_i^A = f_i^p R_i^A$ where 
$R_i^A$ is the EKS98 parameterization 
\cite{Eskola:1998iy,Eskola:1998df}.  The partonic cross
section in Eq.~(\ref{dspnuc}) is the differential of 
Eq.~(\ref{scalingfunctions}) at $k=0$.
Note that the total lepton pair production cross section is equal to the total
$Q \overline Q$ cross section multiplied by the square of the 
lepton branching ratio.

\begin{figure}[htb]
\begin{center}
\includegraphics[width=9.0cm]{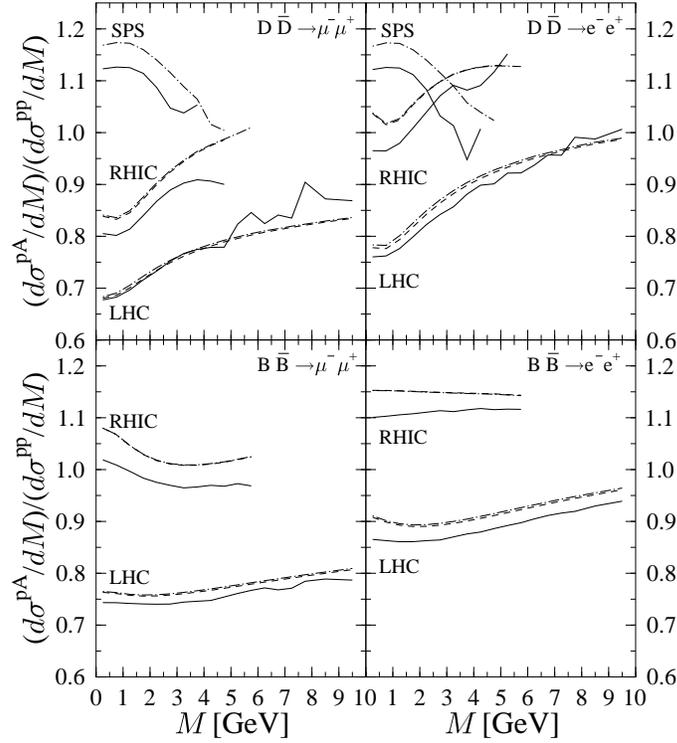} 
\vspace{-0.5cm}
\caption[]{The ratios of lepton pairs from correlated $D \overline D$ and
$B \overline B$ decays in $pA$ to $pp$ collisions at the same energies (solid
curves) compared to the input $R_g^A$ at the
average $x_2$ and $Q$ (dashed)/$\sqrt{\langle Q^2 \rangle}$ (dot-dashed)
of each $M$ bin.  From Ref.~\cite{Eskola:2001gt}.
}
\label{lepvsm}
\end{center}
\end{figure}

We compare the ratios of lepton pair cross sections with
the input $R_g^A$ in Fig.~\ref{lepvsm}. 
The RHIC and LHC results are
integrated over the rapidity intervals appropriate to the PHENIX and ALICE
dilepton coverages.  The ratio is similar to but smaller than $R_g^A$ at all
energies. The higher the energy, the better the agreement.
The ratio always lies below $R_g^A$ because
we include $q \overline q$ annihilation, which decreases with energy, 
but quark and gluon shadowing are different and 
the phase space integration smears the shadowing effect relative to
$R_g^A(\langle x_2 \rangle, \langle Q \rangle)$.  
The calculated ratio deviates slightly more from $R_g^A(\langle
x_2\rangle,\langle Q\rangle)$ for $e^+ e^-$ than for $\mu^+ \mu^-$ 
because the curvature of $R_g^A$ with
$x$ is stronger at larger $x$ and, due to the differences in rapidity
coverage, the average values of $x_2$ are larger for
$e^+ e^-$.

The average $Q^2$ increases with energy and quark mass while the 
average $x_2$ decreases with energy.  At the SPS,
$0.14 \leq \langle x_2 \rangle
\leq 0.32$, $R_g^A$ is decreasing.  
At RHIC,  $0.003 \leq \langle x_2 \rangle \leq
0.012$, $R_g^A$ is increasing quite rapidly. Finally, at the
LHC, $3\times 10^{-5} \leq \langle x_2 \rangle \leq 2\times 10^{-4}$,
$R_g^A$ is almost independent of $x$. The values of
$\langle x_2 \rangle$ are typically larger for electron pairs at
collider energies because the electron coverage is more central than
the muon coverage.  We have only considered $p$Pb collisions but
in Pb$p$ collisions $x_1$ is an order of magnitude or more greater 
than $x_2$ at forward rapidities such as in the ALICE muon spectrometer.  Thus
doing both $pA$ and $Ap$ collisions would greatly increase the $x$ 
range over which the nuclear gluon distribution could be determined.
By assessing $c \overline c$ and $b \overline b$
pairs together, it is possible to investigate the evolution of the nuclear
gluon distribution with $Q^2$, albeit with slightly shifted values of $x$
due to the mass difference.

The importance of such a study as a function of $A$ and $\sqrt{s}$ is clear, 
especially when there is currently little direct
information on the nuclear gluon distribution.  It is desirable for 
the $pp$, $p$Pb, and Pb$p$ measurements to be performed 
at the same energy as the Pb+Pb collisions at 5.5 TeV even though
the relative proton/neutron number is negligible in gluon-dominated processes.
The effect of shadowing is easier
to determine if the $x$ value does not shift with energy, particularly
because the dilepton measurement does not provide a precise determination of 
the heavy quark momenta, only an average.

As shown in Fig.~\ref{lepvsm}, energy systematics would provide a much
clearer picture of the nuclear gluon distribution over the entire $x$
range.  The PHENIX collaboration has already demonstrated that charm
can be measured using single leptons \cite{Adcox:2002cg}.  Perhaps it
might be possible to obtain the bottom cross section the same way as
well at higher $p_T$.  Another handle on the $Q\overline Q$ cross
sections may be obtained by $e\mu$ correlations, since the DY and
quarkonium decays will not contribute. At the LHC, uncorrelated
$Q\overline Q$ pairs can make a large contribution to the
continuum. These pairs, with a larger rapidity gap between the
leptons, are usually of higher mass than the correlated
pairs. However, the larger rapidity gap reduces their acceptance in a
finite detector. Like-sign subtraction also reduces the uncorrelated
background. For further discussion, see
Refs.~\cite{Gavin:ma,Lin:1995pk,Eskola:2001gt}.

\vspace{1.0cm}
\subsection{Cronin effect in proton-nucleus collisions:
a survey of theoretical models}
\label{subsec:accardi}
{\em Alberto Accardi}


\subsubsection{Introduction}
\label{sec:acc_intro}

\begin{figure}[thb]
\begin{center}
\includegraphics[width=6cm]{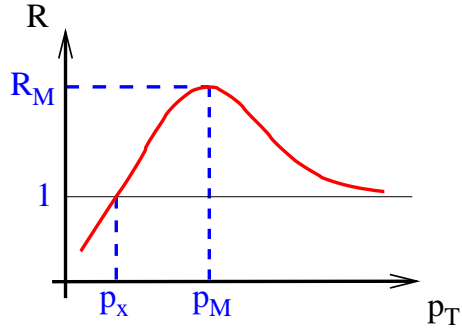} 
\vskip-.3cm
\caption{Definition of $p_\times, p_M,
R_M$.}
 \label{fig:cronineffect}
\end{center}
\end{figure}

In this short note I will compare available theoretical models for the
description of the so-called Cronin effect \cite{Cronin:zm} 
in inclusive hadron spectra in $pA$ collisions. The analysis will be
limited to references containing quantitative predictions for $pA$
collisions
\cite{Wang:1998ww,Zhang:2001ce,Kopeliovich:2002yh,Accardi:2001ih,Vitev:2002pf}.
The observable we study is the {\it Cronin ratio}, $R$, of the inclusive
differential cross sections for proton scattering on two different 
targets, normalized to the respective atomic numbers $A$ and $B$:
\begin{equation*}
    R(p_T) = \frac{B}{A} 
        \frac{d\sigma_{pA}/d^2p_T}{d\sigma_{pB}/d^2p_T} \ .
\end{equation*}
In absence of nuclear effects one would expect $R(p_T)$$=$$1$, 
but for $A>B$ a suppression is measured at small $p_T$, 
and an enhancement at moderate $p_T$ with $R(p_T) \rightarrow 1$ 
as $p_T\rightarrow\infty$.
This behaviour may be characterized by the value of three
parameters: the transverse momentum $p_\times$ at which $R$ crosses unity and the transverse momentum $p_M$
at which $R$ reaches its maximum value $R_M$$=$$R(p_M)$, see
Fig.~\ref{fig:cronineffect}. These {\it Cronin parameters} will be 
studied in Sec.~\ref{sec:predictions}.

The Cronin effect has received renewed interest after the 
experimental discovery at RHIC of an unexpectedly small $R<1$ 
in Au+Au collisions at $\sqrt s $$=$$ 130$ GeV compared to $pp$ collisions at
the same energy \cite{Adcox:2001jp,Reygers:2002nv}. This fact has been
proposed as an experimental signature of large jet quenching,
suggesting that a Quark-Gluon Plasma was created during the collision
\cite{Levai:2001dc,Vitev:2002vr}. 
However, the extrapolation to RHIC and LHC energies of the known Cronin effect
at lower energies is haunted by large theoretical uncertainties
which may make the interpretation such signals unreliable.
More light has been recently shed on this problem by the 
RHIC data on $dA$ collisions at $\sqrt{s}=$ 200 GeV 
\cite{RHICdAu,Adler:2003ii,Adams:2003im,Back:2003ns}.

Since no hot and dense medium is expected to be created in $pA$
collisions, a $pA$ run at the same nucleon-nucleon energy as in $AA$
collisions would be of major importance to test 
the theoretical models and to have reliable baseline spectra for the
extraction of novel physical effects.  
This has been the key in establishing the relevance of
final state effects (jet quenching) as the the cause of the
suppression of the Cronin ratio $R$ in Au+Au collision at RHIC
\cite{RHICdAu,Vitevusermeeting}. This measurement
will also be essential at LHC, where at the accessible Bjorken
$x \lesssim 10^{-3}$ new physics such as gluon saturation may (or may not)
come into play.

A further advantage of $pA$ collisions is the relatively small
multiplicity compared to $AA$ collisions. For this reason, at
ALICE, minijets may be observed at $p_T> 5$ GeV, see Section
\ref{pA_SEC:machine} As I will discuss in Sec.~\ref{sec:predictions},
the Cronin effect on minijet production may then be a further check of
the models.

\subsubsection{The models}

Soon after the discovery of the Cronin effect \cite{Cronin:zm}, it was
realized that the observed nuclear enhancement of the $p_T$-spectra
could be explained in terms of multiple interactions
\cite{Kuhn:1975dq,Lev:1983hh,Lev:1983hh}. The models may be classified
according to the physical object which is rescattering
(the projectile hadron or its partons), and to the ``hardness'' of the
rescattering processes taken into account. Note that a parton is
commonly said to undergo a hard scattering if the 
exchanged momentum is of the order or greater than approximately 1
GeV. However, physically there is no sharp distinction between soft
and hard momentum transfer. Therefore, I prefer to refer to
the so-called two component models of hadron transverse spectra, and
call a scattering which is described by a power-law differential
cross section at large $p_T$ hard, and a scattering whose
cross section is decreasing faster than any inverse power of the
transverse momentum at large $p_T$ soft. 
In Tables~\ref{tab:hadronic} and \ref{tab:partonic},  I provide 
a quick comparison of the hadronic and
partonic rescattering models.

\subsubsection*{Soft hadronic rescattering models 
\cite{Zhang:2001ce,Wang:1998ww}}

These models are based on the pQCD collinearly factorized cross
section for inclusive particle production in $pp$ collisions. 
In order to better describe the hadron transverse momentum
spectra at $\sqrt s\sim20$ GeV, and in the semi-hard region
$p_T\sim 1\dots5$ GeV at higher cms energies, one has to also include an
intrinsic transverse momentum \cite{Owens:1986mp,Field:uq}
for the colliding parton.
The $p_T$ spectrum in a $pA$ collision is obtained from a
Glauber-type model\footnote{Integrations are schematically indicated
with crossed circle symbols. For details see the original
references.}:
\begin{equation}
    \frac{d\sigma^h_{pA}}{d^2p_T} = K \sum\limits_{i,j,k,l}
        F_{i/p} \otimes F_{j/A} 
        \otimes \frac{d\hat\sigma}{d\hat t}(ij\to kl) \, \otimes D^h_k
        \ ,
 \label{pQCD}
\end{equation}
where the proton and nucleus parton distribution functions are,
respectively, 
\begin{align} 
    F_{i/p} = f_{i/p}(x_i,Q^2) \, 
        \frac{e^{-k_{iT}^2/\langle{k_T^2}\rangle_{pA}(b)}}
          {\pi\langle{k_T^2}\rangle_{pA}}  
        \hspace*{.4cm}{\rm{and}}\hspace*{.4cm}
    F_{j/A} = T_A(b) \, f_{j/A}(x_j,Q^2) \, 
        \frac{e^{-k_{jT}^2/\langle{k_T^2}\rangle_{Ap}(b)}}
             {\pi\langle{k_T^2}\rangle_{Ap}}  \ .
 \label{Fsoft}
\end{align}
In Eq.~\eqref{pQCD}, $d\hat\sigma/d\hat t(ij\rightarrow kl)$ is
the pQCD parton-parton cross section, the variables with a hat are
the partonic Mandelstam variables, and a sum over incoming and outgoing
parton flavours is performed.
The proton is considered point-like compared to the target nucleus, 
and the scattering occurs at impact parameter $b$. The nucleus is described
by the Woods-Saxon nuclear thickness function $T_A(b)$.
In Eq.~\eqref{Fsoft}, $f_{i/p(A)}(x,Q^2)$ are the PDFs 
of the proton (nucleus). Isospin imbalance is taken into
account and nuclear shadowing is included by the HIJING 
parametrization \cite{Wang:1991ht}. Partons are assumed to have
an intrinsic transverse momentum with average squared value
$\langle{k_T^2}\rangle_{pA(Ap)}$ and a 
Gaussian distribution. Due to the $k_{i}$ and
$k_j$ integrations a regulator mass $m$$=$$0.8$ GeV has been used
in the pQCD cross section. Finally, $D_k^h(z,{Q'}^2)$ are the fragmentation
functions of a parton $k$ into a hadron $h$ with a fraction $z$ of the
parton momentum.

Soft proton-nucleon interactions are assumed to excite the projectile 
proton wavefunction, so that when the proton interacts with the next
target nucleon its partons have a broadened intrinsic 
momentum. Each proton rescattering is assumed to contribute to
the intrinsic momentum broadening in the same way so that: 
\begin{equation}
    \langle{k_T^2}\rangle_{pA}(b,\sqrt s) 
        = \langle{k_T^2}\rangle_{pp} + \delta \, h_A(b,\sqrt s) \ ,
 \label{ktbroad_acc}
\end{equation}
where $\langle{k_T^2}\rangle_{pp}$ is the parton intrinsic momentum needed to
describe hadron transverse spectra in $pp$ collisions, 
$\delta$ is the average momentum squared acquired in each
rescattering, and 
\begin{equation}
    h_A(b,\sqrt s) = \left\{ \begin{array}{ll}
        \displaystyle \nu_A(b,\sqrt s)-1 
            & {\rm\ if\ \ } \displaystyle \nu_A-1 \leq n \\
        \displaystyle n 
            & {\rm\ if\ \ } \displaystyle \nu_A-1 > n 
        \end{array} \right.
 \label{ha}
\end{equation}
represent the average number of collisions which
are effective in broadening the intrinsic momentum. Both models assume
$h_A$ to be a function of the number of proton-nucleon collisions
$\nu_a(b,\sqrt s) = \sigma_{pp}(\sqrt s) T_A(b)$, with $\sigma_{pp}$
the nucleon-nucleon inelastic cross section. However, Ref.~\cite{Wang:1998ww} 
takes $n$$=$$\infty$, while Ref.~\cite{Zhang:2001ce} assumes an upper limit
$n$$=$$4$ justified in terms of a proton dissociation mechanism: after
a given number of interactions, the proton is so excited that it can no
longer interact as a whole with the next nucleon. 
I will call the first model simply {\it soft} and the
second {\it soft-saturated}. 
In both models, target nucleons do not rescatter, so that 
\begin{equation*}
    \langle{k_T^2}\rangle_{Ap}(b,\sqrt s) 
        = \langle{k_T^2}\rangle_{pp} \ .
\end{equation*}
Further differences between the models are
related to the choices of the hard scales, of the $K$-factor which
simulates NLO contributions to the parton cross section, and to the
parametrizations of $\langle{k^2_T}\rangle_{pp}$ and $\delta$ 
(see Table~\ref{tab:hadronic}). 

\begin{table}[tbp]
\begin{center}
\caption{ Parameters of the soft hadronic rescattering models \cite{Zhang:2001ce}.}
\begin{footnotesize}
\vskip0.2cm
\begin{tabular}{|l|c|c|c|c|c|c|c|}
\hline
 \ model 
        & \!\!hard scales {\scriptsize (GeV)}\!\!
        & \!\!\!$K$\!\! 
        & regul. 
        & proton intrinsic $k_T$ {\scriptsize (GeV$^2$)}
        & \!\!n\!\! 
        & average $k_T$-kick {\scriptsize (GeV$^2$)}
        & nPDF
        \\ 
\hline
 \!Soft 
        & $Q=Q'=p_T$ 
        & \!\!2\!\!
        & \!\!0.8 {\scriptsize GeV}\!\!
        & \!\!$\langle{k_T^2}\rangle_{pp} = 
                1.2 + \!0.2 \alpha_s(q^2)q^2 \, ^\dagger\!$ \!\!
        & \!\!$\infty$\!\! 
        & \!$\delta(Q) = 0.255$ 
        {\small $\frac{\ln^2(Q/{\rm GeV})}{1+\ln^{\ \!}(Q/{\rm GeV})}$}\!\!  
        & \!\!\!HIJING\!\!\!
        \\ 
\hline
 \!Soft-sat. \!\!\!
        & \!\! $Q\!=\!${\normalsize$\frac{p_T}{2z_c}$}
        $;\,Q'\!=\!\!${\normalsize$\frac{p_T}{2} ^\flat$}\!\!
        & \!\!1\!\!
        & \!\!0.8 {\scriptsize GeV}\!\!
        & $\langle{k_T^2}\rangle_{pp} 
               = F(\sqrt s) \ ^\natural$  
        & \!4\!\! 
        & $\delta = 0.4$ \!
        & \!\!\!HIJING\!\!\!
\\ 
\hline
\end{tabular}
\end{footnotesize}
\parbox{15cm}{\footnotesize
$^\dagger$ $q^2 = 2\hat s \hat t \hat u / (\hat s + \hat t 
+ \hat u)^2 $. Parametrization chosen to best reproduce $pp$ data. \\
$^\natural$ No explicit parametrization is given. 
Values of $\langle{k_T^2}\rangle_{pp}$ extracted from a
``best fit'' to $pp$ data, see Fig.15 of Ref.~\cite{Zhang:2001ce}. \\
$^\flat$ These scales used in the computations of Table~\ref{tab:results}.
In Ref.~\cite{Zhang:2001ce} $Q={p_T}/{2}$ and $\,Q'={p_T}/{2z_c}$.
}
 \label{tab:hadronic}
\end{center}
\vskip-.6cm
\end{table}

\subsubsection*{Soft partonic rescatterings: the colour dipole model \cite{Kopeliovich:2002yh}}

In this model, the particle production mechanism 
is controlled by the {\it coherence length} $l_c $$=$$ \sqrt s/(m_N k_T)$,
where $m_N$ is the nucleon mass and $k_T$ the transverse momentum of
the parton which fragments in the observed hadron. 
Depending on the value of $l_c$, three different calculational schemes
are considered.
{\it (a)} In fixed target experiments at low energy (e.g., at SPS), 
where $l_c \ll R_A$, 
the projectile partons interact incoherently with target nucleons.
High-$p_T$ hadrons are assumed to originate
mainly from a projectile parton which experienced a hard 
interaction and whose transverse momentum was broadened 
by soft parton rescatterings. The parton is then put on-shell by a
single semihard scattering computed in factorized pQCD. 
This scheme I will discuss in detail below.
{\it (b)} At LHC, where the c.m. energy is very large and $l_c \gg
R_A$, the partons interact coherently with the target nucleons and high-$p_T$
hadrons are assumed to originate from radiated gluons. Parton
scatterings and gluon radiation are computed in the light-cone dipole
formalism in the target rest frame. 
{\it (c)} At intermediate energies, like RHIC, an interpolation is made
between the results of the low- and high-energy regimes discussed above.
All the phenomenological parameters needed in this model are fixed in 
reactions other than from $pA$ collisions, and in this sense the model
is said to be parameter-free. 

In the short coherence length scheme, pQCD factorization is
assumed to be valid and Eq.~\eqref{pQCD} is used 
with parton masses $m_g = 0.8$ GeV and $m_q = 0.2$ GeV for, viz.,  gluons
and quarks. Moreover, 
\begin{align*} 
    F_{i/p} = 
        f_{i/p} \big( x_i+{\textstyle \frac{\Delta E}{x_a E_p}},Q^2 \big) \, 
        \frac{dN_i}{d^2k_{iT}}(x,b)
        \hspace*{.4cm}{\rm and}\hspace*{.4cm}
    F_{j/A} = T_A(b) \, f_{j/A}\big( x_j,Q^2 \big) \,
        \frac{dN_j^{(0)}}{d^2k_{jT}}(x,b) \ .
\end{align*}
Parton rescatterings are computed in terms of the
propagation of a $q\bar q$ pair through the target nucleus.
The final parton transverse momentum distribution
$dN_i/d^2k_{iT}$ is written as:
\begin{align}
    \frac{dN_i}{d^2k_{iT}} & = \!
        \int \! d^2r_1d^2 r_2\,e^{i\,\vec k_T\,(\vec r_1 - \vec r_2)} 
        \left[ \frac{\langle{k_0^2}\rangle}{\pi}\,
        e^{-\frac12(r_1^2+r_2^2) \langle{k_0^2}\rangle}
        \right] 
        \left[ e^{-\frac12\,\sigma^N_{\bar qq}(\vec r_1-\vec r_2,x)
        \,T_A(b)} \right] 
        = \frac{dN^{(0)}}{d^2k_{T}}
        + O\big( \sigma^N_{\bar qq}T_A\big) \, .
 \label{dNdkt}
\end{align}
The first bracket in Eq.~\eqref{dNdkt} represents the contribution of the
proton intrinsic momentum. The second bracket 
is the contribution of soft parton 
rescatterings on target nucleons, expressed through the
phenomenological dipole cross section. For a quark,
$
    \sigma_{\bar qq}^N(r_T,x)=\sigma_0\,
        \left[ 1-\exp\big(-\frac{1}{4}\,r_T^2\,Q_s^2(x)\big) \right]
$ 
with $Q_s $$=$$ 1\,{\rm GeV}(x/x_0)^{\lambda/2}$ 
(the parameters ar given in Table~\ref{tab:partonic}). 
For a gluon, $\sigma_{gg}^N $$=$$ 9/4 \, \sigma_{\bar qq}^N$ is used. The expansion of 
Eq.~\eqref{dNdkt} to zeroth order in $\sigma_{\bar qq}^N$ 
gives the intrinsic $k_T$ distribution, $dN^{(0)}/d^2k_{T}$, 
of the nucleon partons. The
first order term represents the contribution of one-rescattering
processes, and so on. Partons from the target nucleus are assumed not
to rescatter because of the small size of the projectile. 
Energy loss of the
projectile partons is taken into account by a shift 
of their fractional momentum proportional to the energy of the
radiated gluons, given by the product of the average
mean path length $\Delta L$ and  the energy loss per unit length $dE/dz$ 
\cite{Johnson:2000ph,Johnson:2001xf}.
Since nuclear shadowing effects are computed in the dipole
formalism, see Eq.~\eqref{dNdkt}, 
parton distribution functions in the target are modified only to
take into account antishadowing at large $x$ according to the EKS98
parametrization \cite{Eskola:1998df}.

By Fourier transforming the dipole cross section we find that the
transverse momentum distribution of single parton-nucleon
scattering is Gaussian, which justifies the classification of this
model as ``soft''. However, the single distributions are not just
convoluted to obtain a broadening proportional to the average number of
rescatterings. Indeed, in the second bracket, the rescattering
processes have a Poisson probability distribution. As a result, the nuclear
broadening of the intrinsic momentum is smaller than the product of
the average number of rescatterings and the single scattering
broadening; perhaps a dynamical explanation of the assumption 
that $n \,{\scriptstyle\lneqq}\, \infty$ in Eq.~\eqref{ha},
used in the Soft-saturated model \cite{Zhang:2001ce}.

\begin{table}[tbp]
\begin{center}
\caption{Parameters of the soft partonic rescattering model (at short-$l_c$)
and of the hard partonic rescattering models.}
\begin{footnotesize}
\vskip0.2cm
\begin{tabular}{|l|c|c|c|c|c|c|c|}
\hline 
 \ model 
        & \!\!hard scales\!\!
        & \!\!$K$\!\!
        & \!\!regulators (GeV)\!\!
        & \!\!intr. $k_T$\!\!
        & \!\! $dE/dz$ \!\!
        & \!\!dipole cross-sect.\!\!
        & \!\!nPDF\!\!
        \\ \hline 
 \!Col. dip. \!\!\! 
        & \!\!$Q=Q'=p_T$\!\! 
        & \!\! \scriptsize$\otimes$ \!\! 
        & \!\!$m_g$=0.8, $m_q$=0.2\!\!
        & \!\!as Soft mod.\!\!
        & \!\!\!\! -2.5\,GeV/fm\!\!
        & \!\!$\sigma_0$=23\,mb, $\lambda$=0.288, $x_0$=3$\cdot 10^{-4}$\!\!
        & \!\!EKS98$^\dagger$\!\! 
        \\ \hline 
 \!Hard AT
        & \!\!$Q=Q'=p_T^*$\!\!
        & \!\! 2 \!\! 
        & \!\!$\mu$ free param.\!\!
        & \!\!no\!\!
        & \!\!no\!\!
        & \!\!computed from pQCD\!\!
        & \!\!no\!\!
        \\ \hhline{||--------||}
 \!Hard GV
        & \!\!$Q=Q'=p_T$\!\! 
        & \!\! \scriptsize$\otimes$ \!\! 
        & \!\!$\mu$=0.42$^\flat$\!\!
        & \!\!\!as Soft mod.\!\!
        & \!\!no\!\!
        & \!\!----\!\!
        & \!\!\!EKS98\!\!\!
\\ \hline 
\end{tabular}
\end{footnotesize}
\parbox{15cm}{\footnotesize
$^\star$ $Q$$=$$\mu$ in Ref.~\cite{Accardi:2001ih}.\ \
$^\dagger$ Only at large $x_j$ (EMC effect).\ \
$^\otimes$ Factors out in the Cronin ratio. \\
$^\flat$ $\mu$ determines only the typical momentum transfer in elastic 
rescatterings.}
 \label{tab:partonic}
\end{center}
\vskip-.6cm
\end{table}

\subsubsection*{Hard partonic rescattering model \cite{Accardi:2001ih,Vitev:2002pf}}
\label{sec:hardresc}

The model of Ref.~\cite{Accardi:2001ih}, hereafter labeled ``hard AT'', 
assumes that parton rescatterings are responsible for the Cronin
enhancement. Up to now, it includes only 
semihard scatterings, i.e., scatterings 
described by the pQCD parton-parton cross section. It
 is the generalization to an arbitrary number of hard parton
rescatterings of the early models of 
Refs.~\cite{Kuhn:1975dq,Lev:1983hh,Lev:1983hh,Kastella:ru} and of 
the more recent Refs.~\cite{Wang:1996yf,Wang:2001cy}, which are 
limited to a single hard rescattering. 
As shown in Ref.~\cite{Lev:1983hh}, considering only one rescattering 
may be a reasonable assumption at low energy to describe 
the gross features of the Cronin effect, but already
at RHIC energies this might be insufficient to determine the
Cronin peak $R_M$ \cite{Accardi:2001ih}. The AT model 
assumes that the S-matrix for a collision of $n$ partons from the
projectile on $m$ partons from the target is  
factorizable in terms of S-matrices of parton-parton
elastic-scattering. It also assumes generalized pQCD factorization
\cite{Qiu:2001hj,Fries:2002dn}. 
The result is a unitarized cross section,
as discussed in Refs.~\cite{Accardi:2001ih,Braun:2002kg}:
\begin{align}
    \frac{d\sigma^h_{pA}}{d^2p_T} = \sum\limits_{i}
        f_{i/p} \otimes \frac{d N_{i/A}}{d^2k_T} \, \otimes D^h_i
        + \sum\limits_{j}
        f_{j/A} \, T_A \otimes \frac{d N_{j/p}}{d^2k_T} \, \otimes D^h_j
        \ .
 \label{hardresc}
\end{align}
The first term accounts for multiple semihard scatterings of partons
in the proton on the nucleus. In the second term, partons of the nucleus
are assumed to undergo a single scattering, with 
${dN_{j/p}}/{d^2k_T}=\sum_i f_{i/p} \otimes \sigma_{i/N}^{\rm hard}$.  Nuclear
effects are included in $dN_{i/A}/d^2k_T$, the average transverse
momentum distribution of a proton parton that suffered {\it at least}
one semihard scattering.  In impact parameter space it reads
\begin{align}
    \frac{d N_{i/A}}{d^2k_T}(b) = \int \frac{d^2r}{4\pi}
        e^{\,-i \vec{k}_T \cdot \vec r} 
        \left[  e^{\,-\sigma_{i/N}^{\rm hard}(r)T_A(b)} - 
        e^{\,-\sigma_{i/N}^{\rm hard} T_A(b)}  \right] \ ,
 \label{Harddip}
\end{align}
where unitarity is explicitly implemented at the
nuclear level, as discussed in Ref.~\cite{Braun:2002kg}. 
In Eq.~\eqref{Harddip}, 
$\sigma_{i/N}^{\rm hard}(r) = K \sum_j \int d^2 p 
        \left[ 1 - e^{\,-i \vec p \cdot \vec r} \,\right]
        \frac{d \hat\sigma}{d \hat t}$ 
        {\small \!\!$(ij \rightarrow ij) \otimes f_{j/p}$}. 
Moreover, $\omega_i \equiv \sigma_{i/N}^{\rm hard} T_A(b)$ is
identified with the target opacity to the parton propagation.
Note that $\sigma_{i/N}^{\rm hard}(r) \propto r^2$ as $r\rightarrow 0$ and
$\sigma_{i/N}^{\rm hard}(r) \rightarrow \sigma_{i/N}^{\rm hard}$ as $r\rightarrow \infty$. This,
together with the similarity of Eqs.~\eqref{dNdkt} and \eqref{Harddip}, 
suggests the interpretation of
$\sigma_{i/N}^{\rm hard}(r)$ as a {\it hard dipole cross section},
accounting for hard parton rescatterings on target nucleons 
analogous to what
$\sigma_{q\bar q}^N$ does for soft parton rescatterings.
Note that no nuclear effects on PDF's are included. However,
shadowing is partly taken into account by the multiple scattering
processes.

To regularize the IR divergences of the pQCD cross section, a small mass
regulator, $\mu$, is introduced in the parton propagators. It is
considered a free parameter of the model that sets the scale at
which pQCD computations break down. As a consequence of 
the unitarization of the interaction, due to the inclusion of  
rescatterings, 
both $p_\times$ and $p_M$ are almost insensitive to $\mu$
\cite{Accardi:2001ih}. Therefore, these two quantities are considered 
reliable predictions of the model\footnote{This result is
very different from the conclusion  that $p_\times \propto
p_0$, Ref.~\cite{Wang:2001cy}, based on a
single-rescattering approximation. Hence $p_\times$ cannot be used to ``measure'' the onset of hard
dynamics as proposed in that paper.}. Note, however, 
that they both depend on the
c.m. energy $\sqrt s$ and on the pseudorapidity $\eta$. 
On the other hand, $R_M$ is quite sensitive to the
IR regulator. This sensitivity may be traced back to the inverse-power
dependence of the target opacity $\omega_i$ on $\mu$, 
$\omega_i\propto 1/\mu^a$, where $a>2$ is energy and
rapidity dependent. The divergence of $\omega_i$ as $\mu\rightarrow 0$
indicates the need for unitarization of the parton-nucleon
cross section and deserves further study. Therefore, $\mu$ can be here
considered only as an effective scale which simulates nonperturbative
physics \cite{Eskola:1999fc,Iancu:2002xk}, the nonlinear evolution of
target PDFs \cite{Eskola:2002yc} and physical effects so far
neglected, e.g., collisional and radiative energy losses \cite{enloss}.

Another model which implements hard partonic rescatterings,
labeled ``hard GV'', is found in Ref.~\cite{Vitev:2002pf}. This model is
discussed in more detail in the Chapter on jet physics, see
Ref.~\cite{Vitev:2002aa}. In the hard GV model, the transverse momentum
broadening of a parton which experiences semihard rescatterings is
evaluated, using Eq.~\eqref{Harddip}, to be
\begin{align}
  \langle k_T^2 \rangle_{\rm hard} = \omega \mu^2 
    \ln\left(1+c\frac{p_T^2}{\mu^2}\right)
  \ . 
 \label{GVkt2} 
\end{align}
The IR regulator is fixed to $\mu = 0.42$ GeV and identified with the medium
screening mass. It represents the 
typical momentum kick in each elastic rescattering of a hard parton. 
The factor $c$ and the
constant term 1 are introduced in order to obtain no broadening for
$p_T\approx 0$, as required by kinematic considerations.
The average value of the opacity in the
transverse plane, $\omega\approx(0.4/{\rm{fm}})R_A$ (where $R_A$ is
the nuclear radius), and the factor $c/\mu^2=0.18$ are fixed in order to
reproduce the experimental data at $\sqrt s =27.4$ GeV 
and $\sqrt s=38.8$ GeV \cite{data400lab}. With these values the logarithmic 
enhancement in Eq.~\eqref{GVkt2} is of order 1 for $p_T\approx3$ GeV.
Note that $\omega$ and $c$ are assumed to be independent of $\sqrt s$
and $\eta$. Finally, the transverse spectrum is computed by using
Eq.~\eqref{pQCD} and adding to the semihard broadening of
Eq.~\eqref{GVkt2} the 
intrinsic momentum of the projectile partons, 
$\langle k_0^2 \rangle=1.8$ GeV$^2$:
\begin{equation*}
  \langle k_T^2 \rangle_{pA} = \langle k_T^2 \rangle_{pp} 
    + \langle k_T^2 \rangle_{\rm hard} \ ,
\end{equation*}
with shadowing and antishadowing corrections to target partons from
the EKS98 parametrization \cite{Eskola:1998df}.

\subsubsection{Predictions and conclusions}
\label{sec:predictions}

In Table~\ref{tab:results}, I list the values of the Cronin parameters
$p_M$ and $R_M$ computed in various models of $pA$ collisions at
$\sqrt s=27.4$ GeV (for low-energy experiments), 200 GeV (RHIC) and
5.5 TeV (LHC). The targets considered in the Cronin ratio are also
listed.  I do not include $p_\times$ in the tables since in almost all
models $p_\times \sim 1$~GeV, independent of energy, which lies at the
border of validity of the models.  Uncertainties in the model
calculations are included when discussed in the original references,
see the table footnotes.  In the case of the soft-saturated model, the
uncertainty due to the choice of shadowing parametrization is
illustrated by giving the results obtained with no shadowing, and with
the HIJING parametrization \cite{Wang:1991ht}.  Using the ``new"
HIJING parametrization \cite{Li:2001xa} would change the mid-rapidity
results only at LHC energies, where a 15\% smaller Cronin peak would
be predicted \cite{LevaiPC}\footnote{Note however that the new
parametrization \cite{Li:2001xa}, which predicts a much larger gluon
shadowing at $x\lesssim10^{-2}$ than the ``old" one
\cite{Wang:1991ht}, seems ruled out by data on the Sn/Ca $F_2$ ratio
\cite{Eskola:2002us,Eskola:2001ms}.}.  
In the case of ``hard AT" rescattering, the major theoretical
uncertainty is in the choice of $\mu$, as discussed at the end of 
Sec.~\ref{sec:hardresc}.
In the table, two choices are presented: {\it (a)} an
energy-independent value $\mu=1.5$ GeV, which leads to an increasing
Cronin effect with energy; 
{\it (b)} $\mu$ is identified with the IR cutoff $p_0$ discussed in 
Ref.~\cite{Eskola:2002kv}, in the context of a leading order pQCD
analysis of $pp$ collisions. That analysis found $p_0$
to be an increasing function of $\sqrt s$. By performing a 
simple logarithmic fit to the values extracted from data in
Ref.~\cite{Eskola:2002kv}, 
we find $\mu = p_0(\sqrt s) = 0.060 + 0.283 \log(\sqrt s)$, which
decreases the Cronin effect with energy. 
Note that a scale increasing with $\sqrt s$ also appears
naturally in the so-called ``saturation models" for hadron production in
$AA$ collisions \cite{Eskola:1999fc,Kharzeev:2000ph,Accardi:2001wh}.
In the ``hard GV" model at the LHC, the remnants of the Cronin effect at $p_T
\sim 3$ GeV are overwhelmed  by shadowing and the calculation 
is not considered reliable in this region. The value $R_M=1.05$ at
$p_T\simeq 40$ in Ref.~\cite{Vitev:2002pf} is a result of
antishadowing in the EKS98 parametrization and is not related to
multiple initial state scatterings.

\begin{table}[tbp]
\begin{center}
\caption{Comparison of models of the Cronin effect at $\eta=0$.}
\vskip0.2cm
\begin{small}
\begin{tabular}{cl|ccc|ccc}
 \multirow{2}{2.1cm}{\centerline{$\sqrt s$}} 
        & \multirow{2}{2cm}{\centerline{model}} 
        & \multicolumn{3}{c|}{charged pions} 
        & \multicolumn{3}{c}{partons} 
        \\ 
 && $p_M$ (GeV)  & $R_M$ & Ref. 
        & $p_M$ (GeV) & $R_M$ & Ref. 
\\ \hline
 \multirow{5}{2.1cm}{\begin{center}
        27.4 GeV \\
        $p$W$\big/$$p$Be 
        \\data \cite{data400lab}
        \end{center}} 
 & Soft & 4.0  & 1.55$^\star$ \hspace*{0.95cm} & \cite{Wang:1998ww} 
        &  &  &  \\ 
 & Soft-saturated & 4.5$^\star$; 4.4$^\odot$ & 1.46$^\star$; 1.46$^\odot$ 
        & \cite{LevaiPC} 
        & 5.1$^\star$; 5.1$^\odot$ & 1.50$^\star$; 1.51$^\odot$ 
        & \cite{LevaiPC}  \\
 & Color dipole & 4.5 & 1.43$\pm$0.08$^\otimes$ & \cite{Kopeliovich:2002yh} 
        &  &  &  \\
 & Hard AT &  &  & 
   & 6 $\pm$ 0.8\,$^\dagger$ & 1.1$^\flat$;\ 1.4$^\natural$ &  \\ 
 & Hard GV & 4 & 1.4 & \cite{Vitev:2002pf}
   &  &  &   
\\ \hline 
 \multirow{5}{2cm}{\begin{center}
        200 GeV $p$Au$\big/$$pp$
        \end{center}} 
 & Soft  & 3.5 & 1.35$\pm$0.2$^\ddagger$ & \cite{Wang:1998ww} 
        &  &  & \\ 
 & Soft-saturated & 2.9$^\star$; 2.7$^\odot$ & 1.15$^\star$; 1.47$^\odot$ 
        & \cite{LevaiPC} 
        & 4.4$^\star$; 4.2$^\odot$ & 1.29$^\star$; 1.70$^\odot$ & 
        \cite{LevaiPC}  \\
 & Color dipole & 2.7 & 1.1 & \cite{Kopeliovich:2002yh} 
        &  &  &  \\
 & Hard AT &  &  & 
        & 7$\pm$1\,$^\dagger$  & 1.25$^\flat$;\ 1.2$^\natural$ 
        &  \\
 & Hard GV & 3.0 & 1.3 & \cite{Vitev:2002pf}
        &  &  &   
\\ \hline 
 \multirow{5}{2cm}{\begin{center}
        5500 GeV $p$Pb$\big/$$pp$
        \end{center}} 
 & Soft  & 3.5 & 1.08$\pm$0.02$^\ddagger$ & \cite{WangPC}
        &  &  &  \\ 
 & Soft-saturated & 2.4$^\star$; 2.2$^\odot$ & 0.78$^\star$; 1.36$^\odot$ 
        & \cite{LevaiPC}
        & 4.2$^\star$; 4.2$^\odot$ & 0.91$^\star$; 1.60$^\odot$  
        & \cite{LevaiPC} \\
 & Color dipole & 2.5 & 1.06 & \cite{Kopeliovich:2002yh} 
        &  &  &  \\
 & Hard AT &  &   &  
        & 11$\pm$1.3\,$^\dagger$  & 2.1$^\flat$;\ 1.2$^\natural$ 
        & \\
 & Hard GV & $\approx$ 40$^{\,\lozenge}$ & 1.05$^{\,\lozenge}$ & \cite{Vitev:2002pf}
        &  &  &   
\\ \hline 
\end{tabular}
\end{small}
\parbox{14.9cm}{\footnotesize
$^\star$ With HIJING shadowing \cite{Wang:1991ht}.
$^\odot$ Without shadowing.
$^\otimes$ Error estimated by varying $dE/dz$ within error bars
\cite{Kop02}.\\   
$^\ddagger$ Central value with multiple scattering effects only; the error
estimated by using different shadowing parametrizations. \\ 
$^\dagger$ Numerical errors mainly. \
$^\flat$ Using $\mu=1.5$ GeV. \ $^\natural$ Using 
$\mu=0.060 + 0.283 \log(\sqrt s)$, see text. \\
$^{\lozenge}$ 
Very sensitive to the shadowing parametrization, see text.
} 
 \label{tab:results}
\end{center}
\vskip-.8cm
\end{table}

As discussed in the introduction, experimental reconstruction of
minijets with ALICE may be possible in $pA$ collisions for minijet
transverse momenta $p_T\gtrsim 5$ GeV.  The $p_T$-spectrum of the
partons which  hadronize into the observed minijet may be
obtained by setting $D_i^h $$=$$ \delta(z-1)$ in Eqs.~\eqref{pQCD} and
\eqref{hardresc}. Minijet reconstruction may be very interesting because pQCD
computations suffer from large uncertainties in the determination of
the large-$z$ fragmentation functions,
where they are only loosely constrained by existing data
\cite{Zhang:2002py}. For this reason I also listed 
the Cronin parameters for parton production in
Table~\ref{tab:results}. However, jet reconstruction efficiency should
be accurately evaluated to assess the usefulness of this observable.

Table~\ref{tab:results} shows that there are large theoretical
uncertainties in the extrapolation of the Cronin effect from lower
energies to the LHC. A major source of uncertainty for most of the
models is nuclear shadowing and antishadowing at small
$x$, see Refs.~\cite{Eskola:2002us,Eskola:2001ms} and
Sec.~\ref{pA_SEC:nPDF} for a detailed discussion and comparison of the
existing parametrizations.  For example, the HIJING parametrizations
\cite{Wang:1991ht,Li:2001xa} suggest more gluon shadowing than 
EKS98 \cite{Eskola:1998df} at small $x\lesssim10^{-2}$. At the LHC
the small-$x$ region is dominant at midrapidity and small to
medium $p_T$. On the other hand, the HIJING parametrizations
give less antishadowing than EKS98 at $x\gtrsim10^{-1}$, which is
the dominant region at large $p_T$ for all energies.  At the LHC, all
these effects may lead to up to a factor of 2 uncertainty in the height of
the Cronin peak $R_M$.

In conclusion, a $pA$ run at the LHC is necessary both to test
theoretical models of particle production in a cleaner experimental
environment than $AA$ collisions, and to make reliable extrapolations
to $AA$ collisions, which are the key to disentangling known effects
from new physics. Since, as we have seen, the nuclear effects are
potentially large, it would even be preferable to have a $pA$ run at
the same energy as $AA$. In addition, the study of the $A$ dependence,
or collision centrality, would be interesting since it would change
the opacity of the target -- thus the size of the Cronin effect -- in
a controllable way.  Finally, note that the $\eta$-systematics of the
Cronin effect has so far been considered only in
Ref.~\cite{Accardi:2001ih}.  However, as also discussed in
Ref.~\cite{Vitev:2002aa}, given the large pseudorapidity coverage of
CMS, the $\eta$ dependence might be a very powerful tool to understand
the effect.  It would systematically scan nuclear targets in the
low-$x$ region and would help test proposed models where the
rapidity influences the Cronin effect in different ways.

\vspace{1.0cm}
\subsection{Quarkonium total cross sections}
\label{section531}
{\em Ramona Vogt}

To better understand quarkonium suppression, it is necessary to have a good
estimate of the expected yields.  However, there are still a number of 
unknowns about quarkonium production in the primary
nucleon-nucleon interactions.  In this section, we report the quarkonium 
total cross sections in the CEM \cite{Barger:1979js,Barger:1980mg} 
and NRQCD \cite{Bodwin:1994jh}, both briefly described in 
Section~\ref{section441}.
We also discuss shadowing effects on the $p_T$-integrated
rapidity distributions.

Recall that at leading order, the production cross section of quarkonium state 
$C$ in the CEM is
\begin{eqnarray}
\sigma_C^{\rm CEM} = F_C \sum_{i,j} \int_{4m_Q^2}^{4m_H^2} d\hat s \int dx_1 
dx_2~f_i^p(x_1,Q^2)~f_j^p(x_2,Q^2)~ \hat\sigma_{ij}(\hat s)~\delta(\hat 
s-x_1x_2s)\, \, . \label{sigtil_qxs}
\end{eqnarray} 
In a $pA$ collision, one of the proton parton densities is replaced by that of
the nucleus.  Since quarkonium production in the CEM is gluon dominated, 
isospin is a negligible effect.  However, since shadowing effects on the gluon
distribution are large, it could strongly influence the total cross sections.
We will use the same parton densities and parameters that fit the $Q \overline
Q$ cross section data, given in Table~\ref{qqbtab} in Section~\ref{section431},
to determine $F_C$ for $J/\psi$ and $\Upsilon$ production.  

\begin{figure}[htb]
\setlength{\epsfxsize=0.90\textwidth}
\setlength{\epsfysize=0.35\textheight}
\centerline{\epsffile{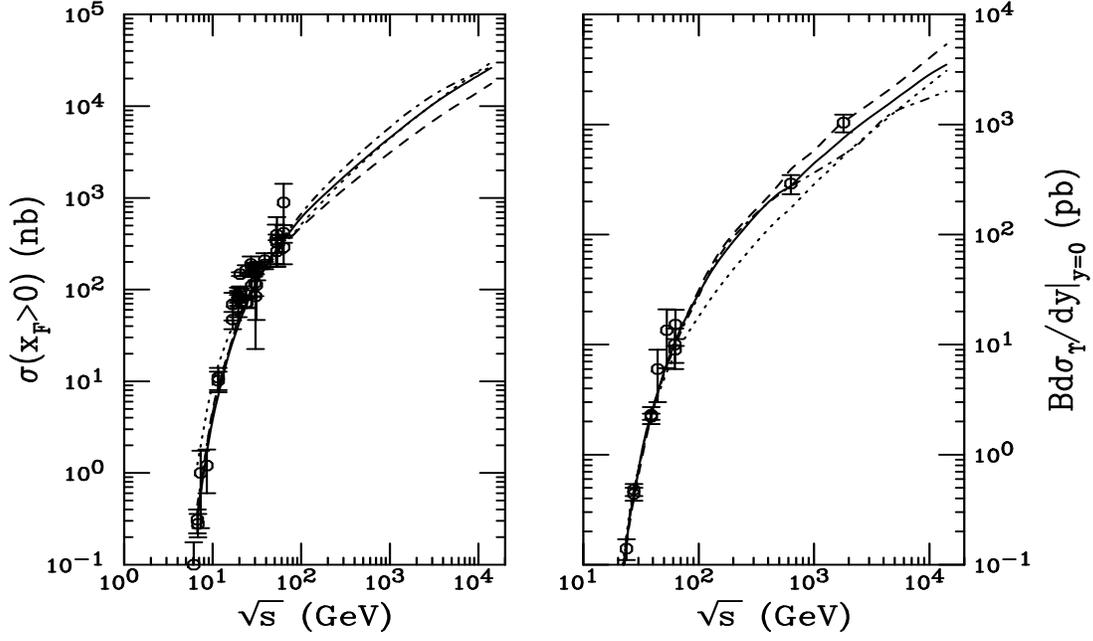}}
\caption{Forward $J/\psi$ (left) and combined $\Upsilon + \Upsilon' +
\Upsilon''$ inclusive (right) cross sections calculated to NLO in the CEM.  On
the left hand side, the
solid curve employs the MRST HO distributions with $m_c = Q/2 = 1.2$ GeV,
the dashed, MRST HO with $m_c = Q = 1.4$ GeV, the dot-dashed, CTEQ 5M with
$m_c = Q/2 = 1.2$ GeV, and the dotted, GRV 98 HO with $m_c = Q = 1.3$ GeV.
On the right hand side, the
solid curve employs the MRST HO distributions with $m_b = Q = 4.75$ GeV,
the dashed, $m_b = Q/2 = 4.5$ GeV, the dot-dashed, 
$m_b = 2Q = 5$ GeV, and the dotted, GRV 98 HO with $m_b = Q = 4.75$ GeV.
}
\label{psiupsdep}
\end{figure}

In the left-hand
side of Fig.~\ref{psiupsdep}, we show the resulting fits to the forward 
$(x_F>0)$ $J/\psi$ total cross sections.  These data also include feeddown from
$\chi_c$ and $\psi'$ decays.  The right-hand plot is the sum of all three
$\Upsilon$ $S$ states, including the branching ratios to lepton pairs.

All the fits are equivalent for $\sqrt{s} = 100$ GeV but differ by up to a 
factor of two at 5.5 TeV.  The high energy $\Upsilon$ results seem to agree 
best with the energy dependence of the MRST calculations with $m_b = 4.75$ GeV
and 4.5 GeV.

The $pp$ cross sections obtained for the individual states are shown in
Tables~\ref{psitab} and \ref{upstab} along with the values of $F_C$ determined
from the fits.  Two values are given for the $\Upsilon$ family.  

\begin{table}[htb]
\caption[]{The production fractions obtained from fitting the CEM cross section
to the $J/\psi$ forward cross sections ($x_F>0$)
and $y=0$ cross sections as a function of
energy.  The parton distribution function (PDF), charm quark mass (in GeV), 
and scales used are the same as those 
obtained by comparison of the $c \overline c$ cross section to the $pp$ data.
The charmonium cross sections (in $\mu$b) 
obtained from the CEM fits for $NN$ collisions at 5.5 TeV are also given.  
The production fractions are then multiplied by the appropriate 
charmonium ratios determined from data.  Note that the $\chi_c$ cross section
includes the branching ratios to $J/\psi$.  The last row gives the NRQCD
cross sections with the parameters from Ref.~\cite{Beneke:1996tk}.}
\label{psitab}
\begin{center}
\begin{tabular}{cccc||cccc} 
PDF & $m_c$ & $Q/m_c$ & $F_\psi$ 
& $\sigma_{J/\psi}^{\rm tot}$ & 
 $\sigma_{J/\psi}^{\rm dir}$ & $\sigma_{\chi_c \rightarrow J/\psi}$ 
& $\sigma_{\psi'}^{\rm dir}$  \\ \hline
MRST HO   & 1.2 & 2 & 0.0144 & 30.6 & 19.0 &  9.2 & 4.3 \\
MRST HO   & 1.4 & 1 & 0.0248 & 20.0 & 12.4 &  6.0 & 2.8 \\
CTEQ 5M   & 1.2 & 2 & 0.0155 & 36.0 & 22.2 & 10.8 & 5.0 \\
GRV 98 HO & 1.3 & 1 & 0.0229 & 32.1 & 19.8 &  9.6 & 4.5 \\ \hline
CTEQ 3L   & 1.5 & 2 &  -     & 83.1 & 48.1 & 27.6 & 13.6 \\ \hline
\end{tabular}
\end{center}
\end{table}
\begin{table}[htbp]
\caption[]{The production fractions obtained from fitting the CEM cross section
to the combined $\Upsilon$ cross sections to muon pairs at $y=0$ as a function 
of energy.  The PDF, bottom quark mass (in GeV), 
and scales used are the same as those 
obtained by comparison of the $b \overline b$ cross section to the $\pi^-p$ 
data.  The direct bottomonium cross sections (in nb) obtained from the CEM 
fits for $NN$ collisions at 5.5 TeV in each case above.  The production 
fractions for the total $\Upsilon$
are multiplied by the appropriate ratios determined from data. In this case,
the $\chi_b$ cross sections do not include any branching ratios to lower
$S$ states.  The last row 
gives the NRQCD
cross sections with the parameters from Ref.~\cite{Beneke:1996tk}.}
\label{upstab}
\begin{center}
\begin{tabular}{ccccc||ccccc} 
PDF & $m_b$ & $Q/m_b$ & $F_{\sum \Upsilon}$ & $F_\Upsilon$
& $\sigma_{\Upsilon}$ & 
 $\sigma_{\Upsilon'}$  & $\sigma_{\Upsilon''}$ 
& $\sigma_{\chi_b(1P)}$ & $\sigma_{\chi_b(2P)}$ \\ \hline
MRST HO   & 4.75 & 1   & 0.000963 & 0.0276 & 188 & 119 & 72 & 390 & 304  \\
MRST HO   & 4.50 & 2   & 0.000701 & 0.0201 & 256 & 163 & 99 & 532 & 414  \\
MRST HO   & 5.00 & 0.5 & 0.001766 & 0.0508 & 128 &  82 & 49 & 267 & 208  \\
GRV 98 HO & 4.75 & 1   & 0.000787 & 0.0225 & 145 &  92 & 56 & 302 & 235  \\ 
\hline
CTEQ 3L   & 4.9  & 2   &   -      & -      & 280 & 155 & 119& 1350 & 1440 \\
\hline
\end{tabular}
\end{center}
\end{table}
The first
is obtained from the combined $S$ state fits shown in the right-hand side of
Fig.~\ref{psiupsdep}, including the branching ratios to lepton pairs, and the
second is that of the total $\Upsilon$ $1S$ state after the branching ratios
have been extracted.  The cross sections given in the tables are, with the 
exception of the total $J/\psi$ cross section including feeddown, all for 
direct production.  To obtain these direct cross sections, we use the 
production ratios given by data and branching ratios to determine the relative
cross sections, as in Refs.~\cite{Gunion:1996qc,Digal:2001ue}.

\begin{figure}[htb]
\setlength{\epsfxsize=0.85\textwidth}
\setlength{\epsfysize=0.40\textheight}
\centerline{\epsffile{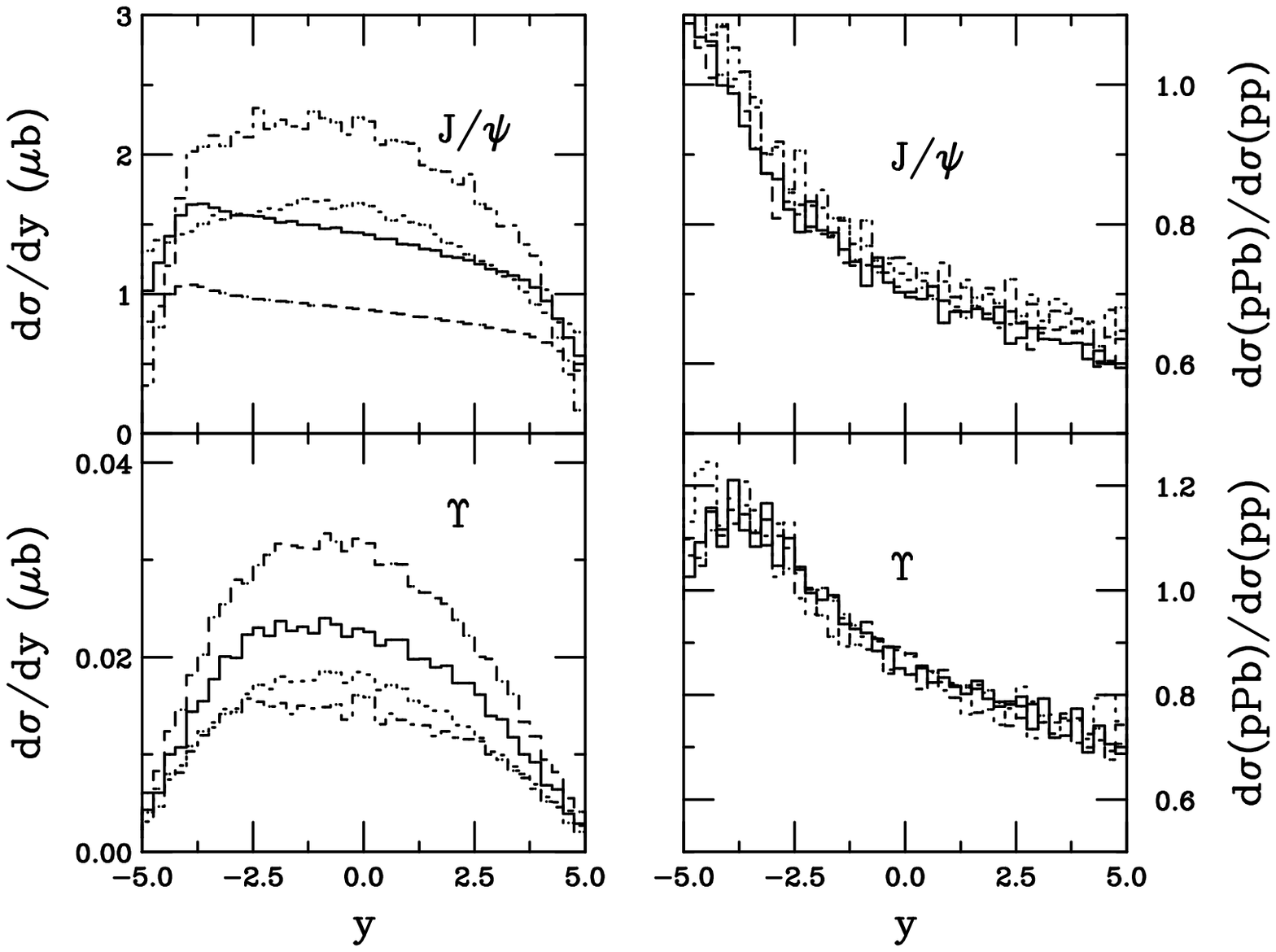}}
\caption{The $J/\psi$ and $\Upsilon$ rapidity distributions in $p$Pb collisions
(left) and the $p$Pb/$pp$ ratios (right).  The $J/\psi$ calculations 
employ the MRST HO distributions with $m_c = Q/2 = 1.2$ GeV (solid),
MRST HO with $m_c = Q = 1.4$ GeV (dashed), CTEQ 5M with
$m_c = Q/2 = 1.2$ GeV (dot-dashed) and GRV 98 HO with $m_c = Q = 1.3$ GeV
(dotted).  The $\Upsilon$ calculations employ the MRST HO distributions with 
$m_b = Q = 4.75$ GeV (solid), $m_b = Q/2 = 4.5$ GeV (dashed),  
$m_b = 2Q = 5$ GeV (dot-dashed), and GRV 98 HO with $m_b = Q = 4.75$ GeV
(dotted).
}
\label{psiupsydep}
\vspace{0.5cm}
\end{figure}

The direct $J/\psi$ and $\Upsilon$ rapidity distributions in $p$Pb interactions
at 5.5 TeV/nucleon are shown in Fig.~\ref{psiupsydep} for all the fits along 
with the $p$Pb/$pp$ ratios at the same energy.  
Note how the average rapidity 
in the $p$Pb distributions is shifted to negative rapidities due to the low 
$x$ shadowing at positive rapidity.  The `corners' in the $J/\psi$ rapidity
distributions at $|y|\sim 4$ are equivalent to $x \sim 10^{-5}$, the lowest
$x$ for which the MRST and CTEQ5 densities are valid.  The $p$Pb/$pp$ ratios
show antishadowing at large negative rapidities, large $x_2$, and $30-40$\%
shadowing at large positive rapidities for the $\Upsilon$ and $J/\psi$
respectively (small $x_2$).  At $y\sim 0$, the effect is $\sim 30$\% for
$J/\psi$ and $\sim 20$\% for $\Upsilon$.  We plot $p$Pb/$pp$ here but the
Pb$p$/$pp$ ratio would simply be these results reflected around $y=0$.

The NRQCD total cross sections, calculated with the CTEQ3L densities 
\cite{Lai:1994bb}, are also shown in Tables~\ref{psitab} and \ref{upstab}.  The
parameters used for the NRQCD calculations were determined for fixed-target
hadroproduction in Ref.~\cite{Beneke:1996tk} with $m_c = 1.5$ GeV, $m_b = 4.9$
GeV and $Q = 2m_Q$.  Calculations of shadowing effects predicted for NRQCD
can be found in Ref.~\cite{Emel'yanov:1999bn}.  The dependence on the shadowing
parameterization is also discussed.  It was found that the EKS98 
parameterization \cite{Eskola:1998iy,Eskola:1998df} predicted weaker shadowing
than the HPC parameterization \cite{HPC}, mainly because of the larger gluon
antishadowing in the EKS98 parameterization \cite{Emel'yanov:1999bn}.  The
predicted shadowing effects were similar for the CEM and NRQCD approaches.

Another approach to quarkonium production has been presented recently.
Ref.~\cite{Hoyer:1998ha} argues that the existing quarkonium data suggest the
presence of a strong color field, $\Gamma$, 
surrounding the hard $Q \overline Q$
production vertex in hadroproduction.  This field is assumed to arise from the 
DGLAP evolution of the initial colliding partons.  The effect of rescatterings
between the quark pair and this comoving field was shown to be in qualitative
agreement with the main features of the $pp$ and $pA$ data \cite{Hoyer:1998ha}.
Quantitative predictions within this model are difficult because the form of 
the field $\Gamma$ is unknown but some predictions assuming an isotropic and
transversely polarized field were given in
Ref.~\cite{Marchal:2000wd}\footnote{Thanks to P. Hoyer and S. Peigne for
comments on their work. See also \cite{Hoyer:2002zz}.}.

So far we have not discussed nuclear absorption in cold matter.  The $A$ 
dependence of $J/\psi$, $\psi'$ and the $\Upsilon$ $S$ states have all been
measured in $pA$ interactions.  (See Ref.~\cite{Vogt:cu} for a compilation of
the data.)  However, there is still no measurement of the $P$ state $A$
dependence.  The CEM makes no distinction between singlet and octet $Q 
\overline Q$ production so that {\it e.g.}\ the $J/\psi$, $\psi'$ and $\chi_c$
$A$ dependencies should all be the same.  The NRQCD approach, on the other
hand, predicts that direct $\psi'$ and $J/\psi$ are $60-80$\% octet but the
$\chi_c$ states are 95\% singlet \cite{Vogt:2001ky}.  If octets and singlets
are absorbed differently, the $\chi_c$ $A$ dependence should be quantitatively
different than the quarkonium $S$ states, see Ref.~\cite{Vogt:2001ky} for a 
detailed discussion at fixed-target energies.  In addition, the hard 
rescattering approach leads to a larger $\chi_{c1}$ $A$ dependence at 
high $x_F$ than for the $\chi_{c2}$ \cite{Hoyerinprep}.  Thus, measuring the
$\chi_c$ $A$ dependencies could help distinguish the underlying quarkonium
production mechanism.

NA60 \cite{NA60} and HERA-B \cite{herab} will measure the $\chi_c$ $A$
dependence for the first time at $\sqrt{s}= 29.1$ and 41.6 GeV
respectively.  At the LHC, it may also be possible to extract the $\chi_b$
$A$ dependence in $pA$ interactions as well. Comparing the $A$ dependence of
all quarkonium states at the LHC with earlier results will also provide a 
test of the energy dependence of the absorption mechanism, suggested to be
independent of energy \cite{Abreu:1999nn}.  Since the absorption
mechanism depends on impact parameter, centrality measurements of the $A$ 
dependence could also be useful in the comparison to $AA$ interactions.

\vspace{1.0cm}
\subsection{$p_T$ broadening in quarkonium production}
\label{section532}
{\em Ramona Vogt}

The broadening of the quarkonium transverse momentum distribution was first
observed by NA3 \cite{Badier:1983dg}.  Additional data at similar energies
showed that the change in the $J/\psi$ $p_T$ slope with $A$ was 
larger than for the Drell-Yan slope \cite{Bordalo:1987cr}.  
The $\Upsilon$ $p_T$ distribution
was shown to broaden still more than the $J/\psi$ \cite{Alde:1991sw}.  

The $p_T$
broadening effect was explained as multiple elastic scattering of the initial
parton before the hard collision \cite{Gavin:tw,Blaizot:1988hh,Hufner:wz}.  The
difference between the $J/\psi$ and Drell-Yan broadening can be attributed to
the larger color factor for gluons $(J/\psi)$ than for quarks (Drell-Yan).
The difference between $\Upsilon$ and $J/\psi$ broadening was more difficult to
explain.  The hard rescattering model of quarkonium production 
\cite{Hoyer:1998ha,Marchal:2000wd} is able to describe the effect 
\cite{Hoyer:1997yf}.  By Coulomb rescattering in the nucleus, the $Q \overline
Q$ pair acquires additional transverse momentum $k_T$ with an upper
limit given by $m_Q$.  Indeed, for $k_T > m_Q$ the color octet $Q
\overline Q$ pair can be `probed' and cannot rescatter as a gluon any
longer.  Since $m_b > m_c$, the $\Upsilon$ transverse momentum
broadening should be larger than that of the $J/\psi$
\cite{Hoyer:1997yf}\footnote{Thanks to P. Hoyer and S. Peigne for
comments on their work. See also \cite{Hoyer:2002zz}.}.

In this section, we investigate the combined effect of shadowing and transverse
momentum broadening on $J/\psi$ and $\Upsilon$ production.  The double 
differential distribution for $Q \overline Q$ pair production in $pp$ 
collisions is
\begin{eqnarray}
\label{stdfact}
s^2\, \frac{d^2\sigma_{pp}(s)}{dt du} =
\sum_{i,j} \int \frac{dx_1}{x_1} \frac{dx_2}{x_2}  d^2k_{T 1} d^2k_{T 2} 
g_p(k_{T 1}^2) g_p(k_{T 2}^2) f_i^p(x_1,Q^2)
f_j^p(x_2,Q^2) {\cal J}_K(\hat s) \hat s^2
\frac{d^2\hat \sigma_{ij}(\hat s)}{d\hat t d\hat u} 
\end{eqnarray}
where ${\cal J}_K$ is a kinematics-dependent Jacobian and $\hat 
s^2 d^2\hat \sigma_{ij}(\hat s)/d\hat t d\hat u$ is the partonic cross
section.  These partonic cross sections are difficult to calculate 
analytically beyond LO and expressions only
exist near threshold for $\hat 
s^2 d^2\hat \sigma_{ij}(\hat s, \hat t, \hat u)/d\hat t d\hat u$, see
{\it e.g.}\ Ref.~\cite{Kidonakis:2001nj}.  
However, the exclusive $Q \overline Q$ pair NLO \cite{Mangano:jk} code 
calculates double differential distributions numerically.

The bare quark $p_T$ distributions have often been sufficient to
describe inclusive $D$ meson production.  However, the $Q \overline Q$
pair distributions, particularly the pair $p_T$ and the azimuthal
angle distributions at fixed target energies, are broader than can be
accounted for by the bare quark distributions without fragmentation
and intrinsic transverse momentum $k_T$.  Heavy quark fragmentation,
parameterized from $e^+ e^- \rightarrow H X$, which reduced the
average momentum of the heavy quark in the final state hadron, is
included.  This effect, along with momentum broadening with $\langle
k_T^2
\rangle = 1$ GeV$^2$, can describe the $Q \overline Q$ measurements 
\cite{Frixione:1994nb}.  We include the broadening here but not 
fragmentation since, in
the CEM, both the $Q$ and $\overline Q$ hadronize together to make the 
quarkonium state, inconsistent with independent fragmentation.

The intrinsic $k_T$ of
the initial partons has been used successfully to describe the
$p_T$ distributions of other processes such as Drell-Yan \cite{Field:uq} 
and hard photon production \cite{Owens:1986mp}.  
Some of the low $p_T$ Drell-Yan data have
subsequently been described by resummation to all orders but the inclusion of
higher orders has not eliminated the need for this intrinsic $k_T$.  

The implementation of intrinsic $k_T$ in the $Q \overline Q$ code in
Ref.~\cite{Mangano:jk} is not the same as in
other processes because divergences are canceled numerically.  Since 
including additional
numerical integrations would slow this process, the $k_T$ kick is added in the
final, rather than the initial, state. 
In Eq.~(\ref{stdfact}), the Gaussian $g_p(k_T^2)$,
\begin{eqnarray}
g_p(k_T^2) = \frac{1}{\pi \langle k_T^2 \rangle_{pp}} \exp(-k_T^2/\langle 
k_T^2 \rangle_{pp}) \, \, ,
\label{intkt}
\end{eqnarray}
with $\langle k_T^2 \rangle_{pp} = 1$ GeV$^2$ \cite{Frixione:1994nb}, 
assumes that the $x$ and $k_T$ dependencies completely
factorize.  If this is true, it does not matter whether the $k_T$ dependence
appears in the initial or final state, modulo some caveats.  
In the code, the $Q \overline Q$ system
is boosted to rest from its longitudinal center of mass frame.  Intrinsic 
transverse
momenta of the incoming partons, $\vec k_{T 1}$ and $\vec k_{T 2}$, are chosen
at random with $k_{T 1}^2$ and $k_{T 2}^2$ distributed according to
Eq.~(\ref{intkt}).   A second transverse boost out of the pair rest frame
changes the initial transverse momentum of
the $Q \overline Q$ pair, $\vec p_T$ to $\vec p_T 
+ \vec k_{T 1} + \vec k_{T 2}$.  The initial
$k_T$ of the partons could have alternatively been given to the entire
final-state system, as is essentially done if applied in the initial state,
instead of to the $Q \overline Q$ pair.  There is no difference if the
calculation is only to LO but at NLO a light parton may also appear in
the final state.  In Ref.~\cite{Frixione:1997ma}, 
it is claimed that the difference 
between adding the $k_T$ to the initial or final state is rather small 
if $k_T^2 \leq 2$ GeV$^2$.  However, this may not turn out to be the case
for nuclei \cite{Vogt:2001nh}.

The average intrinsic $k_T$ is expected to increase in $pA$ interactions.  This
broadening is observed in Drell-Yan \cite{Bordalo:1987cr,Alde:1991sw}, 
$J/\psi$ \cite{Badier:1983dg}, and
$\Upsilon$ \cite{Alde:1991sw} production and has been used to explain high 
$p_T$ pion
production in nuclear interactions \cite{Wang:1998hs}.  We follow the 
formulation of Ref.~\cite{Wang:1998hs}, similar to 
Refs.~\cite{Gavin:tw,Blaizot:1988hh,Hufner:wz} for $J/\psi$ and Drell-Yan
production, where the $k_T$ broadening in $pA$ interactions is
\begin{eqnarray}
\langle k_T^2 \rangle_{pA} = \langle k_T^2 \rangle_{pp} + 
(\langle \nu \rangle -1) \Delta^2(Q) \, \, 
\label{ktbroad}
\end{eqnarray}
with $ \langle k_T^2 \rangle_{pp} = 1$ GeV$^2$.
The number of collisions in a $pA$ interaction, $\nu$,
averaged over impact parameter, is \cite{Chiappetta:1987ca,Gavin:1996ss}
\begin{eqnarray}
\langle \nu \rangle = \sigma_{NN} \frac{\int d^2b \, T_A^2(b)}{\int d^2b \,
T_A(b)} = \frac{3}{2} \sigma_{NN} \rho_0 R_A 
\label{avenu}
\end{eqnarray}
where $T_A(b) = \int dz \rho_A(b,z)$ is the nuclear profile function, 
$\sigma_{NN}$ is the inelastic nucleon-nucleon
scattering cross section, $\rho_0$ is the central nuclear density, and $R_A$ is
the nuclear radius.  Our calculations assume $R_A = 1.2
A^{1/3}$ and $\rho_0 = 0.16/$fm$^3$.
The strength of the nuclear broadening, $\Delta^2$, depends on $Q$, the 
scale of the interaction \cite{Wang:1998hs}
\begin{eqnarray}
\Delta^2(Q) = 0.225 \frac{\ln^2(Q/{\rm GeV})}{1 + 
\ln(Q/{\rm GeV})} {\rm GeV}^2 \, \, .
\label{deltasq}
\end{eqnarray}
Thus $\Delta^2$ is larger
for $b \overline b$ production than $c \overline c$ production.  This
empirically reflects the larger $k_T$ broadening of 
the $\Upsilon$ \cite{Alde:1991sw}
relative to the $J/\psi$ \cite{Badier:1983dg}.  We evaluate 
$\Delta^2(Q)$ at $Q = 
2m_Q$ for both charm and bottom production.
We find $(\langle \nu \rangle - 1) \Delta^2(Q) = 0.35$ GeV$^2$
for $c \overline c$ and 1.57 GeV$^2$ for $b\overline b$ production 
in $pA$ collisions at $b = 0$ and $A = 
200$. We can change the centrality by changing $\langle \nu \rangle - 1$.

\begin{figure}[htb]
\vspace{0.5cm}
\setlength{\epsfxsize=0.95\textwidth}
\setlength{\epsfysize=0.2\textheight}
\centerline{\epsffile{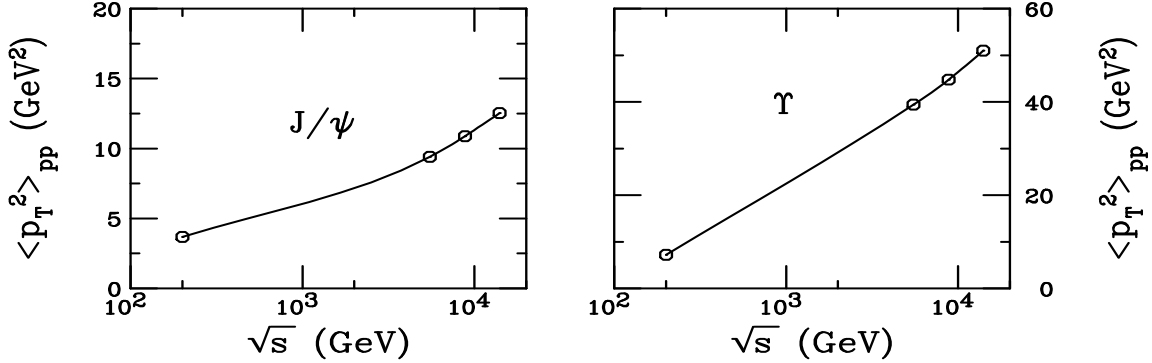}}
\caption{The energy dependence of $\langle p_T^2 \rangle_{pp}$ for $J/\psi$
(left) and $\Upsilon$ (right) production assuming $\langle k_T^2 \rangle_{pp} 
= 1$ GeV$^2$.  The open circles are the calculated energies at $\sqrt{s} = 0.2,
5.5, 8.8$, and 14 TeV.
}
\label{pt2pp}
\end{figure}

In Fig.~\ref{pt2pp}, we show the increase in $\langle p_T^2 \rangle$ for $pp$
collisions at heavy ion collider energies using the MRST parton densities with
$m_c = 1.2$ GeV and $m_b = 4.75$ GeV from Table~\ref{qqbtab}.  
The average 
$p_T^2$ is obtained by integrating over all $p_T$ and rapidity.  Increasing
$m_Q$ increases $\langle p_T^2 \rangle_{pp}$.  At these energies, the 
difference between  $\langle p_T^2 \rangle_{pp}$ with and without $\langle 
k_T^2 \rangle_{pp} = 1$ GeV$^2$ is small, a factor of $\sim 1.3$ for $J/\psi$
and $\sim 1.05$ for $\Upsilon$ at 5.5 TeV.

\begin{figure}[bht]
\setlength{\epsfxsize=0.85\textwidth}
\setlength{\epsfysize=0.30\textheight}
\centerline{\epsffile{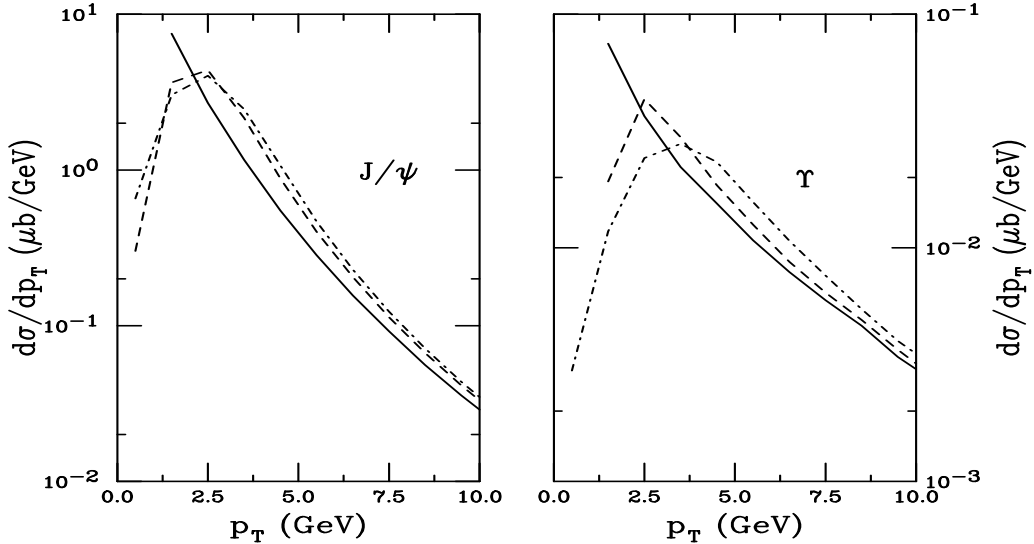}}
\vspace{-0.2cm}
\caption{The $p_T$ distributions with $x_F > 0$ for $J/\psi$
(left) and $\Upsilon$ (right) production in $pA$ interactions at 5.5 
TeV/nucleon.  The
solid curve is the bare distribution, the dashed curve employs $\langle k_T^2
\rangle_{pp} = 1$ GeV$^2$, and the dot-dashed curve uses a 
$\langle k_T^2 \rangle_{pA}$
appropriate for Pb.
}
\label{ptbroad}
\end{figure}

Figure~\ref{ptbroad} compares the bare distributions, $\langle k_T^2
\rangle_{pp} = 1$ GeV$^2$, and $\langle k_T^2
\rangle_{p{\rm Pb}}$ calculated using Eq.~(\ref{ktbroad}), integrated over
all rapidity.
Without intrinsic transverse momentum, the value of the cross section in the
lowest $p_T$ bin is less than zero due to incomplete cancelation of 
divergences, as discussed in Ref.~\cite{Gavai:1994gb}.  This point is left
out of the plots.  It is interesting to note that, while shadowing is a small
effect at large $p_T$, the average $p_T^2$ increases with $A$, even without
any broadening included.  This arises because shadowing is a 25-50\% effect on
$J/\psi$ production at low $p_T$, increasing with $A$, and a 10-20\% effect on
the $\Upsilon$.  The resulting increase in $\langle p_T^2 \rangle$ from $pp$
to $p$Pb is $\sim 1.2$ for $J/\psi$ and $\sim 1.13$ for $\Upsilon$.  Including
 $\langle k_T^2 \rangle_{pp} = 1$ GeV$^2$ increases $\langle p_T^2 \rangle$ 
over all $p_T$ as well, but only by $\sim 1.15$ for both $J/\psi$ and 
$\Upsilon$.  Including the $A$ dependence of the broadening as in 
Eq.~(\ref{ktbroad}) gives about the same increase in $\langle p_T^2 \rangle$
as the bare distributions.

Note that if a $p_T$ cut was imposed, {\it e.g.} $p_T > 5$ GeV, $\langle p_T^2
\rangle$ increases significantly but the effect of broadening is greatly
reduced.  In fact, $\langle p_T^2 \rangle$ actually decreases with increasing
$\langle k_T^2 \rangle_{pA}$ when such a cut is imposed.  This feature is
clear from an inspection of Fig.~\ref{ptbroad}.

\begin{figure}[hbt]
\vspace{0.5cm}
\setlength{\epsfxsize=0.95\textwidth}
\setlength{\epsfysize=0.23\textheight}
\centerline{\epsffile{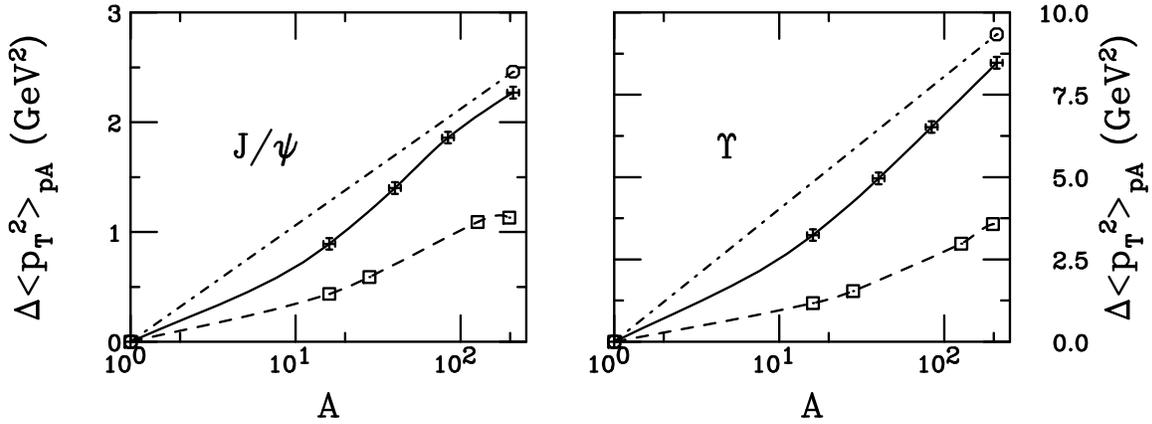}}
\caption{The $\sqrt{s}$ and $A$ dependencies of $\Delta \langle p_T^2 
\rangle_{pA}$ for $J/\psi$
(left) and $\Upsilon$ (right) production in $pA$ interactions.  The
dashed curve is at $\sqrt{s} = 200$ GeV/nucleon, the solid at 5.5 TeV/nucleon 
and the dot-dashed at 8.8 TeV/nucleon.  We use $A=$O, Si, I and Au at 200 
GeV/nucleon, $A=$O, Ar, Kr, and Pb at 5.5 TeV/nucleon and $A=$Pb at 8.8 
TeV/nucleon.
}
\label{pt2pa}
\end{figure}

The energy and $A$ dependence of $\Delta \langle p_T^2 \rangle_{pA} =
\langle p_T^2 \rangle_{pA} - \langle p_T^2 \rangle_{pp}$ is given in 
Fig.~\ref{pt2pa} for $\sqrt{s} = 200$ GeV/nucleon at RHIC as well as 5.5 and
8.8 TeV/nucleon at the LHC.
We find that $\Delta \langle p_T^2 \rangle_{pA}$ for $J/\psi$ and $\Upsilon$
increase by factors of $\sim 3$ and $\sim 5.3$ over $\sqrt{s} = 19.4$ GeV
\cite{Badier:1983dg} and 38.8 GeV \cite{Alde:1991sw} respectively.  
Thus, in this calculation,
the increase in $\Delta \langle p_T^2 \rangle_{pA}$ with $\sqrt{s}$ is stronger
for $\Upsilon$ than $J/\psi$.  Such a phenomenological parameterization should
soon be tested at RHIC for the $J/\psi$.

\clearpage


\section{NOVEL  QCD PHENOMENA IN $pA$ COLLISIONS AT THE LHC}
\label{pA_SEC:noveleffects}
{\em Leonid~Frankfurt and  Mark~Strikman}




\subsection{Physics motivations}
\label{aintro}

An extensive discussion of the novel $pA$ physics at the LHC was
presented in Ref.~\cite{:2001qv}. Here we summarize few key points of
this analysis and extend it to consider effects of the small $x$
high-field regime.

The proton-ion collisions at the LHC will be qualitatively different
from those at fixed target energies.  In soft interactions two prime
effects are a strong increase of the mean number of ``wounded''
nucleons - from $\sim$ 6 to $\sim$ 20 in $pA$ central collisions (due
to increase of $\sigma_{\rm{inel}}(NN)$) and a 50 $\%$ increase of
the average impact parameters in $pp$ collisions (as manifested in the
shrinkage of the diffractive peak in the elastic $pp$ collisions with
increasing energy) making it possible for a proton to interact, often
simultaneously, with nucleons at the same longitudinal coordinate but
different impact parameters.  At the same time, the geometry of $pA$
collisions at the LHC should be closer to that in classical mechanics
due to a strong increase of the $pp$ total cross section and smaller
fluctuations of the interaction strength in the value of the effective
$NN$ cross section. (The dispersion of the fluctuations of the cross
section is given by the ratio of the inelastic and elastic diffractive
cross sections at $t=0$, which is expected to be a factor of $3\dots4$
smaller at the LHC relative to RHIC.) Among other things
this will lead to a much weaker increase of the inelastic coherent
diffractive cross section with $A$ relative to fixed target energies
\cite{Frankfurt:1993qi}.  In hard interactions, the prime effect is a
strong increase of the gluon densities at small $x$ in both projectile
and ion. Gluon densities can be measured at the LHC down to $x \propto
10^{-6}\dots10^{-7} $ where pQCD will be probed in a new domain
\cite{:2001qv}.  As a consequence of blackening of the interaction of
partons from the proton with nuclear partons at small $x_A$, one should
expect a disappearance of low $p_T$ hadrons in the proton
fragmentation region and, hence, a strong suppression of
nonperturbative QCD effects\footnote{A similar phenomenon of the low
$p_T$ suppression of another quantity, the nuclear light cone wave
function at small $x$, is discussed in the saturation models
\cite{Mueller:2001fv,Iancu:2002xk}.}.  Thus the high $p_T$ proton
fragmentation region is a natural place to search for unusual QCD
states of quark-gluon matter which should be long lived due to the
Lorentz time dilation.  In spite of a small coupling constant, gluon
densities of heavy ions become larger over a wide range of $x$ and $Q^2$
and impact parameters than that permitted by the conservation of
probability within the leading-twist approximation
\cite{Frankfurt:1995jw,:2001qv,Mueller:2001fv,Iancu:2002xk,Frankfurt:2000ty}.
Thus a decrease of $\alpha_s$ with $Q^2$ is insufficient for the
applicability of leading-twist pQCD to hard processes in the small $x$
regime where gluon fields may be strong enough for the nonperturbative QCD
vacuum to become unstable and to require modification of QCD dynamics
including modification of the QCD evolution equations.  The
possibility for a novel QCD regime and therefore new phenomena are
maximized in this QCD regime\footnote{It is important to distinguish
between parton densities which are defined within the leading-twist
approximation only \cite{Feynman} and may increase with energy forever
and physical quantities - cross sections/structure functions - whose
increase with energy at a given impact parameter is restricted by
probability conservation.}. At the same time interpretation of these
novel effects would be much more definitive in $pA$ collisions than in
heavy ion collisions.

New classes of strong-interaction phenomena in the short-distance
regime, near the light cone, which could be associated with
perturbative QCD or with the interface of the perturbative and
nonperturbative QCD regions are much more probable in $pA$ than $pp$
collisions. The parton and energy densities may be high enough to
``burn away'' the nonperturbative QCD vacuum structure in a cylinder
of radius $\sim 1.2$~fm and length $\sim 2 R_A$, leaving behind only a
dense partonic fluid governed by new QCD dynamics with small
$\alpha_s$ but large effective short-distance couplings. If there is a
large enhancement of hard-collision, gluon-induced processes, then
there should be an enhancement of hard multi-parton collisions, like
production of several pairs heavy-flavor particles \cite{:2001qv}.
The study of the $x$, $p_T^2$ and $A$-dependence of multi-parton
interactions, and of forward charm and bottom production (in the
proton fragmentation region) should be especially useful for
addressing these questions.

We estimate the boundary of the kinematical regime for the new QCD
dynamics to set in in $pA$ collisions based on the LO formulae for the
dipole cross section and the probability conservation.  For 
central impact parameters with $A\sim 200$ and $Q^2\sim 25$~GeV$^2$,
we find $x_A\leq 5\cdot 10^{-5}$ for quarks and $x_A\leq 10^{-3}$ for
gluons \cite{Frankfurt:2000ty}.

\subsection{Measurements of  nuclear parton densities at ultra small x}

\noindent
One of the fundamental issues in high energy QCD is the dynamics of
hard interactions at small $x$. Depending on the $Q^2$, one
investigates either leading-twist effects or strong color
fields. Since the cross sections for hard processes increase roughly
linearly with $A$ (except at very small $x$ and relatively small $Q^2$
where the counting rates are high anyway), even short runs with
nuclear beams will produce data samples sufficient to measure parton
distributions with {\it statistical } accuracy better than 1$\%$
practically down to the smallest kinematically allowed $x$: $x \sim
10^{-7} $ for quarks and $x \sim 10^{-6} $ for gluons
\cite{Alvero:1998cb}. The systematic errors on the ratios of nuclear
and nucleon structure functions are also expected to be small since
most of these errors cancel in the ratios of the cross sections, cf.
Ref.~\cite{Arneodo:1996qa}. The main limitations on such studies are
the forward acceptance and the transverse momentum resolution of the
LHC detectors.

There are several processes which could be used to probe the small $
x$ dynamics in $pA$ collisions. They include the Drell-Yan pairs,
dijets, jet + photon, charm production, diffractive exclusive 3-jet
production, and multijet events.  It is worth emphasizing that, for
these measurements, it is necessary to detect jets, leptons and
photons at rapidities which are close to the nucleon rapidity. In this
$y$-range, the accompanying soft hadron multiplicities are relatively
small, leading to a soft particle background comparable to that in
$pp$ scattering.  In fact, for $|y_{\rm max}-y| \leq 2-4$ the
background level is likely to be significantly smaller than in $pp$
collisions due to suppression of the leading particle production in
$pA$ scattering, see Ref. \cite{:2001qv}.

Measurements of dimuon production appear to be more feasible at
forward angles than hadron production due to small muon
energy losses. In the kinematics where the leading-twist dominates,
one can study both the small-$x$ quark distributions at relatively
small $p_T$,  $p_T(\mu^+\mu^-) \ll M(\mu^+
\mu^-)$, and the gluon distribution at $p_T(\mu^+\mu^-) \sim
M(\mu^+\mu^-)$ \cite{Catani:2000jh}.  Note also that in the forward
region, the dimuon background from charm production is likely to be
rather small since, in average, muons carry only 1/3 of the charm
decay momentum.

\subsection{Signals for onset of the black body limit}
\label{sec3}

A fast increase of nPDFs for heavy nuclei with energy  contradicts
probability conservation for a wide range of parton momenta and
virtualities for the LHC energy range. On the contrary, the rapid,
power-like growth of the nucleon structure function observed at HERA
may continue forever because of the diffuse edge of a nucleon
\cite{Frankfurt:2001nt}.  At present, new QCD dynamics in the regime
of high parton densities where multi-parton interactions are inhibited
remains a subject of lively debate. Possibilities include a pre-QCD
black-body regime for the structure functions of heavy nuclei
\cite{Gribov:1968gs}, a QCD black-body regime for hard processes with
heavy nuclei \cite{Frankfurt:2001nt}, saturation of parton densities
at a given impact parameter \cite{Mueller:2001fv}, a color glass
condensate \cite{Iancu:2002xk}, turbulence and related scaling laws.
To estimate the pattern of the novel phenomena we shall use the
formulae of the black-body regime because they are generic enough
to illustrate the expectations of the onset of new QCD dynamics within
the existing theoretical approaches\footnote{In the black-body
regime, a highly virtual probe interacts at the same time with several
high $k_T$ partons from the wave function of a hadron
\cite{Frankfurt:2001nt}.  This is qualitatively different from the
asymptotic freedom limit where interactions with only one highly
virtual parton are important.  Hence the interacting parton is far
from being free when probed by an object with a large coherence
length, $l_c=1/2m_Nx$. }.

High parton density effects can be enhanced in nuclei because of their
uniform density and because it is  possibile to select scattering at 
central impact parameters.  Indeed, in the small $x$ regime, a hard
collision of a parton of the projectile nucleon with a parton of the
nucleus occurs coherently with all the nucleons at a given impact
parameter. The coherence length $l_{\mbox c}\approx 1/2m_{N}x$ by far
exceeds the nuclear size. In the kinematic regime accessible at the
LHC, $l_{\mbox c}$ can be up to $10^5$ fm in the nuclear rest frame. 
Here the large coherence length regime can be
visualized in terms of the propagation of a parton through high
density gluon fields over much larger distances than is possible with
free nucleons. In the Breit frame, this corresponds to the fact that
small $x$ partons cannot be localized longitudinally within the
Lorentz-contracted thickness of the nucleus.  Thus low $x$ partons
from different nucleons overlap spatially, creating much larger parton
densities than in the free nucleon case. This leads to a large
amplification of  multi-parton hard collisions, expected at small
$x$ in QCD, see e.g. Refs~\cite{Arneodo:1996qa,:2001qv}.  The eikonal
approximation \cite{Mueller:wy,Mueller:st} and constraints from the
probability conservation \cite{Frankfurt:1995jw,Abramowicz:1995hb}
indicate that new QCD dynamics should be revealed at significantly
larger $x$ in nuclei than in the proton and with more striking effects.
Recently, extensive studies of hard diffraction were performed at HERA.
For review see Ref.~\cite{Abramowicz:1998ii}.
Information obtained in these studies allows to evaluate the
leading-twist nuclear shadowing which turns out to be large both in
the gluon and quark channels and significant up to very large
$Q^2$. For a recent discussion see Ref.~\cite{Frankfurt:2000ty}.  This
leading-twist shadowing tames the growth of the nPDFs at small $x$.
However, even such large shadowing is still insufficient to restore
probability conservation, leaving room for violation of the
leading-twist approximation. This violation may be more serious than
merely an enhancement of higher twist effects since the entire
interaction picture can qualitatively change. The common wisdom that a
small $\alpha_s$ guarantees asymptotic freedom, confirmed
experimentally in many hard processes, is violated at small $x$. This
phenomenon can be considered in terms of effective violation of
asymptotic freedom in this limit. This regime is expected to be
reached in heavy nuclei at $x$ values at least an order of magnitude
larger than in the proton. Hence in the LHC kinematics, the region
where novel QCD phenomena are important will extend at least two
orders of magnitude in $x_A$ over a large range of virtualities.

\noindent
{\bf The space-time picture of the black-body limit in $pA$
collisions} can be seen best in the rest frame of the nucleus. A
parton belonging to the proton emits a hard gluon (virtual photon)
long before the target and interacts with the target in a black regime
releasing the fluctuation, a jet/$q\bar q$ (a Drell-Yan pair).
This leads to a qualitative change in the
picture of $pA$ interactions for the partons with
$x_p$ and $p_T$ satisfying the condition that
\begin{equation}
x_A={4 p_T^2\over x_{p}s_{NN}} ,
\label{limit}
\end{equation}
is in the black-body kinematics for the resolution scale $p_T\leq
p_T^{\rm bbl}(x_A)$. Here $ p_T^{\rm bbl}(x_A)$ is the maximum
$p_T$ at which the black-body approximation is applicable.
Hence by varying $x_A$ for fixed $x_p$ and looking for the the maximum
value of $p_T^{\rm bbl}$ we can determine $p_T^{\rm bbl}(x_p)$.
In the kinematics of the LHC, $Q^2\approx 4(p_T^{\rm bbl})^2$ can
be estimated by using formulae derived in Ref.~\cite{Frankfurt:1995jw}.
At $x_p\sim 0.3$
$Q^2$ may be as large as $(p_T^{\rm bbl}(x_p))^2=15$~GeV$^2$ .
All the partons with such $x_p$ will obtain $p_T{\rm (jet)}\sim
p_T^{\rm bbl}(x_p)$, leading to multi-jet production. The black
body regime will extend down in $x_p$ with increasing $\sqrt{s_{NN}}$.
For the LHC with $p_T \leq 3$~GeV$/c$, this regime may cover the whole
region $x_p \geq 0.01$ where of the order ten partons reside. Hence,
in this limit, most of the final states will correspond to
multi-parton collisions. For $p_T \leq 2$~GeV$/c$ the region extends
to $x_p \geq 0.001$. At the LHC, collisions of such partons correspond
to central rapidities.  Dynamics of the conversion of high-$p_T$
partons with similar rapidities to hadrons is a collective effect
requiring special consideration.

{\bf Inclusive observables.} For  the total cross
sections, logic similar to that employed  in 
$\gamma^*$ - nucleus scattering \cite{Frankfurt:2001nt} should be
applicable. In particular, for the dimuon total cross section,
\begin{equation}
{d\sigma(pA \to \mu^+\mu^- + X)\over dx_Adx_p}= {4\pi\alpha_{\rm
em}^2\over 9}{K(x_A,x_p,M^2)\over M^2} F_{2p}(x_p,Q^2) {M^2\over
6\pi^2} 2\pi R_A^2 \ln\left({x_0\over x_A}\right),
\end{equation}
for large but not too large dimuon masses, $M$. In particular we
estimate $M^2({\rm bbl})(x_A=10^{-7})\approx 60$~GeV$^2$. Here the
$K$-factor has the same origin as in the leading-twist case but it
should be smaller since it originates from gluon emissions only from
the parton belonging to the proton. We define $x_0$ as the maximum $x$
for which the black body limit is valid at the resolution scale $M$.
Hence the expected $M^2$ dependence of dimuon production is
qualitatively different than in $pp$ scattering where scattering at
large impact parameters may mask the contribution from the black-body
limit. In this limit, the $x_A$ dependence of the cross section becomes
weak.

The study of the dimuon $p_T$ distribution may also signal the onset
of the black-body regime. As in the case of leading partons in 
deep inelastic scattering \cite{Frankfurt:2001nt}, a broadening of the
dimuon $p_T$ distribution is expected in the black-body limit
compared to the DGLAP expectations. See Ref.~\cite{Gelis:2002fw} for a
calculation of this effect in the color glass condensate model.

At still lower $x_A$, the black-body formalism will probably
overestimate the cross section because the interaction the sea quarks
and gluons in the proton will also be in the black-body limit.

The onset of the black body regime will also lead to important changes
in hadron production such as a much stronger drop of the $x_F$
spectrum in the proton fragmentation region accompanied by $p_T$
broadening and enhancement of hadron production at smaller rapidities,
see the discussion in section \ref{a.4.4}.

It is worth emphasizing that onset of the black-body limit for ultra
small $x_A$ will places limits on using forward kinematics for PDF
measurements at larger $x_A$.  At the very least it would be necessary
to satisfy the condition $p_T({\rm jet})\gg p_T^{\rm bbl}(x_p)$.
Indeed, as soon as large enough $x_p$ are used for the nPDF
measurements one would have to take into account that such collisions
with nuclear partons are always accompanied by significant $p_T$
broadening, due to the interaction with ``the black component'' of the
nuclear wave function. At a minimum, this would lead to significant
broadening of the $p_T$ distribution of the produced hard system.  One
can also question the validity of the leading-twist expansion for the
$p_T$-integrated cross sections.  To investigate these phenomena, the
ability to make measurements at fixed $x_A$ and different $x_p$ would
be very important.

\subsection{Mapping  the three-dimensional  nucleon  parton structure}

Systematic studies of hard inclusive processes during the last two
decades have led to a reasonably good understanding of the single
parton densities in nucleons. However, very little is known about
multiparton correlations in nucleons. These correlations can provide
critical new insights into strong interaction dynamics, as well as
discriminate between models of the nucleon. Such correlations may be
generated, for example, by fluctuations in the transverse size of the
color field in the nucleon leading, via color screening, to correlated
fluctuations of the gluon and quark densities.

QCD evolution is a related source of correlations since selection of a
parton with a given $x$ and $Q^2$ may lead to a local enhancement of the
parton density in the transverse plane at different $x$ values.  Also,
practically nothing is known about possible correlations between the
transverse size of a particular configuration in the nucleon and the
longitudinal distribution of partons in this configuration.

\subsubsection{Multi-jet production and double parton distributions}
It was recognized already more than two decades ago
\cite{Goebel:1979mi} that the increase of parton densities at small
$x$ leads to a strong increase in the probability of $NN$ collisions
where two or more partons from each nucleon undergo pairwise
independent hard interactions.  Although multijet production through
double-parton scattering was investigated in several experiments
\cite{Akesson:1986iv,Abe:1997bp,Abe:1997xk} at $pp$ and $p\bar p$
colliders, the interpretation of the data was hampered by the need to
model both the longitudinal and the transverse partonic correlations
simultaneously.  Studies of $pA$ collisions at the LHC will provide a
feasible opportunity to study the longitudinal and transverse partonic
correlations in the nucleon separately as well as to check the
validity of the underlying picture of multiple collisions.

The simplest case of a multi-parton process is double parton
collisions.  Since the momentum scale $p_T$ of a hard interaction
corresponds to much smaller transverse distances, $\sim 1/p_T$, in
coordinate space than the hadronic radius, in a double-parton
collision the two interaction regions are well separated in transverse
space. Also, in the center of mass frame, pairs of partons from the
colliding hadrons are located in pancakes of thickness less than $
(1/x_1 +1 /x_2)/p_{\rm cm}$. Thus two hard collisions occur almost
simultaneously as soon as $x_1$ and $ x_2$ are not too small. Hence
there is no cross talk between two hard collisions.  A consequence is
that the different parton processes add incoherently in the cross
section. The double parton scattering cross section, proportional to
the square of the elementary parton-parton cross section, is therefore
characterized by a scale factor with dimension of inverse length
squared. The dimensional quantity is provided by nonperturbative
input, namely by the multi-parton distributions. In fact, because of
the localization of the interactions in transverse space, the two
pairs of colliding partons are aligned in such a way that the
transverse distance between the interacting target partons is
practically the same as the transverse distance between the projectile
partons. The double parton distribution,
$\Gamma(x,x',b)$, is therefore a function of two
momentum fractions and of the transverse distance, $b$. Although
$\Gamma$ also depends on the virtualities of the partons, $Q^2$ and $Q'^2
$, to make the expressions more compact, this $Q^2$ dependence is not
explicitly expressed.  Hence the double-parton scattering cross
section for the two ``2 $\to$ 2'' parton processes $\alpha$ and
$\beta$ in an inelastic interaction between hadrons (or nuclei) $a$
and $b$ can be written as:
\begin{eqnarray}
\sigma_D(\alpha,\beta)&=&{m\over2}\int\Gamma_a(x_1,x_2;b)\hat{\sigma}_{\alpha}
(x_1,x_1')
\hat{\sigma}_{\beta}(x_2,x_2')\Gamma_b(x_1',x_2';b)dx_1dx_1'dx_2dx_2'd^2b
\label{1}
\end{eqnarray}
\par\noindent where $m=1$ for indistinguishable and $m=2$ for
distinguishable parton processes. Note that though the factorization
approximation of Eq.~(\ref{1}) is generally accepted in the analyses
of the multijet processes and appears natural based on the geometry of
the process, no formal proof exists. As we will show below, the study
of the $A$-dependence of this process will be a stringent test of this
approximation.

To simplify the discussion, we neglect small non-additive effects in
the parton densities, a reasonable approximation for $0.02\leq x \leq
0.5$. In this case, we have to take into account only $b$- space
correlations of partons in individual nucleons.

There are thus two different contributions to the double-parton
scattering cross section
($A\equiv a, p \equiv b$):
$\sigma_D=\sigma^1_D+\sigma^2_D$. The first one,
$\sigma_1^D$, the interaction of  two partons from  the same nucleon,
is the same for nucleons and nuclei except for the enhancement of the parton flux. The
corresponding cross section is
\begin{equation}
\sigma^1_D=\sigma_D\int d^2BT_A(B)= A\sigma_D^{NN},
\label{sigma1}
\end{equation}
\par\noindent
where
\begin{equation}
T_A(B)=\int_{-\infty}^{\infty} dz \rho_A(r), {\rm and} \int T_A(B)d^2B=A
\label{TB}
\end{equation}
\par\noindent is the nuclear thickness as a function of the impact
parameter $B$ of the hadron-nucleus collision.

The contribution to $\Gamma_A(x_1',x_2',b)$ from partons in different
target nucleons, $\sigma^2_D$, can be calculated {\it solely} from the
geometry of the process by observing that the nuclear density does not
change within the transverse scale $\left<b\right> \ll R_A$. It
rapidly increases with $A$ as $\int T_A^2(B)d^2B$. Using information
from the CDF double scattering analysis \cite{Abe:1997bp,Abe:1997xk}
on the mean transverse separation of partons in a nucleon, one finds
that the contribution of the second term should dominate in $pA$
collisions: $\sigma^2_D/\sigma^1_D\approx 0.68 (A/12)^{0.39}$ for
$A\ge 40$ \cite{Strikman:2001gz}.  Hence one expects a stronger than
$\propto A$ increase of multijet production in $pA$ collisions at the
LHC. Measurements with a range of nuclei would probe the double-parton
distributions in nucleons and also check the validity of QCD
factorization for such processes. Factorization appears natural but is so
far not derived in pQCD.  An important application of this process
would be the investigation of transverse correlations between the
nuclear partons in the shadowing region.  A study of these
correlations would require selection of both nuclear partons in the
shadowing region, $x_A\leq x_{\rm sh} \sim 10^{-2}$.\footnote{The
$A$-dependence of the ratio of $\sigma^2_D/\sigma^1_D$ in the
kinematics where only one of the nuclear partons has $ x_A\leq x_{\rm
sh}$ is practically the same as for the case when both nuclear partons
have $x\geq x_{\rm sh}$.}

As discussed in section \ref{sec3}, partons with sufficiently large
$x_p$, satisfying Eq.~(\ref{limit}), are expected to interact with
small $x$ partons in the nucleus with a probability of order one.  As
a result, a hard collision of partons with sufficiently large $x_p$
and $x_A\geq 0.01 $ will be accompanied by production of one or more
minijets in the black-body regime with $p_T \sim p_T^{\rm
bbl}(x_A)$. Most sufficiently fast partons from the proton will
generate minijets, leading to a strong suppression of the cross section of
events with only two jets.

Detectors with sufficiently forward rapidity acceptance could detect
events originating from triple parton collisions at $x_A\geq
x_{\rm sh}$ and large enough $p_T$ where $p_T$ broadening related to
the blackbody limit effects are sufficiently small. These triple
parton collisions provide stringent tests of the hard interaction
dynamics as well as provide additional information on two-parton
correlations and unique information on triple-parton correlations.

Other opportunities with multi-jets include:
\begin{itemize}
\item Probing correlations between partons in the nucleus at high
densities, i.e. for $x_{1A}, x_{2A} \ll 10^{-3}$, which would
provide qualitatively new information about the dynamics of nuclear
shadowing   and  the presence of possible new parton condensates.
\item Studying the accompanying soft hadron production (cf. the
discussion in section~\ref{a.1.5}), which would measure the transverse
size of a proton configuration containing partons $x_{1}$ and $x_{2}$. In
particular, the production of 4 jets with two of the jets at large
$x_p$, $x_{1}+x_{2} \geq 0.5$, may provide another way to look for
point-like configurations in nucleons, see section \ref{3jetsec} and
Ref.~\cite{:2001qv} for discussion of other possibilities.
\end{itemize}

\subsubsection{Proton-ion collisions probe  transverse nucleon structure}
\label{a.1.5}

As we discuss in section \ref{3jetsec}, in some subclass of events the
distribution of constituents in the initial proton may be unusually
local in the transverse (impact parameter) plane when the proton
collides with the ion. If this is so, its effective cross section per
nucleon will be greatly reduced, perhaps all the way to the
perturbative-QCD level. If the effective cross section of such a
point-like configuration goes below 20 mb, there will be an
appreciable probability that it can penetrate through the center of a
lead ion and survive, leading to a strongly enhanced diffractive yield
of the products of the point-like configuration in collisions with
heavy ions.

Not only might the properties of the final-state collision products
depend upon the nature of the transverse structure of the proton
primary on arrival at the collision point, but even the conventional
parton distributions may also be affected. For example, let us
consider a nucleon as a quark and small diquark connected by a narrow
QCD flux-tube. It should be clear that if the flux-tube is at right
angles to the collision axis at arrival, then the valence partons will
have comparable longitudinal momentum or $x$. On the other hand, if
the flux tube is parallel to the direction of motion, then one of the
valence systems will have very large $x$ and the other very
small. This happens, because in this case, the internal longitudinal
momenta of the valence systems, in the rest frame of the projectile
proton, are in opposite directions. Therefore the smallness of the
configuration is correlated with the joint $x$-distribution of its
constituents. This kind of nonfactorization may be determined by the
study of perturbative QCD processes such as dilepton, direct-photon,
or dijet production as a function of the centrality\footnote{The
impact parameter dependence is accessible in the $pA$ collisions via a
study of nuclear fragmentation into various channels.  For an
extensive discussion see Ref.~\cite{:2001qv}.}, soft hadron
multiplicity, and $A$.  Naturally such studies would also require
investigation of soft hadron production as a function of impact
parameter. At LHC energies it may differ quite strongly from fixed
target energies.

One possible kinematics where a strong correlation is expected is when
a parton with large $x\geq 0.6$, is selected in the proton. The
presence of such a parton requires three quarks to exchange rather
large momenta. Hence one may expect that these configurations have a
smaller transverse size and therefore interact with the target with a
smaller effective cross section, $\sigma_{\rm eff}(x)$. Suggestions
for such $x$ dependence on the size are widely discussed in the
literature.  Using a geometric (eikonal type) picture of $pA$
interactions as a guide and neglecting (for simplicity) shadowing
effects on nPDFs, one can estimate the number of wounded nucleons,
$\nu(x,A)$, in events with a hard trigger (Drell-Yan pair,
$\gamma$-jet, dijet...)  as a function of $\sigma_{\rm eff}$
\cite{Frankfurt:cv}:
\begin{equation}
\nu (x,A)= 1 + \sigma_{\rm eff}(x)
{A-1 \over A^2}\int T_A^2(B)~d^2B,
\label{nu}
\end{equation}
where the nuclear density per unit area $T_A(B)$ is defined in
Eq.~(\ref{TB}).  At the LHC, for average inelastic $p$Pb
collisions and $\sigma_{\rm eff} \sim \sigma_{\rm{inel}}(pp)$,
Eq.~(\ref{nu}) leads to $\nu \approx $ 10, somewhat larger than the
average number of wounded nucleons in $pA$ collisions due to the
selection of more central impact parameters in events with a hard
trigger.

A decrease of the effective cross section for large $x$, say, by a
factor of 2, would result in a comparable drop of the number of
particles produced at central rapidities as well as in a smaller
number of nucleons produced in the nuclear fragmentation region.

\subsubsection{$A$-dependence of the particle production}
\label{a.4.4}

\paragraph*{Proton fragmentation region.}

\noindent
The $A$-dependence of hadron production in the proton fragmentation
region remains one of the least understood aspects of hadron-nucleus
interactions.  Practically all available data are inclusive and
correspond to energies where the inelastic $NN$ cross section is about
three times smaller than at the LHC. They indicate that the cross
section is dominated by the production of leading particles at large
impact parameters where the projectile interacts with only one or two
nucleons of the nucleus. As a result, very little information is
available about hadron production at the central impact parameters
most crucial for the study of $AA$ collisions. Theoretical predictions
for this region are also rather uncertain.

In eikonal-type models, where the energy is split between several soft
interactions, one may expect a very strong decrease of the leading
particle yield.  The dependence is expected to be exponential with
path length through the nucleus: the mean energy is attenuated
exponentially. On the other hand, if the valence partons of the
projectile do not lose a significant amount of their initial momentum,
as suggested by pQCD motivated models, see the
review in Ref.~\cite{Baier:2000mf}, the spectrum of leading particles
may approach a finite limit for large $A$ and central impact
parameters \cite{Berera:1996ku}. Indeed, in this case, the leading
partons will acquire significant transverse momenta and will not be
able to coalescence back into leading baryons and mesons, as seemingly
happens in nucleon-nucleon collisions. As a result, they will fragment
practically independently, leading to much softer longitudinal
momentum distributions for mesons and especially for baryons
\cite{Berera:1996ku}.

\begin{figure}[htp]
\centerline{\hbox{
\epsfig{figure=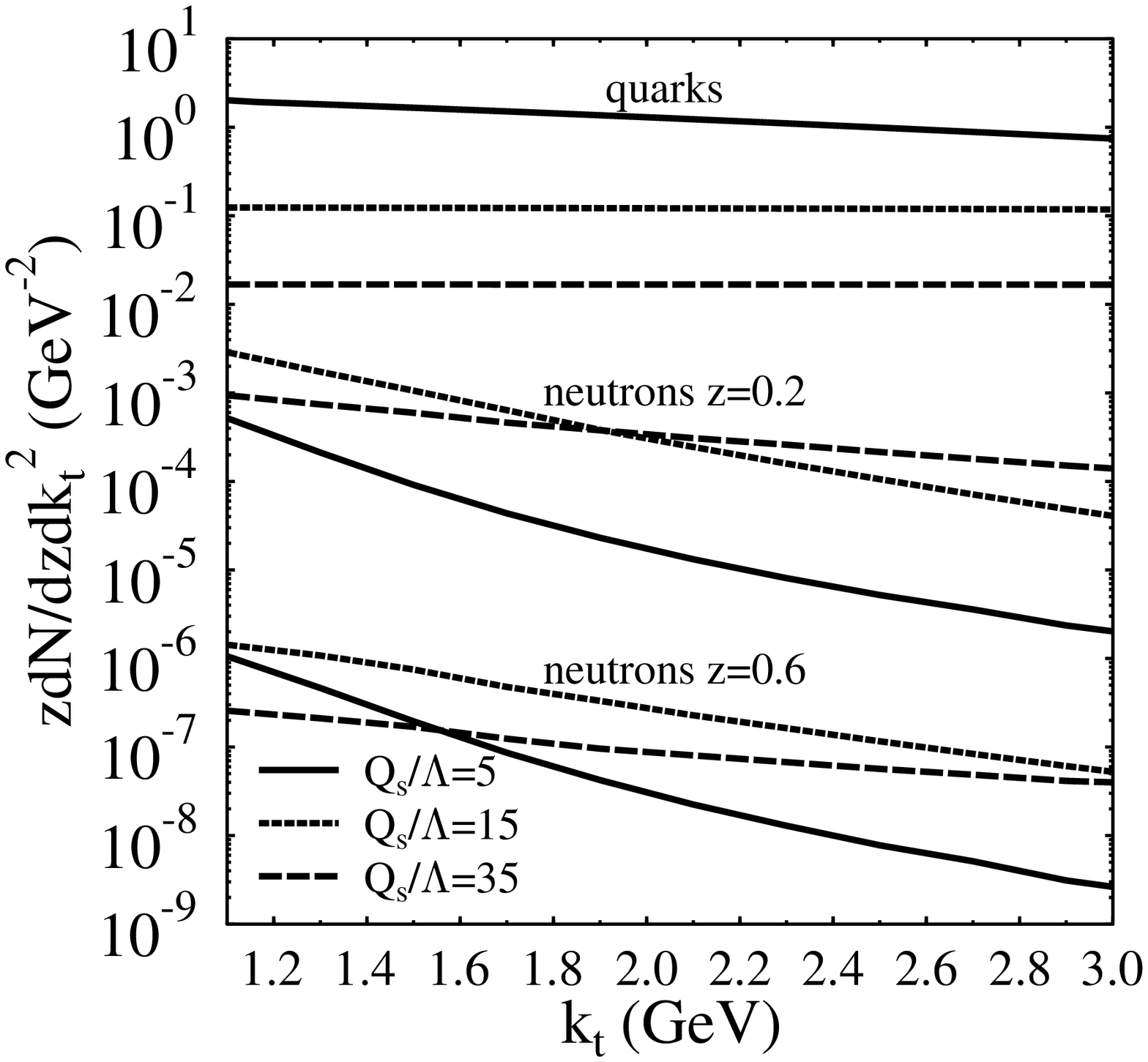,height=8cm}
\epsfig{figure=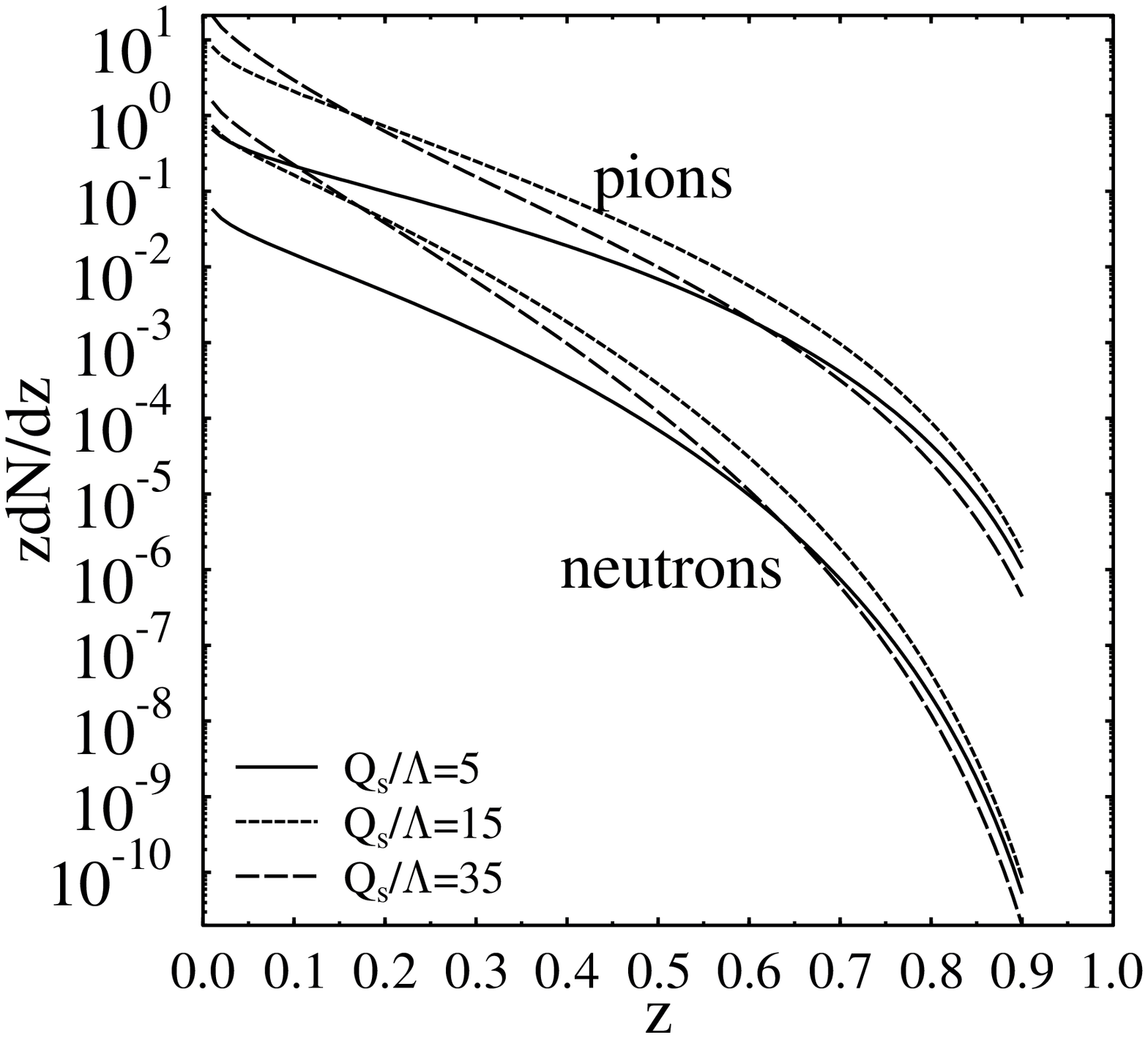,height=8cm}}}
\vspace*{-1cm}
\caption{Left: Transverse momentum distribution of neutrons (in fact,
$n/2+\bar{n}/2$) from the breakup of an incident proton at various
longitudinal momentum fractions $z$ and target saturation momenta
$Q_s$ (bottom six curves). The top three curves depict the underlying
quark distributions. Right: Longitudinal distributions of
$n/2+\bar{n}/2$ and $\pi^0$, integrated over $k_T$. From
Ref. \cite{Dumitru:2002wd}.}  
\vspace{-0.3cm}
\label{novelphys:fig1}
\end{figure}

We discussed previously that the interaction strength of fast partons
in the nucleon ($x_p \geq 10^{-2}$ for LHC) with heavy nuclei at
central impact parameters may approach the black-body limit.  In this
scenario, all these partons will acquire large transverse momenta
$\sim p_T^{\rm bbl}(x_A)$,\footnote{Hence in the black-body limit,
$p_T$ broadening of the partons should be much larger than at low
energies where it is consistent with QCD multiple rescattering model
\cite{Baier:2000mf}. See Ref.~\cite{DFS} for a discussion of matching 
these two regimes.}  leading to a very significant $p_T$ broadening of
the leading hadron spectrum \cite{Dumitru:2002qt}. Probably the most
feasible way to study this effect is measurement of the $p_T$
distribution of leading hadrons (neutrons, $K_L,\pi^0$) in central
collisions. These hadrons will closely follow the $p_T$ of the leading
quarks \cite{Dumitru:2002wd}.  The results of the calculation
\cite{Dumitru:2002wd} using the color glass condensate model for
modeling $p_T$ broadening are shown in Fig.~\ref{novelphys:fig1} for
different values of the saturation parameter $Q_s$.  It is worth
emphasizing that strong $p_T$ broadening, increasing with $x_F$,
clearly distinguishes this mechanism of leading hadron suppression
from the soft physics effect of increasing the total $pp$ cross
section by nearly a factor of three over fixed target energies.

\paragraph*{Central region.}
\noindent
The Gribov-Glauber model, including inelastic screening effects but
neglecting the final state interactions between the hadrons produced
in different Pomeron exchanges, as well as neglecting the interactions
of these hadrons with the nuclear target, predicts that, at high
energies and rapidities $y $ such that $|y_p -y| \gg 1$, the inclusive
$pA$ spectrum should be proportional to that of $pN$ scattering:
\begin{equation}
{d \sigma^{p+A \to h+X}(y, p_T)\over dy~ d^2p_T}= A
 {d \sigma^{p+N \to h+X}(y, p_T)\over dy~ d^2p_T} .
\label{a13}
\end{equation}
Since $\sigma_{\rm{inel}}(pA)/\sigma_{\rm{inel}}(pN) \propto
A^{2/3}$, Eq.~(\ref{a13}) implies that the multiplicity of particles
produced in the inelastic $pA$ collisions should increase as $\propto
A^{1/3}$. Data at fixed target energies do not contradict this
relation but the energy is too low for an unambiguous interpretation.
Effects of splitting the energy between multiple soft $NN$
interactions are much larger in $AA$ collisions than in $pA$
collisions so checking whether a similar formula is valid for $AA$
collisions would require energies much higher than those available at
RHIC.

The opposite extreme is to assume that parton interactions in the
black body regime become important in a large range of $x_p\geq
10^{-2}$.  In this case, multiple scatterings between the partons,
their independent fragmentation together with associated QCD radiation
(gluon bremsstrahlung), may produce most of the total entropy and
transverse energy and will enhance hadron production at central
rapidities relative to the Gribov-Glauber model.

\paragraph*{Nuclear  fragmentation region.}
\noindent
For the rapidities close to that of the nucleus, there are indications
of slow hadron rescatterings, leading to increased nucleon
multiplicity for $1.0 \geq p_N \geq 0.3$~GeV compared to no final
state reinteractions, a factor of $\sim 2$ for heavy nuclei
\cite{Frankfurt:mk}.  As discussed above, it is doubtful that the
Gribov-Glauber picture is correct at LHC energies, so that significant
deviations from Eq.~(\ref{a13}) maybe expected.  It is likely that a
colored quark-gluon system with rapidities close to $y_A$ will be
excited along a cylinder of radius $\sim$ 1 fm and length $2R_A$,
resulting in unusual properties of hadron production in the nuclear
fragmentation region, including production of exotic multi-quark,
multi-gluon states.

We emphasize that a quantitative understanding of particle production in
$pA$ collisions in this phase-space region is a prerequisite for
understanding the corresponding phenomenology in $AA$ collisions.

\subsection{A new hard QCD phenomenon: proton diffraction into three jets}
\label{3jetsec}
\subsubsection{Introduction}

During the last ten years, a number of new hard small-$x$ phenomena
have been observed and calculated in QCD based on the QCD
factorization theorems for inclusive DIS and hard exclusive
processes. These phenomena include: (i) observation of a fast increase
of the PDFs with energy at HERA, reasonably described within the QCD
evolution equation approximation\footnote{For a comprehensive review
of the HERA data on inclusive and exclusive processes and the relevant
theory, see Ref.~\cite{Abramowicz:1998ii}.}; (ii) discovery of color
transparency in pion coherent dissociation into two high-$p_T$ jets
\cite{Aitala:2000hc}, consistent with the predictions of
Refs.~\cite{Frankfurt:it,Frankfurt:2000jm}; (iii) observation of
various regularities in exclusive vector-meson electroproduction
at HERA induced by longitudinally polarized photons and
photoproduction of mesons with hidden heavy flavor consistent with the
predictions of Ref.~\cite{Brodsky:1994kf}. These processes, in a wide
kinematic range of small $x$, have provided effective ways to study
the interaction of small colorless dipoles with hadrons at high
energies and to study the hadron wave functions in their minimal Fock
space configurations. Here we outline new QCD phenomena \cite{:2001qv}
which may become important at the LHC because their cross sections are
rapidly increasing with energy.

A fast increase of hard diffractive cross sections with energy,
predicted within the leading-twist (LT) approximation, would be at
variance with unitarity of the $S$-matrix for the interaction of
spatially small wave packet of quarks and gluons with a hadron target
at a given impact parameter.  This new phenomenon may reveal itself at
sufficiently small $x$ and/or sufficiently heavy nuclei. In the proton
structure functions, this physics is masked by a significant
contribution to the hard collisions of the nucleon periphery. Thus,
unitarity of the $S$-matrix does not preclude an increase of nucleon
structure functions as fast as $\propto {\ln (1/x)}^{3}$
\cite{Frankfurt:2001nt}. Hence no dramatic signals for the breakdown
of the DGLAP regime are expected in this case.  On the contrary, a
breakdown of the LT approximation in hard diffractive processes off
nuclei leads to drastic changes in the cross section dependence on $A$,
incident energy and jet $p_T$.

\subsubsection{Three-jet exclusive diffraction}

A nucleon (meson) has a significant amplitude to be in a configuration
where the valence partons are localized in a small transverse area
together with the rest of the partons. These configurations are
usually referred to as {\it minimal} Fock space configurations -
$\left|3q \right>$ .  Experimental evidence for a significant
probability for these configurations in mesons include the significant
decay amplitude for $\pi\to {\rm leptons}$ as well as the significant
observed diffractive vector meson electroproduction cross sections
and the process $\pi+A\to 2\,{\rm jets} +A$ \cite{Aitala:2000hc}.  In
the proton case, the amplitude of the three quark configuration can be
estimated within QCD-inspired models and, eventually calculated in
lattice QCD.  Note that knowledge of this amplitude is important for
the unambiguous calculation of proton decay within Grand Unification
models.

\begin{figure}
\begin{center}
\includegraphics[width=10.5cm]{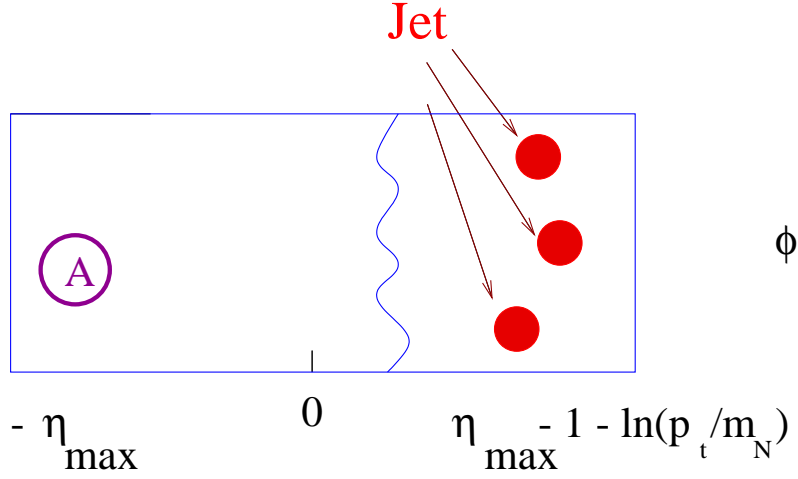}
\vspace{-2.5cm}
\caption{LEGO plot for diffraction of proton into three jets.
}
\label{fig:nuc_3jets}
 \end{center}
\end{figure}

Hadrons, in small size configurations with transverse size $d$,
interact with a small cross section $\propto d^2 xG_N(x,Q^2)$, where
$Q^2\sim 10/d^2$.  The factor $xG_N(x,Q^2)$ leads to a fast increase
of the {\it small } cross section with energy. In the case of
``elastic scattering'' of such a proton configuration off another
proton, this three-quark system with large relative momenta should
preferentially dissociate diffractively into a system of three jets
with large transverse momenta $p_{Ti} \sim \pi/d$, where $d$ is
the transverse size of the minimal configuration. The kinematics of
this process is presented in the LEGO plot of Fig.~\ref{fig:nuc_3jets}
for the case of coherent scattering off a nucleus.

The 3-jet production cross section can be evaluated based on the kind
of QCD factorization theorem deduced in
Ref.~\cite{Frankfurt:2000jm}. The cross section is proportional to the
square of the gluon density in the nucleon at $x\approx {M^2_{\rm 3\,
jets}/ s}$ and virtuality $\sim (1-2) p_T^2$ \cite{Frankfurt:1998eu}.
The distribution over the fractions of the longitudinal momentum
carried by the jets is proportional to the square of the light cone
wave function of the $\left |3q \right>$ configuration, $\psi(z_1,
z_2, z_3, p_{T1}, p_{T2}, p_{T3})$ and large transverse momenta where
\begin{equation}
\psi(z_1, z_2, z_3, p_{T1}, p_{T2}, p_{T3})\propto z_1z_2z_3
\sum_{i\neq j} {1\over p_{Ti}^2}{1\over p_{Tj}^2} .
\end{equation}

The proper renormalization procedure accounting for cancellation of
infrared divergences is implied. Otherwise the light cone gauge
$p_{N,\mu}A_{\mu}=0$, should be used to describe the diffractive
fragmentation region where there are no infrared divergences. Here
$p_{N}$ is the momentum of the diffracted proton. The summation over
collinear radiation is included in the definition of the jets.

Naively, one can hardly expect that exclusive jet production would be
a leading-twist effect at high energies since high-energy hadrons
contain a very large number of soft partons in addition to three
valence quarks. However application of the QCD factorization theorem
demonstrates that in high-energy exclusive processes these partons
are hidden in the structure function of the target.  Hence proton
diffraction into three jets provides important information about the
short-distance quark structure of the proton and also provides unique
information about the longitudinal momentum distribution in the $\left
|3q \right>$ configuration at high $p_T$.  The amplitude of the
process can be written as
\begin{equation}
A=\left[-{\vec \Delta}
\psi(z_1, z_2, z_3, p_{T1}, p_{T2}, p_{T3})\right]\left[{2\pi^2\over 3}
 \alpha_s(xG(x,Q^2,t))\right].
\label{skewed}
\end{equation}
Here  ${\vec \Delta}=\sum^3_{i=1}{\partial^2\over\partial p_i^2 }$. 
In Eq.~(\ref{skewed}) we simplify the expression by substituting a
skewed gluon density reflecting the difference in the light-cone fractions
of the gluons attached to the nucleon by the gluon density which is
numerically  a reasonable approximation for rough estimates.
An additional
factor of 2 in the numerator  relative to the pion allows
 two gluons to be attached to different pairs of quarks.
The $p_T$ and longitudinal momentum  fraction, $z_i$, dependence of the
cross section is given by
\begin{eqnarray}
\frac{d\sigma}{dz_{1} dz_{2} dz_{3}
d^{2}p_{T1}  d^{2}p_{T2} d^{2}p_{T3}}=
 c_N |\alpha_{s} xG(x,Q^{2})|^{2}
\frac{|\phi_{N}(z_{1},z_{2},z_{3})|^{2}}{|p_{T1}|^{4}
|p_{T2}|^{4} |p_{T3}|^{4}}F^2_{2g}(t) \nonumber \\
\delta(\sum p_{Ti}-\sqrt{-t})
\delta(\sum z_{i}-1).
\end{eqnarray}
The coefficient $c_N$ is calculable in QCD and
$\phi_{N}(z_{1},z_{2},z_{3})=z_1z_2z_3$.  The two-gluon form factor of
a nucleon is known to some extent from hard diffractive processes
\cite{Frankfurt:2002ka}.  A numerical estimate of the three-jet
production cross section with $p_T$ of one of the jet greater than a
given value and integrated over all other variables gives, for
$p_T\geq 10$~GeV$/c$,
\begin{equation}
\sigma(pp\to 3\,{\rm jets} + p)\propto \frac{|\alpha_s xG(x,Q^2)|^{2}}
{p_T^{8}} \propto 10^{-6}\dots 10^{-7} \left(\frac{10\,{\rm GeV}}{p_{T}}\right)^{8}\,{\rm mb}
\end{equation}
for LHC energies.  An urgent question is whether this cross section
will continue to grow up to LHC energies, as assumed here based on
extrapolations of $G_N(x,Q^2)$ to $x \sim 10^{-5}$.  The probability
of the $\left |3q \right>$ configuration is estimated using a
phenomenological fit to the probability of configurations of different
interaction strengths in a nucleon, cf.
Refs.~\cite{Frankfurt:1998eu,Frankfurt:1994hf}.

A study of the same process in $pA$ collisions would provide an
unambiguous test of the dominance of hard physics in this process. The
cross section should grow with $A$ as ${A^{2}\over R_A^2}
{G_A^2(x,Q^2)/A^2 G_N^2(x,Q^2)}$. Accounting for the skewedness of the
gluon distribution will lead to some enhancement in the absolute value
of the cross section but will not strongly influence the
$A$-dependence.

Our estimates indicate that for the relevant LHC kinematics, $x \sim
10^{-5}$ and $Q^2_{\rm eff}\sim p_T^2$, nuclear shadowing in the gluon
channel is a rather small correction, reducing the $A$-dependence of the
cross section by $A^{0.1}$, leading to $\sigma(pA\rightarrow 3\, {\rm jets} +A
)\propto A^{1.25}$.  The $A$-dependence of total inelastic cross
section is $\propto A^{0.7}$. Hence we expect that the counting rate per
inelastic interaction will be enhanced in $pA$ collisions for
heavy nuclear target by a factor $\propto A^{0.55}$.  At the same
time, the background from soft diffraction should be strongly
reduced. Indeed, if we estimate the $A$-dependence of soft diffraction
based on the picture of color fluctuations, which provides a good
description of total cross section of the inelastic diffraction off
nuclei at fixed target energies \cite{Frankfurt:1993qi}, we find
$\sigma_{pA}^{\rm{inel.~diff.}} \propto A^{0.25}$.  Scattering off
nuclei also has an obvious advantage in terms of selecting coherent
processes since practically all inelastic interactions with nuclei
would lead to the emission of neutrons at 0$^o$.

A competing process is proton diffraction into 3 jets off the
nuclear Coulomb field. It follows from the theorem proved in
Refs.~\cite{Frankfurt:2000jm,Frankfurt:2002gb} that to leading
order in $\alpha_s$ the dominant contribution is given in the
Weizs\"acker-Williams representation by transverse photon interactions
with external quark lines. The amplitude of the process is
\cite{Frankfurt:2002gb}:
\begin{equation}
{-e^2\psi_N(z_i,p_{Ti})\over q^2-t_{\rm min}}Z F_A(t){2s \over M^2(\rm 3jet)}
\sum_i{e_i(\vec p_{Ti}\cdot \vec{q_T})\over z_i}D^{-3/2}(p_{Ti}),
\end{equation}
where $e_i$ is the electric charge of the quark in the units of the
electron charge, $q^2$ is the virtuality of the exchanged photon and
$D(p_T^2)$ is the renormalization factor for the quark Green's
function.

In the kinematics where LT dominates the QCD mechanism for 3-jet
production, the electromagnetic process is a small correction.
However if the screening effects (approach to the black body limit)
occurs in the relevant $p_T$ range, the electromagnetic process may
compete with the QCD contribution.  In the case of nuclear breakup,
the electromagnetic contribution remains a relatively modest
correction, even in the black body limit.  This is primarily because
the cross section for electromagnetic nuclear breakup does not have a
singularity at $q^2=0$.  Note also that, in the case of QCD
interactions, one can consider processes with a large rapidity gap
which have the same $z_i$ and $p_{Ti}$ dependence as the coherent
process but with a significantly larger cross section.

Another interesting group of hard processes is proton diffraction into
two high $p_T$ jets and one collinear jet. These processes are, in
general, dominated at high energies by collisions of two nucleons at
large impact parameters (otherwise multiple soft interactions would
destroy the coherence).  Hence the $A$-dependence of the dominant term
should be similar to that of
$\sigma_{pA}^{\rm{inel.~diff.}}$. However fluctuations with cross
sections of $\sim 40$~mb would have a much stronger $A$-dependence,
$\sim A^{0.7}$. Thus the study of the $A$-dependence of coherent
diffraction can filter out the smaller than average
components in the nucleons at LHC energies.

\subsubsection{Violation of the LT approximation and new QCD strong interactions}
The energy dependence of cross section in the LT approximation follows
from the properties of the target gluon distribution.  Hence for the
virtualities probed in 3-jet production, the energy dependence is
$\propto x^2G^2(x,Q^2_{\rm eff}) \propto s^{0.8}$ for $Q^2_{\rm
eff}\sim 100 -1000$~GeV$^2$ and $10^{-5}\leq x \leq 10^{-2}$.  In the
black-body limit, the inelastic diffractive processes are strongly
suppressed and hence the energy dependence of the 3-jet production
cross section integrated over $p_T\geq p_{T{\rm min}}\sim$a few GeV
should slow down since only scattering off the grey edge of the
nucleus (nucleon) will effectively survive.  For $Q^2$ relevant to
this process, the black-body limit may only be an interesting
hypothesis for scattering off heavy nuclei.  It would manifest itself
in a strong decrease of the $A$-dependence of 3-jet production at
the $p_T$ corresponding to the black-body limit.  At sufficiently
small $x$ for a heavy nuclear target, $d\sigma(pA\to 3{\rm jets}
+A)/ dp_T^2 \propto A^{1/3}$, in striking contrast with the
LT expectations, $\propto A^{1.2}$.

\subsection{Conclusions}
In conclusion, studies of $pA$ collisions at the LHC will reveal
strong interactions in the high field domain over an extended rapidity
region.  New phenomena will be especially prominent in the proton
fragmentation region but will extend to the central region and the
nuclear fragmentation region as well.


\clearpage

\subsection*{Acknowledgements}
The convenors (KJE, JWQ, WG) thank the contributing authors for fruitful
collaboration. 

\noindent The editor (KJE) is grateful to R. Vogt for many useful
remarks in finalizing the manuscript.

\noindent The following sources of funding are acknowledged:

\noindent~~~ Academy of Finland,

Project 50338: K. J. Eskola, H. Honkanen, V. J. Kolhinen;

Project 80385: V. J. Kolhinen;

\noindent~~~ Alexander von Humboldt Foundation: R. Fries;

\noindent~~~ CICYT of Spain under contract AEN99-0589-C02: N. Armesto;


\noindent~~~ European Commission IHP program,
Contract HPRN-CT-2000-00130: A. Accardi;


\noindent~~~ German Israeli Foundation: L. Frankfurt;

\noindent~~~ Israeli Science Foundation, founded by the 
Israeli Academy of Science and Humanities,

BSF grant $\#$ 9800276: Yu. Kovchegov;


\noindent~~~ Marie Curie Fellowship of the European Community programme TMR,

Contract HPMF-CT-2000-01025: C.A. Salgado; 


\noindent~~~ United States Department of Energy,
 
Contract No. DE-AC03-76SF00515: S.J. Brodsky;

Grant No. DE-FG02-86ER-40251: G. Fai, X.~f. Zhang;

Grant no. DE-FG02-96ER40945: R. Fries;

Grant No. DE-FG03-97ER41014: Yu. Kovchegov;

Grant No. DE-FG02-87ER40371: J.~w. Qiu;

Contract No. DE-FG02-93ER40771: M. Strikman;

Contract No. DE-AC03-76SF00098: R. Vogt;


\noindent~~~ Universidad de C\'ordoba: N. Armesto.

\end{document}